\pdfoutput=1
\documentclass[twoside,11pt]{PhDthesisPSnPDF}

\usepackage{caption}%Options for caption
\usepackage{textcomp}%some fonts
\usepackage[list=true]{subcaption}
\usepackage{paralist,setspace}
\usepackage{enumerate,enumitem}
\usepackage{amsthm}
\usepackage{graphicx}  % needed for figures
\usepackage{amsmath,amsfonts,bbm,MnSymbol}   % for math
\usepackage{amsfonts}
\usepackage{comment}
\usepackage{color}
\usepackage[makeroom]{cancel}

\renewcommand{\[}{\begin{equation}}
\renewcommand{\]}{\end{equation}}
\renewcommand{\Re}{\mathfrak{Re}}
\renewcommand{\Im}{\mathfrak{Im}}
\newcommand{\ket}[1]{|#1\rangle}
\newcommand{\bra}[1]{\langle#1|}
\newcommand{\braket}[2]{\langle#1|#2\rangle}
\newcommand{\pro}[2]{|#1\rangle\langle#2|}
\newcommand{\mean}[1]{\langle#1\rangle}
\newcommand{\abs}[1]{|#1|}
\newcommand{\ov}[1]{\overline{#1}}
\newcommand{\tr}{\mathrm{tr}}
\renewcommand{\det}[1]{|#1|}
\newcommand{\norm}[1]{\left\lvert#1\right\rvert}

\newcommand{\Ref}[1]{{REF}{(#1)}}

\newcommand{\note}[1]{|\emph{#1}{|}}

\newcommand{\A}{\alpha}
\newcommand{\B}{\beta}
\renewcommand{\C}{\gamma}

\newcommand{\R}{\hat{\rho}}
\renewcommand{\G}{{\cal{G}}}
\newcommand{\PI}{\bra{\psi}\hat{P}_i\ket{\psi}}

\newcommand{\HS}{\mathcal{H}}
\newcommand{\CF}{\mathbb{C}}
\newcommand{\RF}{\mathbb{R}}
\newcommand{\de}{\mathrm{d}\epsilon}
\newcommand{\bd}{\boldsymbol{d}}
\newcommand{\bA}{\boldsymbol{\hat{A}}}
\newcommand{\bb}{\boldsymbol{b}}
\newcommand{\bg}{\boldsymbol{\gamma}}

\newcommand{\dif}[1]{\mathrm{d}{#1}}
\newcommand{\be}{\boldsymbol{\epsilon}}
\newcommand{\bdeps}{\mathrm{d}\!\!\;\boldsymbol{\epsilon}}

\newcommand{\vectorization}[1]{\mathrm{vec}{[#1]}}

\newtheorem{definition}{Theorem}[chapter]
\newtheorem{theorem}[definition]{Theorem}

\newtheorem{example}[definition]{Example}

\theoremstyle{definition}

\ifpdf
    \pdfcatalog { /PageMode (/UseOutlines)
                  /OpenAction (fitbh)  }
\fi

\title{Gaussian quantum metrology and space-time probes}

\ifpdf
  \author{\href{pmxdd@nottingham.ac.uk}{Dominik \v Safr\'anek}}
\collegeordept{}
\university{}
  \crest{\includegraphics[width=4cm]{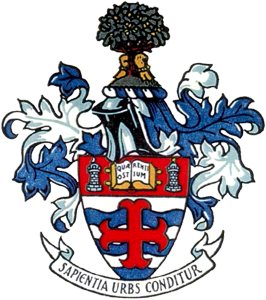}}

\else
  \author{Dominik \v Safr\'anek}
  \collegeordept{School of Mathematical Sciences}
  \university{University of Nottingham}
  \crest{\includegraphics[width=8cm]{UoNlogo.png}}
\fi
\renewcommand{\submittedtext}{Thesis submitted to the University of Nottingham \\ for the degree of Doctor of Philosophy}
\degree{}
\degreedate{September 2016}

\setcitestyle{square}
\begin{document}
%\language{english}

% sets line spacing
\renewcommand\baselinestretch{1.2}
\baselineskip=18pt plus1pt

%: ----------------------- generate cover page ------------------------

\maketitle  % command to print the title page with above variables

\frontmatter

\begin{abstracts}\addcontentsline{toc}{chapter}{Abstract}
In this thesis we focus on Gaussian quantum metrology in the phase-space formalism and its applications in quantum sensing and the estimation of space-time parameters. We derive new formulae for the optimal estimation of multiple parameters encoded into Gaussian states. We discuss the discontinuous behavior of the figure of merit -- the quantum Fisher information. Using derived expressions we devise a practical method of finding optimal probe states for the estimation of Gaussian channels and we illustrate this method on several examples. We show that the temperature of a probe state affects the estimation generically and always appears in the form of four multiplicative factors. We also discuss how well squeezed thermal states perform in the estimation of space-time parameters. Finally we study how the estimation precision changes when two parties exchanging a quantum state with the encoded parameter do not share a reference frame. We show that using a quantum reference frame could counter this effect.
\end{abstracts}

\begin{acknowledgements}\addcontentsline{toc}{chapter}{Acknowledgements}
When I started my PhD studies I felt enthusiastic, excited, determined. I was full of expectations, I was maybe a little childish. And definitely naive. Now I feel confident. This really changes you. And although the change may not be always pleasant, it is definitely worth pursuing. Today I feel the world is open to me. It is full of possibilities and these possibilities are there to take. Just raise your hand. This change I feel inside wasn't for free but I had many people to help me.

I thank Mehdi Ahmadi, who is probably the nicest guy ever, who taught me how to talk to people and that to be humble is really important. That was a great time together doing great research. I thank Antony Richard Lee who had so much enthusiasm and plain humble honesty that you could cut it and give it away to anyone who needs it. Antony, you also helped me in a time of dire need, in a time of crisis. I really thank you for that. I don't know if it could turn out so nicely if you weren't there. I thank Stephen Mumford and other people in the philosophy department at the University of Nottingham. You really opened my eyes. These discussions brought me to question everything I took for granted in physics. I thank Jorma Louko for his great support when I really didn't know what to do next. That advice was excellent. I thank Karishma Hathlia who helped me so much with writing, this thesis included. Keep up the good work, I know you can make it. I thank Tanja Fabsits who helped me so much to grow personally and taught me how to edit and write so it reads well. I hope to meet you again. I thank Ivette Fuentes for teaching me how to present myself, how deal with people, how to sell my work, how to be professional, and how to write. You were a great supervisor, teaching me exactly the things I was lacking the most while allowing my other skills to grow. Finally, I would like to thank everyone I encountered on this incredible journey, especially the people in our office. It has been a great time guys!

\begin{figure}[b!]
\raggedleft
\includegraphics[width=0.5\linewidth]{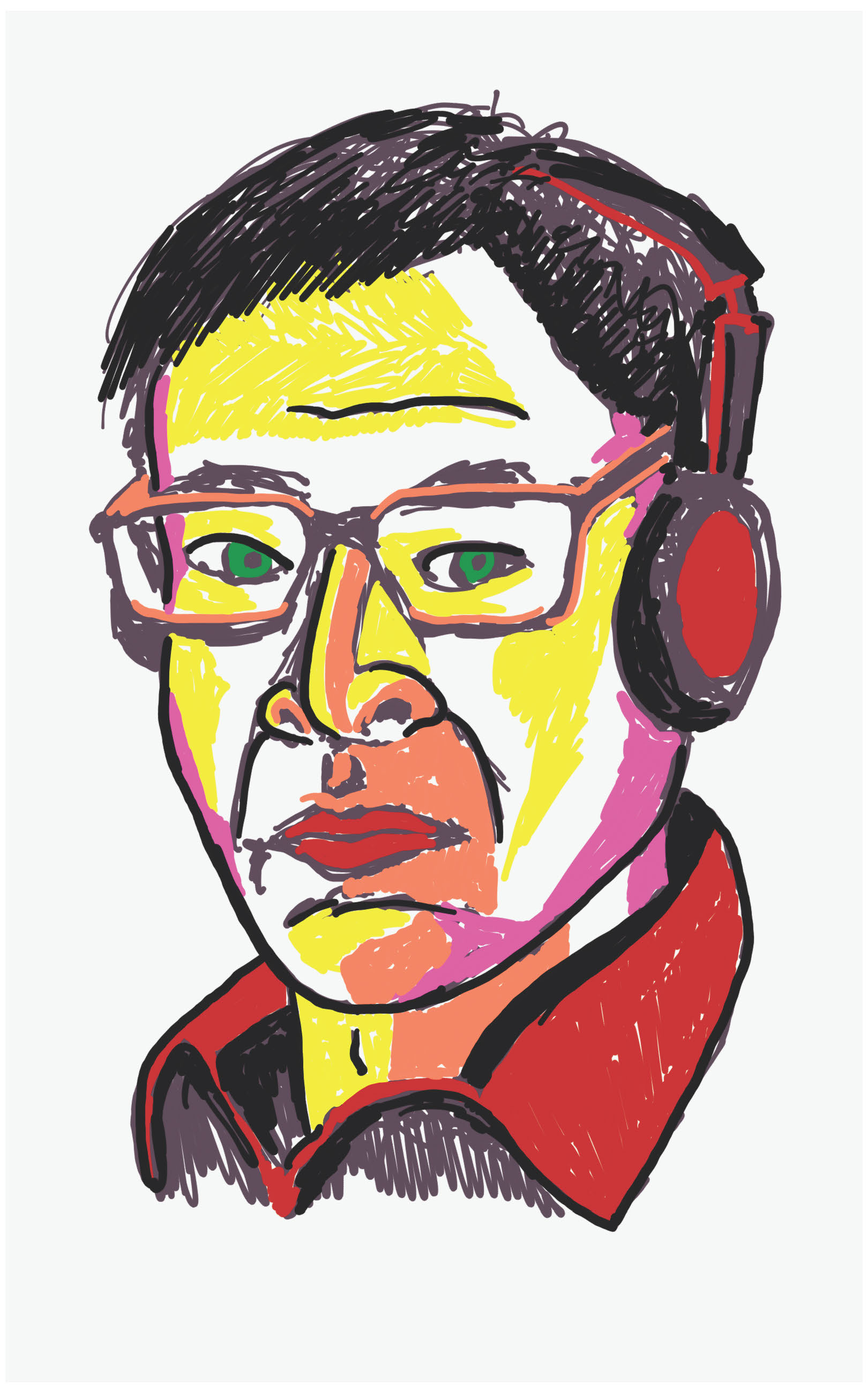}\caption*{\hfill Dominik \v{S}afr\'anek,\\\hfill Vienna, May 2016}
\end{figure}
\end{acknowledgements}

%: ----------------------- contents ------------------------

\setcounter{secnumdepth}{3} % organisational level that receives a numbers
\setcounter{tocdepth}{3}    % print table of contents for level 3
\tableofcontents            % print the table of contents

% the main text starts here with the introduction, 1st chapter,...
\mainmatter

\renewcommand{\chaptername}{} % uncomment to print only "1" not "Chapter 1"

\chapter*{Publications}\label{chapter:publications}
\addcontentsline{toc}{chapter}{Publications}

This thesis is partially based on work presented in the following publications.

\begin{enumerate}[leftmargin=*]
	\item Quantum parameter estimation with imperfect reference frames~\cite{safranek2015quantum}
	\item Quantum parameter estimation using multi-mode Gaussian states~\cite{Safranek2015b}
	\item Ultimate precision: Gaussian parameter estimation in flat and curved spacetime~\cite{Safranek2015a}
	\item Optimal probe states for the estimation of Gaussian unitary channels~\cite{safranek2016optimal}
\end{enumerate}

%Chapters introduction: 1)Little history about the specific chapter (only) 2) Structure of the chapter 3) Good pedagogical references - books and reviews

%%%%%%%%%%%%%%%%%%%%%%%%%%%%%%
\chapter*{Introduction}\addcontentsline{toc}{chapter}{Introduction}
%%%%%%%%%%%%%%%%%%%%%%%%%%%%%%

The aim of this thesis is to provide an elegant and useful basis for future quantum technology and to take small but significant steps towards experimental testing of theories in the overlap of quantum mechanics and general relativity. The importance and future impact of quantum technologies has been recognized not only by governments but also by several large financially-savvy corporations. The world is heading towards the second quantum evolution. Quantum technologies will serve as platform for secure communication, quantum computers will offer highly parallel computations which for certain tasks outperform classical computers,
%could not be met with classical computers in certain tasks,
%quantum simulation will offer a way how to make chemistry experiments safely and cheaply in a comfortable zone of a living room
quantum simulation will allow for safe and inexpensive modeling of chemistry experiments in the comfort of one's own living room, quantum clocks will provide precision in global positioning system which will lead to precise construction and possibly even space-controlled transportation, hand-held devices will be able to measure small distortions in gravity, lasers will be able to measure slight changes in the atmosphere allowing us to predict
extreme weather conditions and ultimately save lives.
%where the next tornado hits

We are still in the beginning though. Currently only a few genuinely practical applications have been developed. On the other hand, the recent amazing discovery of gravitational waves has demonstrated that such technology is feasible~\cite{abbott2016observation}. The new generation of quantum enhanced gravitational wave detectors have already delivered improvement by a factor of $2$~\cite{Demkowicz2013a} and soon we will hear about more such astonishing achievements.

%Despite its practical experimental goals, this work is entirely theoretical.

In this thesis we develop practical tools for the optimal estimation of special class of quantum states -- called Gaussian states -- which are relatively easy to prepare and manipulate in experiments, and thus can serve as an effective building block for this new generation of quantum sensors. We use these tools to show which states to use for which tasks, we show the optimal states. In essence, we give a prescription of how to build the core of such quantum sensors. Moreover, we study how such sensors perform in the estimation of space-time parameters such as proper time, Schwarzschild radius, amplitude of a gravitational wave, or proper acceleration. Finally, we use the powerful tools of quantum metrology to show how to overcome certain issues in distance-communication or distance-sensoring when the reference frame of a sensor and the observer can easily become misaligned. This could prove particularly useful for a new generation of space-based experiments such as eLISA~\cite{amaro2012low} where the detector is in deep space while the operator who reads the data stays on Earth.

This thesis is structured as follows: in Part I of the thesis we introduce and overview tools which have been developed in previous literature, although we believe that some results included have not been published before. For example, we describe the discontinuity of the quantum Fisher information and show that the Bures metric in general does not coincide with the quantum Fisher information matrix, we derive the form of a general Gaussian unitary in the phase-space formalism, the full parametrization of two- and three-mode Gaussian states, and formulae for arbitrary order of the continuous Bogoliubov transformations. Part II of the thesis consists entirely of original work.

In the first chapter~\ref{chap:QM} we overview the necessary tools of quantum metrology that we will use in the main part. The second~\ref{chap:GS} chapter focuses on Gaussian states. In particular we introduce the phase-space formalism of Gaussian states, Gaussian transformation, parametrization of Gaussian states, and state-of-the-art quantum metrology in the phase-space formalism. In the third chapter~\ref{chap:operations_in_QFT} we quickly overview how space-time distortions affect quantum states and show that such transformation are indeed Gaussian transformations. In the fourth chapter~\ref{chap:QM_GS} we develop new expressions for the optimal estimation of Gaussian states. Namely we focus on the quantum Fisher information as the figure of merit. We also use the derived expressions to devise a method for finding optimal probe states for Gaussian channels and we unravel how different characteristics of a probe state affect the estimation precision. The fifth chapter~\ref{chap:QFT_metrology} shows an application of quantum metrology of Gaussian states in quantum field theory in curved space-time. We provide general formulae which show how squeezed thermal states perform as probes for channels that encode space-time parameters.
The last chapter~\ref{chap:reference_frames} is focused on how the estimation precision changes when two parties, one which encodes a parameter into a quantum state and one which decodes, do not share a common reference frame. We show that sending a quantum reference frame in the communication channel could significantly improve the precision with which the parameter is decoded. Finally, we discuss open problems and suggest possible future directions, and we conclude with some short remarks. The table of frequently used notation can be found in appendix~\ref{app:notations}.

\chapter*{Part I}
\addcontentsline{toc}{chapter}{Part I}

%%%%%%%%%%%%%%%%%%%%%%%%%%%%%%
\chapter{Introduction to quantum metrology}\label{chap:QM}
%%%%%%%%%%%%%%%%%%%%%%%%%%%%%%

%Mehdi's article introduction, then my notes introduction
Metrology is the science of measurement. Metrology aims to determine the highest possible precision in measuring parameters of a physical system. It also provides tools to reach that limit of precision. Since measurement plays a central role in quantum physics and cannot be taken out of consideration, measurement theory in quantum theory becomes especially significant.

The groundwork on quantum metrology has been set up by Holevo and Helstrom in their seminal papers\cite{Holevo2011a,meystre1977quantum}. They developed and used many tools from probability theory and statistics and applied it to quantum systems. Their aim was to improve communication protocols with a particular interest in the problem of aligning measurement basis. This problem arises when two parties who wish to communicate cannot do so, because the information the first party sends is encoded with respect to a certain reference frame and the second party does not know what this reference frame is. Put simply, the direction that the first party calls the right direction can be called the left by the other party. If two parties cannot agree on what is left or right, it is impossible for the second party to decode received information. To align a basis between two parties, A and B, some information must be exchanged between the two. This can be achieved in different ways. It has been noted that exchanging information encoded in a quantum state with properties such as superposition, entanglement, and squeezing can be much more effective than exchanging classical information. In fact, the number of qubits needed to reach a given amount of alignment scales as the square root of the number classical bits needed for the same task. %Check
Such a significant advantage is the manifestation of advantages of quantum physics.

The problem of aligning two measurement basis can be viewed as a problem of estimating an angle in which one basis is rotated with respect to the other. As mentioned before, information about this angle is encoded in a quantum state sent from party A to party B. For the first party the task then is to encode the angle efficiently into the quantum state, while for the second party it is to estimate this parameter by measuring the received quantum state. This is a typical metrological setting.

Because aligning measurement basis and estimating angles is essentially the same task, results in the theory of aligning measurement basis can be translated to the theory of estimating angles or phases. But this means that using quantum states to estimate angles, or in fact any parameter~\cite{Giovannetti2004a,Giovannetti2006a}, can yield far better results than any classical method. In other words, the ability to construct devices using quantum systems could lead to a significant improvement in sensing. With this in mind it is no surprise that the field of quantum metrology has begun to grow rapidly. This is illustrated in various review articles and books~\cite{Giovannetti2011a,Toth2014a,Paris2009a,kok2010introduction}.

This chapter is structured as follows: first we give a brief overview on different areas of theoretical quantum metrology. We then quickly delve into the local estimation theory, which is the most developed part of quantum metrology and the main focus of this thesis. We introduce some mathematical results of the current state-of-the-art quantum metrology, its interpretation and use, and its connection to statistics.

\section{Overview}

In this section we sketch the structure quantum metrology. Quantum metrology can be divided into following subfields:

\begin{itemize}[leftmargin=*]
\item \emph{Discrete problems}: With a given input from a discrete set of elements (usually a finite set of quantum states), and usually some a priori probability distribution over this set, the task is to determine which element of the set has been sent. Examples are:
    \begin{itemize}
    \item \emph{Quantum state discrimination}: The task to discriminate between quantum states using an appropriate measurement basis. The method aims to minimize the probability that our guess about the input state is wrong, or more generally, to minimize the cost function. However, using this method that we are never entirely certain that our guess is correct.
    \item \emph{Unambigious discrimination}: The task is, again, to discriminate between quantum states. It differs from the previous method in a way that we choose our measurement basis that allows some measurement results to give an indefinite answer about the state. Put simply, in numerous cases the measurement results do not give us any information about the state we received (so we cannot discriminate), however, after receiving some result we can be certain that our guess is correct.
    \end{itemize}
\item \emph{Continuous problems}: The task is to estimate a parameter or parameters encoded in a quantum state, while the type of dependence of the state on the parameter is usually known. This is exactly the same as to discriminate between a continuum of quantum states. Examples of two different approaches such estimation are:
    \begin{itemize}
    \item \emph{Maximum likelihood estimation}: Given a set of measurement results, MLE assumes that the best guess for the parameter is such that maximizes the probability of receiving those measurement results.
    \item \emph{Bayesian estimation}: Given a set of measurement results, Bayesian estimation assumes that the best guess for the parameter is such that minimizes a cost function.
    \end{itemize}
    Also, it is important to point out that continuous estimation problems divide into two subfields:
    \begin{itemize}
    \item \emph{Global (Bayesian) estimation theory}: Global estimation theory provides general methods how to deal with the estimation of a parameter about what we have not any a priori knowledge about the distribution over the values the parameter can take. The measurement basis do not depend on the parameter we are trying to estimate. The figure of merit in this is the cost function we are trying to minimize. However, this is usually not an easy task. Quantum state tomography can be also viewed as an example of the global estimation theory, where the number of parameters to be fitted is equal to the dimension of the Hilbert space of a quantum system.
    \item \emph{Local (frequentist) estimation theory}: If we know that the parameter is localized around certain approximate value, we may move to the local estimation theory. This theory gives us an optimal measurement which depends on that approximate value, and helps us to estimate the parameter in the shortest amount of time/with minimal resources. The optimal measurement are usually written to be dependent on the unknown parameter we want to estimate -- but in that case we take the previously mentioned approximate value around which the real value is localized.
    \end{itemize}
\end{itemize}

In the following we will focus only on the continuous problems and the local estimation theory.

%Metrology is the science of measurements. In this text, we will deal mostly with the measurements in quantum mechanics, but the basic ideas come from the statistics and the classical information theory. In short, quantum Metrology tells us what is the best precision with what we can measure something, and what is the best strategy to reach that limit of precision. The usual task of quantum metrology is to consider two parties. Alice, which sends some quantum states to Bob, and Bob is trying to measure the incoming states to determine which states has been send or what is the parameter encoded in the states. Bob usually also have some additional information about the problem which makes his guessing easier. This may be the knowledge of probability with what are some states sent, or a knowledge of a particular dependence of the state on the parameter he wants to estimate.  Although we say Bob is trying to guess the best answer, what is the best usually depends on what he is trying to do with that information. This is why he often chooses a cost function, which gives a certain weight to certain results, and he chooses different strategies. The theory is far from complete and is still rapidly evolving, so let me just sketch the current structure as it is now with a little commentary. More details about the quantum estimation in general can be found for example in \cite{Paris2009a,Kok2011a}.

\section{Continuous problems in quantum metrology}

%estimator
Let us consider $N$ copies of the same density matrix $\R_\epsilon$ dependent on the same parameter. The task is to estimate the parameter $\epsilon$ which comes from an uncountable set, usually an interval. %For simplicity we also use a single parameter, multiple parameters will be considered later in section\ref{QMsec:multi}.
First we measure each matrix separately in an appropriate measurement basis. That way we obtain $N$ measurement outcomes $(x_1,\dots,x_N)$. Then we choose an estimator which will give us an estimate of the parameter considering the measurement outcomes. Mathematically, an estimator $\hat{\epsilon}$ is a function which maps the set of possible outcomes into the interval where the parameter lies,
\[
\hat{\epsilon}:(x_1,\dots,x_N)\longrightarrow\hat{\epsilon}(x_1,\dots,x_N).
\]

%different estimators
Properly chosen estimator will approximate the real value of the parameter after several measurements, $\hat{\epsilon}(x_1,\dots,x_N)\approx\epsilon$. Such an appropriate estimator is called consistent. By definition, when number of measurements $N$ goes to infinity, the value of the consistent estimator converges to the real value $\epsilon$. Another class of important estimators are locally unbiased estimators for which the overestimated value and underestimated value balance each other, i.e., such an estimator gives a correct value on average. Defining $p(x|\epsilon)$ as the probability distribution of obtaining outcome $x$ given the value $\epsilon$, the locally unbiased estimator is defined by
\[
\mean{\hat{\epsilon}}:=\int \dif{x}\ p(x|\epsilon)\hat{\epsilon}(x)=\epsilon.
\]
As we will show later, mean squared error of such an estimator is bounded below via the Cram\'er-Rao bound~\eqref{eq:Cramer_Rao}. The last type of estimators are called efficient estimators. Such estimators saturate the minimal value given by the Cram\'er-Rao bound. Asymptotically efficient estimators saturate the Cram\'er-Rao bound in the limit of large $N$. Although efficient estimators might not always exist, asymptotically efficient always do. Examples of asymptotically efficient estimators are the Bayes estimator and the maximum likelihood estimator\cite{meystre1977quantum,BraunsteinCaves1994a,barndorff2000fisher}. For more information about the non-quantum estimation theory see for example \cite{lehmann2006theory,amari2007methods}.

\begin{example}\label{ex1}
\emph{
We assume $\epsilon\in[0,\frac{\pi}{2}]$. Let us have $N$ copies of the state $\ket{\psi}=\cos\epsilon\ket{0}+\sin\epsilon\ket{1}$. To estimate the parameter, we may decide to measure in the computational basis $\{\ket{0},\ket{1}\}$. Since the probability of outcome $0$ is $p_0=\cos^2\epsilon$, a good choice of an estimator is $\hat{\epsilon}(x_1,\dots,x_N)=\arccos{\sqrt{\frac{N_0}{N}}}$, where $N_0$ is a number of times we receive the measurement result $0$. This estimator is clearly consistent. Note that we could have chosen a completely different estimator, for example $\hat{\epsilon}(x_1,\dots,x_N)=N_0+N+4$, but such an estimator is not consistent and does not give an appropriate estimate.
}
\end{example}

\section{Local estimation theory}

%CR bound
Local estimation theory enters the parameter estimation in its latest stage, i.e., when the parameter is localized around certain known approximate value. Then methods which are the most effective for that approximate value are used. For example, optimal measurements will be different for different values as well as optimal probe states for channels encoding the unknown parameter. Since local estimation theory enters in the latest stage of estimation, it also provides the ultimate limit of precision with what we can estimate the parameter. This is given by the Cram\'er-Rao bound\cite{BraunsteinCaves1994a,Paris2009a,jarzyna2015true}, also elegantly proven in~\cite{kok2010introduction}. This is the lower bound on the mean squared error of any locally unbiased estimator $\hat{\epsilon}$ with certain regularity conditions and reads
\[\label{eq:Cramer_Rao}
\mean{\Delta\hat{\epsilon}^2}\geq\frac{1}{N F(\epsilon)}.
\]
For the full statement of the theorem see appendix~\ref{app:CR}. $F(\epsilon)$ is a quantity called the Fisher information which we will define in Eq.~\eqref{eq:the_Fisher_information}, and $N$ is the number of measurements performed on $N$ identical copies of the same quantum state. $\mean{\Delta\hat{\epsilon}^2}$ is the mean squared error of the estimator defined as
\[
\mean{\Delta\hat{\epsilon}^2}=\int \dif{x_1}\dots\dif{x_N}(\hat{\epsilon}(x_1,\dots,x_N)-\epsilon)^2p(x_1|\epsilon)\dots p(x_N|\epsilon).
\]
The Cram\'er-Rao bound says that in average our guess $\hat{\epsilon}(x_1,\dots,x_N)$ cannot be closer to the real value $\epsilon$ than the value given by inverse of the Fisher information and the number of measurements performed. Although Cram\'er-Rao bound is entirely general and holds for any parameter-dependent probability distribution $p(x|\epsilon)$, in quantum physics $p(x|\epsilon)$ is the probability of obtaining the measurement result $x$ given a density matrix $\R_\epsilon$. Mathematically, assuming we are going to perform a measurement (positive-operator valued measure -- POVM) in basis $M=\{M_x\}_x$, $\sum_x M_x=I$, $M_x\geq0$, where $M_x$ is an element of the POVM (while in the special case of projective measurement the operator $M_x$ is a projector onto the Hilbert space given by the eigenvalue $x$), this probability is defined as
\[\label{def:prob_distribution}
p(x|\epsilon)=\tr[M_x\R_\epsilon].
\]
The Fisher information is then defined as
\[\label{eq:the_Fisher_information}
F(\epsilon):=\int\!\!\dif{x}\ \frac{\big(\partial_\epsilon p(x|\epsilon)\big)^2}{p(x|\epsilon)}.
\]
$\partial_\epsilon$ denotes the partial derivative with respect to $\epsilon$, $\partial_\epsilon p(x|\epsilon)=\tr[M_x\partial_\epsilon\R_\epsilon]$, and integral goes over all values of $x$ such that $p(x|\epsilon)>0$. The Fisher information measures how much information a random variable with by the probability distribution $p(x|\epsilon)$ carries about the parameter $\epsilon$. %In quantum physics the probability distribution is given by the combination of the density matrix and the measurement to be performed, as shown by Eq.~\eqref{def:prob_distribution}.
The above definition of the Fisher information comes from the proof of the Cram\'er-Rao bound. In the countable space of possible outcomes (the sample space) the integral is exchanged for the sum over all possible measurement outcomes.

%Detail about the QFI
Strictly speaking, the Fisher information should have been written as $F(\R_\epsilon,M)$, to reflect the fact that the Fisher information depends on a particular structure of density matrix and the choice of measurement to be performed. Here, however, we use the simple notation $F(\epsilon)$ as is common in the literature.

\begin{example}
\emph{Using definition~\eqref{eq:the_Fisher_information}, we calculate the Fisher information for example~\ref{ex1} to be equal to 1. The Cram\'er-Rao bound for the estimation of $\epsilon$ reads $\mean{(\Delta\hat{\epsilon})^2}\geq\frac{1}{N}$.
}
\end{example}

%achievability of Cramer Rao bound
%\cg{something, especially related to the problem of POPs}

%quantum Fisher information
The Fisher information is a measure how much information about the unknown parameter $\epsilon$ can be extracted from the quantum state given a choice of the measurement basis. However, one can rather ask how much information the quantum state itself yields about the parameter. In other words, how much information is in principle extractable from the quantum state. For that reason one can define the quantum Fisher information, which is obtained by maximizing the Fisher information over all possible measurements $M$,
\[\label{def:QFI}
H(\epsilon):=\underset{M}{\mathrm{sup}}\ F(\R_\epsilon,M),
\]
where $\mathrm{sup}_{M}$ denotes supremum. This definition naturally implies $F(\epsilon)\leq H(\epsilon)$, and from the Cram\'er-Rao bound we immediately obtain the quantum Cram\'er-Rao bound,
\[\label{def:quantum_Cramer_Rao}
\mean{\Delta\hat{\epsilon}^2}\geq\frac{1}{N H(\epsilon)}.
\]

%different definitions of the QFI
Definition~\eqref{def:QFI} is one of many possible definitions of the quantum Fisher information and it does not provide an effective formula to calculate such quantity. It is also not clear whether the maximum can be saturated with some optimal measurement under which the Fisher information will be equal to the quantum Fisher information. For a single parameter estimation it is, in fact, possible by doing a projective measurement~\cite{Paris2009a}, a feature which does not hold in general when estimating multiple param-
eters. This optimal measurement might not be unique, however, one of the possible optimal measurements is given by projectors $P_x$ constructed from eigenvectors of the symmetric logarithmic derivative $L_\epsilon$. The symmetric logarithmic derivative is an operator defined as a solution to operator equation
\[\label{def:SLDsolution}
\frac{L_\epsilon\R_\epsilon+\R_\epsilon L_\epsilon}{2}=\partial_\epsilon\R_\epsilon.
\]
An alternative and completely equivalent definition~\cite{Paris2009a} of the quantum Fisher information is then obtained by inserting projectors $P_x\equiv M_x$ into Eq.~\eqref{def:prob_distribution} and evaluating Eq.~\eqref{eq:the_Fisher_information}\footnote{Here it is important to point out that although the symmetric logarithmic derivative $L_\epsilon$ in essence depends on the unknown parameter $\epsilon$, so does its spectral decomposition, the derivative of~\eqref{def:prob_distribution} is still given by $\partial_\epsilon p(x|\epsilon)=\tr[P_x\partial_\epsilon\R_\epsilon]$, and not by $\partial_\epsilon p(x|\epsilon)=\tr[\partial(P_x\R_\epsilon)]$. This is because $\partial_\epsilon p(x|\epsilon)$ measures the change of the probability distribution $p(x|\epsilon)$ when $\epsilon$ is slightly varied while the measurement basis $P_x$ is fixed. But how to reconcile with the fact that $P_x$ seems to be $\epsilon$-dependent? In practice, in the local estimation theory the symmetric logarithmic derivative is evaluated at the approximate value $\epsilon_{\mathrm{app}}$ of the parameter, and the optimal measurement basis $P_x$ is fixed at this approximate value. Therefore, $P_x$ does not depend on the real value $\epsilon$.}  (see appendix~\ref{app:CR}), which yields
\[\label{def:H_using_L}
H(\epsilon)=\tr[\partial_\epsilon\R_\epsilon L_\epsilon]=\tr[\R_\epsilon L_\epsilon^2].
\]
The above definition must be the same as the definition given by Eq.~\eqref{def:QFI} because all Fisher informations are upper bounded by $\tr[\R_\epsilon L_\epsilon^2]$ (see~\cite{Paris2009a}), and as mentioned before a special pick of the measurement will results in the Fisher information to be equal to $\tr[\R_\epsilon L_\epsilon^2]$.
%From the symmetric logarithmic derivative it is also possible to construct an unbiased estimator  which saturates the Cram\'er-Rao bound in a single-shot experiment (N=1)~\cite{Paris2009a}, a feature which does not hold in general when estimating multiple parameters.

\begin{example}
\emph{Why the name symmetric logarithmic derivative? From the definition~\eqref{def:SLDsolution} it is clear why it is called symmetric. For the other part let us for simplicity assume that $L_\epsilon$ and $\R_\epsilon$ commute (so we can sum the symmetric part of the definition) and that $\R_\epsilon$ is full rank operator (so $\R_\epsilon^{-1}$ exists). Under those conditions it is easy to find a solution to Eq.~\eqref{def:SLDsolution}, $L_\epsilon=\partial_\epsilon\log{\R_\epsilon}$.
}
\end{example}

Solving equation~\eqref{def:SLDsolution} is not easy in general. However, the solution has been found~\cite{hubner1993computation} for the case when the spectral decomposition of the density matrix is known,
\begin{equation}\label{SLD}
L_\epsilon=2\!\!\!\!\sum_{\substack{k,l \\p_k+p_l> 0}}\!\!\!\!\frac{\langle \psi_k|\partial_{\epsilon}\R_{\epsilon}|\psi_{l}\rangle}{p_k+p_l} |\psi_{k}\rangle\langle\psi_{l}|,
\end{equation}
where the vectors  $\{|\psi_k\rangle\}$ are the eigenvectors of $\R_{\epsilon}$, i.e. $\R_{\epsilon}=\sum_k p_k|\psi_k\rangle\langle\psi_k|$. The summation involves only elements for which $p_i+p_j>0$. The quantum Fisher information is then
\begin{equation}\label{QFI}
H(\epsilon)=2\!\!\!\!\sum_{\substack{k,l \\p_k+p_l> 0}}\!\!\!\!\frac{\left|\langle \psi_k|\partial_{\epsilon}\R_{\epsilon}|\psi_{l}\rangle\right|^2}{p_k+p_l}.
\end{equation}

\begin{example}\label{ex:pure}
\emph{Now we can derive an elegant formula for the quantum Fisher information of pure states. Assuming $\R_\epsilon=\pro{\psi}{\psi}$ we sum over all elements in Eq.~\eqref{QFI}. It is important not to forget eigenvectors with zero eigenvalue (there will be $\mathrm{dim}\HS-1$ many of them). Then we use the Parseval identity and the normalization condition $\braket{\psi}{\psi}=1$. By differentiating the normalization condition twice we find that $\braket{\psi}{\partial_{\epsilon}\psi}$ is purely imaginary. Using this property we finally derive \[\label{eq:QFI_pure}
H(\epsilon)=4(\braket{\partial_{\epsilon}\psi}{\partial_{\epsilon}\psi}-\norm{\braket{\psi}{\partial_{\epsilon}\psi}}^2).\]
%%%%%Another way SLD...
A relatively easier alternative route is to prove that $L_\epsilon=2(\pro{\psi}{\partial_{\epsilon}\psi}+\pro{\partial_{\epsilon}\psi}{\psi})$ solves Eq.~\eqref{def:SLDsolution} and then to use Eq.~\eqref{def:H_using_L}. The third option for deriving this formula is through the Uhlmann fidelity, see example~\ref{ex:pure2}.
}
\end{example}

\section{Estimating channels: encoding operations, probes, and uncertainty relations}

In the previous section we introduced a formalism which can be used to find the ultimate limit of precision with what a parameter encoded in a quantum state can be estimated. In here, we will introduce a common metrological scenario aimed to estimate quantum channels.

Estimating channels in quantum metrology has numerous stages. These stages can be depicted as follows,
\[\label{scheme_for_estimating_channels}
\xrightarrow{preparation}\R_0\xrightarrow{channel(\epsilon)}\R_\epsilon\xrightarrow{measurement}(x_1,\dots,x_N)\xrightarrow{estimation}{\hat{\epsilon}(x_1,\dots,x_N)\approx\epsilon}
\]
First, the probe state $\R_0$ is prepared. This probe state is fed into a channel which encodes the unknown parameter $\epsilon$. The structure of the channel is usually known. For example, it is known that it is a phase-changing channel. However, it is not known how much phase-change is introduced by that channel. It is the phase we are trying to estimate.
%Such an example represents the gravitational wave detector LIGO\Cit{}

After the parameter is encoded, an appropriate measurement basis is chosen. Repeating this procedure on $N$ identical states $\R_\epsilon$ we obtain a statistics of measurement results. Those results are then turned into an estimate of the parameter through an estimator $\hat{\epsilon}$.

The task of quantum metrology is then three-folded: First, it is finding the optimal state $\R_0$ for probing the channel, i.e., the state which is the most sensitive to the channel. Within the local estimation theory, this is usually done by maximizing the quantum Fisher information under some fixed condition on the probe state. For example, finding the best probe state given a fixed amount of energy. The second task of quantum metrology is to find the optimal measurement. The optimal measurement is such that it produces a statistics of the measurement results which is the most informative about the parameter, i.e., statistics which leads to the lowest mean squared error on the parameter we want to estimate. An optimal measurement is given by the condition that the Fisher information for that particular measurement is equal to the quantum Fisher information. One of the optimal measurements can be always found by diagonalizing the symmetric logarithmic derivative. Third task of quantum metrology is to choose an appropriate estimator, which gives an appropriate meaning to the estimate with respect to the real value $\epsilon$. Two obvious choices are previously mentioned MLE estimator, which gives the value has the highest probability to produce the measurement results, or the Bayes estimator, which minimizes the risk that the estimated value is far away from the real value.

This thesis will focus on the mathematical formalism of the first stage, as well as finding optimal Gaussian probe states, both being discussed in the next chapters. The other two stages will not be discussed in detail, but we will point out certain directions when appropriate. Now we will present some basic results about the channel estimation.

Assuming the encoding operation is a unitary in an exponential form, or more precisely one-parameter unitary group, $\R_\epsilon=e^{-i\hat{K}\epsilon}\R_0e^{+i\hat{K}\epsilon}$, where $\hat{K}$ is a Hermitian operator, the quantum Fisher information is a constant independent on the parameter we want to estimate. This serves as a good check when calculating the quantum Fisher information for such channels. Note, however, that the symmetric logarithmic derivative still depends on the parameter, and so does the optimal POVM. Channels represented by one-parameter unitary group are very common. Examples include phase-changing channels, mode-mixing channels, squeezing channels, or a displacing channel.

Because of the unitarity of the channel, pure states remain pure after the parameter is encoded. We can use the result of example~\ref{ex:pure} and derive the quantum Fisher information for pure states undergoing such unitary channels,
\[\label{pureK}
H(\epsilon)=4\bra{\psi_0}\Delta\hat{K}^2\ket{\psi_0}=:4\mean{\Delta\hat{K}^2},
\]
where $\Delta\hat{K}:=\hat{K}-\bra{\psi_0}\hat{K}\ket{\psi_0}$. This interesting result shows that the quantum Fisher information scales quadratically with the encoding operator $\hat{K}$. This means, if the encoding is twice as fast, the quantum Fisher information is four times bigger, and the mean squared error with what we can estimate the parameter is $\frac{1}{4}$ of the previous mean squared error. Also, the right hand side is essentially the variance of the observable $\hat{K}$. But this immediately says that the best probe states are such which maximize the variance in the observable $\hat{K}$, which is the generator of translations in the parameter we want to estimate. For example, the encoding operator can be the Hamiltonian and the encoded parameter time. Because of identity~\eqref{pureK}, the best time-probes are such which have the widest distribution in energy. These are called the GHZ states and play an important role in the estimation theory.

For one measurement ($N=1$) and SI units ($\hbar\neq 1$), the quantum Cram\'er-Rao bound~\eqref{eq:Cramer_Rao} gives an interesting relation for the pure states,
\[
\mean{\Delta\hat{K}^2}\mean{\Delta\hat{\epsilon}^2}\geq\frac{\hbar^2}{4}.
\]
The quantum Cram\'er-Rao bound thus represents a type of the Heisenberg uncertainty relations. However, this is not between two observables as it is usually considered, but between one observable and one parameter. If for example the encoding operator is the Hamiltonian and the parameter time, this inequality states that the mean squared error on the time estimate multiplied by the square root of the variance in energy of the quantum state cannot go below $\frac{\hbar}{2}$ in a single-shot experiment.

The equivalent formulae for the mixed states exist and can be found for example in~\cite{Paris2009a}. However, the quantum Fisher information for mixed states is not equal to four times the variance as it is for pure states. In fact, it is always lower. For more details on the Heisenberg uncertainty relations, as well as on its connection to the speed of evolution of quantum states, see for example~\cite{kok2010introduction,taddei2013quantum}.

\section{Classical and Heisenberg scaling}

As briefly mentioned in the previous section, finding optimal probe states usually encompasses fixing a certain parameter, usually the mean energy of the probe state, or equivalently, the mean number of particles in the probe state. In the beginning of this chapter we assumed we have $N$ identical copies of the same state dependent on the parameter we want to estimate, i.e., in total we have $\R_\epsilon^{\otimes N}$. A single measurement of such large state can extract as much information as $N$ measurements on the $N$ identical subsystems. For a single-parameter estimation, if each subsystem represents one particle, there is no difference between measuring all of these particles at the same time, or each particle separately.\footnote{Nevertheless it is important to note that entangled measurements can improve the estimation of multiple parameters~\cite{fujiwara1995quantum}.} The improvement on the precision of the estimated parameter then scales as central limit theorem dictates for any identical and identically distributed variables. That is why the quantum Fisher information scales as $H(\epsilon)\sim N$ for such states, called the shot-noise limit. However, one can consider an alternative input state with the same `cost' which performs much better. If the probe state is an entangled state such as GHZ state, the task is no longer equivalent to measuring each particle separately and central limit theorem does not apply. The quantum enhancement is possible. Such states can then scale as $H(\epsilon)\sim N^2$, called the Heisenberg limit.

\begin{example}\label{ex:Heisenberg_scaling}
\emph{Consider the phase-changing encoding operator $e^{-i\hat{N}\epsilon}$, where $\hat{N}$ is the total number operator. We can use results of example~\ref{ex:pure} or equation~$\eqref{pureK}$ to calculate the quantum Fisher information for the probe states. First, $\R_0=\R_{0s}^{\otimes N}$, where $\R_{0s}=\pro{\psi_{0s}}{\psi_{0s}}$ is the pure state defined as $\ket{\psi_{0s}}=\frac{1}{\sqrt{2}}(\ket{0}+\ket{1})$. Second, the GHZ probe state $\R_0=\pro{\psi_0}{\psi_0}$, where $\ket{\psi_0}=\frac{1}{\sqrt{2}}(\ket{0,\dots,0}+\ket{1,\dots,1})$. Although both states have the same mean energy $\frac{N}{2}$, the quantum Fisher information of the separable state achieves the shot-noise limit, $H(\epsilon)=N$, in contrast to the entangled state which achieves the Heisenberg limit, $H(\epsilon)=N^2$.
}
\end{example}

Note however, despite its name, whether the Heisenberg limit is or is not a fundamental limit depends on the particular definition one chooses to use. For the definitions which define the scaling of the quantum Fisher information with respect to the mean energy of the state, or with respect to the maximum energy of the state, a sequences of states can be found to achieve super-Heisenberg scaling~\cite{luis2004nonlinear,boixo2008quantum}. Constructing such states is particularly easy in the Fock space, since its infinite-dimensional structure allows for states to have an arbitrarily large variance in energy while having an arbitrarily low mean value of energy. We will discuss this issue in more detail in section~\ref{sec:Heisenberg_limit_Gaussian}. On the other hand, when one considers the scaling with respect to the amount of resources needed to prepare such a probe state, the Heisenberg limit is indeed the fundamental limit~\cite{zwierz2010general}.

%One may encounter the classical and Heisenberg scaling of the quantum Fisher information. To show what it is about, let me just bring the ideas from the beginning. We assumed we have $N$ identical copies of the same state dependent on the parameter we want to estimate. I.e., in total we have $\R_\epsilon^{\otimes N}$. A single measurement of such large state is equivalent to the $N$ measurements on the $N$ subsystems, that is why the quantum Fisher information for such state scales as $H(\epsilon)\sim N$, called the classical scaling. However, one can consider an alternative input state with the same `cost', which performs much better. The cost is usually considered the total number of particles used in the experiment. The best probe states with the same cost are usually highly entangled and they perform as $H(\epsilon)\sim N^2$ which is called the Heisenberg scaling. Preparing such states is very important for estimating the parameter and thus creating them experimentally would be a huge step allowing for many new technologies.

\section{Geometry of estimation and multi-parameter estimation}\label{QMsec:multi}

We have considered only one parameter to be estimated so far. But there are tasks where it is important to estimate multiple parameters. These scenarios include for example simultaneous estimation of a phase and the decoherence or estimation of multiple spins pointing in different directions. The theory outlined in previous sections can be naturally generalized to multi-parameter estimation. However, there are some problems which are connected to the impossibility of estimating the parameters simultaneously. This is tied to the fact that optimal measurements for estimation of different parameters do not necessarily commute. Higher precision in one parameter induces a trade-off on the precision in others. For that reason, it is not even clear what figure of merit to maximize. Whether to maximize the total variance on the parameters, which gives each parameter the same importance, a weighted sum of variances, or a covariance between the two parameters. For an introduction to the multi-parameter estimation see for example~\cite{Paris2009a,szczykulska2016multi}.

Assuming the density matrix depends on a vector of parameters $\boldsymbol{\epsilon}=(\epsilon_1,...,\epsilon_n)$, in the analogy of Eq.~\eqref{def:SLDsolution} we define symmetric logarithmic derivatives,
\[
\frac{L_{i}\R_{\be}+\R_{\be} L_{i}}{2}=\partial_{i}\R_{\boldsymbol{\epsilon}}.
\]
We use a simplified notation $\partial_{i}\equiv\partial_{\epsilon_i}$.
The quantum Fisher information matrix is then a symmetric positive or symmetric positive semi-definite matrix given by
\[\label{def:Information_matrix}
H^{ij}(\boldsymbol{\epsilon})=\tr\left[\frac{L_{i}L_{j}+L_{j}L_{i}}{2}\R_{\be}\right],
\]
from which it is possible to derive a multi-parameter equivalent of Eq.~\eqref{QFI},
\begin{equation}\label{QFI_multi}
H^{ij}(\be)=2\!\!\!\!\sum_{\substack{k,l \\p_k+p_l> 0}}\!\!\!\!\frac{\Re(\langle \psi_k|\partial_i\R_{\be}|\psi_{l}\rangle\langle \psi_l|\partial_j\R_{\be}|\psi_{k}\rangle)}{p_k+p_l},
\end{equation}
where $\Re$ denotes the real part. Performing $N$ identical measurements on $N$ identical copies of a quantum state the multi-parameter quantum Cram\'er-Rao bound reads
\[
\mathrm{Cov}[\hat{\boldsymbol{\epsilon}}]\geq\frac{1}{N}{H}^{-1}(\boldsymbol{\epsilon}),
\]
where $\mathrm{Cov}[\hat{\boldsymbol{\epsilon}}]=\mean{\hat{\epsilon}_i\hat{\epsilon}_j}-\mean{\hat{\epsilon}_i}\mean{\hat{\epsilon}_j}$ is the covariance matrix of the locally unbiased estimator vector $\hat{\be}$ (i.e., matrix with variances of single parameters on the diagonal and correlation coefficients being the non-diagonal elements), and $H^{-1}(\boldsymbol{\epsilon})$ the inverse of the matrix defined in Eq.~\eqref{def:Information_matrix}. The above equation should be understood as an operator inequality. It states that $\mathrm{Cov}[\hat{\boldsymbol{\epsilon}}]-\frac{1}{N}{H}^{-1}$ is a positive semi-definite or a positive definite matrix.

In contrast to the one-parameter scenario for which an optimal measurement can be always found, it is not always possible to find the optimal measurement the multi-parameter quantum Cram\'er-Rao bound, i.e., it is not always possible to find a measurement for which the Fisher information matrix $\Big($defined as $F^{ij}(\be)=\int\!\!\dif{x}\ \frac{\partial_{i} p(x|\be) \partial_{j} p(x|\be)}{p(x|\be)}$$\Big)$ equals the quantum Fisher information matrix. This is because projectors $P_x^{(i)}$ from spectral decompositions of different symmetric logarithmic derivatives do not necessarily commute which is a general problem in the multi-parameter estimation.
Several advancements in the attainability of the multi-parameter bound are reviewed in~\cite{szczykulska2016multi}.

The quantum Fisher information matrix is connected to an important statistical measure called the Bures metric~\cite{Bures1969a}. To define this metric we first introduce the Bures distance. The Bures distance is a measure of distinguishability between two quantum states $\R_{1,2}$ and is defined through the Uhlmann fidelity~\cite{Uhlmann1976a}
\[\label{def:Uhlmann_Fidelity}
\mathcal{F}({\R}_{1},{\R}_{2})\,:=\,\big(\tr\sqrt{\sqrt{{\R}_{1}}\,{\R}_{2}\,\sqrt{{\R}_{1}}}\big)^{2}
\]
as
\[\label{def:bures_distance}
d_B^2(\R_1,\R_2)=2\big(1-\sqrt{\mathcal{F}(\R_1,\R_2)}\big).
\]
The Bures distance gives rise to the Bures metric $g^{ij}$ which measures the amount of distinguishability of two close density matrices in the coordinate system $\be$ through the definition for the line element,
\[\label{eqn:bures}
d_B^2(\R_{\boldsymbol{\epsilon}},\R_{\boldsymbol{\epsilon}+\boldsymbol{\mathrm{d}\epsilon}})=\sum_{i,j}g^{ij}(\be)\mathrm{d}\epsilon_i\mathrm{d}\epsilon_j.
\]

It is usually thought and it is mentioned in~\cite{Paris2009a} that the quantum Fisher information matrix~\eqref{def:Information_matrix} is a multiple of the Bures metric. Although usually true, this is not always the case. This belief is based on derivations of explicit formulae for the Bures metric in~\cite{hubner1992explicit} and \cite{sommers2003bures}. The first derivation assumes that the density matrix is invertible, while the second lacks particular details concerning the treatment of problematic points. Moreover, both derivations are finding expressions for %seemingly coordinate-independent\footnote{Although the expression looks coordinate-independent, the Bures metric is always defined in certain coordinates, which in our case is the vector of parameters $\be$.}
infinitesimal distance $d_B^2(\R,\R+\mathrm{d}\R)$ which is ill-defined at the boundary of the convex space of density matrices (when $\R^{-1}$ does not exist), because certain choices of $\mathrm{d}\R$ can cause $\R+\mathrm{d}\R$ not to be a density matrix anymore. In general, however, for parameterized quantum states $\R_{\be}$ in which a slight change in the parameter $\be$ results in an eigenvalue of the density matrix to vanish (or, equivalently, results in an eigenvalue to `pop out'), there is an extra term in the Bures metric %induced by the Bures distance
which needs to be accounted for. As we show in appendix~\ref{app:discontinuity_of_QFI}, the quantum Fisher information matrix is connected to the Bures metric through relation
\[\label{eq:bures_metric}
g^{ij}(\be)=\frac{1}{4}\Big(H^{ij}(\be)+2\!\!\!\!\!\!\sum_{k:\,p_k(\be)=0}\!\!\!\!\!\!\partial_{i} \partial_{j}p_k(\be)\Big).
\]
By $k:\,p_k(\be)=0$ we mean that the sum goes over all eigenvalues $p_k$ such that their value vanishes at point $\be$. From the above relation we can see that the (four times) Bures metric and the quantum Fisher information matrix do not coincide only at certain points $\be$, at which an eigenvalue vanishes. When change of the parameter does not result in the change of purity, for example when the operation encoding $\be$ is a unitary operation, the (four times) Bures metric and the quantum Fisher information matrix are identical. It is worth noting that the Hessian $\mathcal{H}_k^{ij}(\be):=\partial_{i} \partial_{j}p_k(\be)$ is a positive or a positive semi-definite matrix, because $p_k$ reaches the local minimum at point $\epsilon$ such that $p_k(\epsilon)=0$. Therefore the following matrix inequality holds,
\[
4g\geq H
\]
and $4g= H$ if and only if for all $\be$ and $k$ such that $p_k(\be)=0$, $\mathcal{H}_k(\be)=0$.

Expression~\eqref{eq:bures_metric} has a surprising interpretation. It can be shown from Eq.~\eqref{QFI} that even for analytical functions $\R_\epsilon$ the quantum Fisher information can be discontinuous at points $\be$ for which $p_k(\be)=0$. This discontinuity is however removable. It is possible to redefine these points in a way which makes the quantum Fisher information matrix continuous in the following sense: for $\R_{\be}\in C^{(2)}$ every element of the redefined matrix $H_c^{ij}$ is a continuous function in parameter $\epsilon_i$ while all other parameters $\epsilon_k, k\neq i$ are kept fixed. The same holds for the parameter $\epsilon_j$. Such a redefinition leads exactly to the expression defined by the Bures metric,
\[\label{eq:connection_between_Hc_and_H}
H_c^{ij}(\be)=4g^{ij}(\be)=H^{ij}(\be)+2\!\!\!\!\!\!\sum_{k:\,p_k(\be)=0}\!\!\!\!\!\!\partial_{i} \partial_{j}p_k(\be).
\]
In the end, it is important to point out that although the continuous version of the quantum Fisher information seems to have nicer properties, the quantum Cram\'er-Rao bound holds only for the (possibly discontinuous) quantum Fisher information matrix $H$. For more details see appendices~\ref{app:CR} and~\ref{app:discontinuity_of_QFI}.

Combining the above equation with Eq.~\eqref{eqn:bures} we obtain the expression for the continuous quantum Fisher information matrix in terms of fidelity,
\[\label{eq:QFI_matrix_using_fidelity}
\sum_{i,j}H_c^{ij}\mathrm{d}\epsilon_i\mathrm{d}\epsilon_j
=8\big(1-\sqrt{\mathcal{F}(\R_{\boldsymbol{\epsilon}},\R_{\boldsymbol{\epsilon}+\boldsymbol{\mathrm{d}\epsilon}})}\big).
\]
This means that we can calculate the continuous quantum Fisher information matrix by expanding the Uhlmann fidelity to the second order in infinitesimal parameters. For a single parameter we can simply write
\[\label{QFI_using_fidelity}
H_c(\epsilon)=8\lim_{\mathrm{d}\epsilon\rightarrow0}\frac{1-\sqrt{\mathcal{F}(\R_\epsilon,\R_{\epsilon+\mathrm{d}\epsilon})}}{\mathrm{d}\epsilon^{2}}.
\]

\begin{example}\label{ex:pure2}
\emph{
In the case of pure states $H$ equals $H_c$ because the purity does not change. Therefore, using Eq.~\eqref{QFI_using_fidelity} we can derive the quantum Fisher information for pure states one more time. The result should coincide with the results of example~\ref{ex:pure}.
}
\end{example}

The next chapter will introduce the current state-of-the art quantum metrology on Gaussian states. We build on these results and take them even further in chapter~\ref{chap:QM_GS} in which we derive new formulae for the parameter estimation and we develop a new and effective formalism for finding optimal Gaussian probe states. We will also discuss the issue of discontinuity of the quantum Fisher information matrix in the context of Gaussian states. Chapter~\ref{chap:reference_frames} then sheds light on the quantum metrology in the context of quantum reference frames. There we will show how having misaligned or imperfect reference frames affects the estimation precision.

%%%%%%%%%%%%%%%%%%%%%%%%%%%%%%
\chapter{Gaussian states}\label{chap:GS}
%%%%%%%%%%%%%%%%%%%%%%%%%%%%%%

%opening advertisement
Gaussian states are of great use in experimental quantum physics, mainly because they combine several useful properties. They are relatively straightforward to prepare and handle, especially in optical systems~\cite{andersen201530}, and they are resistant to decoherence~\cite{demkowicz2013fundamental}. Although they resemble some properties of classical fields, they also exhibit quantum phenomena such as entanglement and thus can be used for quantum information protocols, for instance quantum teleportation~\cite{furusawa2007quantum} and quantum cryptography~\cite{jouguet2013experimental,huang2016long}. Moreover, Gaussian states are also simple to handle mathematically via elegant phase-space formalism.

%references
There has been extensive literature published on Gaussian states. Let us mention for example lecture notes on continuous variable quantum information~\cite{ferraro2005gaussian}, PhD thesis on entanglement of Gaussian states~\cite{adesso2007entanglement} or Gaussian channels~\cite{schafer2013information}, and review articles on Gaussian states~\cite{wang2007quantum,Weedbrook2012a,Adesso2014a}. In this introductory chapter we focus on aspects of continuous variable quantum information directly related to finding optimal probe states for the estimation of Gaussian channels.

%content
This chapter is organized as follows: we first introduce the Fock space of a bosonic field which is necessary to define Gaussian states. We introduce the phase-space formalism of Gaussian states and Gaussian channels. In particular, we provide symplectic matrices in the real and the complex phase-space formalism for the most common Gaussian unitary channels. We introduce basic Gaussian states and fully parametrize one-, two-, and three-mode Gaussian states. This parametrization will be used in section~\ref{sec:estimation_of_channels} to find the optimal Gaussian probe states for the estimation of Gaussian channels. Finally, we give an overview on the current state-of-the-art in the estimation of Gaussian states in the phase-space formalism. Later in chapter~\ref{chap:QM_GS} we build on these results and derive new easy-to-use formulae.

\section{Fock space of a bosonic field}\label{sec:Fock_space}
%introduction to bosons, then construction of the fock space, annihilation and creation operators with Eq.1, construction of a number basis on an example of two particles
Bosons are particles which follow Bose-Einstein statistics and are characterized by an integer spin. Examples include photons -- particles of light, W and Z bosons -- particles mediating the weak interaction, phonons -- excitations of a vibrational field, or Cooper pairs -- bound states of electrons responsible for super-conductivity. An important property of bosons is that their statistics gives no restriction on the number of indistiguishable particle occupying the same quantum state. That is why a quantum description of many such particles offer a rich structure, described by a bosonic Fock space.

%Construction of the Fock space
Let $\HS$ be a single particle Hilbert space. The (bosonic) Fock space is the direct sum of the symmetric tensor powers of the Hilbert space $\HS$,
\[
F(\HS)=\bigoplus_{n=0}^\infty S\left(\HS^{\otimes n}\right).
\]
$\HS^{\otimes 0}=\CF$ and $S$ is the operator which symmetrizes the Hilbert space, i.e., $S\left(\HS^{\otimes n}\right)$ consists of such states $\ket{\psi}\in\HS^{\otimes n}$ which are completely symmetric with respect to the exchange of particles.

%basis of the Fock space
Assuming $\HS=\mathrm{span}\{\ket{\psi_1},\ket{\psi_2},\dots\}$, we can construct an elegant way of how to write a basis of the Fock space. $F(\HS)=\mathrm{span}\{\ket{n_1^{(1)},n_2^{(2)},\dots}\}_{n_1,n_2,\dots}$ where $n_1^{(1)}$ denotes that there are $n_1$ particles in the state $\ket{\psi_1}$, $n_2^{(2)}$ denotes that there are $n_2$ particles in the state $\ket{\psi_2}$ and so forth. Because the particles are indistinguishable, exchanging any two particles in the same state of the single-particle Hilbert space $\HS$ should not change the full state in the Fock space. Hence a pure state in the Fock space can be fully described only by the number of particles in each single-particle state and this notation is consistent. Any other state in the Fock space can be described as a linear combination of these number vectors. Construction of the number basis will be clarified in the following example.

%Example on two-mode Fock space
\begin{example}\label{ex:twomode_Fock_space}
\emph{
Let $\HS=\mathrm{span}\{\ket{\uparrow},\ket{\rightarrow}\}$ be a Hilbert space a polarized photon with $\ket{\uparrow}$ representing a vertically polarized photon and $\ket{\rightarrow}$ representing a horizontally polarized photon. The Fock space is
\[
F(\HS)=\CF\oplus S\left(\HS\right) \oplus S\left(\HS\otimes\HS\right)\oplus S\left(\HS\otimes\HS\otimes\HS \right)\oplus\cdots,
\]
where $S(\HS)=\HS$ denotes the single particle Hilbert space, $S(\HS\otimes\HS)$ the symmetrized two particles Hilbert space, $S(\HS\otimes\HS\otimes\HS)$ the symmetrized three particles Hilbert space. Number states are then constructed as
\begin{align}
\ket{00}&=1\oplus\boldsymbol{0}\oplus\boldsymbol{0}\oplus\boldsymbol{0}\oplus\cdots,\nonumber\\
\ket{10}&=0\oplus\ket{\uparrow}\oplus\boldsymbol{0}\oplus\boldsymbol{0}\oplus\cdots,\nonumber\\
\ket{01}&=0\oplus\ket{\rightarrow}\oplus\boldsymbol{0}\oplus\boldsymbol{0}\oplus\cdots,\nonumber\\
\ket{20}&=0\oplus\boldsymbol{0}\oplus\ket{\uparrow}\otimes\ket{\uparrow}\oplus\boldsymbol{0}\oplus\cdots,\\
\ket{11}&=0\oplus\boldsymbol{0}\oplus\frac{1}{\sqrt{2}}(\ket{\uparrow}\otimes\ket{\rightarrow}+\ket{\rightarrow}\otimes\ket{\uparrow})\oplus\boldsymbol{0}\oplus\cdots,\nonumber\\
\ket{02}&=0\oplus\boldsymbol{0}\oplus\ket{\rightarrow}\otimes\ket{\rightarrow}\oplus\boldsymbol{0}\oplus\cdots,\nonumber\\
\ket{30}&=0\oplus\boldsymbol{0}\oplus\boldsymbol{0}\oplus\ket{\uparrow}\otimes\ket{\uparrow}\otimes\ket{\uparrow}\oplus\cdots,\nonumber\\
\ket{21}&=0\oplus\boldsymbol{0}\oplus\boldsymbol{0}\oplus
\frac{1}{\sqrt{3}}(\ket{\uparrow}\otimes\ket{\uparrow}\otimes\ket{\rightarrow}+\ket{\uparrow}\otimes\ket{\rightarrow}\otimes\ket{\uparrow}+\ket{\rightarrow}\otimes\ket{\uparrow}\otimes\ket{\uparrow})\oplus\cdots\nonumber\\
&\cdots\nonumber
\end{align}
Clearly, these vectors are linearly independent and any vector of the Fock space can be written as their linear combination. They form a basis of the Fock space. The zero vector in the Fock space commonly denoted as $0$ is defined as
\[
0=0\oplus\boldsymbol{0}\oplus\boldsymbol{0}\oplus\boldsymbol{0}\oplus\cdots.\\
\]
}
\end{example}

%Annihilation and creation operators
Each Fock space can be equipped with a set of annihilation and creation operators which are necessary to define Gaussian states. We assign one annihilation $\hat{a}_i$ and one creation operator $\hat{a}_i^\dag$ to each single-particle basis vector $\ket{\psi_i}$. The action of these field operators is to either annihilate or to create a particle in the state $\ket{\psi_i}$,
\begin{subequations}
\begin{align}
\hat{a}_i\ket{\dots,n_{i-1},n_i,n_{i+1},\dots}&=\sqrt{n_i}\ket{\dots,n_{i-1},n_i-1,n_{i+1},\dots},\\
\hat{a}_i\ket{\dots,n_{i-1},0,n_{i+1},\dots}&=0,\\
\hat{a}_i^\dag\ket{\dots,n_{i-1},n_i,n_{i+1},\dots}&=\sqrt{n_i+1}\ket{\dots,n_{i-1},n_i+1,n_{i+1},\dots}.
\end{align}
\end{subequations}
The field operators satisfy the commutation relations $[\hat{a}_i,\hat{a}_j^\dag]=\delta_{ij}\mathrm{id}$, where $\delta_{ij}$ denotes Kronecker delta and $\mathrm{id}$ denotes the identity element of the algebra.

\begin{example}\label{ex:building_a_Fock_basis}
\emph{
Having defined the action of creation operators, it is clear that all basis vectors introduced in example~\ref{ex:twomode_Fock_space} can be written as
\[
\ket{n_1^{(1)}n_2^{(2)}}=\frac{(\hat{a}_1^\dag)^{n_1}}{\sqrt{n_1!}} \frac{(\hat{a}_2^\dag)^{n_2}}{\sqrt{n_2!}} \ket{00}.
\]
Later we will focus on quantum field theory where even more compact notation is used. This is due to the fact that there are often infinitely many modes, making it impossible to write write them all into one long vector. The vacuum state will be denoted as $\ket{0}:=\ket{00}$ and the state of $n_1$ particles in the first mode as $\ket{n_1^{(1)}}:=\ket{n_1^{(1)} 0}$.
}
\end{example}
%\Miss{How to construct Fock basis out of the creation and annihilation operators, say that the creation and annihilation operators are actually maps from higher symmetrized hilbert space to smaller.}

%Definition of the symplectic form
Now we will switch to the more elegant notation which will be appropriate for an effective description of Gaussian states. Assuming there is a finite number of basis vectors $\ket{\psi_i}$ -- from now on called modes -- of the single particle Hilbert space $\HS=\mathrm{span}\{\ket{\psi_1},\dots,\ket{\psi_N}\}$, we collect their associated annihilation and creation operators into a vector $\bA:=(\hat{a}_1,\dots,\hat{a}_N,\hat{a}_1^\dag,\dots,\hat{a}_N^\dag)^T$. The commutation relations between the operators can be written in compact form,
\[\label{def:commutation_relation}
[\bA^{i},\bA^{j\dag}]=K^{ij}\mathrm{id}\quad\!\Rightarrow\quad\! K=
\begin{bmatrix}
I & 0 \\
0 & -I
\end{bmatrix},
\]
where $I$ denotes the identity matrix. This equation also defines matrix $K$ to which we will later refer to as to the symplectic form. This matrix has numerous useful properties, namely $K^{-1}=K^\dag=K$ and $K^2=I$.

\section{Gaussian states in the phase-space formalism}

%characteristic function
Quantum states are usually described by a positive semi-definite operator called the density matrix $\R$, however, for bosonic systems an alternative and completely equivalent description exists which is particularly useful for a description of Gaussian states. Given a state $\R$ we define \emph{the symmetric characteristic function} as
\[
\chi(\boldsymbol{\xi})\,=\,\mathrm{tr}[\R\,\hat{D}(\boldsymbol{\xi})],
\]
where $\hat{D}(\boldsymbol{\xi})\,=\,e^{\boldsymbol{\hat{A}}^{\dag}K\boldsymbol{\xi}}$ is the \emph{Weyl displacement operator} with the variable of the form $\boldsymbol{\xi}\,=\,\boldsymbol{\gamma}\oplus\overline{\boldsymbol{\gamma}}$. Gaussian states are those whose characteristic function is, by definition, of Gaussian form, i.e.,
\[
\chi(\boldsymbol{\xi})\,=\,e^{-\frac{1}{4}\boldsymbol{\xi}^{\dag}\sigma\boldsymbol{\xi}-i\,{\bd}^{\dag}K\boldsymbol{\xi}}.
\]

%covariance matrix
In the analogy of classical probability theory, Gaussian states are completely described by the first and the second statistical moments $\bd$ and $\sigma$ of the field. The \emph{displacement vector} $\bd$ and the \emph{covariance matrix} $\sigma$ are defined as~\cite{Weedbrook2012a},
\begin{subequations}\label{def:covariance_matrix}
\begin{align}
\bd^i&=\mathrm{tr}\big[\R\boldsymbol{\hat{A}}^i\big],\\
\sigma^{ij}&=\mathrm{tr}\big[\R\,\{\Delta\boldsymbol{\hat{A}}^i,\Delta\boldsymbol{\hat{A}}^{j{\dag}}\}\big].
\end{align}
\end{subequations}
The density operator $\hat\rho$ specifies the state of the field and $\{\!\cdot,\cdot\!\}$ denotes the anti-commutator, and $\Delta\boldsymbol{\hat{A}}:=\boldsymbol{\hat{A}}-\bd$.

%properties and structure of the covariance matrix
From the definition~\eqref{def:covariance_matrix} we can observe the following structure of the first and second moments:
\[\label{def:first_and_second_moments}
\bd=
\begin{bmatrix}
\boldsymbol{\tilde{d}} \\ \overline{\boldsymbol{\tilde{d}}}
\end{bmatrix},\quad
\sigma\,=\,\begin{bmatrix}
X & Y \\
\overline{Y} & \overline{X}
\end{bmatrix},
\]
where \emph{bar} denotes the complex conjugation. The covariance matrix is a positive-definite Hermitian matrix, $\sigma^\dag=\sigma$, i.e., $X^\dag=X$ and $Y^T=Y$, and further satisfying~\cite{simon1994quantum}
\[\label{eq:sigma_K_positivity}
\sigma+K\geq0
\]
which is a consequence of the commutation relations~\eqref{def:commutation_relation}.

%Tracing of the covariance matrix
Sometimes we can be interested in a subsystem of the Gaussian states. In the density matrix formalism, the density matrix of a subsystem is obtained by the partial tracing, i.e., tracing over all states we are not interested in. Partial tracing in the covariance matrix formalism is very simple. Tracing over modes we are not interested in is done simply by taking away the rows and columns of the covariance matrix and elements of the displacement vector associated with those modes.

%The real form
We emphasise that Eq.~\eqref{def:covariance_matrix} defines the complex form of the covariance matrix, which is defined by the anti-commutator of annihilation and creation operators. Most authors use the real form, which is defined in terms of position and momenta operators. Defining vector of position and momenta operators %\newline
$\boldsymbol{\hat{Q}}:=(\hat{x}_1,\dots,\hat{x}_N,\hat{p}_1,\dots,\hat{p}_N)^T$, where $\hat{x}_i:=\frac{1}{\sqrt{2}}(\hat{a}^\dag_i+\hat{a}_i)$, $\hat{p}_i:=\frac{i}{\sqrt{2}}(\hat{a}^\dag_i-\hat{a}_i)$, the real form displacement and the real form covariance matrix are defined as
\begin{subequations}\label{def:covariance_matrix_real}
\begin{align}
\bd_{\Re}^i&=\mathrm{tr}\big[\R\boldsymbol{\hat{Q}}^i\big],\\
\sigma_{\Re}^{ij}&=\mathrm{tr}\big[\R\,\{\Delta\boldsymbol{\hat{Q}}^i,\Delta\boldsymbol{\hat{Q}}^j\}\big],
\end{align}
\end{subequations}
where $\Delta\boldsymbol{\hat{Q}}:=\boldsymbol{\hat{Q}}-\bd_{\Re}$.

%Other notations
Other notations exist which adds to the confusion in literature. One other common example includes different ordering of the quadrature operators, $\boldsymbol{\hat{Q}}:=(\hat{x}_1,\hat{p}_1,$$\hat{x}_2,\hat{p}_2,$$\dots,$ $\hat{x}_N,\hat{p}_N)^T$. In this thesis we will consistently use the complex form unless specified differently in concrete examples. This is because the complex form expose the inner symmetries in more detail than the real form and because some formulae and matrices are much more elegantly expressed in the complex form. Also, the complex form is generally easier to work with at a small cost of admitting complex numbers. For more information about the real form and its connection to the complex form see appendix~\ref{app:real_covariance}, or~\cite{Arvind1995a,Adesso2014a}.

\section{Gaussian unitaries and symplectic geometry}

%Gaussian unitary
Gaussian transformation is an transformation which maps Gaussian states into Gaussian states. Gaussian unitary is a Gaussian transformation which is represented by a unitary operator, i.e., it transforms Gaussian state $\R$ into Gaussian state $\R'=\hat{U}\R\hat{U}^\dag$. All such operators can be generated via an exponential map with the exponent at most quadratic in the field operators~\cite{Weedbrook2012a},
\[\label{def:Gaussian_unitary}
\hat{U}=\exp\big(\tfrac{i}{2}\bA^\dag W \bA+\bA^\dag K \bg\big),
\]
where $W$ is a Hermitian matrix of the form following the same structure as the covariance matrix~\eqref{def:first_and_second_moments},
\[\label{def:W_for_Gaussian_unitary}
W=\begin{bmatrix}
X & Y \\
\ov{Y} & \ov{X}
\end{bmatrix},
\]
$\bg$ a complex vector of the form $\bg=(\tilde{\bg},\ov{\tilde{\bg}})^T$, and $K$ is the matrix defined in Eq.~\eqref{def:commutation_relation}. In the case that $W=0$, the Gaussian operator~\eqref{def:Gaussian_unitary} corresponds to the~\emph{Weyl displacement operator} $\hat{D}(\tilde{\bg})$, while for $\bg=0$ we obtain other Gaussian transformations such as the phase-changing operator, one- and two-mode squeezing operators, or mode-mixing operators depending on the particular structure of $W$. For more details see section~\ref{sec:list_of_Gaussian_unitaries}.

\subsection{Transformation of the first and the second moments}
Under the unitary channel~\eqref{def:Gaussian_unitary} the first and the second moments transform according to rule
\[\label{def:transformation}
\bd'=S\bd+\bb,\ \ \sigma'=S\sigma S^\dag,
\]
where, as we prove in appendix~\ref{app:derivation_of_S_and_b},
\[\label{eq:S_and_b}
S=e^{iKW},\ \
\bb=\Big(\!\int_0^1e^{iKWt}\mathrm{d}t\!\Big)\ \!\bg.
\]
The above identities together with transformation relations~\eqref{def:transformation} are central for the effective description of Gaussian states. They allow us to transform the density matrix description of Gaussian states to the phase-space formalism, which immensely simplify every calculation. In the density matrix formalism, Gaussian states can be usually written only in terms of Taylor series in operators, while in the phase-space formalism they are represented by one vector and one matrix.

\subsection{Symplectic group and the Lie algebra}
%symplectic group
The matrix $S$ from Eq.~\eqref{eq:S_and_b}, called the \emph{symplectic matrix}, has the same structure as $W$ and satisfies the relation
\[\label{def:structure_of_S}
S=
\begin{bmatrix}
\A & \B \\
\ov{\B} & \ov{\A}
\end{bmatrix},\ \ SKS^\dag=K.
\]
These two properties define the complex representation of the real symplectic group $Sp(2N,\mathbb{R})$. Note that transformations for which $\B=0$ are called passive, while transformation with $\B\neq0$ are called active. This is because symplective matrix representing a passive transformation commutes with the total number operator, $\hat{N}=\sum_{i+1}^N\hat{a}_i^\dag\hat{a}_i$, i.e., states before and after passive transformation contain the same mean number of bosons. In contrast, active transformation either create or annihilate particles. %passive vs active transformations

%Lie algebra
When describing quantum metrology on Gaussian states, the Lie algebra associated the symplectic group will prove to be very useful. The complex form of the Lie algebra associated with the real symplectic group $Sp(2N,\mathbb{R})$ is defined by properties
\[\label{def:P_1}
P=
\begin{bmatrix}
R & Q \\
\ov{Q} & \ov{R}
\end{bmatrix},\ \ PK+KP^\dag=0.
\]
The second property implies that $R$ is skew-Hermitian, $R^\dag=-R$, and $Q$ is symmetric (and complex in general), $Q^T=Q$. Note that we used the definition of the Lie algebra more common to mathematical literature~\cite{helgason1979differential}: for any $t\in\RF$, $S=e^{Pt}$ is symplectic. Some authors~\cite{Arvind1995a,deGosson2006a} define the Lie algebra by the property for any $t\in\RF$, $S=e^{iPt}$ is symplectic matrix leading to $PK-KP^\dag=0$. This is rather a cosmetic difference and does not affect any of the results of this thesis.

%Properties of symplectic group
\begin{example}
\emph{\emph{Properties of the symplectic group $Sp(2N,\mathbb{R})$.}
\begin{itemize}[leftmargin=*]
\item The symplectic group $Sp(2N,\mathbb{R})$ is connected, non-compact, simple Lie group.
\item Both defining properties~$\eqref{def:structure_of_S}$ are necessary to define the real symplectic group. This group is a subgroup of the the more general pseudo-unitary group~\cite{Mostafazadeh2004a} which is defined only by the second property, $U(N,N)=\{S\in GL(2N,\mathbb{C})|SKS^\dag=K\}$.
\item Eqs.~\eqref{def:structure_of_S} can be rewritten in two useful ways. The first one is actually identical to the definition of the Bogoliubov transformations used in the quantum field theory in curved space-time~\cite{Birrell1984a},
\begin{subequations}\label{def:Bogo_id}
\begin{align}
\A\A^\dag-\B\B^\dag&=I,\label{def:first_Bogo_id}\\
\A\B^T&=\B\A^T.\label{def:second_Bogo_id}
\end{align}
\end{subequations}
Eq.~\eqref{def:first_Bogo_id} implies that $\A^{-1}$ always exists. We can therefore define $\C:=\A^{-1}\B$ and obtain even simpler form,
\begin{subequations}
\begin{align}
\A^{-1}(\A^{-1})^\dag+\C\C^\dag&=I,\\
\C&=\C^T.
\end{align}
\end{subequations}
\item Let $S$ be a symplectic matrix. $\A$ is unitary $\Leftrightarrow$ $\B=0$ $\Leftrightarrow$ $S$ is unitary. \emph{Proof:} $\B=0\Rightarrow\A$: follows trivially from Eq.~\eqref{def:first_Bogo_id}. $\B=0\Leftarrow\A$: For unitary $\alpha$ Eq.~\eqref{def:first_Bogo_id} implies $\B^\dag\B=0$. Sequence of implications follows:
    $$
    \B^\dag\B=0\ \Rightarrow\ \forall\ket{\psi},\ |\!|\B\ket{\psi}|\!|^2=\bra{\psi}\B^\dag\B\ket{\psi}=0\ \Rightarrow\ \forall\ket{\psi},\ \B\ket{\psi}=0\ \Rightarrow\ \B=0.
    $$
    $\B=0$ $\Leftrightarrow$ $S$ follows the same logic. \qed
\item $\mathrm{dim}(Sp(2N,\mathbb{R}))=2N^2+N$. \emph{Proof:} The dimension of a matrix group is the same as the dimension of the associated Lie algebra defined in Eq.~\eqref{def:P_1}. Because number of real parameters needed to fully characterize the element of the Lie algebra $P$ is $N^2$ for the skew-hermitian matrix $R$ and $N^2+N$ for the symmetric matrix $Q$ respectively, the dimension of the Lie algebra is $2N^2+N$.\qed
\item $\mathrm{det}(S)=1$. \emph{Proof:} A simple proof\footnote{This simple proof is attributed to Jan Kohlrus.} involves transforming the symplectic matrix into the real form using appendix~\ref{app:real_covariance}, and applying a property of the Pfaffian, $\mathrm{pf}(S_\Re\Omega S_\Re^T)=\mathrm{pf}(\Omega)\mathrm{det}(S_\Re)$. A complicated proof can be found in~\cite{Arvind1995a}. \qed
\item $S^T\in Sp(2N,\mathbb{R})$, $S^{-1}=KS^\dag K\in Sp(2N,\mathbb{R})$, $K\in Sp(2N,\mathbb{R})$.
\item Any symplectic matrix can be decomposed using Euler's decomposition. For more details see section~\ref{sec:parametrization_of_Gaussian_states}.
\end{itemize}}
\end{example}

%the real form properties AND for more information see arvind
All definitions and properties in this section can be of course rewritten in terms of the real representation of the real symplectic group. For details see appendix~\ref{app:real_covariance} or~\cite{Arvind1995a}.

\subsection{A list of Gaussian unitaries}\label{sec:list_of_Gaussian_unitaries}
In this section we provide a list of basic Gaussian unitaries. We parametrize one-mode and two-mode Gaussian unitaries and provide their symplectic matrices.

%Weyl displacement operator
%\subsubsection{The Weyl displacement operator}
The simplest Gaussian unitary which acts only on the displacement vector and leave the covariance matrix invariant is previously mentioned displacement operator \[\label{eq:Weyl_displacement_operator}
\hat{D}(\tilde{\bg})=\exp\big(\bA^\dag K \bg\big),
\]
where $\bg=(\tilde{\bg},\ov{\tilde{\bg}})^T$. According to Eq.~\eqref{def:transformation}, this operator acts as $\bd'=\bg$, $\sigma'=\sigma$. Other transformations described in this section will characterized by $\bg=\boldsymbol{0}$ and will act as $\bd'=S\bd$, $\sigma'=S\sigma S^\dag$.

%one-mode unitaries
%\subsubsection{One-mode unitary operators}
For one-mode states ($N=1$), the Hermitian matrix $W$ in Eq.~\eqref{def:W_for_Gaussian_unitary} describing the Gaussian unitary can be fully parametrized as
\[\label{eq:W1}
W=\begin{bmatrix}
-\theta & i r e^{i \chi} \\
-i r e^{-i \chi} & -\theta
\end{bmatrix}.
\]
For $r=0$ and $\bg=\boldsymbol{0}$ the Gaussian unitary~\eqref{def:Gaussian_unitary} represents a one-mode phase-shift $\hat{R}(\theta)=\exp(-i\theta\hat{a}^\dag \hat{a})$. We will denote its correspondent symplectic matrix derived using Eq.~\eqref{eq:S_and_b} as $S=R(\theta)$. Choosing $\theta=0$ instead, we obtain one-mode squeezing at angle $\chi$, $\hat{S}(r,\chi)=\exp(-\frac{r}{2}(e^{i\chi}\hat{a}^{\dag2}-e^{-i\chi}\hat{a}^{2}))$. Squeezing at angle zero will be denoted as $\hat{S}(r)$ and its symplectic matrix equivalent will be denoted as $S(r)$.

%two-mode unitaries
%\subsubsection{Two-mode unitary operators}
In the analogy with one-mode Gaussian channels, for two-mode states ($N=2$) we parametrize the Hermitian matrix $W$ as
\[\label{eq:W2}
W=\begin{bmatrix}
-\theta_1 & -i\theta_{B}e^{i \chi_{B}}  & i r_1 e^{i \chi_1} & i r_{T} e^{i \chi_{T}} \\
i\theta_{B}e^{-i \chi_{B}} & -\theta_2 & i r_{T} e^{i \chi_{T}} & i r_2 e^{i \chi_2} \\
-i r_1 e^{-i \chi_1} & -i r_{T} e^{-i \chi_{T}} & -\theta_1 & i\theta_{B}e^{-i \chi_{B}}\\
-i r_{T} e^{-i \chi_{T}} & -i r_2 e^{-i \chi_2} & -i\theta_{B}e^{i \chi_{B}} & -\theta_2
\end{bmatrix}.
\]
Setting all parameters apart from $\theta_1$ to zero, the Gaussian unitary~\eqref{def:Gaussian_unitary} represents the one-mode phase-shift operator $\hat{R}_1(\theta_1)=\exp(-i\theta_1\hat{a}_1^\dag \hat{a}_1)$, and we write $S=R_1(\theta_1)$. Similarly, for $\theta_2$ we have $S=R_2(\theta_2)$. Setting all parameters apart from $\theta_{B}$ and $\chi_{B}$ to zero, we obtain the general mode-mixing channel $\hat{B}(\theta_{B},\chi_{B})=\exp(\theta_{B}(e^{i\chi_{B}}\hat{a}_1^\dag\hat{a}_2-e^{-i\chi_{B}}\hat{a}_2^\dag\hat{a}_1))$, where $\chi_{B}$ represents the angle of mode-mixing. For $\chi_{B}=0$ we obtain the usual beam-splitter with transmissivity $\tau=\cos^2\theta_{B}$, denoted $\hat{B}(\theta_{B})$. Following the same logic, parameters $r_1$ and $r_2$ represent the one-mode squeezing of the first and the second mode as defined in the previous section, denoted $\hat{S}_1(r_1,\chi_1)$, $\hat{S}_2(r_2,\chi_2)$, and parameter $r_{T}$ represents the two-mode squeezing at angle $\chi_{T}$, $\hat{S}_{T}(r_{T},\chi_{T})=\exp(-r_{T}(e^{i\chi_{T}}\hat{a}_1^\dag\hat{a}_2^\dag-e^{-i\chi_{T}}\hat{a}_1\hat{a}_2))$.

%multi-mode unitaries
%\subsubsection{Multi-mode unitary operators}
Multi-mode channels ($N\geq3$) can be obtained generalizing the same parametrization which has been used in~\eqref{eq:W2}. This essentially means there are not any other Gaussian unitary operators other than phase-changing, mode-mixing and single- and two-mode squeezing channels and their combinations. Number of parameters needed for fully parametrize a Gaussian unitary is $2N^2+N$ for the symplectic matrix and $2N$ for the displacement vector, thus $2N^2+3N$ in total.

\subsubsection{The phase-space representation of Gaussian unitaries}\label{sec:The_phase_space_Gaussian_unitaries}

%How to transform between different forms
Now we provide a list of the introduced Gaussian unitaries in both the complex or the real form matrices defined by Eq.~\eqref{def:covariance_matrix}, Eq.~\eqref{def:covariance_matrix_real} respectively. Symplectic matrices in most other commonly used notations are obtained by rearranging some rows and columns of either complex or the real form of the symplectic matrix. For example, the symplectic matrix in the real form~\eqref{def:covariance_matrix_real} given by `$xxpp$' vector transforms into `$xpxp$' form given by $\boldsymbol{\hat{Q}}:=(\hat{x}_1,\hat{p}_1,\hat{x}_2,\hat{p}_2)^T$ as
\[
S_\Re=\begin{bmatrix}
S_{x_1x_1} & S_{x_1x_2}  & S_{x_1p_1} & S_{x_1p_2} \\
S_{x_2x_1} & S_{x_2x_2} & S_{x_2p_1} & S_{x_2p_2} \\
S_{p_1x_1} & S_{p_1x_2} & S_{p_1p_1} & S_{p_1p_2} \\
S_{p_2x_1} & S_{p_2x_2} & S_{p_2p_1} & S_{p_2p_2}
\end{bmatrix}\ \longrightarrow\
S_{\Re,xpxp}=\begin{bmatrix}
S_{x_1x_1} & S_{x_1p_1}  & S_{x_1x_2} & S_{x_1p_2} \\
S_{p_1x_1} & S_{p_1p_1} & S_{p_1x_2} & S_{p_1p_2} \\
S_{x_2x_1} & S_{x_2p_1} & S_{x_2x_2} & S_{x_2p_2} \\
S_{p_2x_1} & S_{p_2p_1} & S_{p_2x_2} & S_{p_2p_2}
\end{bmatrix}
\]
In addition, it is often convenient to consider one-mode operations acting on a multi-mode state. One-mode operations which leave the other modes invariant are easily lifted into multi-mode operations by adding identities onto suitable places as illustrated on Eq.~\eqref{eq:phase_operator}.\\

%A list of Gaussian unitaries
\noindent
\emph{Rotation/phase-change} $\hat{R}(\theta)=\exp(-i\theta\hat{a}^\dag \hat{a})$, $\hat{R}_1(\theta)=\exp(-i\theta\hat{a}_1^\dag \hat{a}_1)$,
\begin{flalign}\label{eq:phase_operator}
R(\theta)&=\begin{bmatrix}
e^{-i\theta} & 0 \\
0 & e^{i\theta}
\end{bmatrix},
\quad R_\Re(\theta)=\begin{bmatrix}
\cos\theta & \sin\theta \\
-\sin\theta & \cos\theta
\end{bmatrix},&&\\
\quad R_1(\theta)&=\begin{bmatrix}
e^{-i\theta} & 0  & 0 & 0 \\
0 & 1 & 0 & 0 \\
0 & 0 & e^{i\theta} & 0 \\
0 & 0 & 0 & 1
\end{bmatrix},
\quad
R_{1\Re}(\theta)=\begin{bmatrix}
\cos\theta & 0  & \sin\theta & 0 \\
0 & 1 & 0 & 0 \\
-\sin\theta & 0 & \cos\theta & 0 \\
0 & 0 & 0 & 1
\end{bmatrix}.&&\nonumber
\end{flalign}
\emph{One-mode squeezing} $\hat{S}(r,\chi)=\exp(-\frac{r}{2}(e^{i\chi}\hat{a}^{\dag2}-e^{-i\chi}\hat{a}^{2}))$,
\begin{flalign}\label{eq:squeezing_operator}
S(r,\chi)&=\begin{bmatrix}
\cosh r & -e^{i\chi}\sinh r \\
-e^{-i\chi}\sinh r & \cosh r
\end{bmatrix},\\
\quad S_\Re(r,\chi)&=\begin{bmatrix}
\cosh r-\cos\chi\sinh r & -\sin\chi\sinh r \\
-\sin\chi\sinh r & \cosh r+\cos\chi\sinh r
\end{bmatrix}.&&
\end{flalign}
\emph{Mode-mixing} $\hat{B}(\theta,\chi)=\exp(\theta(e^{i\chi}\hat{a}_1^\dag\hat{a}_2-e^{-i\chi}\hat{a}_2^\dag\hat{a}_1))$,
\begin{flalign}\label{eq:mode_mixing_operator}
B(\theta,\chi)&=\begin{bmatrix}
\cos\theta & e^{i\chi}\sin\theta  & 0 & 0 \\
-e^{-i \chi}\sin\theta & \cos\theta & 0 & 0 \\
0 & 0 & \cos\theta & e^{-i \chi}\sin\theta \\
0 & 0 & -e^{i \chi}\sin\theta & \cos\theta
\end{bmatrix},&&\\
\quad B_\Re(\theta,\chi)&=\begin{bmatrix}
\cos\theta & \cos\chi\sin\theta & 0 & -\sin\chi\sin\theta \\
-\cos\chi\sin\theta & \cos\theta & -\sin\chi\sin\theta & 0 \\
0 & \sin\chi\sin\theta & \cos\theta & \cos\chi\sin\theta\\
\sin\chi\sin\theta & 0 & -\cos\chi\sin\theta & \cos\theta
\end{bmatrix}.&&\nonumber
\end{flalign}
\emph{Two-mode squeezing} $\hat{S}_{T}(r,\chi)=\exp(-r(e^{i\chi}\hat{a}_1^\dag\hat{a}_2^\dag-e^{-i\chi}\hat{a}_1\hat{a}_2))$,
\begin{flalign}\label{eq:twomode_squeezing_operator}
S_T(r,\chi)&=\begin{bmatrix}
\cosh r & 0  & 0 & -e^{i\chi}\sinh r \\
0 & \cosh r & -e^{i\chi}\sinh r & 0 \\
0 & -e^{-i\chi}\sinh r & \cosh r & 0 \\
-e^{-i\chi}\sinh r & 0 & 0 & \cosh r
\end{bmatrix},&&\\
S_{T\Re}(r,\chi)&=\begin{bmatrix}
\cosh r & -\cos \chi\sinh r  & 0 & -\sin \chi\sinh r \\
-\cos \chi\sinh r & \cosh r & -\sin \chi\sinh r & 0 \\
0 & -\sin \chi\sinh r & \cosh r & \cos \chi\sinh r \\
-\sin \chi\sinh r & 0 & \cos \chi\sinh r & \cosh r
\end{bmatrix}.&&\nonumber
\end{flalign}

\section{Common Gaussian states}\label{sec:common_Gaussian_states}
In this section we introduce the most common Gaussian states. As we will see in section~\eqref{sec:parametrization_of_Gaussian_states}, characteristics of all other Gaussian states are mixtures of characteristics of these basic ones. In that sense the following list is complete.

%thermal state
\subsection{Thermal state}\label{sec:thermal_state}
The simplest Gaussian state is the thermal state. Assuming the single particle Hilbert space is spanned by $N$ states -- modes, each mode is characterized by the energy $E_i$ of the state $\ket{\psi_i}$. We assume that each mode is thermally populated, i.e., number of particles in each mode is given by the thermal distribution, $\R_{{\mathrm{th}}i}=\frac{1}{Z}\mathrm{exp}(-\frac{E_i}{kT}\hat{n}_i)$, where $\hat{n}_i=\hat{a}_i^\dag\hat{a}_i$ denotes the number operator associated with mode $i$, $k$ is the Boltzmann constant, and $Z=\tr[e^{-\frac{E_i}{kT}\hat{n}_i}]$ defines the partition function.%~\cite{Weedbrook2012a}
The full thermal state is then a tensor product of the thermal states of each mode, $\R_{\mathrm{th}}=\hat{\rho}_{\mathrm{th}1}\otimes\cdots\otimes\hat{\rho}_{\mathrm{th}N}$. The displacement vector of the thermal state is equal to zero and the covariance matrix in both complex and the real form is a diagonal matrix,
\[
\bd=\boldsymbol{0}, \quad \sigma_{\mathrm{th}}=\mathrm{diag}(\lambda_1,\dots,\lambda_N,\lambda_1,\dots,\lambda_N).
\]
$\lambda_i=\coth(\frac{E_i}{2kT})$ are called symplectic eigenvalues for reasons described in the next section. They can be also expressed in terms of the mean number of thermal bosons, $\lambda_i=1+2n_{{\mathrm{th}}i}$, where $n_{{\mathrm{th}}i}:=\tr[\hat{n}_i\R_{\mathrm{th}}]$.
%Generally, each mode could have different temperature, $\lambda_i=\coth(\frac{\omega_i\hbar}{2kT_i})$.
Larger temperatures and smaller energies correspond to larger symplectic eigenvalues. For each $i$, $\lambda_i\geq1$ and $\lambda_i=1$ for $T=0$. Thermal state corresponding to $T=0$ is the lowest-energy state called vacuum and is described by the identity matrix $\sigma=I$.
%isothermal (isotropic) states?

%coherent state
\subsection{Coherent state}
A Gaussian state characterized only by its displacement vector is the coherent state,
\[
\ket{\A}=e^{-\frac{\abs{\A}^2}{2}}\sum_{n=0}^\infty \frac{\A^n}{\sqrt{n!}}\ket{n}.
\]
Coherent state is an eigenvector of the annihilation operator, $a\ket{\A}=\A\ket{\A}$. Coherent states typically describe beams of light emitted by a laser~\cite{zhang1990coherent}. Mathematically, coherent state can be created by the action of the Weyl displacement operator~\eqref{eq:Weyl_displacement_operator} on the vacuum (thus an equivalent name would be a single-mode displaced vacuum), $\ket{\A}=\hat{D}(\A)\ket{0}$. The first and the second moments can be easily derived using Eq.~\eqref{def:transformation},
\[
\bd=(\A,\ov{\A})^T, \quad \sigma=I.
\]

%squeezed state
\subsection{Single-mode squeezed state}
Squeezed state is created by an action of the squeezing operator~\eqref{eq:squeezing_operator} on the vacuum, $\ket{S(r,\chi)}=S(r,\chi)\ket{0}$. For $\chi=0$ this state takes the form~\cite{kok2010introduction}
\[
\ket{S(r)}=\frac{1}{\sqrt{\cosh\abs{r}}}\sum_{n=0}^\infty\frac{\sqrt{(2n)!}}{n!}\left(\frac{-r}{2\abs{r}}\right)^n\tanh^n\abs{r}\ket{2n}
\]
Such states for example from a laser light by going through an optical parametric oscillator~\cite{breitenbach1997measurement,Lvovsky2014squeezed}.
The first and the second moments are
\[
\bd=\boldsymbol{0}, \quad \sigma=S(r,\chi)S^\dag(r,\chi)=S(2r,\chi).
\]

%two-mode squeezed state
\subsection{Two-mode squeezed state}
Two-mode squeezed states are entangled two-mode states created by an action of the two mode squeezing operator~\eqref{eq:twomode_squeezing_operator} on the vacuum, $\ket{S_T(r,\chi)}=S_T(r,\chi)\ket{0}$. For $\chi=0$ this state takes the form~\cite{kok2010introduction}
\[
\ket{S_T(r)}=\frac{1}{\cosh\abs{r}}\sum_{n=0}^\infty\left(\frac{-r}{\abs{r}}\right)^n\tanh^n\abs{r}\ket{n,n}
\]
Physically, two-mode squeezed states are prepared by sending squeezed and anti-squeezed state (squeezed with the negative squeezing) through a beam-splitter. The first and the second moments are
\[
\bd=\boldsymbol{0}, \quad \sigma=S_T(r,\chi)S_T^\dag(r,\chi)=S_T(2r,\chi).
\]

\begin{example}
\emph{
It is easy to show that tracing over one-mode of a two-mode squeezed state $\ket{S_T(r)}$ leaves us with a thermal state.
}
\end{example}

\section{Number of particles in a Gaussian state}
For some applications it is useful to know the mean number of particles in a Gaussian state or the mean energy of a Gaussian state. For example, in quantum metrology we are usually interested how well the sensitivity of a Gaussian probe state scales with its energy. Calculating this quantity is very simple when using the complex form of the covariance matrix. Defining the mean number of particles in mode $i$, $0\leq i\leq N$, as $n_i:=\tr[\hat{a}_i^\dag \hat{a}_i\R]$ we can use the definition of the covariance matrix~\eqref{def:covariance_matrix} and the commutation relations~\eqref{def:commutation_relation} to derive
\[
n_i=\frac{1}{2}\big(\sigma^{ii}+2\ov{d}^id^i-1\big).
\]
The mean energy of the probe state is then $\mean{E}=\sum_{i=1}^Nn_iE_i$ where $E_i$ is the energy of a particle in mode $i$. The mean number of particles in a Gaussian state can be calculated as
\[\label{eq:mean_number_of_particles}
n:=\sum_{i=1}^Nn_i=\frac{1}{2}\left(\frac{1}{2}\tr[\sigma]+\bd^\dag \bd-N\right).
\]

\section{Williamson's decomposition of the covariance matrix}\label{sec:Williamson's_decomposition}
%in the previous section we saw that we can construct covariance matrix by multiplying symplectic matrices. The opposite is also true, the covariance matrix can be always decomposed using symplectic matrices.
%Symplectic eigenvalues for one and two-mode state

%Williamson's theorem
In the section~\ref{sec:common_Gaussian_states} we have illustrated that covariance matrices can be constructed by applying symplectic matrices on diagonal matrix. In this section will show that the opposite is also true. We introduce a theorem which is crucial for understanding structure of Gaussian states, and which will be later used for a parametrization of Gaussian states. This is the Williamson's decomposition of the positive definite matrices. According to the Williamson's theorem~\cite{Williamson1936a,deGosson2006a,Simon1998a}, any positive-definite matrix can be diagonalized by symplectic matrices of the form introduced in Eq.~\eqref{def:structure_of_S},
\[\label{def:Williamson's_decomposition}
\sigma=SDS^\dag.
\]
$D$ is the diagonal matrix consisting of \emph{symplectic eigenvalues},\\
$D=\mathrm{diag}(\lambda_1,\dots,\lambda_N,\lambda_1,\dots,\lambda_N)$.

%calculating symplectic eigenvalues and symplectic matrix
%\subsection{Calculating symplectic eigenvalues and the symplectic matrix}
Symplectic eigenvalues can be found by solving the usual eigenvalue problem for the matrix
\[\label{def:A}
A:=K\sigma,
\]
where $K$ is the symplectic form defined by commutation relations~\eqref{def:commutation_relation}. Eigenvalues of $A$ always appear in pairs. If $\lambda_i$ is an eigenvalue of $A$, then also $-\lambda_i$ is an eigenvalue of the same operator. The symplectic spectrum is then defined as a collection of the positive eigenvalues of $A$. In other words, $\lambda_i$ is a symplectic eigenvalue of $\sigma$ if and only if it is positive and $\pm\lambda_i$ are the eigenvalues of the operator $A$. Combining Eqs.~\eqref{def:structure_of_S} and~\eqref{def:Williamson's_decomposition} we find $\tr[A]=\tr[A^3]=0$. This together with the expansion of determinant gives analytical formulae for the symplectic eigenvalues of a single mode Gaussian state,
\[\label{eq:singlemode_symplectic_eigenvalue}
\lambda=\sqrt{\mathrm{det}(A)}=\frac{1}{\sqrt{2}}\sqrt{\tr[A^2]},
\]
and of a two-mode Gaussian state,
\[\label{eq:twomode_symplectic_eigenvalues}
\begin{split}
\lambda_{1,2}&=\frac{1}{2}\sqrt{\tr[A^2]\pm\sqrt{(\tr[A^2])^2-16\mathrm{det}(A)}}\\
&=\frac{1}{2}\sqrt{\tr[A^2]\pm\sqrt{4\tr[A^4]-(\tr[A^2])^2}}.
\end{split}
\]
Diagonalizing symplectic matrices $S$ can be found for example by a method described in~\cite{Simon1998a}.

%connection with examples and purity of a Gaussian state
%\subsection{Symplectic eigenvalue and purity of a Gaussian state}
We have seen in the previous section that the thermal state was represtented only by symplectic eigenvalues. These eigenvalues were connected with purity of the state. Temperature equal to zero -- symplectic eigenvalues equal to one -- results in vacuum, which is a pure state. Higher temperature lead to a mixed state. But that is true not only for a thermal state, but any Gaussian state. Every symplectic eigenvalue of a Gaussian state is larger than one, $\lambda_i\geq1$, which is a consequence of Eq.~\eqref{eq:sigma_K_positivity}. Purity of a Gaussian state can be calculated as
\[
\mu(\hat{\rho})=\prod_{i=1}^{N}\lambda_{i}^{-1}=\frac{1}{\sqrt{\mathrm{det}(A)}}.
\]
A Gaussian state is pure if all symplectic eigenvalues are equal to one. We say that mode $i$ is pure if $\lambda_i=1$.

\section{Parametrization of Gaussian states}\label{sec:parametrization_of_Gaussian_states}

%Euler's decomposition
%\subsection{Euler's decomposition of the symplectic matrix}
In this section we will use the Williamson's decomposition to fully parametrize Gaussian states of a given number of modes. But to do that, we need to fully parametetrize symplectic matrices first. Any symplectic matrix~\eqref{def:structure_of_S} can be decomposed using Euler's decomposition~\cite{Arvind1995a,Weedbrook2012a} as
\[\label{def:S_decomposition}
S=
\begin{bmatrix}
U_1 & 0 \\
0 & \ov{U}_1
\end{bmatrix}
\begin{bmatrix}
\cosh{M_{\boldsymbol{r}}} & -\sinh{M_{\boldsymbol{r}}} \\
-\sinh{M_{\boldsymbol{r}}} & \cosh{M_{\boldsymbol{r}}}
\end{bmatrix}
\begin{bmatrix}
U_2 & 0 \\
0 & \ov{U}_2
\end{bmatrix},
\]
where $U_1$ and $U_2$ denote unitary matrices, and $M_{\boldsymbol{r}}=\mathrm{diag}(r_1,\dots,r_N)$ is the diagonal matrix of the squeezing parameters. This shows that any symplectic matrix can be decomposed into two passive operations and one active, which is consisted of single mode squeezers. This is important from an experimental point of view because that means there does not need to be any direct two mode squeezing operation as long as there are single mode squeezers and beam splitters.

%parametrization of Gaussian states
With a full parametrization of unitary matrices $U_1$ and $U_2$, one can use this decomposition to fully parametrize the covariance matrix via Eq.~\eqref{def:Williamson's_decomposition}. Moreover, since the displacement vector is fully parametrized by its elements, we have a full parametrization of Gaussian states. Note, however, that some parameters may not add any additional complexity and can be removed. This is a consequence of the fact that in Eq.~\eqref{def:Williamson's_decomposition} some parts of (the decomposition of) $U_2$ vanish, because they commute with the diagonal matrix $\mathrm{diag}(\lambda_1,\dots,\lambda_N)$. Since the parametrizations of unitary matrices up to $N=3$ are known, we can explicitly write the most general single-, two-, and three-mode Gaussian states.

%single mode
%\subsection{General one-mode Gaussian state}
The most general one-mode Gaussian state is the one-mode squeezed rotated displaced thermal state~\cite{Weedbrook2012a},
\[\label{eq:general_1mode_state}
\hat{\rho}_0=\hat{D}(\tilde{\bg})\hat{R}(\theta)\hat{S}(r)\rho_{\mathrm{th}}(\lambda)\hat{S}^\dag(r)\hat{R}^\dag(\theta)\hat{D}^\dag(\tilde{\bg}),
\]
where the variable in the Weyl displacement operator $\hat{D}(\tilde{\bg})$ is of the form $\tilde{\bg}=\norm{{d}}e^{i\phi_d}$. The first and the second moments of this state are
\[
\bd=(\tilde{\bg},\ov{\tilde{\bg}})^T,\quad \sigma=R(\theta)S(r)D(\lambda)S(r)^\dag R^\dag(\theta),
\]
where $D(\lambda)=\mathrm{diag}(\lambda,\lambda)$.

%two mode
%\subsection{General two-mode Gaussian state}
Applying the parametrization of the general $2\times2$ unitary matrix to Eq.~\eqref{def:S_decomposition} we find the most general two-mode Gaussian state,
\[\label{eq:general_2mode_state}
\begin{split}
\hat{\rho}_0=&\hat{D}(\tilde{\bg})\hat{R}_1(\phi_1)\hat{R}_2(\phi_2)\hat{B}(\theta_2)\hat{R}_{\mathrm{as}}(\psi_2)\hat{S}_1(r_1)\hat{S}_2(r_2)\\
&\hat{R}_{\mathrm{as}}(\psi_1)\hat{B}(\theta_1)\hat{\rho}_{\mathrm{th}}(\lambda_1,\lambda_2)(\ \cdot\ )^\dag,
\end{split}
\]
where we define $\hat{R}_{\mathrm{as}}(\psi):=\hat{R}_1(\psi)\hat{R}_2(-\psi)$ and $\tilde{\bg}=(\norm{{d}_1}e^{i\phi_{d1}},\norm{{d}_2}e^{i\phi_{d2}})$. The displacement vector and the covariance matrix are obtained in analogy to the single mode state by removing `hats' while the displacement operator affects only the displacement vector, $\bd_0=(\tilde{\bg},\ov{\tilde{\bg}})^T$.

%three mode
%\subsection{General three-mode Gaussian state}
A general $3\times3$ unitary matrix can be fully parametrized using the $C\!K\!M$ matrix (Cabibbo--Kobayashi--Maskawa~\cite{chau1984comments}). Assuming $\hat{B}_{ij}(\theta,\chi)$ is the mode-mixing operation between modes $i$ and $j$ (3-mode generalizations of Eq.~\eqref{eq:mode_mixing_operator}), $\hat{B}_{ij}(\theta):=\hat{B}_{ij}(\theta,0)$ the beam-splitter operation respectively, we can define $C\!K\!M$ operator as
\[
\hat{C\!K\!M}(\theta_1,\theta_2,\theta_3,\chi_1):=\hat{B}_{23}(\theta_1)\hat{B}_{13}(\theta_2,\chi_1)\hat{B}_{12}(\theta_3).
\]
We also denote a collection of single mode rotations and a collection single mode squeezers as
\begin{subequations}
\begin{align}
\hat{R}(\phi_1,\phi_2,\phi_3)&:=\hat{R}_1(\phi_1)\hat{R}_2(\phi_2)\hat{R}_3(\phi_3),\\ \hat{S}(r_1,r_2,r_3)&:=\hat{S}_1(r_1)\hat{S}_2(r_2)\hat{S}_3(r_3).
\end{align}
\end{subequations}
The most general three-mode Gaussian state is
\[
\begin{split}
\hat{\rho}_0=&\hat{D}(\tilde{\bg})\hat{R}(1,\phi_1,\phi_2)\hat{C\!K\!M}(\theta_1,\theta_2,\theta_3,\chi_1)\hat{R}(\phi_3,\phi_4,\phi_5)\hat{S}(r_1,r_2,r_3)\\
&\hat{R}(1,\phi_6,\phi_7)\hat{C\!K\!M}(\theta_4,\theta_5,\theta_6,\chi_2)\R_{\mathrm{th}}(\lambda_1,\lambda_2,\lambda_3)(\ \cdot\ )^\dag,
\end{split}
\]
where $\tilde{\bg}=(\norm{{d}_1}e^{i\phi_{d1}},\norm{{d}_2}e^{i\phi_{d2}},\norm{{d}_3}e^{i\phi_{d3}})$.

%Number of parameters
The number of parameters $\#(N)$ needed to fully parametrize $N$-mode Gaussian states is $5$ for a one-mode state, $14$ for a two-mode state, and $27$ for a three-mode state. In general the following formula holds,
\[\label{eq:number_of_parameters}
\#(N)=2N^2+3N.
\]
Interestingly, this means that number of parameters needed to fully parametrize a Gaussian state is the same as the number of parameters needed to fully parametrize a Gaussian unitary~\eqref{def:Gaussian_unitary}. We can prove this expression by studying properties of the displacement and the covariance matrix~\eqref{def:first_and_second_moments}. Because the covariance matrix is a Hermitian matrix its sub-block $X$ is also a Hermitian matrix and its sub-block $Y$ is ad (complex) symmetric matrix. But Hermitian matrices of size $N\times N$ are fully parametrized by $N^2$ parameters and symmetric matrices are fully parametrized by $N^2+N$ parameters, i.e., the covariance matrix is fully parametrized by $2N^2+N$ parameters. The displacement vector is parametetrized by $N$ absolute values of the displacement and $N$ phases. Summed up, this gives Eq.~\eqref{eq:number_of_parameters}.

Pure Gaussian states are characterized by a significantly smaller number of parameters,
\[\label{eq:number_of_parameters_pure}
\#_{\mathrm{pure}}(N)=N^2+3N.
\]
This comes from the the Euler's decomposition~\eqref{def:S_decomposition} of the symplectic matrix and the Williamson's decomposition of the covariance matrix~\eqref{def:Williamson's_decomposition}. In case of pure states all symplectic eigenvalues are equal to one, and the unitary matrix $U_2$ commutes with the diagonal matrix representing the vacuum state. Therefore, we have $N^2$ parameters needed to parametrize the unitary matrix $U_1$, $N$ squeezing parameters, and $2N$ parameters of the displacement vector. Summed up, this gives Eq.~\eqref{eq:number_of_parameters_pure}.

\section{State-of-the-art quantum metrology in the phase-space formalism}\label{sec:state_quantum_metrology}

%introduction
In this section we review state-of-the-art methods of quantum metrology in the phase-space formalism. %For our contribution see chapter~\Ref{my quantum metrology}.

In the first chapter we introduced several formulae for the quantum Fisher information. However, expressions introduced there were only for states represented by a density matrix. On the other hand, as we illustrated in section~\ref{sec:common_Gaussian_states}, density matrices of Gaussian states can be usually expressed only in terms of relatively complicated infinite series. This is why calculating the quantum Fisher information -- the figure of merit of the local quantum estimation -- has been quite a difficult task for Gaussian states until recently. This has changed when new expressions using the phase-space formalism have been derived.

The first leap in deriving general formulae has been taken by Pinel et al.~\cite{Pinel2012a}, who found an expression for the quantum Fisher information for pure states, i.e., for the states which are pure at point $\epsilon$ and remain pure even if the $\epsilon$ slightly changes. The same year Marian and Marian found the formula for the fidelity between one-mode and two-mode Gaussian states~\cite{Marian2012a}, which allowed for the derivation of the general formula for the one-mode state~\cite{Pinel2013b}. Also, Spedalieri et al.~found a formula for the fidelity between one pure and one mixed Gaussian state~\cite{Spedalieri2013a}, from which one can derive a slightly more general formula for pure states, i.e., for the states which are pure at the point $\epsilon$ but the small change in $\epsilon$ introduces impurity. A different path has been followed by Monras~\cite{Monras2013a}, who connected the quantum Fisher information to the solution of the so-called Stein equation. Using this approach, he derived the quantum Fisher information for a generalization of the pure states called iso-thermal states, and a general formula for any multi-mode Gaussian state in terms of an infinite series. Using the previous result, Jiang derived a formula~\cite{Jiang2014a} for the Gaussian states in exponential form and simplified a known formula for pure states. Quite recently, Gao and Lee derived an exact formula~\cite{Gao2014a} for the quantum Fisher information for the multi-mode Gaussian states in terms of the inverse of certain tensor products, elegantly generalizing the previous results, however with some possible drawbacks, especially in the necessity of inverting relatively large matrices. The last result from Banchi et al.~\cite{Banchi2015a} provides a very elegant expression for the quantum Fisher information for multi-mode Gaussian states written in terms of inverses of certain super-operators.

The original results has been been published in many different notations. We translate all of them into the complex form, although we mention the real form version in some examples. In the following, we will give details on these results which relate to our work introduced later in the thesis.

%single mode QFI
%\subsection{The quantum Fisher information for single-mode states}
The simplest case of a Gaussian state is a single mode Gaussian state. Making the identification $\hat{\rho}_{1}\rightarrow(\boldsymbol{d}_{1},\sigma_{1})$ the Uhlmann fidelity between two one-mode states~\cite{Marian2012a} is given by
\[
\mathcal{F}_{1}(\hat{\rho}_{1},\hat{\rho}_{2})\,=\,2\,\frac{e^{-(\boldsymbol{d}_1-\boldsymbol{d}_2)^{\dag}(\sigma_{1}+\sigma_{2})^{-1}(\boldsymbol{d}_1-\boldsymbol{d}_2)}}{\sqrt{\Delta+\Lambda}-\sqrt{\Lambda}},
\]
where $\Delta=\det{{\sigma}_{1}+{\sigma}_{2}}$, $\Lambda=\det{{\sigma}_{1}+{K}}\det{{\sigma}_{2}+{K}}$, and $\det{\cdot}:=\mathrm{det}[\cdot]$ denotes determinant. One can use this formula and the connection between the Uhlmann fidelity and the Quantum Fisher information~\eqref{QFI_using_fidelity}, expand the determinants in the small parameter $\de$, and derive the quantum Fisher information for a single mode state~\cite{Pinel2013b},%\footnote{Note that for $\det{A}=1$, i.e., when the state is pure, the second term is undefined and should not be accounted for.}
\[\label{eq:one_mode_quantum_fisher_information}
H(\epsilon)=\frac{1}{2}\frac{\mathrm{tr}\Big[\big(A^{-1}\dot{A}\big)^{2}\Big]}{1+\det{A}^{-1}}
+\frac{1}{2}\,\frac{\det{A}^{-1}\mathrm{tr}[A^{-1}{\dot{A}}]^{2}}{1-\det{A}^{-2}}+2\dot{\bd}^\dag\sigma^{-1}\dot{\bd},
\]
where $A:=K\sigma$ for the complex form.

%pure states/isothermal=isotropic states
%\subsection{Pure states}
A very elegant expression for the quantum Fisher information can be derived for pure states. Taking a different approach for finding this quantity -- solving equations for the symmetric logarithmic derivative~\eqref{def:SLDsolution} -- has been taken in~\cite{Monras2013a}. This equation translate into the Stein equation\footnote{Stein equation for $X$, $X-F X F^\dag=W$, is a discrete-time Lyapunov equation~\cite{Bhatia2006a}.} which can be solved in terms of infinite series~\cite{Bhatia2006a}. This series has been evaluated for \emph{isothermal} (also called \emph{isotropic}) which are defined as states with all eigenvalues being equal, $\lambda_1=\cdots=\lambda_N=\lambda$. The quantum Fisher information reads
\[\label{eq:nu_pure}
H(\epsilon)=\frac{\lambda^{2}}{2(1+\lambda^{2})}\mathrm{tr}\big[(A^{-1}\dot{A})^2\big]+2\dot{\bd}^\dag\sigma^{-1}\dot{\bd}.
\]
As noted in~\cite{Jiang2014a}, using $\sigma^{-1}=\frac{1}{\lambda^2}K\sigma K$ this the expression can be further simplified,
\[\label{eq:nu_pure_simplified}
H(\epsilon)=-\frac{1}{2(1+\lambda^{2})}\mathrm{tr}\big[\dot{A}^2\big]+\frac{2}{\lambda^2}\dot{\bd}^\dag A K\dot{\bd}.
\]
For pure states we take $\lambda=1$.

%Note however, that these two formulae hold only when small variation in $\epsilon$ does not change the purity of the state, i.e., only the case when $\lambda$ is independent of $\epsilon$ up to the second order in the small parameter $\de$, $\lambda(\epsilon+\de)=\lambda(\epsilon)+\mathcal{O}(\de^3)$. More general case can be derived using the Uhlmann fidelity between one mixed and one pure state~\cite{Spedalieri2013a}. We are going to address this issue later in section~\ref{sec:pure_states_QM}.

%Limit formula
%\subsection{Limit expression for multi-mode states}
For some applications, an exact expression for the quantum Fisher information is not necessary. It can be easier to numerically obtain an approximate value of this quantity. The same method used to find the expression for pure states can be also used for general mixed states, however, in terms of an infinite sum. The quantum Fisher information for any number of modes reads
\[\label{eq:Monras_QFI}
H(\epsilon)=\frac{1}{2}\tr\big[\dot{\sigma}Y\big]+2\dot{\bd}^\dag\sigma^{-1}\dot{\bd},
\]
where $Y=-\sum_{n=0}^\infty (K\sigma)^{-n}\dot{(\sigma^{-1})}(\sigma K)^{-n}$. The limit converges if and only if all symplectic eigenvalues are larger than one, i.e., when all modes are mixed.

%Multi-parameter multi-mode mixed-states formula
%\subsection{Multi-parameter estimation of multi-mode mixed states}
A similar method of solving the equation for the symmetric logarithmic derivative has been used~\cite{Gao2014a} to derive an exact formula for estimation multiple parameters for multi-mode mixed Gaussian states. This also generalizes the single parameter results of~\cite{Monras2013a}. The original formula is written in terms of tensor elements. However, we notice the result can be expressed in an elegant matrix form. The quantum Fisher information matrix for Gaussian state $(\bd,\sigma)$ can be calculated as
\begin{subequations}\label{eq:mixed_QFI}
\begin{align}
H^{ij}(\be)&=\frac{1}{2}\vectorization{\partial_i\sigma}^\dag\mathfrak{M}^{-1}\vectorization{\partial_j\sigma}+2\partial_i\bd^\dag\sigma^{-1}\partial_j\bd,\\
\mathfrak{M}&=\ov{\sigma}\otimes\sigma-K\otimes K,
\end{align}
\end{subequations}
where $\otimes$ denotes the Kronecker product, $\vectorization{\cdot}$ is a vectorization of a matrix, and $\partial_i\equiv\partial_{\epsilon_i}$. Again, this formula holds only for states for which all symplectic eigenvalues are larger than one. The symmetric logarithmic derivative reads
\[
\mathcal{L}_i=\Delta \bA^\dag\mathcal{A}_i\Delta \bA-\frac{1}{2}\tr[\sigma\mathcal{A}_i]+2\Delta\boldsymbol{A}^\dag\sigma^{-1}\partial_i{\boldsymbol{d}},
\]
where $\Delta \bA:=\bA-\bd$, $\vectorization{\mathcal{A}_i}:=\mathfrak{M}^{-1}\vectorization{\partial_i\sigma}$. The quantum Fisher information matrix is then defined as $H^{ij}(\boldsymbol{\epsilon})=\frac{1}{2}\tr[\hat{\rho}\{\mathcal{L}_i,\mathcal{L}_j\}]$. For the full derivation of the above matrix formulae and the real form version see appendix~\ref{app:mixed_state}. Note that although the above multi-mode formula encompasses all previous formulae, it may be harder to use. For example, calculating the quantum Fisher information of a single-mode state with Eq.~\eqref{eq:mixed_QFI} requires inverting $4\times4$ matrix $\mathfrak{M}$, while Eq.~\eqref{eq:one_mode_quantum_fisher_information} only requires inverting $2\times2$ matrix $A$.

%Braunstein multi-mode formula
The last formula we will present here is again the expression for multi-mode mixed Gaussian state~\cite{Banchi2015a}. Defining a super-operator $S_Y(X):=YXY$, the quantum Fisher information matrix reads,
\[
H^{ij}(\be)=\frac{1}{2}\tr\big[\partial_i\sigma(S_\sigma-S_K)^{-1}\partial_j\sigma\big]+2\partial_i\bd^\dag\sigma^{-1}\partial_j\bd.
\]
Despite a very elegant form this expression seems slightly impractical for actual mathematical calculations. This is because the task of inverting the super-operator $S_\sigma-S_K$ leads to the same task as before -- solving the Stein equation.

All formulae for mixed states introduced here suffer of the same problem - they cannot be applied to states which have at least one symplectic eigenvalue equal to one.\footnote{For example, this procedure sets the term $\frac{1}{2}\,\frac{\det{A}^{-1}\mathrm{tr}[A^{-1}{\dot{A}}]^{2}}{1-\det{A}^{-2}} $ in Eq.~\eqref{eq:one_mode_quantum_fisher_information} that is undefined for $\det{A}=1$ to zero.} These are exactly cases where the continuous quantum Fisher information and the quantum Fisher information might not coincide, as shown by Eq.~\eqref{eq:connection_between_Hc_and_H}. It turns out that in cases where this happens the solution is to use the regularization procedure~\eqref{eq:regularization_procedure}. We discuss this problem in sections~\ref{sec:when_Williamson's_decomposition} and~\ref{sec:problems_at_pops}. Then, in analogy of the expression for a single mode Gaussian state, we derive the quantum Fisher information two-mode Gaussian states. We also simplify the limit formula~\eqref{eq:Monras_QFI} and provide an estimate of the remainder of the series. Finally, we derive an elegant and useful expression for the quantum Fisher information for the case when the symplectic decomposition of the covariance matrix is known.

%%%%%%%%%%%%%%%%%%%%%%%%%%%%%%
\chapter{Operations in quantum field theory and state-of-the-art in estimating space-time parameters}\label{chap:operations_in_QFT}
%%%%%%%%%%%%%%%%%%%%%%%%%%%%%%

With the enormous success of quantum theory the question arose how to combine this theory with special and general relativity and whether such theory is even possible. The first attempts were performed by Klein~\cite{klein1926quantentheorie} and Gordon~\cite{gordon1926comptoneffekt} who came up with an idea of deriving an equation of motion in a similar way to the Schr\"odinger equation -- simply by exchanging energy and momenta for its respective operators in the energy-momentum relation. This led to the Klein-Gordon equation which we now use to describe scalar fields of spinless particles. After numerous interpretational problems -- especially with the notion of particle -- quantum field theory was born. One of the most precise theories we have today successfully predicted and confirmed anomalous magnetic dipole moments, hyperfine splitting of energy levels of a hydrogen atom, and the quantum Hall effect. Assuming that the space is not necessarily flat has led to further generalization of the theory called quantum field theory in curved space-time. This theory attempts to describe quantum fields in large velocities and accelerations and on scales where gravity plays a role. The famous predictions of this theory are: Hawking radiation~\cite{Hawking1974a} which says that particles can escape an enormous black hole potential behind the Schwarzchild horizon, the Unruh effect~\cite{Unruh1979a} which illustrates that an accelerating observer sees more particles than an inertial observer, and the dynamical Casimir effect~\cite{moore1970quantum} which shows particles can be created between two moving mirrors. Predictions of this theory, so far, have only been confirmed in analogue systems~\cite{Wilson2011a}. Despite the practical success of this theory, it is not believed to be the final theory. This is simply because the theory attempts to describe quantum fields propagating on a fixed space-time. But gravity, which gives rise to the space-time, is itself provided by other quantized fields. These quantum properties of gravity are expected to have observable effects on either very small scales or in high energies and will be described by a future theory of quantum gravity.

An excellent although quite concise text on quantum field theory in curved space-time has been written by Birrel and Davies~\cite{Birrell1984a}. A more mathematical approach can be found in~\cite{ashtekar1975quantum} and a more pedagogical approach in~\cite{wald1994quantum}.

This chapter is organized as follows: we first summarize the quantization of the Klein-Gordon field while omitting mathematical technicalities that can be found in the references above. Then we introduce Bogoliubov transformations which can describe how different observers perceive the field and how the field evolves. However, these transformations are not suitable for the description of continuous evolution. For that reason we follow on~\cite{Bruschi2013b,Bruschi2013a} and show how the equations of motion for continuous transformations are constructed. Such equations are usually difficult to solve exactly and perturbation methods need to be used. Previous works considered only the first order correction to the solution of continuous Bogoliubov transformations in the small parameter of interest. We derive a general prescription on how to calculate these coefficients to any order which can later be used for more precise approximation of the quantum Fisher information. Finally, we overview the current state-of-the-art of quantum metrology applied in the estimation of space-time parameters.

\section{Quantization of the Klein-Gordon field}

%geometry
The space-time in general relativity is described by a smooth manifold equipped with patches of \emph{local coordinates}. Put simply, manifolds are objects which when viewed from a sufficiently small region resemble the flat space. Local coordinates are then a mathematical description of how an observer measures space and time in this sufficiently small region. In these local coordinates $x^\mu=(t,\boldsymbol{x})$, where $t$ represents time and $\boldsymbol{x}$ position, we define a line element
\[\label{def:metric}
\mathrm{d}s^2=g^{\mu\nu}\mathrm{d}x_\mu \mathrm{d}x_\nu,
\]
where $g^{\mu\nu}$ are the elements of the metric tensor and $\mu,\nu=0,1,2,3$.

The simplest example of particles living on the manifold are spin-0 particles which can be described by either real or complex scalar field $\phi$. The massless scalar field $\phi$ obeys the Klein-Gordon equation~\cite{Birrell1984a},
\[\label{eq:klein_gordon_equation}
\nabla^{\mu}\nabla_{\mu}\,\phi(t,\boldsymbol{x})=0.
\]
The operator $\nabla_{\mu}$ is the covariant derivative defined with respect to the metric tensor $\boldsymbol{g}$ in local coordinates $(t,\boldsymbol{x})$.

%field operator, quantization
When space-time $\boldsymbol{g}$ admits global or asymptotic time-like killing vector field, it is possible to quantize the field. This Killing vector field then splits the set of linearly independent solutions -- modes -- of Klein-Gordon Eq.~\eqref{eq:klein_gordon_equation} to either positive frequency modes $u_{\boldsymbol{k}}$ or negative frequency modes $\ov{u}_{\boldsymbol{k}}$. Because Eq.~\eqref{eq:klein_gordon_equation} is a linear equation, the full solution is a linear combination of these positive and negative frequency modes,
\[\label{eq:general_solution_to_KG}
\phi(t,\boldsymbol{x})=\int\!\!\!\mathrm{d}\boldsymbol{k}\ \ a_{\boldsymbol{k}}u_{\boldsymbol{k}}(t,\boldsymbol{x})+\ov{a}_{\boldsymbol{k}}\ov{u}_{\boldsymbol{k}}(t,\boldsymbol{x}).
\]
Following the standard quantization procedure~\cite{Birrell1984a}, the coefficients $a_{\boldsymbol{k}}$, $\ov{a}_{\boldsymbol{k}}$ are lifted into the annihilation and creation operators $a_{\boldsymbol{k}}\rightarrow\hat{a}_{\boldsymbol{k}}$, $\ov{a}_{\boldsymbol{k}}\rightarrow\hat{a}_{\boldsymbol{k}}^\dag$. In simplified terms, mode $u_{\boldsymbol{k}}$ is usually associated with a particle, the creation operator $\hat{a}_{\boldsymbol{k}}^\dag$ creates this particle, and the annihilation operator $\hat{a}_{\boldsymbol{k}}$ annihilates this particle. However, it is important to point out that this view is very tricky and in general the notion of particle in quantum field theory is still unsettled~\cite{Birrell1984a}.

We will restrict ourselves to the real Klein-Gordon field quantized in a finite space (for example a box or a cavity) in $1+1$ dimensions. In that scenario the spectrum of modes is discrete and the field operator constructed from Eq.~\eqref{eq:general_solution_to_KG} takes the form of an infinite sum
\[\label{eq:field_expansion}
\hat{\phi}(t,\boldsymbol{x})=\sum_n \hat{a}_n u_n(t,\boldsymbol{x})+\hat{a}_n^\dag \ov{u}_n(t,\boldsymbol{x}).
\]
$n$ labels the mode and annihilation and creation operators satisfy the same commutation relations as introduced in Eq.~\eqref{def:commutation_relation}. The Fock space describing the system of relativistic bosons is constructed identically to the non-relativistic case from section~\ref{sec:Fock_space}.

\section{Bogoliubov transformation}

%different set of modes
The field expansion~\eqref{eq:field_expansion} is not unique. It can be written in different basis of solutions to the Klein-Gordon equation denoted $\{v_n\}_n$. We collect both sets of modes into a (possibly infinite) vectors $\boldsymbol{u}=(u_1,u_2,\dots)$, $\boldsymbol{v}=(v_1,v_2,\dots)$. Solutions to the Klein-Gordon equation form a linear vector space. Therefore every solution $u_n$ can be expressed as a linear combination of modes $v_n$, $\ov{v}_n$,
\[\label{eq:bogoliubov_transformation_definition}
\begin{bmatrix}
\boldsymbol{u} \\ \overline{\boldsymbol{u}}
\end{bmatrix}=
S
\begin{bmatrix}
\boldsymbol{v} \\ \overline{\boldsymbol{v}}
\end{bmatrix},
\]
where matrix
\[\label{def:Bogoliubov_S}
S=
\begin{bmatrix}
\A & \B \\
\overline{\B} & \overline{\A}
\end{bmatrix}
\]
is called the Bogoliubov transformation and $\A$ and $\B$ are the Bogoliubov coefficients.

%use of Bogoliubov transformations
Bogoliubov transformations are important in the quantum field theory in curved space-time, because they relate solutions to the Klein-Gordon equation written in one local coordinates to a different local coordinates. Some problems can be very hard to solve in certain coordinates, while they can be relatively easy in specially picked ones. Bogoliubov transformation can be used, for example, to derive many results of quantum field theory in curved space-time such as black-hole evaporation~\cite{Hawking1974a,hawking1975particle,fabbri2005modeling}, the Unruh effect~\cite{takagi1986vacuum,Unruh1979a}, and the dynamical Casimir effect~\cite{moore1970quantum}. They also used to model the time evolution of quantum states, and quantum states from the point of view of different observers.

%deriving Bogoliubov identities
The field operator~\eqref{eq:field_expansion} can be written in two different vector forms,
\[\label{eq:two_field_expansion}
\hat{\phi}(t,x)=
\begin{bmatrix}
\ov{\boldsymbol{u}} \\ \boldsymbol{u}
\end{bmatrix}^\dag
\cdot\begin{bmatrix}
\hat{\boldsymbol{a}} \\ \hat{\boldsymbol{a}}^\dag
\end{bmatrix}=
\begin{bmatrix}
\ov{\boldsymbol{v}} \\ \boldsymbol{v}
\end{bmatrix}^\dag
\cdot\begin{bmatrix}
\hat{\boldsymbol{b}} \\ \hat{\boldsymbol{b}}^\dag
\end{bmatrix}
\]
Combining Eqs.~\eqref{eq:two_field_expansion} and~\eqref{eq:bogoliubov_transformation_definition} we derive transformation relations between two sets of annihilation and creation operators,
\[\label{eq:transformation_of_field}
\begin{bmatrix}
\hat{\boldsymbol{b}} \\ \hat{\boldsymbol{b}}^\dag
\end{bmatrix}=
\ov{S}^\dag\begin{bmatrix}
\hat{\boldsymbol{a}} \\ \hat{\boldsymbol{a}}^\dag
\end{bmatrix}.
\]
Both sets of annihilation and creation operators must obey the commutation relations~\eqref{def:commutation_relation} which gives a condition on the matrix $S$,
\[
SKS^\dag=K.
\]
This is equivalent to
\begin{subequations}
\begin{align}
\A\A^\dag-\B\B^\dag&=I,\\
\A\B^T&=\B\A^T,
\end{align}
\end{subequations}
known as Bogoliubov identities. But these conditions are exactly the defining conditions of the symplectic group, Eqs.~\eqref{def:structure_of_S} and~\eqref{def:Bogo_id} respectively. In other words, Bogoliubov transformations are identical to Gaussian unitary transformations and they transform Gaussian states into Gaussian states. This is why the phase-space description of Gaussian states and can be easily applied in the quantum field theory in curved space-time which will be used in following chapters.

%Expressions for Bogoliubov transformations in terms of the inner product
The Bogoliubov coefficients can be calculated using the (pseudo-)inner product of the Klein-Gordon equation~\cite{Crispino2008} which depends on the metric~\eqref{def:metric}
as
\begin{subequations}\label{eq:basic_bogos}
\begin{align}
\alpha_{mn}&\,=\, \big(u_{m},v_{n}\big)\big|_{\Sigma},\\
\beta_{mn}&\,=\, \big(u_{m},\overline{v}_{n}\big)\big|_{\Sigma}.
\end{align}
\end{subequations}
These coefficients encode the information of a transformation between two sets of solutions of the Klein-Gordon equation on a given time-like hypersurface $\Sigma$.

\section{Continuous Bogoliubov coefficients}

%Why continuous transformations
The Bogoliubov transformations in Eq.~\eqref{eq:basic_bogos} are defined only for a fixed time, they will not in general be suitable to describe the continuous evolution of a quantum state. In particular, they are not suitable for describing the evolution of a quantum state which transform an initial state at time $\tau_{0}$ to a final state at time $\tau$. In this section we utilize perturbative methods and derive expressions for continuous Bogoliubov transformations up to any order in the expansion in a small parameter $\epsilon$. The method was formerly described in~\cite{Bruschi2013b,Bruschi2013a}, however, previous work considered only the first order correction. Here we derive expressions for perturbative coefficients up to any order.

%Construction of continuous Bogoliubov transformations
To construct the continuous Bogoliubov transformations consider a function $h(\tau)$ which parametrizes the physical scenario, for example a cavity moving through a curved space-time, a passing gravitational wave, or an accelerating cavity. For example, $h(\tau)$ could describe the proper acceleration of the cavity. The Bogoliubov co-efficients, which describe the evolution of a quantum state inside of the cavity, would be then dependent on this function as $\A[h(\tau)]$ and $\B[h(\tau)]$. The matrices $\A[h(\tau)]$ and $\B[h(\tau)]$ represent the Bogoliubov transformation from one set of modes at time $\tau_{0}$ to a new set of modes at time $\tau$. After the Bogoliubov transformation has been applied, the evolution of the field is governed by the free Hamiltonian $H_f[h(\tau)]$. By ``free'' here we mean that it governs the evolution the state would undergo if it is left alone, without adding any extra energy into the system, for example without adding any further acceleration.  For example, for a non-interacting field the free Hamiltonian this is given by the mode frequencies defined as $\Omega_{j}[h(\tau)]$. In other words, at time $\tau$, each mode is characterized by a frequency which depends, for example, on the acceleration $h(\tau)$. To construct the continuous transformations, consider a successive combination of transformations composed of: i) An initial transformation at time $\tau=\tau_{0}$, $\mathbin{_{o}\A}:=\A[h(\tau_{0})]$ and $\mathbin{_{o}\B}:=\B[h(\tau_{0})]$, ii) Evolution under the ``free" Hamiltonian in the new modes for some time $\Delta\tau$, iii) Applying the inverse of the initial transformation for the coefficients $\mathbin{_{o}\boldsymbol{\alpha}}$ and $\mathbin{_{o}\boldsymbol{\beta}}$. Combining these successive transformations and letting each interval $\Delta\tau\rightarrow 0$, while keeping the \emph{total} proper time fixed, we can construct a continuous Bogoliubov transformation. Following this procedure, we can derive the following initial value problem for the \emph{total} continuous Bogoliubov transformation~\cite{Bruschi2013b},
\[\label{eqn:exact_continuous_bogo_ODE}
\mathrm{d}_{\tau}S(\tau,\tau_{0})=iKH[h(\tau)] S(\tau,\tau_{0}),\quad S(\tau_{0},\tau_{0})=I.
\]
Here, we have defined the effective Hamiltonian of the system as
\[\label{def:Hamiltonian_continuous_bogos}
H=\begin{bmatrix}
A & B \\
\overline{B} & \overline{A}
\end{bmatrix}.
\]
The form of Eq.~\eqref{eqn:exact_continuous_bogo_ODE} is to be expected from the Lie group structure of symplectic transformations, i.e., matrix $iKH$ is an element of the Lie algebra associated with the symplectic group and satisfy Eq.~\eqref{def:P_1}. Similar results have been pointed out before in the quantum field theory literature~\cite{Brown2013a,Bruschi2013a}.

%submatrix form
Taking advantage of the block structure of the symplectic matrices~\eqref{def:Bogoliubov_S}, Eq.~\eqref{eqn:exact_continuous_bogo_ODE} can be rewritten as two coupled equations,
\begin{subequations}
\label{eqn:exact_continuous_bogo_ODE_alpha_beta}
\begin{align}
\mathrm{d}_{\tau}\A&= +i\,\Big(A\A+B\overline{\B}\Big),\quad\A(\tau_{0},\tau_{0})=I,\\
\mathrm{d}_{\tau}\overline{\B}&= -i\,\Big(\overline{A}\,\overline{\B}+\overline{B}\A\Big),\quad\B(\tau_{0},\tau_{0})=0.
\end{align}
\end{subequations}
Eq.~\eqref{eqn:exact_continuous_bogo_ODE} or Eq.~\eqref{eqn:exact_continuous_bogo_ODE_alpha_beta} give a concrete recipe to determine the symplectic transformation induced by the Bogoliubov transformations of our quantum field theory. If those Bogoliubov transformations describe the evolution of the state, once the matrices $\A$ and $\B$ are determined one can use the full symplectic transformation $S$ to evolve the moments of the Gaussian state via Eq.~\eqref{def:transformation}.

%Example on estimation of proper acceleration
\begin{example}
\emph{
The Hamiltonian~\eqref{def:Hamiltonian_continuous_bogos} depends on the particular method of construction of `gluing' together infinitesimal fixed-time Bogoliubov transformation and free-time evolution. Our construction can for example describe the previously mentioned case when the function $h(\tau)$ describes the proper acceleration of an observer. Then the effective Hamiltonian reads,
\begin{subequations}\label{eq:accelerating_cavity_Hamiltonian}
\begin{align}
A&= \mathbin{_{o}\A^{\dag}}\Omega\mathbin{_{o}\A}+\mathbin{_{o}\B^{T}}\Omega\mathbin{_{o}\overline{\B}}, \\
B&= \mathbin{_{o}\A^{\dag}}\Omega\mathbin{_{o}\B}+\mathbin{_{o}\B^{T}}\Omega\mathbin{_{o}\overline{\A}}.
\end{align}
\end{subequations}
Above, we have also introduced the free Hamiltonian given by the frequency matrix $\Omega:=\mathrm{diag}(\Omega_{1}[h(\tau)],\Omega_{2}[h(\tau)],\ldots)$. In different scenarios, such as slowly moving cavities in Schwarzchild space-time, different methods of construction are more viable for describing the evolution of the physical system. This then leads to a different effective Hamiltonian.
}
\end{example}

%General solution
The general solution to Eq.~\eqref{eqn:exact_continuous_bogo_ODE}, Eq.~\eqref{eqn:exact_continuous_bogo_ODE_alpha_beta} respectively, can be formally written as the time-ordered exponential,
\[
S=\mathrm{Texp}\left(i\int_{\tau_0}^\tau KH\ \mathrm{d}\tau\right).
\]
However, this solution is usually difficult to compute exactly and ODE methods, such as solving order by order in a parameter, have to be used. In later chapters we will use this perturbative method to estimate the ultimate limits of precision in estimating space-time parameters.

%ODE method
\subsection{Perturbative method of finding the continuous Bogoliubov coefficients}\label{sec:perturbative_method}
Let us assume the function $h(\tau)$ introduced in the previous subsection can be factorized as $h(\tau)=\epsilon f(\tau)$, where $\epsilon$ is a fixed constant -- a small parameter, and $f(\tau)$ is another function independent of $\epsilon$. Because the effective Hamiltonian $H[h(\tau)]$ depends on time only through the function $h(\tau)$, setting $\epsilon=0$ will make this Hamiltonian time-independent. We define this time-independent part of the effective Hamiltonian as $H_0:=H[0]$.
%For convenience, we also define the time-independent part of the Hamiltonian (which will later serve as the free Hamiltonian) as $H_0:=H(\tau_0,\tau_0)$.
We look for the solution of Eq.~\eqref{eqn:exact_continuous_bogo_ODE} in the form
\[\label{def:tildeS}
S=e^{iKH_0(\tau-\tau_0)}\tilde{S},\quad
\tilde{S}=\begin{bmatrix}
\tilde{\A} & \tilde{\B} \\
\overline{\tilde{\B}} & \overline{\tilde{\A}}
\end{bmatrix}.
\]
This is of course only a mathematical substitution which we could have chosen differently. However, this choice of substitution ensures that for $\epsilon=0$ we retrieve the exact solution of the free-time evolution instead of just perturbative expansion of such evolution. The Eq.~\eqref{eqn:exact_continuous_bogo_ODE} transforms into
\[\label{eqn:exact_continuous_bogo_ODE2}
\mathrm{d}_{\tau}\tilde{S}(\tau,\tau_{0})=iV(\tau,\tau_{0}) \tilde{S}(\tau,\tau_{0}),\quad \tilde{S}(\tau_{0},\tau_{0})=I,
\]
where we have defined the new effective Hamiltonian,
\[
V(\tau,\tau_{0}):=e^{-iH_0(\tau-\tau_0)}KH[h(\tau)]e^{iH_0(\tau-\tau_0)}-KH_0.
\]
Matrix $iV$ is again the element of the Lie algebra associated with the symplectic group and satisfy Eq.~\eqref{def:P_1}. This provides an internal structure of the matrix $V$,
\[
V=\begin{bmatrix}
V_{11} & V_{12} \\
\overline{V}_{12} & \overline{V}_{11}
\end{bmatrix},\ \ V_{11}=V_{11}^\dag,\ \ V_{12}=-V_{12}^T.
\]
From now on we assume every matrix $M$ has a perturbative expansion in $\epsilon$,
\[
M=\sum_{k=0}^\infty M^{(k)}\epsilon^k.
\]
Clearly from the definition $H_0=H^{(0)}$ and $V^{(0)}=0$. By inserting expansions of matrix $\tilde{S}$ and matrix $V$ to Eq.~\eqref{eqn:exact_continuous_bogo_ODE2} we obtain a set of differential equations,
\begin{subequations}
\begin{align}
\sum_{k=0}^\infty \mathrm{d}_\tau \tilde{\A}^{(k)}\epsilon^k=i\sum_{k=1}^\infty V_{11}^{(k)}\epsilon^k+i\sum_{k,l=1}^\infty\left(V_{11}^{(l)}\tilde{\A}^{(k)}+V_{12}^{(l)}\ov{\tilde{\B}}^{(k)}\right)\epsilon^{k+l},\\
\sum_{k=0}^\infty \mathrm{d}_\tau \tilde{\B}^{(k)}\epsilon^k=i\sum_{k=1}^\infty V_{12}^{(k)}\epsilon^k+
i\sum_{k,l=1}^\infty\left(V_{11}^{(l)}\tilde{\B}^{(k)}+V_{12}^{(l)}\ov{\tilde{\A}}^{(k)}\right)\epsilon^{k+l}.
\end{align}
\end{subequations}
Rearranging terms in infinite summations gives
\begin{subequations}
\begin{align}
\sum_{k=0}^\infty \mathrm{d}_\tau \tilde{\A}^{(k)}\epsilon^k&= iV_{11}^{(1)}\epsilon+i\sum_{k=2}^\infty \left(V_{11}^{(k)}+\sum_{l=1}^{k-1}\left(V_{11}^{(l)}\tilde{\A}^{(k-l)}+V_{12}^{(l)}\ov{\tilde{\B}}^{(k-l)}\right)\right)\epsilon^{k},\\
\sum_{k=0}^\infty \mathrm{d}_\tau \tilde{\B}^{(k)}\epsilon^k&= iV_{12}^{(1)}\epsilon+
i\sum_{k=2}^\infty\left(V_{12}^{(k)}+\sum_{l=1}^{k-1}\left(V_{11}^{(l)}\tilde{\B}^{(k-l)}+V_{12}^{(l)}\ov{\tilde{\A}}^{(k-l)}\right)\right)\epsilon^{k}.
\end{align}
\end{subequations}
Finally, by comparing coefficients of different powers of $\epsilon$ we obtain recursive formulae for an arbitrarily high coefficient of a continuous Bogoliubov transformation,
\[
\begin{split}\label{eqns:general_solution_arbitrarily_high_order}
\tilde{\A}^{(0)}(\tau,\tau_0)&= I,\\
\tilde{\B}^{(0)}(\tau,\tau_0)&= 0,\\
\tilde{\A}^{(1)}(\tau,\tau_0)&= i\int_{\tau_0}^{\tau}\!\!\!\mathrm{d}t\ V_{11}^{(1)}(t,\tau_0),\\
\tilde{\B}^{(1)}(\tau,\tau_0)&= i\int_{\tau_0}^{\tau}\!\!\!\mathrm{d}t\ V_{12}^{(1)}(t,\tau_0),\\
\tilde{\A}^{(k)}(\tau,\tau_0)&= i\int_{\tau_0}^{\tau}\!\!\!\mathrm{d}t\ \left(V_{11}^{(k)}+\sum_{l=1}^{k-1}\left(V_{11}^{(l)}\tilde{\A}^{(k-l)}+V_{12}^{(l)}\ov{\tilde{\B}}^{(k-l)}\right)\right)(t,\tau_0),\\
\tilde{\B}^{(k)}(\tau,\tau_0)&= i\int_{\tau_0}^{\tau}\!\!\!\mathrm{d}t\ \left(V_{12}^{(k)}+\sum_{l=1}^{k-1}\left(V_{11}^{(l)}\tilde{\B}^{(k-l)}+V_{12}^{(l)}\ov{\tilde{\A}}^{(k-l)}\right)\right)(t,\tau_0).
\end{split}
\]
Expressions for $\tilde{\A}^{(0)}$ and $\tilde{\B}^{(0)}$ have been obtained from the initial condition in Eq.~\eqref{eqn:exact_continuous_bogo_ODE2}. The full solution to the continuous Bogoliubov transformation is obtain by combining the above equations and Eq.~\eqref{def:tildeS}.

%example on accelerating cavities
\begin{example}\label{ex:accelerated_cavity}
Accelerating cavity first introduced in~\cite{Bruschi2013b}.
\emph{The scenario is the following. Assume a quantum state inside of a non-moving cavity. Starting at proper time $\tau_0=0$, the cavity goes through a period $\tau$ of the proper acceleration $a$ (as measured in the centre of the cavity) and period $\tau$ of retardation $-a$, stopping again at time $2\tau$. The proper length of the cavity $L=1$ is considered constant during the whole procedure. We are going to expand in the small parameter $a$, i.e. $\epsilon\equiv a$.
%Hamiltonian
The effective Hamiltonian is given by Eqs.~\eqref{def:Hamiltonian_continuous_bogos} and~\eqref{eq:accelerating_cavity_Hamiltonian}. The function is given by $h(t)=af(t)$, where $f(t)=1$ for $\tau_0\leq t\leq\tau$ and $f(t)=-1$ for $\tau\leq t\leq2\tau$. The frequencies of the free Hamiltonian $H_f=\mathrm{diag}(\Omega_1[a f(\tau)],\Omega_2[a f(\tau)],\dots,\Omega_1[a f(\tau)],\Omega_2[a f(\tau)])$ and the Bogoliubov coefficients which transforms the cavity from the still state to the accelerating state with proper acceleration $a$ are given by
\begin{subequations}
\begin{align}
\Omega_n[a f(t)]&\,=\,\omega_n+\mathcal{O}(a^2),\quad \omega_n=\frac{n\pi}{L},\\
\mathbin{_{o}\A}_{mn}&\,=\, 1+\mathcal{O}(a^2),\ &m=n\nonumber\\
&\,=\, \frac{(-1+(-1)^{m+n})\sqrt{mn}}{(m-n)^3\pi^2}a+\mathcal{O}(a^2),\ &m\neq n\\
\mathbin{_{o}\B}_{mn}&\,=\, \frac{(1-(-1)^{m+n})\sqrt{mn}}{(m+n)^3\pi^2}a+\mathcal{O}(a^2),&
\end{align}
\end{subequations}
The continuous coefficients up to the first order are obtained combining Eqs.~\eqref{eqns:general_solution_arbitrarily_high_order} and Eq.~\eqref{def:tildeS},
\begin{subequations}\label{eq:bogos_acceleration}
\begin{align}
\alpha_{mn}(a)&\,=\, e^{i\omega_n2\tau}+\mathcal{O}(a^2),\ &m=n\nonumber\\
&\,=\, -\frac{8i\sqrt{mn}}{(m-n)^3\pi^2}e^{\frac{1}{2}i\pi(m+n-2m\tau+6n\tau)}\sin{\tfrac{(m+n)\pi}{2}}\sin^2{\tfrac{(m-n)\pi\tau}{2}}\,a+\mathcal{O}(a^2),\ &m\neq n\\
\beta_{mn}(a)&\,=\, -\frac{8i\sqrt{mn}}{(m+n)^3\pi^2}e^{\frac{1}{2}i\pi(m+n-2m\tau+2n\tau)}\sin{\tfrac{(m+n)\pi}{2}}\sin^2{\tfrac{(m+n)\pi\tau}{2}}\,a+\mathcal{O}(a^2).&
\end{align}
\end{subequations}
}
\end{example}

%This will be used this later
The method introduced here can be used to derive the Bogoliubov transformations depending on the arbitrarily space-time parameter, not only the proper acceleration. In chapter~\ref{chap:QFT_metrology} we show how to derive the quantum Fisher information for estimating such parameters encoded by a general Bogoliubov transformations into squeezed and two-mode squeezed thermal states.

%\section{Overview on the current state-of-the-art metrology in quantum field theory}
\section{State-of-the-art quantum metrology for estimating space-time parameters}\label{sec:state_QM_in_qft}
%then check metrology in QFT and do some overview - our group: ant, ivette, nico, david, mehdi, and the chinese guys - check who they cite

%introduction in QFT+quantum info
After successful efforts in combining quantum physics and general relativity on macroscopic scales leading to infamous black-hole evaporation~\cite{Hawking1974a,hawking1975particle,fabbri2005modeling}, the Unruh effect~\cite{takagi1986vacuum,Unruh1979a}, and the dynamical Casimir effect~\cite{moore1970quantum}, the focus shifted to studying relativistic quantum effects from a more analytical perspective of quantum information. Pioneering articles in this newly established field focused on entanglement and its generation due to either gravity or non-inertial motion~\cite{alsing2002lorentz,pachos2002generation,Peres2004quantum_information,fuentes2005alice,alsing2006entanglement} and on quantum information protocols such as quantum teleportation~\cite{alsing2003teleportation,alsing2004teleportation} and communication channels~\cite{Peres2004quantum_information}. Relativistic quantum information was booming. It has been discovered that relativistic motion could serve as a source for generating quantum gates~\cite{friis2012quantum,Bruschi2013relativistic_motion,martin2014quantum}. To complement studies on entanglement, it was investigated how motion affects other quantum information figures of merit such as quantum discord~\cite{datta2009quantum}. Moreover, it has been shown that continuous variable methods could prove fruitful in relativistic quantum information~\cite{adesso2012continuous}.
%Introducing meaning of measurements, decoherence, entanglement, entropy, and communication channels in quantum field theory~\cite{Peres2004quantum_information},
One of the latest achievements was the introduction of quantum metrology into quantum field theory. As quantum metrology provides the limits of precision in measuring parameters, its methods can be used to validate the predictions in quantum field theory in curved space-time,  and also to demonstrate whether measuring space-time parameters is achievable with current and future technology.

Quantum metrology was first applied to the scenario in which a quantum state in a cavity is used estimate the temperature of the Unruh-Hawking effect~\cite{aspachs2010optimal}. This work showed that Fock states have an advantage as probes over Gaussian states. It was followed in~\cite{wang2014quantum} where a pair of Unruh-DeWitt detectors were used instead of cavity modes, and it was found that the limit of precision depends on an effective coupling strength and a longer interaction time. Authors of~\cite{Tian2015a} have evaluated the Fisher information and so, unlike the previous work, were taking realistic measurements into account. They showed that the optimal bound given by the quantum Fisher information could be achieved by the population measurement.

Other applications of quantum metrology include relativistic quantum accelerometers. The first article~\cite{dragan2011quantum} shows how a field in a cavity can be used to distinguish two scenarios that are kinematically indistinguishable: an accelerated cavity as seen by a stationary observer and an accelerated observer looking at a stationary cavity. Authors of article~\cite{Yao2014quantum} have shown how the precision in estimating parameters encoded in a general two-qubit state changes under accelerated motion. Interestingly, they found that the quantum Fisher information converges to a constant as the acceleration goes to infinity. A similar analysis has been made for fermionic cavities~\cite{shamsi2014quantum}. It has been illustrated in~\cite{Ahmadzadegan2014a} that the internal atomic degrees of freedom are sensitive to the atom's spatial trajectory and could serve as a good accelerometer.

Quantum metrology has been also used to study the precision of quantum clocks. It has been shown that the accelerated motion affects the precision in measuring time~\cite{Lindkvist2015a}, and that the quantum Fisher information degrades with higher acceleration. When acceleration is applied coherent states are more robust against the loss of precision, despite the fact that squeezed states perform better than coherent states in the absence of motion. As proposed in~\cite{lindkvist2014twin}, superconducting circuits could perform well in simulating the time dilation. %Also, it has been recognized that finite length of the cavity affects the time dilation -- larger the length, stronger the time dilation.

Quantum metrology has been applied to determine relativistic parameters such as the Schwarzchild radius using squeezed light~\cite{Bruschi2014a,kohlrus2015quantum,kish2016estimating} and the mass of a black hole~\cite{Doukas2014a} using scattering experiments. There are proposals for implementation in analogue gravity, using optical waveguides. Finally, a gravitational wave detector using Bose-Einstein condensates has been proposed~\cite{Sabin2014a}. It has been shown that phonon states which exist on a BEC can be altered by a passing gravitational wave. This is because the gravitational wave effectively changes boundary conditions of the BEC trap, leading to a similar effect as the dynamical Casimir effect. The changed phonon state could be measured and compared to its unchanged counterpart, effectively serving as a sensor for the gravitational wave. The initial proposal could achieve few orders of magnitudes higher precision than the current large interferometers~\cite{Caron1995a,Abbott2004a,Grote2008a} in the ideal case. The follow up article~\cite{Sabin2015a} analyzed the effects of finite temperature and concluded that there is no significant effect on the precision of such a proposed detector. However, what is still not taken into account is decoherence and practical aspects of the preparation of the initial state and the prospect of performing an optimal measurement.

Finally, work has been done on estimating space-time parameters in general, providing useful formulae for the quantum Fisher information for any encoding Bogoliubov transformation. Authors of~\cite{Ahmadi2014a} considered a general Bogoliubov transformation acting on two one-mode squeezed states. They derived the quantum Fisher information for estimating a parameter of such a transformation and applied the derived expression to estimating the proper acceleration of a cavity. Later, the same authors generalized this result to also incorporate the displaced squeezed states and two-mode squeezed states as probes~\cite{Ahmadi2014a}. In article~\cite{Friis2015a} a general procedure for finding the quantum Fisher information for initially pure probe states has been proposed. As this work also applies to non-Gaussian states it can be considered to be more general than previous contributions. However, as this work is done in the density matrix formalism it is impractical for the Gaussian states due to the complicated form of their density matrices.

The work presented in this section has been almost exclusively limited to pure initial states. But in real scenarios one cannot achieve perfect vacuum, nor is it possible to have perfect squeezed and coherent states. Achieving exactly pure states is, in fact, forbidden by the third law of thermodynamics. Probe states are always exposed to thermal fluctuations. In chapter~\ref{chap:QFT_metrology} we provide expressions for the quantum Fisher information for an arbitrary channel acting on one- and two-mode squeezed thermal states, allowing us to study the effects of temperature on estimating space-time parameters.

\chapter*{Part II}
\addcontentsline{toc}{chapter}{Part II}

%%%%%%%%%%%%%%%%%%%%%%%%%%%%%%
\chapter{Quantum metrology on Gaussian states}\label{chap:QM_GS}
%%%%%%%%%%%%%%%%%%%%%%%%%%%%%%

In this chapter we focus on quantum metrology of Gaussian states in the phase-space formalism. This chapter is divided into two sections. In the first section we derive numerous new formulae for the multi-parameter estimation of multi-mode Gaussian states and discuss problems of discontinuity of the figure of merit -- the quantum Fisher information. The first section can be also viewed as a continuation of section~\ref{sec:state_quantum_metrology} in which we overview current state-of-the-art quantum metrology of Gaussian states. In the second section we apply the derived formulae to devise a practical method of finding optimal probe states for Gaussian unitary channels. We use this method to find optimal Gaussian probe states for common Gaussian channels and for some channels which have not been optimized before. We also discuss related issues such as how different parameters of the probe state affect the estimation precision and whether entanglement plays a significant role in quantum metrology.

\section{The quantum Fisher information in the phase-space formalism}\label{sec:new_formulae}% in the phase-space formalism}

In this section we derive new formulae for the quantum Fisher information in the phase-space formalism following known formulae overviewed in section~\ref{sec:state_quantum_metrology}. This section partially consists of results we published in~\cite{Safranek2015b}. Some results are extended, for example, we provide expressions for the multi-parameter estimation. Also, many proofs and derivations are performed in a different and simpler way, mostly based on the matrix form of the quantum Fisher information matrix~\eqref{eq:mixed_QFI}. First we derive the quantum Fisher information matrix for two-mode states. Then we use the quantum Fisher information matrix in terms of the Williamson's decomposition of the covariance matrix, which will beautifully expose the inner structure of this figure of merit. We also simplify the limit formula~\eqref{eq:Monras_QFI} and provide an estimate for the remainder of the series, which is very useful for numerical calculations. Finally we show different versions of the quantum Fisher information for pure states and address the problems of the quantum Fisher information at the points of purity.

\subsection{Two-mode Gaussian states}
In the analogy of the derivation of the quantum Fisher information for single mode Gaussian states~\eqref{eq:one_mode_quantum_fisher_information} we can derive the expression for two-mode Gaussian states. The calculations are more involved than in deriving the single mode case and the following derivation is shortened compared to the original version published in~\cite{Safranek2015b}.

The derivation goes as follows. First, we use the expression for the Uhlmann fidelity between two two-mode Gaussian states derived in~\cite{Marian2012a},
\[\label{eq:fidelity_basic_formula}
\mathcal{F}(\rho_1,\rho_2)=\frac{4e^{-\delta\boldsymbol{d}^{\dagger}\left(\sigma_{1}
+\sigma_2\right)^{-1}\delta\boldsymbol{d}}}{\left(\sqrt{\Gamma}+\sqrt{\Lambda}\right)-\sqrt{\left(\sqrt{\Gamma}+\sqrt{\Lambda}\right)^{2}-\Delta}},
\]
where $\delta\boldsymbol{d}=\boldsymbol{d}_1-\boldsymbol{d}_2$ is a relative displacement and $\Delta,\Gamma,\Lambda$ denotes three determinants defined as
\begin{subequations}\label{eqs:GDL_unpolished}
\begin{align}
\Delta &=\det{\sigma_{1}+\sigma_{2}},\\
\Gamma &=\det{I+K\sigma_{1}K\sigma_{2}},\\
\Lambda &=\det{\sigma_{1}+K}\det{\sigma_{2}+K}.
\end{align}
\end{subequations}
According to Eq.~\eqref{QFI_using_fidelity} expanding fidelity between two close states $\rho_\epsilon$ and $\rho_{\epsilon+\de}$ in the small parameter $\de$ will give us the quantum Fisher information. First we need to expand determimants~\eqref{eqs:GDL_unpolished} in the small parameter $\de$. However doing that directly leads to numerous problems. That is why we rewrite these determimants in terms of the Williamson's decomposition~\eqref{def:Williamson's_decomposition} of the covariance matrix, $\sigma=SDS^\dag$, and expand this decomposition in the small parameter $\de$ instead. An element of the Lie algebra associated with the symplectic group will naturally appear,
\[
P:=S^{-1}\dot{S},
\]
where \emph{dot} denotes the derivative with respect to $\epsilon$. Because $P$ satisfies identities of the Lie algebra~\eqref{def:structure_of_S}, we use these identities to simplify the expression for the expansion of the fidelity and derive
\[\label{GeneralQFISD}
\begin{split}
H&(\epsilon)=\frac{1}{\det{D}-1}\bigg(\det{D}\Big(\tr[P^2]-\tr[D^{-1}KPDKP]\Big)\\
&\!+\!\sqrt{\det{C}}\Big(\tr[(C^{-1}P)^2]\!+\!\tr[(C^{-1}DKP)^2]\!-\!\tr[C^{-1}P^2]\Big)\!\bigg)\\
&+\frac{1}{2}\tr\big[(D+K)^{-1}D^{-1}\dot{D}^2\big]+2\dot{\bd}^\dag\sigma^{-1}\dot{\bd},
\end{split}
\]
where $C=I+D^2$. At this stage, we would like to obtain the quantum Fisher information in terms of the covariance matrix again. Defining $A:=K\sigma$ introduced in Eq.~\eqref{def:A} and using identities
\begin{subequations}
\begin{align}
\tr[(A^{-1}\dot{A})^2]&=2\tr[P^2]-2\tr[D^{-1}KPDKP]+\tr[D^{-1}\dot{D}D^{-1}\dot{D}]\\
\tr[((1+A^2)^{-1}\dot{A})^2]&=2\tr[(C^{-1}DKP)^2]+2\tr[(C^{-1}P)^2]-2\tr[C^{-1}(P)^2]\nonumber\\
&+\tr[(C^{-1}\dot{D})^2].
\end{align}
\end{subequations}
we derive the quantum Fisher information for two-mode Gaussian states,
\[\label{GeneralQFI}
\begin{split}
H(\epsilon)&=\frac{1}{2(\det{A}-1)}\Bigg(\det{A}\tr\Big[\big(A^{-1}\dot{A}\big)^2\Big]+\sqrt{\det{I+A^2}}\tr\Big[\big((I+A^2)^{-1}\dot{A}\big)^2\Big]\\
&+4\big(\lambda_1^2-\lambda_2^2\big
)\bigg(-\frac{\dot{\lambda_1}^2}{\lambda_1^4-1}
+\frac{\dot{\lambda_2}^2}{\lambda_2^4-1}\bigg)\Bigg)+2\dot{\bd}^\dag\sigma^{-1}\dot{\bd}.
\end{split}
\]
The symplectic eigenvalues $\lambda_{1,2}$ can be calculated via Eq.~\eqref{eq:twomode_symplectic_eigenvalues}. The above formula is not directly applicable when one of the modes is pure, i.e., when at least one of the symplectic eigenvalues is equal to $1$. We will address this issue in following sections.

Similarly to the quantum Fisher information for single mode states~\eqref{eq:one_mode_quantum_fisher_information}, formula for two-mode states~\eqref{GeneralQFI} provides the advantage over the general quantum Fisher information for multi-mode states~\eqref{eq:mixed_QFI} because it requires inverting much smaller matrices, i.e., $4\times 4$ matrix $A$ in Eq.~\eqref{eq:one_mode_quantum_fisher_information} matrix as compared to $16\times 16$ matrix $\mathfrak{M}$ in Eq.~\eqref{eq:mixed_QFI}. We will use this computational advantage in chapter~\ref{chap:QFT_metrology} to derive how two-mode Gaussian states perform as probes in estimating space-time parameters.

\subsection{When the Williamson's decomposition is known}\label{sec:when_Williamson's_decomposition}
In section~\ref{sec:Williamson's_decomposition} we have shown that every covariance matrix can be diagonalized using symplectic matrices. In this section we are going to derive an expression for the quantum Fisher information matrix for the case when the Williamson's decomposition of the covariance matrix is known. As we will see in section~\eqref{sec:estimation_of_channels}, such expression can be very useful for example when we are trying to find the optimal probe state for the estimation of Gaussian unitary channels.

%Derivation of the formula
We define matrices $P_i:=S^{-1}\partial_i{S}$ which are elements of the Lie algebra associated with the symplectic group~\eqref{def:structure_of_S},
\[\label{def:P_i}
P_i=
\begin{bmatrix}
R_i & Q_i \\
\ov{Q}_i & \ov{R}_i
\end{bmatrix},\ \ P_iK+KP_i^\dag=0.
\]
Rewriting Eq.~\eqref{eq:mixed_QFI} in terms of the Williamson's decomposition of the covariance matrix, switching to element-wise notation, and simplifying using identities~\eqref{def:structure_of_S} and~\eqref{def:P_i}, we derive an exact expression for the quantum Fisher information matrix of Gaussian states in terms of the Williamson's decomposition of the covariance matrix,
\[\label{eq:multimode_QFI}
\begin{split}
H^{ij}(\be)&=\!\!\!\sum_{k,l=1}^{N}\!\frac{(\lambda_k\!-\!\lambda_l)^2}{\lambda_k\lambda_l\!-\!1}\Re[\ov{R_{i}}^{kl}R_{j}^{kl}]+\frac{(\lambda_k\!+\!\lambda_l)^2}{\lambda_k\lambda_l\!+\!1}\Re[\ov{Q_{i}}^{kl}Q_{j}^{kl}]\\
&+\sum_{k=1}^{N}\frac{\partial_i\lambda_k\partial_j\lambda_k}{\lambda_k^2-1}+2\partial_i\boldsymbol{d}^\dag\sigma^{-1}\partial_j\boldsymbol{d},
\end{split}
\]
where $\Re$ denotes the real part, $R_i=\A^\dag\partial_i{\A}-\overline{\B^\dag\partial_i{\B}}$ is the skew-Hermitian and $Q_i=\A^\dag\partial_i{\B}-\overline{\B^\dag\partial_i{\A}}$ the (complex) symmetric matrix. This formula represents a multi-parameter generalization of the one-parameter result of~\cite{Safranek2015b}. The full derivation can be found in appendix~\ref{app:Williamson's_QFI}.

%Beauty of the formula
Beauty of the above formula lies in the fact that we can see each contribution of the different parts of the Gaussian state to the quantum Fisher information. The first part consists of matrices $R_i$ and $Q_i$. These matrices are proportional to the derivative of the diagonalizing symplectic matrix $S$ which encode squeezing and the orientation of squeezing. If the symplectic matrix $S$ is very sensitive to the changes in $\epsilon$, the derivatives of $S$ are large which leads to the higher quantum Fisher information. In other words, the first part of this equation shows the contribution to the ultimate precision from the changes of squeezing.  Moreover, these terms have constant factors given by symplectic eigenvalues. As we explained in section~\ref{sec:Williamson's_decomposition}, the symplectic eigenvalues $\lambda_i$ encode purity or equivalently temperature of the state. The second contribution is proportional to the changes in the symplectic eigenvalues and thus to the changes in purity or the temperature of the state. The third contribution is proportional to the changes in displacement. Summed up, how well we can estimate the vector of parameters $\be$ is given by how sensitive is the squeezing, purity, and the displacement of a Gaussian to the changes in the vector of parameters $\be$. We will talk more about the effects of temperature in section~\ref{sec:effects_of_temperature}.

%Problems at the points of purity
Symplectic eigenvalues are larger than one, $\lambda_i\geq1$, and $\lambda_i=1$ if and only if the mode is in the pure state. The multi-mode Gaussian state is pure when all symplectic eigenvalues are equal to one. To be able to apply Eq.~\eqref{eq:multimode_QFI} to states which have some eigenvalues equal to one, we have to define these problematic points. There are two possible ways of how to define them. The first choice will give us the quantum Fisher information matrix~\eqref{def:Information_matrix} for which the Cram\'er-Rao bound holds, while the second choice will give us the continuous quantum Fisher information matrix~\eqref{eq:connection_between_Hc_and_H}, which corresponds to the quantity defined by the Bures distance. Recalling the discussion in section~\ref{QMsec:multi} of the first chapter, the quantum Fisher information matrix for states where some of the modes are pure is obtained simply by not summing over terms which are undefined.  Equivalently, to obtain the quantum Fisher information $H$, for $\be$ such that $\lambda_k(\be)=\lambda_l(\be)=1$ we define
\begin{subequations}\label{def:problematic_points0}
\begin{align}
\frac{(\lambda_k\!-\!\lambda_l)^2}{\lambda_k\lambda_l\!-\!1}(\be)\!&:=0,\\
\frac{\partial_i\lambda_k\partial_j\lambda_k}{\lambda_k^2-1}(\be)\!&:=0.
\end{align}
\end{subequations}
Now we look at the second choice of defining these points which will give the continuous quantum Fisher information matrix $H_c$ (see Eq.~\eqref{eq:connection_between_Hc_and_H}). Assuming $\be$ is such that $\lambda_k(\be)=1$ and $\sigma(\be)\in C^{(2)}$, the function $\lambda_k$ must achieve the local minimum at point $\be$ and the Taylor expansion must be of the form $\lambda_k(\be+\mathrm{d}\be)=1+\frac{1}{2}\mathrm{d}\be^T\mathcal{H}_k\mathrm{d}\be+\cdots$, where $\mathcal{H}_k^{ij}=\partial_i\partial_j\lambda_k$ is the positive semi-definite matrix called Hessian. To obtain the continuous quantum Fisher information $H_c$, for $\be$ such that $\lambda_k(\be)=\lambda_l(\be)=1$ we define the problematic terms as the limit of the values given by its neighborhood,
%Symplectic eigenvalues are larger than one, $\lambda_i\geq1$, and $\lambda_i=1$ if and only if the mode is in the pure state. The multi-mode Gaussian state is pure when all symplectic eigenvalues are equal to one. To be able to apply Eq.~\eqref{eq:multimode_QFI} to states which have some eigenvalues equal to one, in analogy of~\cite{Safranek2015b} we define the problematic terms in a way which makes the quantum Fisher information matrix a continuous function. Assuming $\be$ is such that $\lambda_k(\be)=1$ and $\sigma(\be)\in C^{(2)}$, the function $\lambda_k$ must achieve the local minimum at point $\be$ and the Taylor expansion must be of the form $\lambda_k(\be+\mathrm{d}\be)=1+\frac{1}{2}\mathrm{d}\be^T\mathcal{H}_k\mathrm{d}\be+\cdots$, where $\mathcal{H}_k^{ij}=\partial_i\partial_j\lambda_k$ is the positive semi-definite matrix called Hessian. For $\be$ such that $\lambda_k(\be)=\lambda_l(\be)=1$ we define\footnote{The same definition of problematic points can be derived for pure states using an alternative definition of the quantum Fisher information matrix given by Eq.~\eqref{eq:QFI_matrix_using_fidelity}, where the Uhlmann fidelity $\mathcal{F}$ can be calculated using the results of~\cite{Spedalieri2013a}.
\begin{subequations}\label{def:problematic_points}
\begin{align}
\frac{(\lambda_k\!-\!\lambda_l)^2}{\lambda_k\lambda_l\!-\!1}(\be)\!&:=\!\lim_{\mathrm{d}\be\rightarrow 0}\frac{(\lambda_k(\be+\mathrm{d}\be)\!-\!\lambda_l(\be+\mathrm{d}\be))^2}{\lambda_k(\be+\mathrm{d}\be)\lambda_l(\be+\mathrm{d}\be)\!-\!1}=0,\\
\frac{\partial_i\lambda_k\partial_j\lambda_k}{\lambda_k^2-1}(\be)\!&:=\!\lim_{\de_i\rightarrow 0}\lim_{\de_j\rightarrow 0}\lim_{\overset{\de_r\rightarrow 0}{r\neq i,j}}\!\frac{\partial_i\lambda_k(\be+\mathrm{d}\be)\partial_j\lambda_k(\be+\mathrm{d}\be)}{\lambda_k^2(\be+\mathrm{d}\be)-1}=\mathcal{H}_k^{ij}(\be).
\end{align}
\end{subequations}
This construction is identical to the way of how the proof of the continuity of the continuous Fisher information was built (see appendix~\ref{app:discontinuity_of_QFI}, Eq.~\eqref{eq:construction_of_continuous_QFI}). The above definitions~\eqref{def:problematic_points0} and~\eqref{def:problematic_points} allows us to use Eq.~\eqref{eq:multimode_QFI} for any Gaussian state.

%Elegant form
When all symplectic eigenvalues are larger than one, we can define Hermitian matrix $\widetilde{R}_i^{kl}:=\frac{\lambda_k-\lambda_l}{\sqrt{\lambda_k\lambda_l-1}}R_i^{kl}$, symmetric matrix $\widetilde{Q}_i^{kl}:=\frac{\lambda_k+\lambda_l}{\sqrt{\lambda_k\lambda_l+1}}Q_i^{kl}$, and diagonal matrix $L:=\mathrm{diag}(\lambda_1,\dots,\lambda_N)$. We rewrite Eq.~\eqref{eq:multimode_QFI} in an elegant way,
\[\label{eq:exact_multimode_compact}
%\begin{split}
H^{ij}(\be)=\frac{1}{2}\tr\big[\widetilde{R}_i\widetilde{R}_j^\dag+\widetilde{R}_j\widetilde{R}_i^\dag+\widetilde{Q}_i\widetilde{Q}_j^\dag+\widetilde{Q}_j\widetilde{Q}_i^\dag\big]
+\tr\big[(L^2-I)^{-1}\partial_iL\partial_jL\big]+2\partial_i\boldsymbol{d}^\dag\sigma^{-1}\partial_j\boldsymbol{d}.
%\end{split}
\]
%The three contributions consisting of the changes of squeezing, purity and displacement with the small variations in $\be$ we talk about earlier are clearly visible in this formula.
%The quantum Fisher information matrix then consists of three terms. The first one is connected to the change of the orientation and the squeezing of the Gaussian state with small variations in $\be$, the second to the change of purity, and the third to the change of displacement.

\subsection{Regularization procedure}\label{sec:regularization_procedure}
%{short derivation taking into account the formulas in the previous section, (27), relation to pure states and Ant formula multiparameter}
We have seen so far that many expressions for the quantum Fisher information have problems when some of the symplectic eigenvalues is equal to one. Using the knowledge discovered in the previous section we can devise a method how to use the formulae for mixed states~\eqref{eq:one_mode_quantum_fisher_information},~\eqref{eq:mixed_QFI},~\eqref{GeneralQFI},~\eqref{eq:exact_multimode_compact} to calculate the quantum Fisher information matrix for any state. This regularization procedure is the Gaussian version of the regularization procedure described in appendix~\ref{app:discontinuity_of_QFI}, Eq.~\eqref{eq:regularization_procedure_density_matrix}. It goes as follows. First we multiply the covariance matrix by a regularization parameter $\nu>1$, use any expression for the quantum Fisher information matrix of mixed Gaussian states and calculate $H(\be)\equiv H(\R_{\be,\nu})$, $\R_{\be,\nu}\equiv(\bd(\be),\nu\sigma(\be))$, and in the end we perform the limit $\nu\rightarrow 1$. We need to make sure, however, that such method leads to the proper definition of the problematic points given either by Eq.~\eqref{def:problematic_points0} or Eq.~\eqref{def:problematic_points}. Assuming $\lambda_k(\be)=\lambda_l(\be)=1$ and performing the limit, both problematic terms are set to zero, $\lim_{\nu\rightarrow1}\frac{(\nu\lambda_k-\nu\lambda_l)^2}{\nu\lambda_k\nu\lambda_l-1}=\lim_{\nu\rightarrow1}\frac{0}{\nu^2-1}=0$, $\lim_{\nu\rightarrow1}\frac{(\nu\partial_i{\lambda}_k)^2}{(\nu\lambda_k)^2-1}=\lim_{\nu\rightarrow1}\frac{0}{\nu^2-1}=0$. This corresponds to the definition~\eqref{def:problematic_points0} for the quantum Fisher information matrix. To obtain the continuous quantum Fisher information matrix given by definition~\eqref{def:problematic_points} we need to add a Hessian matrix for every $k$ for which $\lambda_k(\be)=1$. Together we write
\[\label{eq:regularization_procedure}
H, H_c(\be)=\lim_{\nu\rightarrow1}H\big(\bd(\be),\nu\sigma(\be)\big)+c\!\!\!\!\!\!\sum_{k:\lambda_k(\epsilon)=1}\!\!\!\!\!\!\mathcal{H}_k(\be),
\]
where $c=0$ corresponds to the quantum Fisher information matrix $H$, and $c=1$ corresponds to the continuous quantum Fisher information matrix $H_c$.

\subsection{Limit formula}
In previous sections we presented exact formulae, however, in some cases an approximate value is enough. Here we will simplify and generalize the limit expression for the quantum Fisher information given by Eq.~\eqref{eq:Monras_QFI} to multi-parameter estimation. Defining matrix $A:=K\sigma$ the limit expression for the quantum Fisher information matrix reads,
\[\label{eq:limit_formula}
H^{ij}(\be)=\frac{1}{2}\sum_{n=1}^\infty\tr\big[A^{-n}\partial_i{A}A^{-n}\partial_j{A}]+2\partial_i\boldsymbol{d}^\dag\sigma^{-1}\partial_j\boldsymbol{d}.
\]
When using approximate methods, it is convenient to estimate the error connected to the approximation. In complete analogy of the proof presented for the estimation of a single parameter~\cite{Safranek2015b}, in appendix~\ref{app:Remainder} we find a bound on the remainder of the series. Defining $R_M^{ij}:=\frac{1}{2}\sum_{n=M+1}^\infty\tr\big[A^{-n}\!\partial_i{A}A^{-n}\!\partial_j{A}]$, we have
\[
|R_M^{ij}|\leq\frac{\sqrt{\mathrm{tr}[(A\partial_iA)^2]}\sqrt{\mathrm{tr}[(A\partial_jA)^2]}}{2\lambda_{\mathrm{min}}^{2(M+1)}(\lambda_{\mathrm{min}}^2-1)},
\]
where $\lambda_{\mathrm{min}}:=\min_{k}\{\lambda_k\}$ is the smallest symplectic eigenvalue of the covariance matrix $\sigma$. The bound shows that the series converges as a geometric series. This is a very fast convergence -- it is enough to take few terms of the infinite summation and the quantum Fisher information will be very well approximated. On the other hand, the bound for the remainder is of no use when some symplectic eigenvalue is equal to one, i.e., when some of the modes is pure. In the case when some symplectic eigenvalue is equal to one, the infinite series~\eqref{eq:limit_formula} converges, but it is not absolutely convergent. Moreover, this converging expression does not give the correct quantum Fisher information. Both of these statements can be checked by a careful analysis of the elements in the series given by Eq.~\eqref{eq:expression_in_terms_of_P} which is explained in more detail in~\cite{Safranek2015b}. Put simply, some terms which should contribute are identically zero as a consequence of the fact that the limit for the smallest eigenvalue $\lambda_{\mathrm{min}}\rightarrow0$ and the upper limit in the infinite summation in Eq.~\eqref{eq:limit_formula} do not commute. The correct expression for the states with at least one symplectic eigenvalue equal to one must be obtained using the regularization procedure~\eqref{eq:regularization_procedure}.

To prove formula~\eqref{eq:limit_formula} we show how it connects to Eqs.~\eqref{eq:multimode_QFI} and~\eqref{eq:mixed_QFI}. Using
\[\label{eq:expression_in_terms_of_P}
\begin{split}
&\tr[A^{-n}\partial_i{A}A^{-n}\partial_j{A}]=2\tr[D^{-n+1}\!K^{-n+1}\!P_{i}D^{-n+1}\!K^{-n+1}\!P_{j}]\\
&-\tr[D^{-n+2}\!K^n\!P_{i}D^{-n}\!K^n\!P_{j})]
-\tr[D^{-n+2}\!K^n\!P_{j}D^{-n}\!K^n\!P_{i})]\\
&+\tr[D^{-n}\partial_i{D}D^{-n}\partial_j{D}]
\end{split}
\]
and changing to element-wise notation, the infinite sum~\eqref{eq:limit_formula} turns out to be geometric series in powers of $\lambda_k$'s which can be evaluated. Then, using $R_i^{kl}=-\ov{R}_i^{lk}$, $Q_i^{kl}=Q_i^{lk}$ which follows from identity~\eqref{def:P_i}, we prove that Eq.~\eqref{eq:limit_formula} simplifies to Eq.~\eqref{eq:multimode_QFI}. To obtain Eq.~\eqref{eq:mixed_QFI}, we use $\sigma^\dag=\sigma$, properties of vectorization $\tr[A^\dag B]=\vectorization{A}^\dag\vectorization{B}$ and properties of Kronecker product~\eqref{id:Kronecker_product_ids} to transform the infinite sum in Eq.~\eqref{eq:limit_formula} into a Neumann series which can be evaluated,
\[
\begin{split}
\sum_{n=1}^\infty\tr\big[A^{-n}\partial_i{A}A^{-n}\partial_j{A}]&=
\vectorization{\partial_i\sigma}^\dag\bigg(\sum_{n=0}^\infty(\ov{A}\otimes A)^{-n}\bigg)\big(\ov{\sigma}\otimes\sigma\big)^{-1}\vectorization{\partial_j\sigma}\\
&=\vectorization{\partial_i\sigma}^\dag(I-\ov{A^{-1}}\otimes A^{-1})^{-1}\big(\ov{\sigma}\otimes\sigma\big)^{-1}\vectorization{\partial_j\sigma}\\
&=\vectorization{\partial_i\sigma}^\dag(\ov{\sigma}\otimes\sigma-K\otimes K)^{-1}\vectorization{\partial_j\sigma}
\end{split}
\]
Combining the above expression with Eq.~\eqref{eq:limit_formula} gives Eq.~\eqref{eq:mixed_QFI}.

\subsection{Pure states}\label{sec:pure_states_QM}

Combining Eq.~\eqref{eq:limit_formula}, the regularization procedure~\eqref{eq:regularization_procedure}, and $A^2(\be)=I$, which holds for pure states, we derive the quantum Fisher information matrix for states which are pure at point $\be$,
\[\label{eq:pure_non-elegant}
H^{ij},H_c^{ij}(\be)=\frac{1}{4}\mathrm{tr}[\sigma^{-1}\partial_i\sigma\sigma^{-1}\partial_j\sigma]+c\sum_{k}\mathcal{H}_k^{ij}+2\partial_i\boldsymbol{d}^\dag\sigma^{-1}\partial_j\boldsymbol{d}.
\]
If either $c=0$ (the definition corresponding to $H$), or if the state remains pure when the vector of parameters $\be$ is slightly varied, i.e., $\mathcal{H}_k(\be)=0$, the above expression reduces to the known formula for pure states given by Eq.~\eqref{eq:nu_pure} which has been derived in~\cite{Pinel2012a}.

Calculating the continuous quantum Fisher information matrix $H_c$ from Eq.~\eqref{eq:pure_non-elegant} requires to calculate derivatives of symplectic eigenvalues. However, it is possible to find an alternative form which avoids that need,
\[\label{eq:pure_elegant}
H_c^{ij}(\be)=\frac{1}{4}\big(2\mathrm{tr}[\sigma^{-1}\partial_i\partial_j\sigma]-\mathrm{tr}[\sigma^{-1}\partial_i\sigma\sigma^{-1}\partial_j\sigma]\big)+2\partial_i\boldsymbol{d}^\dag\sigma^{-1}\partial_j\boldsymbol{d},
\]
which generalizes the result of~\cite{Safranek2015b} to the multi-parameter estimation. To show that Eq.~\eqref{eq:pure_non-elegant} for $c=1$ and Eq.~\eqref{eq:pure_elegant} are identical, one needs to use the Williamson's decomposition of the covariance matrix $\sigma=SDS^\dag$, find~$\mathrm{tr}[\sigma^{-1}\partial_i\partial_j\sigma]$ and $\mathrm{tr}[\sigma^{-1}\partial_i\sigma\sigma^{-1}\partial_j\sigma]$ in terms of $K,P_i,P_{ij}:=S^{-1}\partial_i\partial_jS$ (which gives similar expressions to Eq.~\eqref{eq:expression_in_terms_of_P}), and use identities~\eqref{def:structure_of_S},~\eqref{def:P_i}, $P_{ij}K+P_iKP_j^\dag+P_jKP_i^\dag+KP_{ji}^\dag=0$. Both~\eqref{eq:pure_non-elegant} and~\eqref{eq:pure_elegant} then reduce to the same expression,
\[\label{eq:QFI_almost_pure}
H_c(\epsilon)=\frac{1}{2}\Big(\tr\big[P_iP_j\big]-\tr\big[KP_iKP_j\big]\Big)+\frac{1}{2}\tr[\partial_i\partial_jD].
\]
In analogy of Eq.~\eqref{eq:nu_pure_simplified}, Eqs.~\eqref{eq:pure_non-elegant},~\eqref{eq:pure_elegant} can be further simplified by using $\sigma^{-1}=K\sigma K=AK$ to avoid the necessity of inverting the covariance matrix.

Eq.~\eqref{eq:QFI_almost_pure} is an expression for the quantum Fisher information in terms of the decomposition of the covariance matrix. This expression can be further simplified by expressing $P_i$ in terms of its submatrices. In analogy of Eq.~\eqref{eq:exact_multimode_compact} we write
\[\label{eq:exact_multimode_compact_pure}
%\begin{split}
H^{ij},H_c^{ij}(\be)=\tr\big[{Q}_i{Q}_j^\dag+{Q}_j{Q}_i^\dag\big]
+c\,\tr\big[\partial_i\partial_jL\big]+2\partial_i\boldsymbol{d}^\dag\sigma^{-1}\partial_j\boldsymbol{d}.
%\end{split}
\]

\subsection{Problems at points of purity}\label{sec:problems_at_pops}
%introduction
In section~\ref{sec:when_Williamson's_decomposition}  we showed two possible definitions of the problematic points of the quantum Fisher information and devised the regularization procedure in a way that takes this definition into account. Here we illustrate this difference more concretely on an example and show how different definitions relate to mixed states.

\begin{figure}[t]
\centering
\includegraphics[width=0.6\linewidth]{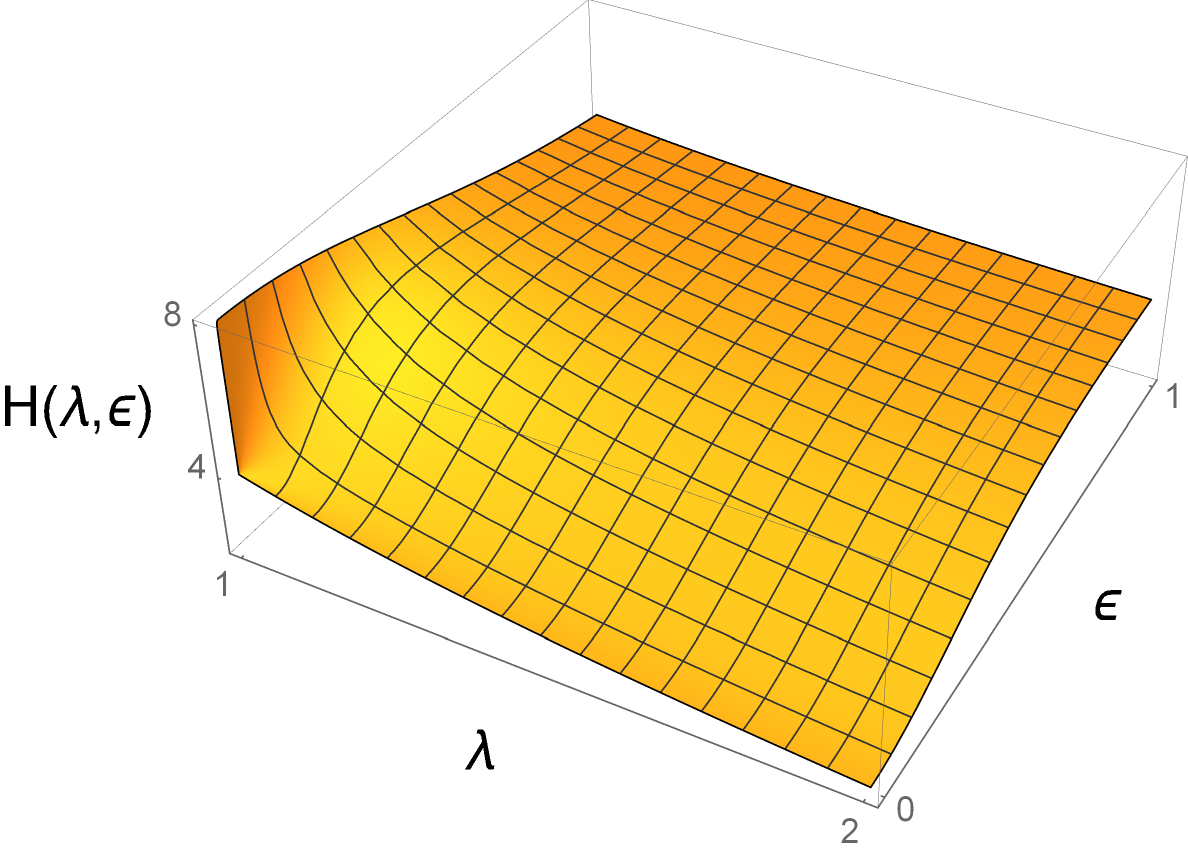}
\caption{The quantum Fisher information for the estimation of the parameter of the single mode state~\eqref{eq:problematic_example}.}\label{fig:problematic_example}
\end{figure}

\begin{example}\label{ex:problematic_example}
\emph{Let $\R$ be a displaced two-mode squeezed thermal state with the covariance matrix $\sigma=S_T(\epsilon)\lambda S_T^\dag(\epsilon)$ and the displacement vector $\bd=(\epsilon,\epsilon,\epsilon,\epsilon)^T$. We assume that experimenter has access only to the first mode, i.e., we trace over the second mode. The resulting state is a single mode state with moments\footnote{We choose $\bd$ non-zero to avoid the problem with identifiability of $\epsilon$. For $\bd=\boldsymbol{0}$ values $\epsilon$ and $-\epsilon$ would produce exactly the same statistics because $\cosh$ appearing in Eq.~\eqref{eq:problematic_example} is an even function. Consequently an experimentalist is not be able to distinguish between the two values, which would suggest the quantum Fisher information should be zero. Such an example favors the first definition of problematic points~\eqref{def:problematic_points0}. However, we wanted to illustrate differences between the two definitions on a less trivial example where the choice of an appropriate definition is not so clear.}
\[\label{eq:problematic_example}
\bd=\begin{bmatrix}
\epsilon \\
\epsilon
\end{bmatrix}, \quad \sigma=\begin{bmatrix}
\lambda\cosh (2\epsilon) & 0 \\
0 & \lambda\cosh (2\epsilon)
\end{bmatrix}.
\]
The quantum Fisher information for the estimation of the parameter $\epsilon$ is obtained either from Eq.~\eqref{eq:one_mode_quantum_fisher_information} or Eq.~\eqref{eq:multimode_QFI},
\[
H(\lambda,\epsilon)=\frac{4\lambda^2\sinh^2(2\epsilon)}{\lambda^2\cosh^2(2\epsilon)-1}
+\frac{4}{\lambda \cosh(2\epsilon)}.
\]
The first term corresponds to the Fisher information gained from the change of purity, while the second term corresponds to the Fisher information gained from the change of the displacement. The graph of this function is shown on Fig.~\ref{fig:problematic_example}. This function of two variables is continuous everywhere apart from the point $(\lambda,\epsilon)=(1,0)$ where it is not defined. There are two reasonable ways of how to define this point.
The first way will ensure that for any $\epsilon$ the quantum Fisher information is continuous in $\lambda$, but it will not be continuous in $\epsilon$ for $\lambda=1$. We define,
\[\label{def:problematic_points0_example}
H(1,0):=\lim_{\lambda\rightarrow1}\lim_{\epsilon\rightarrow0}H(\lambda,\epsilon)=4.
\]
Such a definition corresponds to the choice of the definition of problematic points for the quantum Fisher information, Eq.~\eqref{def:problematic_points0}. The quantum Fisher information is given as the limit of the quantum Fisher information of close mixed states $\R_{\lambda+\mathrm{d}\lambda,\epsilon}$.
The second way will ensure that for any $\lambda$ the Quantum Fisher information is continuous in $\epsilon$, but it will not be continuous in $\lambda$ for $\epsilon=0$. We define
\[
H(1,0):=\lim_{\epsilon\rightarrow0}\lim_{\lambda\rightarrow1}H(\lambda,\epsilon)=8.
\]
Such a definition corresponds to the choice of the definition of problematic points for the continuous quantum Fisher information, Eq.~\eqref{def:problematic_points}. The quantum Fisher information is given as the limit of the quantum Fisher information of close pure states $\R_{\lambda,\epsilon+\de}$.}
\end{example}

%physical arguments for either choice
The two choices of the definition of problematic points correspond to the difference on a more fundamental level. From the physical point of view, the pro for the first definition is that pure states, in fact, do not exist because their existence is forbidden by the third thermodynamical law. There are always some thermal fluctuations which will result for the state not to be pure. According to this argument, the quantum Fisher information for pure state should be calculated as the limit of mixed states. The pro for the second definition is that $\epsilon=0$ is the point of measure zero and thus cannot be ever measured. The real -- measured -- value of $\epsilon$ will be close to zero but never equal. It is important to remind, however, that the right figure of merit for the Cram\'er-Rao bound is the quantum Fisher information given by the first definition of problematic points, Eqs.~\eqref{def:problematic_points0} and \eqref{def:problematic_points0_example} respectively.%\footnote{Another argument for the second choice lies in the fact that in our example the $\epsilon$ is not identifiable, i.e., $\epsilon$ and $-\epsilon$ produce exactly the same statistics because $\cosh$ appearing in Eq.~\eqref{eq:problematic_example} is an even function. As a result an experimentalist cannot distinguish between the two values, therefore the quantum Fisher information should be zero. However, one could easily argue with a different example: instead of $\bd=\boldsymbol{0}$ in Eq.~\eqref{eq:problematic_example} one could choose $\bd=(\epsilon,\epsilon)^T$. This choice suddenly makes $\epsilon$ identifiable and the logic of the argument for both definitions of problematic points can be repeated in the very same manner.}

%problem with two parameters and smoothness of the QFI
Such a problematic behavior of the quantum Fisher information can never be avoided when dealing with pure states which change its purity. The issue becomes especially problematic when we want to analyze the quantum Fisher information depending on two parameters as illustrated on example~\ref{ex:problematic_example}. We formalize this problem in the following theorem,
\begin{theorem}
Let quantum state $\R\equiv(\bd(\epsilon),\sigma(p,\epsilon))$ be a smooth function of two parameters, where $\epsilon$ is the parameter we are going to estimate, and where parameter $p$ is encoded only in the symplectic eigenvalues of the covariance matrix. If there exists a symplectic eigenvalue $\lambda_k(p,\epsilon)$ of the covariance matrix such that $\lambda_k(p,\epsilon)=1$, but $\lambda_k(p+\mathrm{d} p,\epsilon)>1$ and $\partial_{\epsilon\epsilon}\lambda_k(p,\epsilon)\neq 0$, then the multi-parameter Taylor expansion of the form
\[\label{not_existing_form}
\begin{split}
H(\sigma(p+\mathrm{d} p,\epsilon+\mathrm{d} \epsilon))&=H(\sigma(p,\epsilon))+\partial_pH(\sigma(p,\epsilon))\mathrm{d} p+\partial_\epsilon H(\sigma(p,\epsilon))\mathrm{d} \epsilon\\
&+\mathcal{O}(\mathrm{d} p^2,\mathrm{d} p\mathrm{d} \epsilon, \mathrm{d} \epsilon^2)
\end{split}
\]
does not exist for either choice of definition of problematic points.
\end{theorem}
The proof is based on the fact that either $\partial_pH(\sigma(p,\epsilon))=-\infty$ or $\partial_\epsilon H(\sigma(p,\epsilon))=+\infty$ from which the theorem immediately follows. The full proof can be found in appendix~\eqref{app:not_existing_form}. Intuition why this theorem should hold can be also obtained by studying the second term in Eq.~\eqref{eq:H_expanded}, which shows how the quantum Fisher information behave around problematic points.

%outlook
The problem of points of purity is not limited to Gaussian states. As we can see on Fig.~\ref{BobsQFIexample3} or Fig.~\ref{fig:Continuous_QFI}, a non-Gaussian state can suffer of the same discontinuity.
%The different choice of the problematic terms also illustrates that not all formulae for the quantum Fisher information are entirely equivalent. While the first choice of the problematic points~\eqref{def:problematic_points} corresponds to the quantum Fisher information given by Eq.~\eqref{QFI_using_fidelity}, the second choice~\eqref{def:problematic_points2} corresponds to Eq.~\eqref{QFI}. However, these formulae coincide when every eigenvalue of the density matrix $\R$ is either non-zero or it is zero but it remains zero for a small variation in $\epsilon$, i.e. $\lambda(\epsilon)=0$ and $\lambda(\epsilon+\mathrm{d}\epsilon)=0$.
We will encounter this problem again in the next chapter where we are trying to expand the quantum Fisher information at the same time in the small space-time parameter $\epsilon$ and in a small parameter connected to temperature $Z\,=\,e^{-E_{m}/2T}$.

\subsection{Unitary encoding operations}

Let us assume that a single parameter was encoded into an initial Gaussian state $\R_0$ via a Gaussian unitary transformation $\hat{U}_\epsilon$ which forms a one parameter group. We construct this unitary from the general Gaussian unitary~\eqref{def:Gaussian_unitary},
\[\label{def:Gaussian_unitary_epsilon}
\hat{U}_\epsilon=\exp\big(\big(\tfrac{i}{2}\bA^\dag W \bA+\bA^\dag K \bg\big)\epsilon\big).
\]
The final state $\R_\epsilon=\hat{U}_\epsilon\R_0\hat{U}_\epsilon^\dag$ is given by the first and the second moments
\begin{subequations}\label{eq:moments_epsilon}
\begin{align}
\bd_\epsilon&=S_\epsilon\bd_0+\bb_\epsilon,\\
\sigma_\epsilon&=S_\epsilon \sigma_0 S_\epsilon^\dag,
\end{align}
\end{subequations}
where $S_\epsilon=e^{iKW\epsilon}$ and $\bb_\epsilon=\big(\int_0^1e^{iKWt\epsilon}\mathrm{d}t\big)\bg\epsilon$. We have $\dot{\sigma}_\epsilon=S_\epsilon[iKW,\sigma_0]S_\epsilon^\dag$, $\dot{\bd}_\epsilon=S_\epsilon(iKW\bd_0+\bg)$, and the quantum Fisher information is independent of $\epsilon$.

We can use Eq.~\eqref{eq:mixed_QFI} and properties of the Kronecker product~\eqref{id:Kronecker_product_ids} to derive the quantum Fisher information for mixed Gaussian states,
\begin{subequations}\label{eq:mixed_QFI_unitary}
\begin{align}
H(\epsilon)&=\frac{1}{2}\mathrm{vec}\big[[iKW,\sigma_0]\big]^\dag\mathfrak{M}^{-1}\mathrm{vec}\big[[iKW,\sigma_0]\big]
+2(iKW\bd_0+\gamma)^\dag\sigma_0^{-1}(iKW\bd_0+\gamma),\\
\mathfrak{M}&=\ov{\sigma}_0\otimes\sigma_0-K\otimes K.
\end{align}
\end{subequations}
Similarly, the quantum Fisher information for pure Gaussian states is obtained from Eq.~\eqref{eq:pure_non-elegant},
\[
H(\epsilon)=\frac{1}{2}\mathrm{tr}\big[[iKW,\sigma_0]\ \!\sigma_0^{-1}iKW\big]+2(iKW\bd_0+\gamma)^\dag\sigma_0^{-1}(iKW\bd_0+\gamma).\\
\]
The scenario in which the Willimson's decomposition of the covariance matrix is known is discussed in section~\ref{sec:general_method}.

The above formulae can be generalized to multiparameter estimation where parameters are encoded via a Gaussian unitary $\hat{U}_{\be}=\mathrm{exp}\big(\sum_i\hat{G}_i\epsilon_i\big)$. However, in the case when $\hat{G}_i$ do not commute the quantum Fisher information matrix cannot be simplified and in general depends on the vector of parameters $\be$.

\section{Estimation of Gaussian unitary channels}\label{sec:estimation_of_channels}

We have shown in the first chapter that finding optimal probe states for the estimation of quantum channels is one of the main tasks of quantum metrology. In this section we develop a practical method for finding optimal Gaussian probe states and illustrate this method on the estimation of the most common Gaussian unitary channels. This section is partially based on results we published in~\cite{safranek2016optimal}.

\begin{figure}[t!]
\centering
\includegraphics[width=1\linewidth]{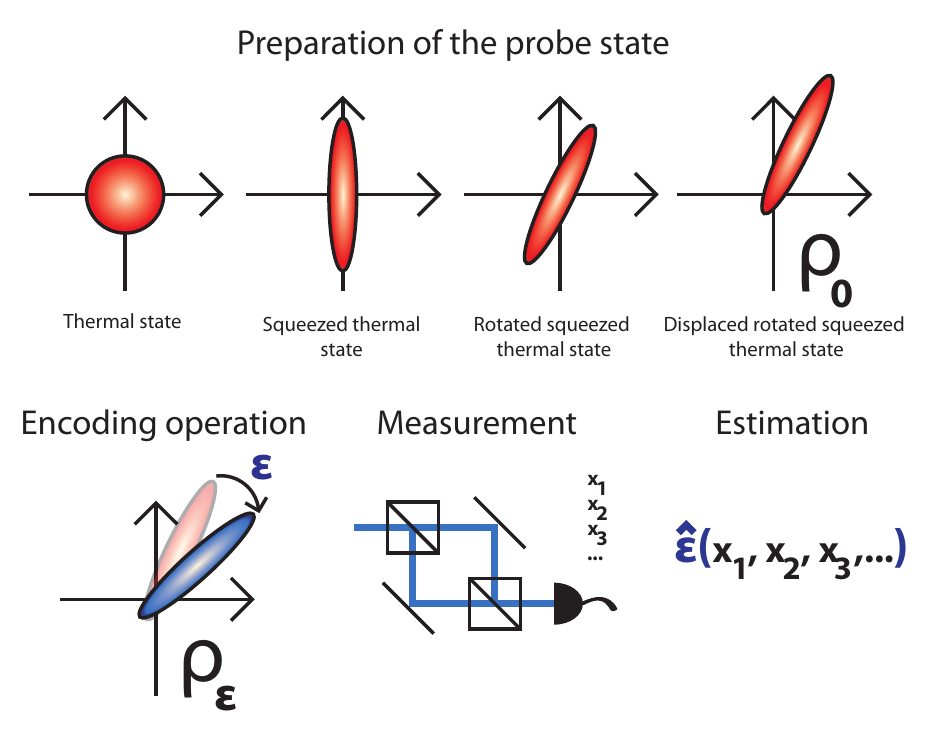}
\caption{Scheme of the usual metrology setting illustrated on a one-mode Gaussian probe state. First, we prepare the state by using various Gaussian operations, then we feed the state into the channel we want to estimate, perform an appropriate measurement, and an estimator $\hat{\epsilon}$ gives us an estimate of the true value of the parameter. In this section we are interested in optimizing over the preparation stage for a given encoding Gaussian unitary channel $\hat{U}(\epsilon)$.
}\label{fig:scheme}
\end{figure}

The typical setup is given by scheme~\eqref{scheme_for_estimating_channels} which is illustrated on Fig.~\ref{fig:scheme} where a single mode Gaussian states is being used to estimate a one-mode Gaussian channel. The probe state is fed into the channel, the channel encodes the parameter on the state of the system and, finally, measurements are performed with the aim of gaining maximal information about the parameter.

Gaussian states are usually not optimal probe states for estimating Gaussian unitary channels. Non-gaussian states such as GHZ states usually perform as better probes. However, previous theoretical studies show that Gaussian states can be still effectively used for the estimation of Gaussian channels such as phase-changing~\cite{Monras2006a,Aspachs2008a,Sparaciari2015a,Sparaciari2015b}, squeezing~\cite{Milburn1994a,Chiribella2006a,Gaiba2009a}, two-mode squeezing and mode-mixing channels~\cite{Gaiba2009a}. Previous studies analyzed specific channels and for each channel \emph{only one} probe state achieving the Heisenberg limit was found. Moreover, Gaussian state metrology is often restricted to pure states. Less attention has been given to thermal states, which are of great relevance in practice. In the laboratory, quantum states can never be isolated from the environment which thermalizes them. In this section we develop a formalism that can effectively be used to study \emph{any} Gaussian probe state for \emph{any} one- and two-mode Gaussian unitary channel. We also develop methods to find \emph{all} optimal Gaussian probe states for these channels. We take advantage of recent progress in the phase-space formalism of Gaussian states. We make use of Euler's decomposition of symplectic matrices~\eqref{def:S_decomposition}, the Williamson's decomposition of the covariance matrix in the complex form~\eqref{def:Williamson's_decomposition}, and the expression for the quantum Fisher information in terms of the Williamson's decomposition of the covariance matrix~\eqref{eq:multimode_QFI}. These techniques enable us to simplify expressions so that formulae can be easily used in practical applications. As an example, we derive optimal states for channels that, to our knowledge, have not been studied before. These are the channels such as generalized mode-mixing, two-mode squeezing, or combining the phase-change and squeezing. Interestingly, we find that in the estimation of two-mode channels, separable states consisting of two one-mode squeezed states perform as well as their entangled counterpart: two-mode squeezed states. This shows that entanglement between the modes does not enhance precision in this case.

Our formalism also enables us to further our understanding of the effects of temperature in probe states. It has been reported in~\cite{Aspachs2008a} that higher temperature in squeezed thermal states can enhance phase estimation, while higher temperature of displaced thermal states is detrimental. We show that the effects of thermalised probe states on the estimation of Gaussian channels are generic, i.e., for all Gaussian unitary channels, temperature effects are always manifested in multiplicative factors of four types. Two of the factors correspond to the ones previously found in~\cite{Aspachs2008a}. The other two factors, that we discovered, show that not only temperature of the probe state, but also temperature difference between different modes of the probe state helps the estimation.

This section is organized as follows: first we present a general framework for finding optimal Gaussian probe states for Gaussian unitary channels. Then we study the effects of temperature on the estimation strategy and show that effects of temperature are generic. We apply our formalism to present concrete examples for one- and two-mode Gaussian unitary channels and we generalize bounds on the ultimate limit of precision of estimation of Gaussian channels found in~\cite{Milburn1994a,Chiribella2006a,Monras2006a,Aspachs2008a,Gaiba2009a}. The previous examples included mixed states, however we design a simplified way of how to find optimal pure probe states for the estimation of an arbitrary Gaussian unitary channel. Finally we discuss the connection between entanglement of Gaussian probe states with the Heisenberg scaling and summarize our findings.

\subsection{General method}\label{sec:general_method}

In this subsection we are going to describe the general method of finding the optimal probe states. Put simply, first we take parametrization of general one- or two- mode Gaussian states and use the formula for the quantum Fisher information in terms of the Williamson's decomposition derived in the previous section. Finally we maximize the quantum Fisher information by choosing appropriate parameters in the parametrization of the probe state.

Let us assume we have full control over the preparation of the initial probe state $\hat{\rho}_0\equiv(\bd_0,\sigma_0)$, with the Williamson decomposition $\sigma_0=S_0D_0S_0^\dag$ of the covariance matrix. The diagonal matrix, $D_0$, represents a thermal state and the symplectic matrices, $S_0$, represent the operations we are going to perform on this thermal state. After the probe state is created, we feed it into a Gaussian channel encoding the parameter we want to estimate.

%Although the method we are going to describe can be easily used to estimate any Gaussian unitary channel~\eqref{def:Gaussian_unitary}, for simplicity we consider only channels with purely quadratic generators represented by a single symplectic matrix $S_\epsilon$. These are obtained from Eq.~\eqref{def:Gaussian_unitary} by setting $\bg=0$.

Using Eqs.~\eqref{def:transformation} we find the final state is given by the first and the second moments
\begin{subequations}\label{eq:moments_epsilon}
\begin{align}
\bd_\epsilon&=S_\epsilon\bd_0+\bb_\epsilon,\\
\sigma_\epsilon&=S_\epsilon S_0D_0S_0^\dag S_\epsilon^\dag.
\end{align}
\end{subequations}
As the covariance matrix appears precisely in the form of the Williamson decomposition, we can use formula~\eqref{eq:multimode_QFI} directly. Applying Eqs.~\eqref{def:structure_of_S}, \eqref{def:P_i}, and \eqref{eq:moments_epsilon}, we derive
\begin{subequations}\label{eq:eq_for_general_method}
\begin{align}
&P=S_0^{-1}P_\epsilon S_0,\label{eq:first_part}\\
&\sum_{k=1}^N\frac{\dot{\lambda_k}^2}{\lambda_k^2-1}=0,\label{eq:second_part}\\
&2\dot{\bd}^\dag\sigma^{-1}\dot{\bd}=2(P_\epsilon\bd_0+S_\epsilon^{-1}\dot{\bb}_\epsilon)^\dag\sigma_0^{-1}(P_\epsilon\bd_0+S_\epsilon^{-1}\dot{\bb}_\epsilon),\label{eq:displacement_part}
\end{align}
\end{subequations}
where we have denoted $P_\epsilon:=S_\epsilon^{-1}\dot{S}_\epsilon$. Due to the unitarity of the channel the symplectic eigenvalues do not change, and the expression~\eqref{eq:second_part} vanishes. This scheme can be used for any Gaussian unitary channel. However, in next sections we are going to study Gaussian unitary channels which form a one-parameter unitary group given by Eq.~\eqref{def:Gaussian_unitary_epsilon}. In that case $P_\epsilon=iKW$, $\dot{\bb}_\epsilon=S_\epsilon\bg$, and the quantum Fisher information is independent of $\epsilon$. The problem of finding the optimal states is thus reduced to finding the optimal parameters from the parametrization of the initial state for a given Gaussian channel represented by a constant matrix $W$ and a constant vector $\bg$.

\subsection{Effects of temperature}\label{sec:effects_of_temperature}

It is interesting to note that the symplectic eigenvalues in Eq.~\eqref{eq:multimode_QFI} appear only in a form of multiplicative factors, independent of other parameters and channels we estimate.

This is particularly interesting from a physical point of view because, as we explained in sections~\ref{sec:thermal_state} and~\ref{sec:Williamson's_decomposition}, the symplectic eigenvalues encode temperature. The symplectic eigenvalue describing a thermal state of the harmonic oscillator with frequency $\omega_k$ is given by $\lambda_k=\coth(\frac{\omega_k\hbar}{2kT})$, or alternatively, $\lambda_k=1+2n_{{\mathrm{th}}k}$ where $n_{{\mathrm{th}}k}$ denotes the mean number of thermal bosons in each mode.

In Eq.~\eqref{eq:multimode_QFI} we can identify four types of multiplicative factors given by symplectic eigenvalues, $\frac{\lambda_k^2}{1+\lambda_k^2}$, $\frac{(\lambda_k+\lambda_l)^2}{\lambda_k\lambda_l+1}$, $\frac{(\lambda_k-\lambda_l)^2}{\lambda_k\lambda_l-1}$, and $\frac{1}{\lambda_k}$. First, let us focus on effects of temperature given by the first three types of factors which multiply matrices $R$ and $Q$. These represents sensitivity of squeezing and orientation of squeezing of the probe state with respect to the channel we estimate.
The first type of factor, $\frac{\lambda_k^2}{1+\lambda_k^2}$, is one of the two to appear for the \emph{isothermal} (sometimes called \emph{isotropic}) states for which all symplectic eigenvalues are equal. This class also encompasses all pure states. Because $1\leq\lambda_k\leq+\infty$, we have $\frac{1}{2}\leq\frac{\lambda_k^2}{1+\lambda_k^2}\leq1$, where the lower bound is attained by pure states and the upper bound by thermal states with infinite temperature. This means that for isothermal states temperature helps the estimation with maximal enhancement of a factor of two, a fact already noted in~\cite{Aspachs2008a}. Next, for mixed multi-mode states we have the second and third type of factors, $\frac{(\lambda_k-\lambda_l)^2}{\lambda_k\lambda_l-1}$ and $\frac{(\lambda_k+\lambda_l)^2}{\lambda_k\lambda_l+1}$. These terms become especially important when there is a large difference between the symplectic eigenvalues. Considering $\lambda_l \rightarrow 1$ we have
\begin{subequations}
\begin{align}
\frac{(\lambda_k-\lambda_l)^2}{\lambda_k\lambda_l-1}&\longrightarrow\lambda_k-1=2n_{{\mathrm{th}}k},\\
\frac{(\lambda_k+\lambda_l)^2}{\lambda_k\lambda_l+1}&\longrightarrow\lambda_k+1=2(n_{{\mathrm{th}}k}+1).
\end{align}
\end{subequations}
Generally, assuming $\lambda_k\gg \lambda_l$ yields
\[
\frac{(\lambda_k-\lambda_l)^2}{\lambda_k\lambda_l-1}\approx\frac{(\lambda_k+\lambda_l)^2}{\lambda_k\lambda_l+1}\approx\frac{2n_{{\mathrm{th}}k}}{2n_{{\mathrm{th}}l}+1}.
\]
This shows that the enhancement by temperature difference is no longer bounded by some fixed value as in the previous case.

If we keep one mode sufficiently cool and the other hot, or if one mode has a high frequency and the other a low frequency, we can, in principle, achieve an infinite enhancement in the estimation of the unknown parameter. In general, states with a large variance in energy, which in this case is in the form of thermal fluctuations, have a higher ability to carry information, and thus can carry more information about the parameter we want to estimate. We will refer to this phenomenon as temperature-enhanced estimation.

We have shown that temperature and temperature difference enhances the first two terms in Eq.~\eqref{eq:multimode_QFI} due to first three types of factors. However, the opposite behaviour is observed in the last term. This last term shows how sensitive the displacement is to the small changes in the parameter of the channel. Factors of the fourth type, $\frac{1}{\lambda_k}$, are hidden in the inverse of the initial covariance matrix in this last term as shown in Eq.~\eqref{eq:displacement_part}. As temperature rises and the symplectic eigenvalues grow to infinity, this factor goes to zero and the precision in estimation diminishes.

Let us look at what these factors mean physically for different probe states. Channels quadratic in the field operators do not affect the displacement of non-displaced states such as squeezed thermal states. This means that the precision in estimation of such channels when using non-displaced states will be affected only by factors of the first three types. When using a squeezed thermal state as a probe, temperature and temperature difference in different modes of this probe will always help the estimation. In contrast, when a displaced thermal state is used as a probe, the effect of quadratic channels on the squeezing of such probes is very minor. In other words, covariance matrix of displaced thermal states is almost unchanged by a quadratic channel and completely unchanged in the case of passive channels. Therefore the first three types of factor play a minor role. Quadratic channels will greatly change the displacement of a displaced thermal state therefore the factor of the last type $\frac{1}{\lambda_k}$ is of great relevance. Higher temperature in displaced thermal states decreases the precision of estimation of quadratic channels. Put simply, it is good to have either a hot squeezed state or a cold displaced state as a probe.

\subsection{Optimal probe states for the estimation of Gaussian unitary channels}
In this subsection we illustrate the general methods of finding optimal probe state for the estimation of parameters of special class of Gaussian unitary channels. These channels will be one- and two- mode  unitary channels generated by a purely quadratic Hamiltonian.

\subsubsection{Estimation of one-mode channels: combining squeezing and phase-change}

%definition of the channel
First we are going to look at the estimation of one-mode Gaussian unitary channels with purely quadratic generators. These channels are fully parametrized by Eq.~\eqref{eq:W1}. We will look on a channel which combines the squeezing and a phase-change which is constructed by substituting $\theta\rightarrow\omega_p\epsilon$, $r\rightarrow \omega_s\epsilon$ in Eq.~\eqref{eq:W1}. The resulting symplectic matrix $S_\epsilon:=e^{iKW}$ then represents an encoding operator $\hat{S}_\epsilon=\exp((-i\omega_p\hat{a}^\dag \hat{a}-\frac{\omega_s}{2}(e^{i\chi}\hat{a}^{\dag2}-e^{-i\chi}\hat{a}^{2}))\epsilon)$. $\omega_p$ and $\omega_s$ are the frequencies with which the state is rotated and squeezed respectively. We assume these frequencies and the squeezing angle $\chi$ are known, so $\epsilon$ is the only unknown parameter we are trying to estimate.

%QFI
Using the general probe state~\eqref{eq:general_1mode_state} we derive the quantum Fisher information,
\[\label{eq:one_mode_channel_QFI}
\begin{split}
H(\epsilon)&=\frac{4\lambda_1^2}{1+\lambda_1^2}\Big(\omega_s^2\big(\cos^2(2\theta+\chi)+\cosh^2(2r)\sin^2(2\theta+\chi)\big)\\
&+\omega_p^2\sinh^2(2r)-\omega_s\omega_p\sin(2\theta+\chi)\sinh(4r)\Big)\\
&+\frac{4\norm{{d}}^2}{\lambda_1}\Big(e^{2r}\big(\omega_s\cos(\theta-\phi_d+\chi)-\omega_p\sin(\theta+\phi_d)\big)^2\\
&+e^{-2r}\big(\omega_s\sin(\theta-\phi_d+\chi)+\omega_p\cos(\theta+\phi_d)\big)^2\Big).
\end{split}
\]
Assuming all $\omega_s,\omega_p,r$ are positive, this function clearly achieves its maximum when $\sin(2\theta+\chi)=-1$, $\sin(\theta-\phi_d+\chi)=1$, and $\sin(\theta+\phi_d)=-1$. For example, these conditions are fulfilled when $\theta=-\frac{\chi}{2}-\frac{\pi}{4}$, $\phi_d=\frac{\chi}{2}-\frac{\pi}{4}$, which leads to
\[\label{eq:one_mode_channel_QFI_max}
H_{\mathrm{max}}(\epsilon)=\frac{4\lambda_1^2}{1+\lambda_1^2}\big(\omega_s\cosh(2r)+\omega_p\sinh(2r)\big)^2+\frac{4\norm{{d}}^2}{\lambda_1}e^{2r}\big(\omega_s+\omega_p\big)^2.
\]
This shows that both displacement and squeezing, if properly oriented, enhance the estimation precision. However, to study what strategy is the best when only a fixed amount of energy of the probe state is available we use the relation for the mean total number of Bosons given by Eq.~\eqref{eq:mean_number_of_particles} and derive
\[\label{eq:mean_number_of_bosons_one}
n=n_{{d}}+n_{\mathrm{th}}+(1+2n_{\mathrm{th}})\sinh^2r,
\]
where $n_{{d}}:=\norm{{d}}^2$ denotes the mean number of Bosons coming from the displacement. Together with the relation $\lambda_1=1+2n_{\mathrm{th}}$ Eq.~\eqref{eq:one_mode_channel_QFI_max} transforms into
\[
\begin{split}
H_{\!\mathrm{max}}(\epsilon)&=\frac{2\Big(\omega_s(2n\!-\!2n_{{d}}\!+\!1)+2\omega_p\sqrt{\!n\!-\!n_{{d}}\!-\!n_{\mathrm{th}}\!}\sqrt{\!n\!+\!1\!-\!n_{{d}}\!+\!n_{\mathrm{th}}\!}\Big)^{\!2}\!\!\!\!}{1\!+\!2n_{\mathrm{th}}(1\!+\!n_{\mathrm{th}})}\\ &+\frac{4n_{{d}}\big(2n\!-\!2n_{{d}}\!+\!1\!+\!2\sqrt{\!n\!-\!n_{{d}}\!-\!n_{\mathrm{th}}\!}\sqrt{\!n\!+\!1\!-\!n_{{d}}\!+\!n_{\mathrm{th}}\!}\big)^2\!\!\!}{(1\!+\!2n_{\mathrm{th}})^2}(\omega_s\!+\!\omega_p)^2
\end{split}
\]
Keeping $n$ fixed, the maximum is achieved when $n_{\mathrm{th}}=n_{{d}}=0$, i.e., when all available energy is invested into squeezing, which coincides with some special cases~\cite{Aspachs2008a,Gaiba2009a}. The quantum Fisher information then reaches the Heisenberg limit,
\[\label{eq:optimized_QFI_squeezing_phase}
H_{\mathrm{max}}(\epsilon)=2\big(\omega_s(2n+1)+\omega_p2\sqrt{n}\sqrt{1+n}\big)^2.
\]
On the other hand, if we decide to invest only into the displacement (which corresponds to the coherent probe state), i.e., $n=n_{{d}}$, we obtain the shot-noise limit $H_{\mathrm{max}}(\epsilon)=2\omega_s^2+4n(\omega_s+\omega_p)^2$.

\begin{figure}[t!]
\centering
\includegraphics[width=0.8\linewidth]{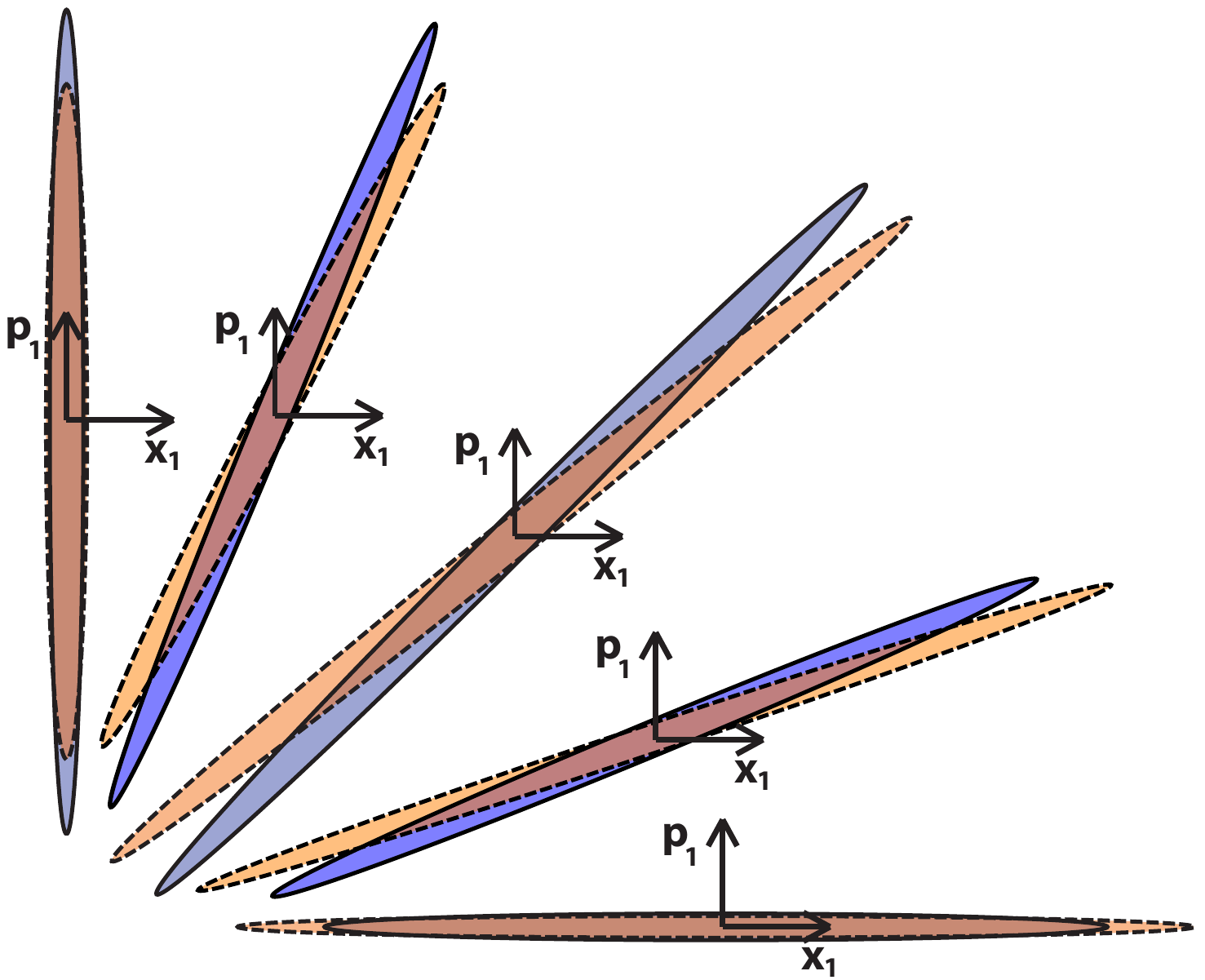}
\caption{Estimation of the one-mode squeezing channel $\hat{S}(\epsilon)$ around point $\epsilon=0$ using various squeezed states. Each squeezed state is represented by a covariance matrix depicted as an ellipse. Each state has the same energy, but the initial rotation varies. The initial squeezing was set to $r=0.8$, the initial displacement $\tilde{d_0}=0$, and the final squeezing $\epsilon=0$~(blue with full line) or $\epsilon=0.1$~(orange with dashed line). The initial rotation from left to right $\theta=0,\frac{\pi}{8},\frac{\pi}{4},\frac{3\pi}{8},\frac{\pi}{2}$. Covariance matrices with $\theta=\frac{\pi}{4}$ can be easily distinguished allowing for the optimal estimation of the parameter $\epsilon$.}\label{fig:three_cov}
\end{figure}

%phase-changing and squeezing channel
The phase changing and squeezing channel are the special cases of the channel introduced above. The phase changing channel is given by $(\omega_p,\omega_s)=(1,0)$ and squeezing channel by $(\omega_p,\omega_s)=(0,1)$. Corresponding quantum Fisher information and optimized quantum Fisher information is given by Eq.~\eqref{eq:one_mode_channel_QFI} and~\eqref{eq:optimized_QFI_squeezing_phase}. It is interesting to note that the optimal probe state for a phase-changing channel is an arbitrary squeezed state achieving the quantum Fisher information of $H(\epsilon)=8n(n+1)$, while for a squeezing channel it is the vacuum squeezed in the $45^\circ$ from the angle from which the channel squeezes. Such state then achieves the Heisenberg scaling $H(\epsilon)=2(2n+1)^2$. For illustration how different squeezed states perform in the estimation of the squeezing channel, see Fig.~\ref{fig:three_cov}.

\subsubsection{Estimation of two-mode channels: mode-mixing and two-mode squeezing channels}

%general scheme
In this section we are going to study the estimation of two-mode Gaussian unitary channels with purely quadratic generators, namely two-mode squeezing and mode-mixing channels, using a wide class of two-mode mixed probe states and the general two-mode pure state. In the analogy with one-mode Gaussian channels, these channels are fully parametrized by Hermitian matrix $W$ given by Eq.~\eqref{eq:W2}.

%probe state
Although analysis with the general probe state~\eqref{eq:general_2mode_state} can be made, the results seem to be too complicated to be used effectively. Also, as the first three operations applied on the thermal state only swap and entangle the symplectic eigenvalues, we do not expect much generality will be lost when not considering them. Moreover, in the case of the isothermal states (which also covers all pure states), such operations do not have any effect. This is why we restrict ourselves to probe states which we write in the covariance matrix formalism as
\begin{subequations}\label{eq:simplified_probe_state}
\begin{gather}
\bd_0=(\tilde{\bg},\ov{\tilde{\bg}})^T,\\
\sigma_0=R_1(\phi_1)R_2(\phi_2)B(\theta)R_{\mathrm{as}}(\psi)S_1(r_1)S_2(r_2)D_0(\cdot)^\dag,
\end{gather}
\end{subequations}
where $\tilde{\bg}=(\norm{{d}_1}e^{i\phi_{d1}},\norm{{d}_2}e^{i\phi_{d2}})$ and $D_0=\mathrm{diag}(\lambda_1,\lambda_2,\lambda_1,\lambda_2)$. Also, it is believed that mixed states cannot improve the
quality of estimation when fixing the energy of the probe state, the optimal states are always pure.\footnote{This belief comes fact that the quantum Fisher information is a convex function on the space of density matrices. However, there are some problems for more detailed discussion see section~\ref{sec:are_pure_optimal}.}%<-\Ref{This "It is believed" is very dodgy}
As Eq.~\eqref{eq:simplified_probe_state} encompasses all pure states, it is enough to use this restricted class of states to find the optimal.

\textbf{Two-mode squeezing channel.}
First we are going to study the optimal states for the estimation of the two-mode squeezing channel $\hat{S}_{T}(\epsilon,\chi)$, assuming the direction of squeezing $\chi$ is known. This channel was introduced in Eq.~\eqref{eq:twomode_squeezing_operator}. Using the state from Eq.~\eqref{eq:simplified_probe_state} we find only two cases which lead to significantly different results. In the first case a beam-splitter is not used ($\theta=0$) in the preparation process, which corresponds to using two simultaneously sent, but non-entangled single-mode squeezed probe states. In the second case the balanced beam-splitter is used ($\theta=\pi/4$), which corresponds to using two-mode squeezed-type probe states. The full expression for the quantum Fisher information is a mixture of these two qualitatively different cases. Defining $\phi_\chi:=\phi_1+\phi_2+\chi$, $\phi_{1\chi}:=\phi_1-\phi_{d2}+\chi$, $\phi_{2\chi}:=\phi_2-\phi_{d1}+\chi$, the quantum Fisher information for the estimation of a two-mode squeezing channel $\hat{S}_{T}(\epsilon,\chi)$ reads
\[
\begin{split}
H(\epsilon)&=2\cos^2(2\theta)\bigg(\!\frac{(\lambda_1\!+\!\lambda_2)^2}{\lambda_1\lambda_2\!+\!1}\big(\!\cos^2\!\!\phi_\chi\cosh^2(\!r_1\!-\!r_2\!)\!+\!\sin^2\!\!\phi_\chi\cosh^2(\!r_1\!+\!r_2\!)\big)\\
&+\frac{(\lambda_1\!-\!\lambda_2)^2}{\lambda_1\lambda_2\!-\!1}\big(\!\cos^2\!\!\phi_\chi\sinh^2(\!r_1\!-\!r_2\!)\!+\!\sin^2\!\!\phi_\chi\sinh^2(\!r_1\!+\!r_2\!)\!\big)\!\bigg)\\
&+4\sin^2(2\theta)\bigg(\frac{\lambda_1^2}{\lambda_1^2+1}\big(\cos^2(\phi_\chi+2\psi)+\sin^2(\phi_\chi+2\psi)\cosh(2r_1)\big)\\
&+\frac{\lambda_2^2}{\lambda_2^2+1}\big(\cos^2(\phi_\chi-2\psi)+\sin^2(\phi_\chi-2\psi)\cosh(2r_2)\big)\bigg)\\
&+\frac{4}{\lambda_1}\Big(e^{2r_1}\big(\norm{{d}_1}\sin\theta\cos(\phi_{2\chi}+\psi)-\norm{{d}_2}\cos\theta\cos(\phi_{1\chi}+\psi)\big)^2\\
&+e^{-2r_1}\big(\norm{{d}_1}\sin\theta\sin(\phi_{2\chi}+\psi)-\norm{{d}_2}\cos\theta\sin(\phi_{1\chi}+\psi)\big)^2\Big)\\
&+\frac{4}{\lambda_2}\Big(e^{2r_2}\big(\norm{{d}_1}\sin\theta\cos(\phi_{2\chi}-\psi)+\norm{{d}_2}\cos\theta\cos(\phi_{1\chi}-\psi)\big)^2\\
&+e^{-2r_2}\big(\norm{{d}_1}\sin\theta\sin(\phi_{2\chi}-\psi)+\norm{{d}_2}\cos\theta\sin(\phi_{1\chi}-\psi)\big)^2\Big).\\
\end{split}
\]

Assuming both $r_1$ and $r_2$ are positive, one of the optimal states is given by setting $\theta=0$, $\phi_1=\phi_2=\frac{\pi}{4}-\frac{\chi}{2}$, $\phi_{d1}=\phi_{d2}=\frac{\pi}{4}+\frac{\chi}{2}$ which corresponds to two one-mode squeezed displaced thermal states, squeezed in the angle of $45^\circ$ from the squeezing angle of the channel. This probe states then leads to
\[\label{eq:max_QFI_nonentangled_twomode}
\begin{split}
H_{\!\mathrm{max}}\!(\epsilon)\!\!&=\!\!\frac{2(\lambda_1\!\!+\!\!\lambda_2)^2\!\!}{\lambda_1\lambda_2\!+\!1}\cosh^2\!(\!r_1\!+\!r_2\!)
+\frac{2(\lambda_1\!\!-\!\!\lambda_2)^2\!\!}{\lambda_1\lambda_2\!-\!1}\sinh^2\!(\!r_1\!+\!r_2\!)\\
&+\frac{4\norm{{d}_2}^2}{\lambda_1}e^{2r_1}
+\frac{4\norm{{d}_1}^2}{\lambda_2}e^{2r_2}.
\end{split}
\]

In contrast, for $r_1\leq0$, $r_2\geq0$, one of the optimal states are given by setting $\theta=\frac{\pi}{4}$, $\phi_\chi=0$, $\psi=\phi_{1\chi}=\phi_{2\chi}=\frac{\pi}{4}$ and leads to
\[\label{eq:max_QFI_entangled_twomodeb}
\begin{split}
H_{\mathrm{max}}(\epsilon)&=\frac{4\lambda_1^2}{\lambda_1^2+1}\cosh^2(2r_1)
+\frac{4\lambda_2^2}{\lambda_2^2+1}\cosh^2(2r_2)\\
&+\frac{2}{\lambda_1}(\norm{{d}_1}-\norm{{d}_2})^2e^{-2r_1}
+\frac{2}{\lambda_2}(\norm{{d}_1}+\norm{{d}_2})^2e^{2r_2}.
\end{split}
\]
This probe state corresponds to the two-mode squeezed displaced thermal state.

However, optimizing over the energy of the probe using the relation for the mean number of particles,
\[\label{eq:mean_number_of_bosons_two}
n=n_{{d}_1}+n_{\mathrm{th}1}+\lambda_1\sinh^2r_1
+n_{{d}_2}+n_{\mathrm{th}2}+\lambda_2\sinh^2r_2,
\]
where $n_{{d}_i}:=\norm{{d}_i}^2$, and $\lambda_i=1+2n_{\mathrm{th}i}$, $i=1,2$, we find that all states perform the best when all energy is invested into squeezing and the squeezing parameters are equal. Both Eq.~\eqref{eq:max_QFI_nonentangled_twomode} and Eq.~\eqref{eq:max_QFI_entangled_twomodeb} then lead to the same Heisenberg limit $H_{\mathrm{max}}(\epsilon)=4(n+1)^2$. To conclude, the optimal probe states for estimating two-mode squeezing channel are either two one-mode squeezed states or a two-mode squeezed state while squeezing is in the angle of $45^\circ$ from the squeezing angle of the channel. This adds to the current knowledge of optimal states for squeezing channels, since until now the research has been focused only on one type of a two-mode squeezing channel and non-entangled probe states~\cite{Gaiba2009a}. Investing all energy into the displacement of the state we obtain the shot-noise limit $H_{\mathrm{max}}(\epsilon)=4(n+1)$. The same shot noise limit is also achieved by any single-mode state used as a probe for the two-mode squeezing channel.

\textbf{Mode-mixing channel.}
Using the probe state from Eq.~\eqref{eq:simplified_probe_state}, and defining $\phi_\chi:=\phi_1-\phi_2+\chi$, $\phi_{1\chi}:=\phi_1+\phi_{d2}+\chi$, $\phi_{2\chi}:=\phi_2+\phi_{d1}-\chi$, the quantum Fisher information for the estimation of a mode-mixing channel $\hat{B}(\epsilon,\chi)$ introduced in Eq.~\eqref{eq:mode_mixing_operator} reads
\[\label{eq:mode_mixing_full}
\begin{split}
H(\epsilon)&=4\sin^2(2\theta)\sin^2\!\!\phi_\chi\bigg(\frac{\lambda_1^2}{\lambda_1^2+1}\sinh^2(2r_1)+\frac{\lambda_2^2}{\lambda_2^2+1}\sinh^2(2r_2)\bigg)\\
&+\frac{2(\lambda_1\!+\!\lambda_2)^2}{\lambda_1\lambda_2\!+\!1}\!\Big(\!\big(\!\cos(\!2\theta\!)\sin\!\phi_\chi\sin(\!2\psi\!)\!-\!\cos\!\phi_\chi\cos(\!2\psi\!)\!\big)^2\!\!\sinh^2(\!r_1\!-\!r_2\!)\\
&+\big(\!\cos(\!2\theta\!)\sin\!\phi_\chi\cos(\!2\psi\!)\!+\!\cos\!\phi_\chi\sin(\!2\psi\!)\!\big)^2\!\!\sinh^2(\!r_1\!+\!r_2\!)\!\Big)\\
&+\frac{2(\lambda_1\!-\!\lambda_2)^2}{\lambda_1\lambda_2\!-\!1}\!\Big(\!\big(\!\cos(\!2\theta\!)\sin\!\phi_\chi\sin(\!2\psi\!)\!+\!\cos\!\phi_\chi\cos(\!2\psi\!)\!\big)^2\!\!\cosh^2(\!r_1\!-\!r_2\!)\\
&+\big(\!\cos(\!2\theta\!)\sin\!\phi_\chi\cos(\!2\psi\!)\!+\!\cos\!\phi_\chi\sin(\!2\psi\!)\!\big)^2\!\!\sinh^2(\!r_1\!+\!r_2\!)\\
&+\frac{1}{2}\cos(\!2\theta\!)\sin(\!2\phi_\chi\!)\sin(\!4\psi\!)\sinh(2r_1)\sinh(2r_2)\!\Big)\\
&+\frac{4}{\lambda_1}\Big(e^{2r_1}\big(\norm{{d}_1}\sin\theta\cos(\phi_{2\chi}+\psi)+\norm{{d}_2}\cos\theta\cos(\phi_{1\chi}+\psi)\big)^2\\
&+e^{-2r_1}\big(\norm{{d}_1}\sin\theta\sin(\phi_{2\chi}+\psi)+\norm{{d}_2}\cos\theta\sin(\phi_{1\chi}+\psi)\big)^2\Big)\\
&+\frac{4}{\lambda_2}\Big(e^{2r_2}\big(\norm{{d}_1}\cos\theta\cos(\phi_{2\chi}-\psi)-\norm{{d}_2}\sin\theta\cos(\phi_{1\chi}-\psi)\big)^2\\
&+e^{-2r_2}\big(\norm{{d}_1}\cos\theta\sin(\phi_{2\chi}-\psi)-\norm{{d}_2}\sin\theta\sin(\phi_{1\chi}-\psi)\big)^2\Big).\\
\end{split}
\]

For positive $r_1$ and $r_2$ one of the optimal states is given by $\theta=0$, $\phi_\chi=\frac{\pi}{2}$, and $\phi_{1\chi}=\phi_{2\chi}=0$,
\[\label{eq:max_QFI_nonentangled_bs}
\begin{split}
H_{\mathrm{max}}(\epsilon)&=\frac{2(\lambda_1\!\!+\!\!\lambda_2)^2\!\!}{\lambda_1\lambda_2\!+\!1}\sinh^2(\!r_1\!+\!r_2\!)
+\frac{2(\lambda_1\!\!-\!\!\lambda_2)^2\!\!}{\lambda_1\lambda_2\!-\!1}\cosh^2(\!r_1\!+\!r_2\!)\\
&+\frac{4\norm{{d}_2}^2}{\lambda_1}e^{2r_1}
+\frac{4\norm{{d}_1}^2}{\lambda_2}e^{2r_2}.
\end{split}
\]
These conditions are fulfilled for example for $\phi_1=\frac{\pi}{4}-\frac{\chi}{2}$, $\phi_{d1}=\frac{\pi}{4}+\frac{\chi}{2}$, $\phi_2=-\frac{\pi}{4}+\frac{\chi}{2}$, $\phi_{d2}=-\frac{\pi}{4}-\frac{\chi}{2}$. This corresponds to two-single mode squeezed displaced thermal states of which the angle of between squeezing of each mode of the probe is given by the angle of the mode-mixing channel $\chi$.

Setting $\theta=\psi=\frac{\pi}{4}$, $\phi_\chi=\frac{\pi}{2}$, and $\phi_{1\chi}=\phi_{2\chi}=-\frac{\pi}{4}$ in Eq.~\eqref{eq:mode_mixing_full} we obtain
\[\label{eq:QFI_entangled_BS}
\begin{split}
H_{\mathrm{max}}(\epsilon)&=\frac{4\lambda_1^2}{\lambda_1^2+1}\sinh^2(2r_1)+\frac{4\lambda_2^2}{\lambda_2^2+1}\sinh^2(2r_2)\\
&+\frac{2}{\lambda_1}\big(\norm{{d}_1}+\norm{{d}_2}\big)^2e^{2r_1}+\frac{2}{\lambda_2}\big(\norm{{d}_1}-\norm{{d}_2}\big)^2e^{2r_2},
\end{split}
\]
This state corresponds to the two-mode squeezed displaced thermal state.

For mode-mixing channels we find a unique phenomenon which does not occur with the squeezing channels, and which can be exploited only when using a beam-splitter in the preparation process. Setting $\lambda_1=\lambda_2=1$, $r_1=r_2=r$, $\theta=\psi=\frac{\pi}{4}$, and $\phi_1+\phi_2+\phi_{d1}+\phi_{d2}=-\frac{\pi}{2}$ in Eq.~\eqref{eq:QFI_entangled_BS}, we derive
\[\label{eq:QFI_universal_BS}
H(\epsilon)=4\sinh^2(2r)+4\Big(\big(\norm{{d}_1}^2+\norm{{d}_2}^2\big)\cosh(2r)+2\norm{{d}_1}\norm{{d}_2}\sinh(2r)\Big).
\]
Any free parameter has not been at this point set to be dependent on the angle of the mode-mixing $\chi$. Also, the leading order here is identical to the energy-optimal probe states. In other words, we have found the universal optimal probe state for the mode-mixing channels $\hat{B}(\epsilon,\chi)$. If we set the initial displacement $\bd_0$ to zero, according to Eq.~\eqref{eq:simplified_probe_state} this probe state becomes the two-mode squeezed vacuum in the direction of $\chi_T=\frac{\pi}{2}$, $\hat{\rho}_{0}%=\hat{B}(\frac{\pi}{4})\hat{R}_{\mathrm{as}}(\frac{\pi}{4})\hat{S}_1(r)\hat{S}_2(r)(\cdot)^\dag
=\hat{S}_{T}(r,\tfrac{\pi}{2})\hat{S}_{T}^\dag(r,\tfrac{\pi}{2})$.

\begin{figure}[t!]
\centering
\includegraphics[width=0.8\linewidth]{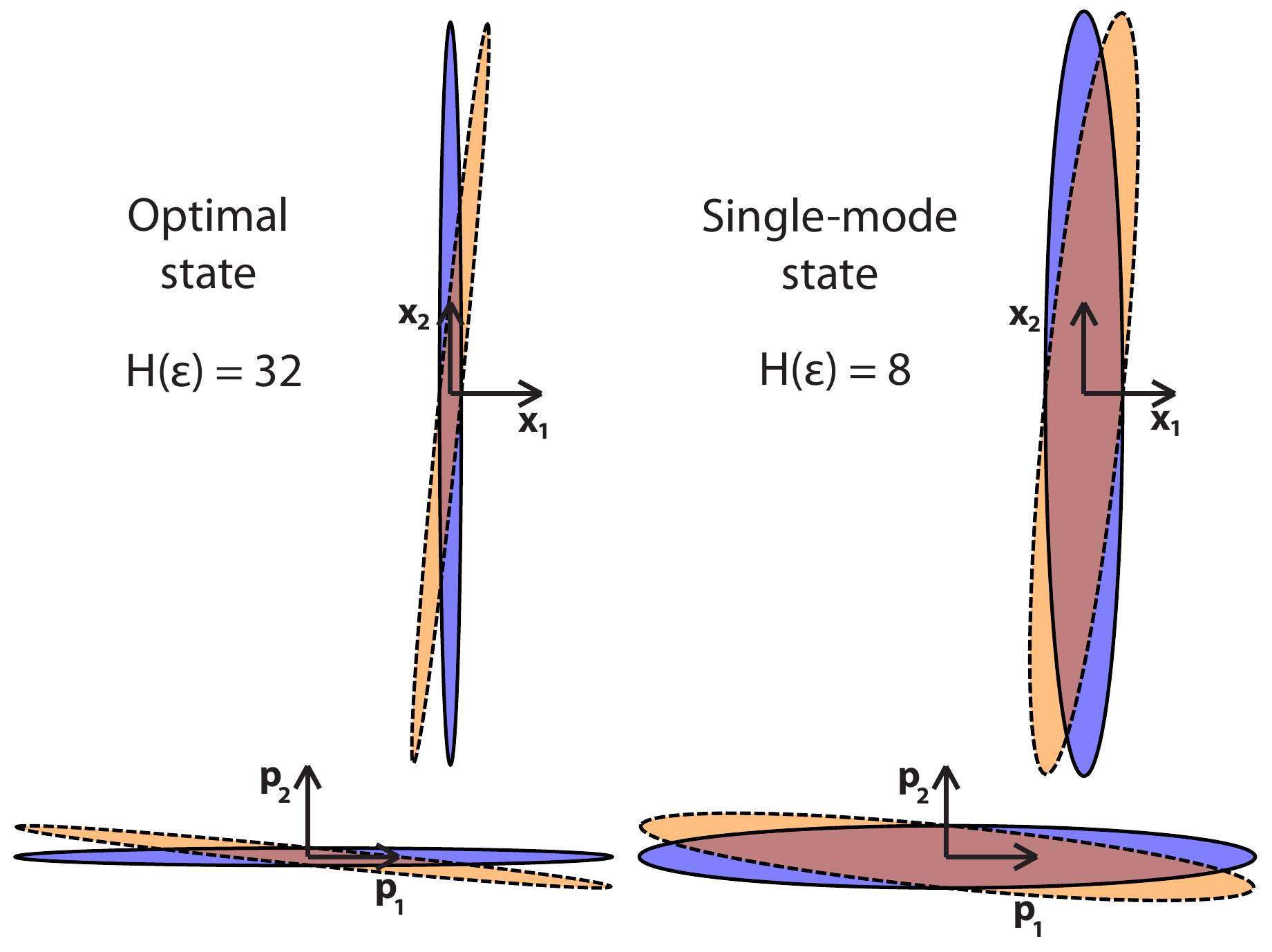}
\caption{Estimation of the beam-splitter $\hat{B}(\epsilon)$ around point $\epsilon=0$ using one of the optimal states $\hat{\rho}=\hat{S}_1(r)\hat{S}_2(-r)\hat{S}_2^\dag(-r)\hat{S}_1^\dag(r)$, and the single-mode state $\hat{\rho}=\hat{S}(r_1)\hat{S}^\dag(r_1)\otimes\ket{0}\bra{0}$, both with the same mean energy $n=2$. We plot the real form marginal covariance matrices showing correlations between positions in the first and the second mode $x_1$ and $x_2$, and momenta $p_1$ and $p_2$ in the real form phase-space, before(blue with full line) and after(orange with dashed line) beam-splitter $\hat{B}(0.1)$ has been applied. There are no correlations between position and momentum. Clearly, the optimal state is more sensitive to the channel allowing the optimal estimation of the parameter $\epsilon$.}\label{fig:one_and_optimal}
\end{figure}

Optimizing over the energy of the probe using the relation for the mean number of particles~\eqref{eq:mean_number_of_bosons_two} we find that all probe states leading to Eqs.~\eqref{eq:max_QFI_nonentangled_bs}, \eqref{eq:QFI_entangled_BS}, and~\eqref{eq:QFI_universal_BS} perform the best when all energy is, as in the case of estimating two-mode squeezing channel, invested into squeezing and the squeezing parameters are equal. All probe states then lead to the same Heisenberg limit $H_{\mathrm{max}}(\epsilon)=4n(n+2)$. In contrast, investing all energy into the displacement of the state we obtain the shot-noise limit $H_{\mathrm{max}}(\epsilon)=4n$. The same shot noise limit is also achieved by any one-mode state used as a probe for the mode-mixing channel. For illustration of how a one-mode state compares to the optimal state see figure~\ref{fig:one_and_optimal}.

\subsection{Discussion on optimality of pure states and squeezed states%Are pure states optimal probes? Are squeezed states?
}\label{sec:are_pure_optimal}

In the previous sections we have studied how mixed states perform as probe states. We have shown that investing some energy into the temperature of the probe state can help the estimation. However, after optimizing over the mean energy we have shown the pure probe states are always optimal. Why this happens can be viewed in the following way: let us assume that we have the set of all states with a fixed mean energy. Mixed states of that set are states on which we (as observers) lack some information. Pure states, where we do not lack any information, should serve as better probes as we know how to retrieve the information about the parameter more accurately. However, the way in which to prove this statement mathematically seems unclear. What we would like to prove is that for any mixed state $\R_0$ there exists a pure state $\ket{\psi_0}$ with the same energy (i.e., the same mean value given by a positive semi-definite operator $\hat{E}$) such that for any encoding channel $\mathrm{C}_\epsilon$ (CTPT map) the quantum Fisher information satisfies $H(\mathrm{C}_\epsilon(\ket{\psi_0}\bra{\psi_0}))\geq H(\mathrm{C}_\epsilon(\rho_0))$. One of states $\ket{\psi_0}$ could be the purification of the state $\R_0$. However the purification of a quantum state lies in a larger Hilbert space than the mixed state itself and the definition of the energy operator may not necessarily extend to this larger Hilbert space. We could extend the operator $\hat{E}$ to the larger Hilbert space in any possible way as long as it reduces to the original $\hat{E}$ on the former Hilbert space. In this way the energy of the purification should be larger unless we define the energy of all states in the extended Hilbert space as zero -- but this seems very arbitrary and unphysical as we cannot access those states in this extended Hilbert space anyway. Therefore the question is: ``For any mixed state $\R_0$ of energy $E$ from Hilbert space $\HS$, does there exist a pure state $\ket{\psi_0}$ from the same Hilbert space and with the same energy as the mixed state such that $H(\mathrm{C}_\epsilon(\ket{\psi_0}\bra{\psi_0}))\geq H(\mathrm{C}_\epsilon(\R_0))$?''

We will not answer this question precisely but we can provide some insight. The quantum Fisher information is a convex function on the set of density matrices~\cite{Toth2014a}, i.e., for any mixed state $\R_0=\sum_{i}p_i\ket{\psi}\bra{\psi}$ and for any encoding unitary $\hat{U}_\epsilon=e^{-i\hat{G}\epsilon}$ the following inequalites are satisfied,
\[\label{eq:convex_QFI}
H\Big(\hat{U}_\epsilon\R_0\hat{U}_\epsilon^\dag\Big)\ \!\leq\ \!\sum_{i}p_iH\Big(\hat{U}_\epsilon\ket{\psi_i}\bra{\psi_i}\hat{U}_\epsilon^\dag\Big)\ \!\leq\ \!\max_{i}\ \! H\Big(\hat{U}_\epsilon\ket{\psi_i}\bra{\psi_i}\hat{U}_\epsilon^\dag\Big).
\]
This shows that for any mixed state there exists a pure state in the same Hilbert space which provides better or equal precision in the estimation of the parameter $\epsilon$. However, this state may not have the same energy as the original mixed state $\R_0$. Our original question thus remains unanswered. On the other hand, for Gaussian states we can use an elegant formula for the quantum Fisher information in terms of the Williamson's decomposition of the covariance matrix~\eqref{eq:multimode_QFI}. We will illustrate the optimality of high energy single mode pure state in the estimation of a unitary channel $\hat{U}_\epsilon$. We will schematically show the maximal scaling with particle number for each element in Eq.~\eqref{eq:multimode_QFI} and then combine these scalings into a single formula. We will assume all particle numbers are large, $n,n_{d},n_{\mathrm{th}}\gg 1$. Eq.~\eqref{eq:mean_number_of_bosons_one} then becomes
\[\label{eq:mean_number_of_bosons_one_large}
n=n_{d}+n_{\mathrm{th}}+(1+2n_{\mathrm{th}})e^{2r}.
\]
Given the Euler's decomposition~\eqref{def:S_decomposition}, elements of the symplectic matrices scale as $S_0\sim e^r$, $S_0^{-1}\sim e^r$. In combination with Eq.~\eqref{eq:first_part} this implies $P\sim e^{2r}$, $Q\sim e^{2r}$, $R\sim e^{2r}$, and $\sigma_0^{-1}\sim \frac{e^{2r}}{2n_{\mathrm{th}}+1}$. Temperature factor scales as $\frac{\lambda^2}{1+\lambda^2}\sim \frac{n_{\mathrm{th}}^2}{n_{\mathrm{th}}^2}$ and the displacement scales as $\bd_0\sim \sqrt{n_{d}}$. Scaling of these quantities and Eq.~\eqref{eq:mean_number_of_bosons_one_large} give us the scaling of the quantum Fisher information,
\[\label{eq:scaling_QFI}
H(\epsilon)\sim C_1\frac{n_{\mathrm{th}}^2}{n_{\mathrm{th}}^2}\left(\frac{n-n_{\mathrm{th}}-n_{d}}{2n_{\mathrm{th}}+1}\right)^2+C_2\frac{n-n_{\mathrm{th}}-n_{d}}{2n_{\mathrm{th}}+1}(\sqrt{n_{d}}+C_3)^2.
\]
Factors $C_1,C_2,C_3$ are real constants which depend on the probe state we use and the channel we probe. Fixing the total mean number of particles $n$ of the probe state we can see that the maximum of the quantum Fisher information is achieved when the number of thermal bosons is the lowest, i.e., when the initial state is pure. This reasoning partially and schematically shows that pure states are indeed optimal states for the estimation of Gaussian unitary channels.

Eq.~\eqref{eq:scaling_QFI} also shows that the optimal probe is either the squeezed state or a mixture of squeezed and displaced state. We assume the encoding channel is generated by a purely quadratic Hamiltonian for which $C_3=0$. Fixing the $n_{\mathrm{th}}=0$ the Eq.~\eqref{eq:scaling_QFI} becomes
\[\label{eq:scaling_QFI}
H(\epsilon)\sim %C_1\left(n-n_{d}\right)^2+C_2(n-n_{d})n_{d}
%\sim C_1\left(n-n_{d}\right)\left(n+(\frac{C_2}{C_1}-1)n_{d}\right).
\left(n-n_{d}\right)\left(C_1n+(C_2-C_1)n_{d}\right).
\]
For $C_2\leq 2C_1$ the maximum is achieved when $n_{d}=0$, i.e., when the probe state is a squeezed state. For $C_2>2C_1$ the maximum is achieved for $n_{d}=\frac{C_2-2C_1}{2(C_2-C_1)}$, i.e., when the probe state is a squeezed displaced state. In all channels studied we had $C_2=2C_1$ when assuming $r_1=r_2$, $\norm{{d}_1}=\norm{{d}_2}$ and $\lambda_1=\lambda_2=1$. As an empirical rule the optimal state is always a squeezed state. However a rigorous mathematical proof is still necessary.

\subsection{Simplified way of finding optimal probes}\label{sec:optimizing_for_pure_state}

We illustrated in the previous section that it is likely that optimal probe states are always pure. If this is the case, we can restrict our search for optimal probe states to pure states and significantly simplify the algorithm.

Covariance matrix for pure states can be always written as $\sigma_0=S_0S_0^\dag$. Considering the Euler's decomposition of the symplectic matrix $S_0$~\eqref{def:S_decomposition}, the covariance matrix of a pure state is fully parametrized by
\[\label{eq:S_0_pure}
S_0=
\begin{bmatrix}
U_1 & 0 \\
0 & \ov{U}_1
\end{bmatrix}
\begin{bmatrix}
\cosh{M_{\boldsymbol{r}}} & -\sinh{M_{\boldsymbol{r}}} \\
-\sinh{M_{\boldsymbol{r}}} & \cosh{M_{\boldsymbol{r}}}
\end{bmatrix}.
\]
The quantum Fisher information can be obtained by combining Eqs.~\eqref{eq:S_0_pure},~\eqref{eq:exact_multimode_compact_pure}, and~\eqref{eq:eq_for_general_method},
\[
H(\epsilon)=2\tr\left[QQ^\dag\right]+2\boldsymbol{v}^\dag\begin{bmatrix}
\cosh{M_{\boldsymbol{2r}}} & \sinh{M_{\boldsymbol{2r}}} \\
\sinh{M_{\boldsymbol{2r}}} & \cosh{M_{\boldsymbol{2r}}}
\end{bmatrix}\boldsymbol{v},
\]
where
\begin{subequations}
\begin{align}
Q&=\cosh{M_{\boldsymbol{r}}}U_1^\dag Q_\epsilon \ov{U}_1\cosh{M_{\boldsymbol{r}}}-\sinh{M_{\boldsymbol{r}}}\ov{U}_1^\dag \ov{Q}_\epsilon U_1\sinh{M_{\boldsymbol{r}}}\nonumber\\
&+\sinh{M_{\boldsymbol{r}}}\ov{U}_1^\dag\ov{R}_\epsilon \ov{U}_1\cosh{M_{\boldsymbol{r}}}-\cosh{M_{\boldsymbol{r}}}U_1^\dag R_\epsilon U_1\sinh{M_{\boldsymbol{r}}},\\
\boldsymbol{v}&=\begin{bmatrix}
U_1^\dag & 0 \\
0 & \ov{U}_1^\dag
\end{bmatrix}
\left(P_\epsilon\bd_0+S_\epsilon^{-1}\dot{\bb}_\epsilon\right),\ \
P_\epsilon:=S_\epsilon^{-1}\dot{S_\epsilon}=\begin{bmatrix}
R_\epsilon & Q_\epsilon \\
\ov{Q}_\epsilon & \ov{R}_\epsilon
\end{bmatrix}.
\end{align}
\end{subequations}
Finding the optimal probe states then reduces to maximizing the quantum Fisher information over the unitary matrix $U_1$, the matrix of squeezing $M_{\boldsymbol{r}}$, and the vector of displacement $\bd_0$. For the estimation of one- two- and three- mode Gaussian channels the unitary $U_1$ can be fully parametrized as discussed in section~\eqref{sec:parametrization_of_Gaussian_states}.

In the case when the encoding operation is a one-parameter unitary group~\eqref{def:Gaussian_unitary_epsilon} with the hermitian matrix $W$ of structure given by Eq.~\eqref{def:W_for_Gaussian_unitary} we have
$P_\epsilon=iKW$, $R_\epsilon=iX$, $Q_\epsilon=iY$, and $\boldsymbol{v}=\big(\tilde{\boldsymbol{v}},\ov{\tilde{\boldsymbol{v}}}\big)^T$, $\tilde{\boldsymbol{v}}=U_1^\dag\big(iX\tilde{\bd}_0+iY\ov{\tilde{\bd}}_0+\tilde{\bg}\big)$.

\subsection{Role of entanglement and the Heisenberg limit}\label{sec:Heisenberg_limit_Gaussian}

As illustrated in previous sections entangled states such as two mode squeezed states do not achieve any advantage over the two single-mode squeezed states in the estimation of two-mode channels. In this section we first show why it is usually thought that entanglement in the probe state is necessary to achieve the Heisenberg limit, and why this reasoning is not applicable in the continuous variable states known as Gaussian states.

Although there are many possible definitions of Heisenberg limit in quantum metrology~\cite{Giovannetti2011a}, in this thesis we adopt the definition where the precision of estimation is compared to the mean energy of the probe state, mean number of Bosons respectively. Sequence of states $\rho_m$ with ever-increasing mean value of energy $\lim_{m\rightarrow\infty}\mean{\hat{E}}_{\rho_m}=\infty$ is said to reach the Heisenberg limit iff there exists a number $c>0$ such that
\[\label{def:limit_Heisenberg}
\lim_{m\rightarrow\infty}\frac{H(\rho_m)}{(\mean{\hat{E}}_{\rho_m})^2}=c.
\]
In the case of a Bosonic system the operator $\hat{E}$ which measures the energy of the probe state is up to a scaling constant identical to the total number operator, $\hat{E}\equiv\hat{N}$.

%Usually, higher energy of the probe state leads to the higher quantum Fisher information.
We first consider a Hilbert space $\HS$ such that for every state $\rho\in\HS$ the quantum Fisher information is bounded by the same value $B_H$, i.e.,
\[\label{eq:bound_H}
\exists B_H>0,\ \ \forall \rho\in \mathcal{H},\ \ H(\rho)\leq B_H.
\]
It is not possible to create a sequence of states from such Hilbert space to achieve the Heisenberg limit, because by definition $\lim_{m\rightarrow\infty}\frac{H(\rho_m)}{(\mean{\hat{E}}_{\rho_m})^2}\leq\lim_{m\rightarrow\infty}\frac{B_H}{\mean{\hat{E}}_{\rho_m}}=0$.
However, we can increase the quantum Fisher information by adding more particles, which corresponds to expanding the Hilbert space. We consider a (fully) separable state
\[\label{eq:construction_separable_states}
\rho_m=\sum_ip_i\rho_i^{(1)}\otimes\rho_i^{(2)}\otimes\cdots\otimes\rho_i^{(m)}\in\mathcal{H}^{\otimes m},
\]
where $\sum_ip_i=1$. Assuming that energy of each added state does not go below certain value, i.e.,
\[\label{eq:bound_E}
\exists B_E>0,\ \ \forall i,\ \forall k,\ \ \mean{\hat{E}}_{\rho_i^{(k)}}\geq B_E,
\]
and using convexity of the quantum Fisher information and additivity under tensoring~\cite{Toth2014a}, we derive
\[\label{eq:proof_separable_states_QFI}
\lim_{m\rightarrow\infty}\!\frac{H(\rho_m)}{(\mean{\hat{E}}_{\rho_m})^2}\leq\!\lim_{m\rightarrow\infty}\!\frac{\sum_{i,k}p_iH(\rho_i^{(k)})}{(\sum_{i,k}p_i\mean{\hat{E}}_{\rho_i^{(k)}}\!)^2}
\leq\!\lim_{m\rightarrow\infty}\!\frac{mB_H}{m^2B_E}=0.
\]
This illustrates that under conditions~\eqref{eq:bound_H} and~\eqref{eq:bound_E}, the construction~\eqref{eq:construction_separable_states} using separable states cannot lead to the Heisenberg limit and entangled states are necessary. This follows the proofs from~\cite{Pezze2009a,Demkowicz2012a,Toth2014a} showing that existence of entanglement in an $m$-qubit state is necessary condition for the scaling of the quantum Fisher information larger than the shot-noise limit.

Although $1$-qubit Hilbert space, from which the $m$-qubit Hilbert space is created, satisfies Eq.~\eqref{eq:bound_H}, such condition is no longer satisfied by the Fock space representing a Bosonic system. There are states in the Fock space, such as squeezed states and coherent states, which can lead to an arbitrarily large precision in the estimation. Therefore proof~\eqref{eq:proof_separable_states_QFI} does not apply anymore and entanglement is not necessary. As shown in previous sections, separable states such as squeezed states can also achieve the Heisenberg limit.

Moreover, in the Fock space it is possible to construct states which achieve arbitrary scaling of the quantum Fisher information with the energy of the probe state. We will illustrate this on the the example of pure states and unitary (not necessarily Gaussian) encoding operation $e^{-i\hat{K}\epsilon}$ where $\hat{K}$ is a Hermitian operator. For pure states the quantum Fisher information is simply the variance in the operator~\eqref{pureK} and the ratio~\eqref{def:limit_Heisenberg} reads
\[
\frac{H(\rho_m)}{(\mean{\hat{E}}_{\rho_m})^2}=\frac{4\mean{\Delta\hat{K}^2}_{\rho_m}}{(\mean{\hat{E}}_{\rho_m})^2}.
\]
But in the infinite dimensional Hilbert space the variance of a Hermitian operator $\hat{K}$ is not bounded by the mean value of energy. Even for the phase estimation for which $\hat{K}=\hat{N}^2$ and $\hat{E}=\hat{N}$ the variance in particle numbers is not related to the mean value of the particle number. We can construct states with arbitrarily large variance and arbitrarily low mean. From such states we can construct a series $\rho_m$ which achieve any scaling $f$,
\[
H(\rho_m)\sim f(\mean{\hat{E}}_{\rho_m}).
\]
However, the states which achieve such extraordinary scaling cannot be Gaussian states. These states could nevertheless achieve extremely high precisions while having a very low energy, being extremely interesting from the experimental point of view.

In comparison to $m$-qubit systems, which use entanglement as a resource, the resources in Bosonic systems are rather highly superposed states spanning across all infinite-dimensional Hilbert space, while entanglement does not play a significant role anymore.

\section{Summary}

In the first section we derived new formulae for the quantum Fisher information matrix of Gaussian states for the estimation of the vector of parameters $\be$. This included expressions in terms of the Williamson's decomposition of the covariance matrix and the limit formula which can be used for effective numerical calculations. We also noted problems when a pure mode changes its purity due to small variations in the vector of parameters $\be$, and devised a regularization procedure to fix them. This also led to a new expression for pure states that takes into account possible changes in purity.

In the second section we took advantage of the derived formulae and we devised a method of finding optimal probe states for the estimation of Gaussian channels based on the phase-space formalism. We also simplified this method by restricting ourselves to pure probe states. We applied this method to the estimation of one- and two-mode Gaussian channels. We found that for every channel we studied the optimal states are either squeezed or two-mode squeezed states. Further, the entanglement of the probe state does not play any significant role, which corresponds to the findings of~\cite{Gaiba2009a,Friis2015a}. This is not in contradiction with some previous studies that show that entanglement is necessary to achieve the Heisenberg limit~\cite{Pezze2009a,Demkowicz2012a}, as assumptions taken there do not apply anymore to the Fock space describing these bosonic systems.

In estimating parameters of phase-changing, one-mode squeezing, mode-mixing, and two-mode squeezing channels ($\hat{R},\hat{S},\hat{B},\hat{S}_T$ respectively), the quantum Fisher information reaches the Heisenberg limits
\begin{subequations}
\begin{align}
H_{R}(\epsilon)&=2\sinh^2(2r)=8n(n+1),\\
H_{S}(\epsilon)&=2\cosh^2(2r)=2(2n+1)^2,\\
H_{B}(\epsilon)&=4\sinh^2(2r)=4n(n+2),\\
H_{S_T}(\epsilon)&=4\cosh^2(2r)=4(n+1)^2,
\end{align}
\end{subequations}
where $r$ denotes the squeezing of one of the modes in the probe state, and $n$ is the mean total number of particles of the probe state.
These results generalize the precision bounds found in~\cite{Milburn1994a,Chiribella2006a,Monras2006a,Aspachs2008a,Gaiba2009a}. Alternatively, if we choose coherent states as probe states, we obtain the shot-noise limits
\begin{subequations}
\begin{align}
H_{R}(\epsilon)&=4n,\\
H_{S}(\epsilon)&=2(2n+1),\\
H_{B}(\epsilon)&=4n,\\
H_{S_T}(\epsilon)&=4(n+1).
\end{align}
\end{subequations}
These are the same limits we find when using any one-mode state to probe two-mode Gaussian channels. In addition, we have shown that non-Gaussian probe states spanning over the full infinite-dimensional Fock space can achieve arbitrarily high scaling of the quantum Fisher information which can be of interest for experimental application.

%Also, probing two-mode channels with one-mode states leads to the same shot-noise limit independent of the type of the probe state.
%This is interesting because that basically means that no matter what form of the one-mode probe state you use to estimate two-mode channel, the result is always the same.

%optimal states, entanglement-{say what we have found, what has been found elsewhere, why our case works even without the entanglement}, precision bounds

Authors of~\cite{Aspachs2008a} showed that the temperature of the probe state may enhance the estimation precision by a factor of two, and authors of~\cite{Gaiba2009a} explored of how temperature acts in the estimation of mode-mixing channels. We demonstrated that effects of temperature are generic. Independent of which Gaussian unitary channel is probed, the effects of temperature always come in multiplicative factors of four types. The first three appear when the channel changes the squeezing or the orientation of squeezing of the probe state. The first one accounts for the absolute number of thermal bosons in each mode and corresponds to the one found in~\cite{Aspachs2008a}. The next two take into account differences between thermal bosons in each mode. Larger differences then lead to higher precision in the estimation, while the enhancement factor scales with the ratio of the number of thermal bosons $\frac{n_{{\mathrm{th}}i}}{n_{{\mathrm{th}}j}}$, for $n_{{\mathrm{th}}i}\gg n_{{\mathrm{th}}j}\gg 0$. The last type of factor is of the form $(2n_{{\mathrm{th}}i}+1)^{-1}$ and appears when the Gaussian channel changes the displacement of the probe state.

We have shown how different aspects of a probe state affect the estimation precision, and have provided a framework that can be effectively used to study optimal probe states for the construction of new-era quantum detectors. In addition to applications for existing gravitational wave detectors~\cite{Abbott2004a,Caron1995a}, our results may be useful for designing new gravimeters~\cite{Snadden1998a,Altin2013a,Sabin2014a}, climate probes~\cite{Tapley2004a}, or for the estimation of space-time parameters~\cite{Danzmann1996a,Everitt2011a,Bruschi2014a}.

%%%%%%%%%%%%%%%%%%%%%%%%%%%%%%
\chapter{Applications in quantum field theory: Estimating effects of space-time}\label{chap:QFT_metrology}
%%%%%%%%%%%%%%%%%%%%%%%%%%%%%%

%introduction
In this chapter we derive the ultimate precision limits with which we can estimate parameters encoded by a general Bogoliubov transformation into one- and two- mode Gaussian probe states. As detailed in chapter~\ref{chap:operations_in_QFT} such transformations can represent a quantum state from the point of view of an accelerated observer, cavities moving in curved space-time, action of an expanding universe on a quantum state, or a gravitational wave passing through a Bose-Einstein condensate. Each such transformation contains parameters % such as proper time, gravitational field strengths, accelerations, Schwarzchild parameter
which, upon correct estimation, could tell us more about heavenly bodies and gravity, how non-inertial observers see quantum particles, or the universe itself. The precise estimation of these parameters can lead to novel applications in gravimeters, space-time probes, and gravitational wave detectors. Moreover, since the predictive power of quantum field theory in curved space-time lies in the overlap of quantum physics and general relativity, measuring such parameters could either validate the theory or it could lead to a new theory of quantum gravity.

Previous work in this direction, as overviewed in section~\ref{sec:state_QM_in_qft}, considered almost exclusively pure probe states. However, in realistic situations probe states are mixed. In this chapter we provide a framework for the computation of optimal precision bounds for mixed single- and two-mode Gaussian states within quantum field theory. This enables the estimation of space-time parameters in the case when the states are initially at some finite temperature.

This chapter is structured as follows: first we introduce a general method in which we will approach the problem. Then we compute the perturbative expression of the quantum Fisher information associated to one- and two-mode mixed Gaussian states as a function of the Bogoliubov coefficients. We compute explicitly the quantum Fisher information for the case where we wish to estimate a state parameter around the value $\epsilon_{0}\,=\,0$. Finally, we apply these results to calculate the quantum Fisher information for the estimation of the proper acceleration using one- and two-mode squeezed thermal states.

\section{General method}
%input state, tracing over, encoding operation, the difference: infinite-dimensional nature, problem with the points of purity

The general method is very similar to the one used for finding the optimal probe states for Gaussian channels: here, however, because of computational complexity we will not optimize over probe states.

%initial state
Let us consider a quantum state represented by a covariance matrix $\tilde{\sigma}_0$ of the field.This covariance matrix is infinite dimensional which reflects the infinite-dimensional nature of states in quantum field theory in curved space-time. It consists of the initial covariance matrix of a mode or modes of interest, $\sigma_0:=\tr_E[\tilde{\sigma}_0]$, and the covariance matrix of the remaining modes of the field (or ``environment'' $E$), $\sigma_{E}:=\tr_{\neg E}[\tilde{\sigma}_{0}]$, which contains no initial correlation with the system modes $\sigma_{0}$. Therefore, the initial state $\tilde{\sigma}_{0}$ is separable in the subsystem-environment bipartition. Its block structure is given as for any Gaussian state by Eq.~\eqref{def:first_and_second_moments},
\[
\tilde{\sigma}_0\,=\,\begin{bmatrix}
X_0 & Y_0 \\
\overline{Y}_0 & \overline{X}_0
\end{bmatrix},
\]
Note this infinite-dimensional matrix takes into account any modes of interest and also the remaining ``environment" modes of the system. We will also assume the initial state has zero initial displacement, $\boldsymbol{d}_0=0$.

%transformation and the final state
The transformation between the initial and final state of the field, when restricted to the Gaussian case, is given by the (infinite-dimensional) Bogoliubov matrix $\tilde{S}$. The covariance of the transformed state is calculated via Eq.~\eqref{def:transformation} and the Bogoliubov matrix $\tilde{S}$ is the same as transformations of the field operators~\eqref{eq:transformation_of_field}, i.e.,
\[
\tilde{S}:=\ov{S}^\dag=\begin{bmatrix}
\ov{\A} & -\ov{\B} \\
-\B & \A
\end{bmatrix}.
\]
Matrix $S$ from which the matrix $\tilde{S}$ is computed can be found by solving Eq.~\eqref{eqn:exact_continuous_bogo_ODE} for example by using the perturbative method described in section~\ref{sec:perturbative_method}.\footnote{Note however that the definition and meaning of the matrix $\tilde{S}$ in this section differs from the definition and meaning of the matrix $\tilde{S}$ in section~\ref{sec:perturbative_method}. While in here the matrix $\tilde{S}$ represents the transformation of the field operators given by Eq.~\eqref{eq:transformation_of_field}, $\tilde{S}$ in section~\ref{sec:perturbative_method} denotes a convenient substitution which has been used for the purposes of the perturbation method and is given by Eq.~\eqref{def:tildeS}.} This perturbation method uses the expansion in the small parameter $\epsilon$. Both $\A=\A(\epsilon)$ and $\B=\B(\epsilon)$ depend on this parameter, but to keep expressions sufficiently short we do not explicitly indicate this dependence. Finally, we describe the transformation from the initial subsystem state $\sigma_{0}$ to a final subsystem state $\sigma(\epsilon)$ via a map $\mathcal{E}$ defined as
\[
\label{eqn:QFI_quantum_channel}
\sigma(\epsilon)=\mathcal{E}[\tilde{\sigma}(\epsilon)]=\tr_{E}\big[{\tilde{S}}(\epsilon)\tilde{{\sigma}}_{0}{\tilde{S}}(\epsilon)^{\dag}\big]
=\begin{bmatrix}
X & Y \\
\ov{Y} & \ov{X}
\end{bmatrix}.
\]
The Bogoliubov transformation between the initial and final states can be viewed as a quantum channel on the space of quantum states.

%general matrix after transformation
The elements of the final matrix $\sigma(\epsilon)$ for an arbitrary Gaussian state are given by
\begin{subequations}
\begin{align}
X^{ij}&\,=\,\sum_{a,b}\Big(\ov{\A}^{ia}\,X_0^{ab}\,\A^{jb}-\ov{\B}^{ia}\,\overline{Y}_{0}^{ab}\A^{jb}-\ov{\A}^{ia}Y_{0}^{ab}\B_{jb}+\ov{\B}^{ia}\overline{X}_{0}^{ab}\B^{jb}\Big),\\
Y^{ij}&\,=\,\sum_{a,b}\Big(-\ov{\B}^{ia}\,\overline{X}_{0}^{ab}\,\ov{\A}^{jb}+\ov{\A}^{ia}\,Y_{0}^{ab}\,\ov{\A}^{jb}+\ov{\B}^{ia}\,\overline{Y}_{0}^{ab}\,\ov{\B}^{jb}-\ov{\A}^{ia}\,X_{0}^{ab}\,\ov{\B}^{jb}\Big).
\end{align}
\end{subequations}
These expressions, coupled with the exact definitions of the quantum Fisher information and Bogoliubov co-efficients, can be used to compute the quantum Fisher information for any state within our quantum field theory framework. The resulting expressions are rather unwieldy and hence we have not written them out explicitly. However, in the following we will write the quantum Fisher information explicitly for the cases where the initial state is one- and two-mode squeezed thermal state and the environment is assumed to be a thermal state. The expressions provided will be perturbative expressions in the small parameter $\epsilon$ we estimate, while the zeroth order will be exact for $\epsilon=0$, i.e., $H^{(0)}=H(0)$. Note that when using the perturbative solution of the continuous Bogoliubov transformations~\eqref{eqns:general_solution_arbitrarily_high_order}, having the transformation matrices expanded up to order $(n)$ in the unknown parameter $\epsilon$ the quantum Fisher information can be expressed up to order $(n-1)$. This is because the quantum Fisher information is a function of at most first derivatives of the transformation matrix $S$, $H(\sigma(\epsilon))=H(\tilde{\sigma}_0,\A,\B,\tfrac{\mathrm{d}}{\mathrm{d}\epsilon}{\A},\tfrac{\mathrm{d}}{\mathrm{d}\epsilon}{\B})$.

%assumptions
In the following we also implicitly assume that the Bogoliubov coefficients have the following property
\begin{equation}
\label{eqn:simplifying_condition}
\frac{\mathrm{d}}{\mathrm{d}\epsilon}\alpha_{jj}\big|_{\epsilon\,=\,0}\,=\,\frac{\mathrm{d}}{\mathrm{d}\epsilon}\beta_{jj}\big|_{\epsilon\,=\,0}\,=\,0.
\end{equation}
This is equivalent to the statement that the first order coefficients of the diagonal $\boldsymbol{\alpha}$ and $\boldsymbol{\beta}$ are zero i.e. $\alpha^{(1)}_{jj}\,=\,\beta^{(1)}_{jj}\,=\,0$. As an example, these assumptions hold when the symplectic operation is symmetric around zero, i.e., $\alpha(\epsilon)=\alpha(-\epsilon)$, $\beta(\epsilon)=\beta(-\epsilon)$, and also for the special case of $\alpha_{mn},\beta_{mn}\in\mathbb{R}$. Physically, this condition means that the channel does not affect the same mode up to the first order in $\epsilon$, i.e., the channel is mostly mode-entangling channel. It is possible to generalise this work to cases where diagonal first order Bogoliubov coefficients are non-zero, however, since Bogoliubov coefficients considered in previous literature~\cite{Bruschi2012a,Bruschi2013b,Friis2013a,Sabin2014a} all satisfy Eq.~\eqref{eqn:simplifying_condition}, we will restrict to such case. In the following sections, we consider that all quantities (matrix and scalar) can be expanded in the form,
\begin{equation}
f(\epsilon)\,=\,f^{(0)}+f^{(1)}\epsilon+f^{(2)}\epsilon^{2}+\mathcal{O}(\epsilon^{3}).
\end{equation}
Throughout this chapter we will be using Planck units $\hbar\,=\,c\,=\,k_{B}\,=\,1$.

\section{Estimating space-time parameters with single-mode Gaussian states}
%general probe matrix
We first compute the quantum Fisher information of a single mode undergoing a Bogoliubov transformation that depends on the physical parameter to be estimated. We consider the following initial state, \begin{equation}
\label{eqn:Single_mode_state}
\sigma_{0}\,=\,\lambda_{m}\,
\begin{bmatrix}
\cosh(2r) & \sinh(2r) \\
\sinh(2r) & \cosh(2r)
\end{bmatrix},\quad\sigma_{E}\,=\,\bigoplus_{j\ne m}\lambda_{j}\,I,
\end{equation}
which corresponds to a single mode squeezed thermal state with squeezing parameter $r$, thermal parameter $\lambda_{m}\ge 1$ and all other modes in a separable thermal state. The temperature of the state, denoted by $T$, is related to the thermal parameter through $\lambda_{m}\,=\,\coth(E_{m}/2T)$ where $E_{m}=\omega_{m}$ is the energy of each mode. Note that for zero temperature, the thermal parameter reduces to $\lambda_{m}\,=\,1$.

Exact elements of the the final state $\sigma(\epsilon)$ can be computed as
\begin{subequations}
\label{eqn:one_mode_XYsummed}
\begin{align}
X^{mm}&=\lambda_{m}\,\Big(\cosh(2r)\,\big(|\alpha^{mm}|^{2}+|\beta^{mm}|^{2}\big)-2\,\mathrm{Re}\big[\alpha^{mm}\overline{\beta}^{mm}\big]\sinh(2r)\Big)\nonumber\\
&+\sum_{a\ne m}\lambda_{a}\,\big(|\alpha^{ma}|^{2}+|\beta^{ma}|^{2}\big),\\
Y^{mm}&=\lambda_{m}\Big(-2\,\cosh(2r)\,\overline{\alpha}^{mm}\overline{\beta}^{mm}+\big(\overline{\alpha}^{mm\ \!2}+\overline{\beta}^{mm\ \!2}\big)\sinh(2r)\Big)-2\,\sum_{a\ne m}\lambda_{a}\overline{\alpha}^{ma}\overline{\beta}^{ma}.
\end{align}
\end{subequations}
We can write this expression as a series expansion in $\epsilon$ around the point $\epsilon_{0}\,=\,0$,
\begin{equation}
\sigma(\epsilon)\,=\,\sigma^{(0)}+\sigma^{(1)}\,\epsilon+\mathcal{O}(\epsilon^{2}).
\end{equation}
It should also be noted that, in general, the covariance matrix elements $X^{(j)mn}$ and $X^{(j)mn}$ will depend on both squeezing, $r$, and the thermal parameters $\lambda_{m}$. We will also denote phases acquired due to free time evolution as $G^{m}\,=\,e^{+i\,\omega_{m}\tau}$ with $\omega_{m}$ the zeroth order contribution to frequency of the mode $m$.

We now proceed to choose specific values for the temperature and squeezing to find analytically the quantum Fisher information in regimes of interest.

\subsection{Zero initial temperature}
We start by considering an initial state with zero temperature. The perturbative expansion of the quantum Fisher information in Eq.~\eqref{eq:one_mode_quantum_fisher_information} needs particular attention. If we consider a state which is initially pure one finds that the denominators in Eqs.~\eqref{eq:one_mode_quantum_fisher_information} vanish. This potentially problematic point can be handled in multiple ways which is discussed in more detail in section~\ref{sec:problems_at_pops}. However, one can make a series expansion of each term and by applying L'H\^{o}pital's rule one obtains a finite result,
\begin{equation}
\label{eqn:one_mode_qfi_zero_temp}
\begin{split}
H_{1}(\epsilon)\,&=\,X^{(2)mm}\,\cosh(2\,r)-\mathrm{Re}[(G^{m})^2\,Y^{(2)mm}]\,\sinh(2r)\\
&+\frac{2}{3}\Big(X^{(3)mm}\,\cosh(2\,r)-\mathrm{Re}[(G^{m})^2\,Y^{(3)mm}]\,\sinh(2\,r)\Big)\,\epsilon+\mathcal{O}(\epsilon^{3})
\end{split}
\end{equation}
For convenience and clarity, we have left the second order covariance matrix elements written in the general form $X^{(2)mm}$ and $Y^{(2)mm}$. This elegant expression builds upon the zeroth order result of~\cite{Ahmadi2014a} and extends it to the linear regime in $\epsilon$. It should be noted that, for the expansion~\eqref{eqn:one_mode_qfi_zero_temp} to be valid, the initial squeezing $r$ and parameter $\epsilon$ must satisfy $e^{2r}\,\epsilon\ll1$.

%\subsection{Small initial temperature}

%A case of physical relevance is that of small temperature. In realistic situations the field is never in the vacuum state.  It is possible to compute analytical formulas in different regimes of interest that depend on the relative magnitude of the temperature parameter and the parameter $\epsilon$. We start by analyzing the case where the temperature is ``small" as compared to $\epsilon$. The thermal parameter $\lambda_{m}$ in this case has the form $\lambda_{m}\,=\,1+2\,Z^{2}+\mathcal{O}(Z^{3})$, with $Z\,=\,e^{-E_{m}/2T}$. We can identify the ratio $Z^{2}/\epsilon^{2}$ as our new expansion parameter. This defines our ``small'' temperature regime as the one where $Z^{2}\ll\epsilon^{2}$. We find the quantum Fisher information takes the expression
%\begin{equation}
%\label{eqn:small_temperature_one_mode_qfi}
%H_1(\epsilon)\,\approx\,H_{1}^{(0)}-4\frac{Z^{2}}{\epsilon^{2}}.
%\end{equation}
%One notes that the main contribution is the zeroth order and zero temperature QFI $H_{1}^{(0)}$ which coincides with the zeroth order contribution from Eq.~\eqref{eqn:one_mode_qfi_zero_temp}. The second term is a small correction.  We have neglected the linear contribution to the zero temperature result for simplicity. Clearly in the regime where $Z^{2}\approx\epsilon^{3}$ both the ``small" temperature correction and the linear zero temperature correction are comparable. In this case both must be accounted for and will have competing effects in the quantum Fisher information.

\subsection{Large initial temperature}

In this case we find that the zeroth order quantum Fisher information for a single mode is identically zero, i.e. $H^{(0)}_{1}\,=\,0$ at any non-zero temperature. This implies that the estimation of the parameter $\epsilon$ around zero is impossible for a one-mode squeezed state with a non-zero temperature. The first non-trivial contribution comes at $\mathcal{O}(\epsilon^{2})$. The result is,
\begin{equation}\label{one:mode:large:temperature:result}
\begin{split}
H^{(2)}_{1}&=\frac{|Y_{mm}^{(2)}|^{2}}{\lambda_{m}^{2}+1}+\frac{(X_{mm}^{(2)})^{2}}{\lambda_{m}^{2}-1}\,-\, \frac{2\,\lambda_{m}^{2}\,X_{mm}^{(2)}\mathrm{Re}[G_{m}^{2}\overline{Y}_{mm}^{(2)}]}{\lambda_{m}^{4}-1}\sinh(4r)\,\\
&+\frac{2\,\lambda_{m}^{2}\Big((X_{mm}^{(2)})^{2}+\mathrm{Re}[G_{m}^{4}(\overline{Y}_{mm}^{(2)})^{2}]\Big)}{\lambda_{m}^{4}-1}\sinh^{2}(2r)
\end{split}
\end{equation}
Clearly the condition $\lambda_{m}>1$ is key and the equation holds in the regime $\epsilon^{2}\ll\lambda_{m}-1$ or, in terms of the previous subsections notation, $\epsilon^{2}\ll Z^{2}$ by which the ``large" temperature regime is defined.

Note that the significant non-smooth difference between the zeroth order of the large temperature case and the small temperature case is given by a contribution from the change of the purity of the state as illustrated on example~\ref{ex:problematic_example}.

\section{Estimating space-time parameters with two-mode Gaussian states}

%general probe matrix
Here we compute the quantum Fisher information for thermal two-mode states with non-degenerate thermal parameters (i.e., the frequencies of the two modes are different). The initial state is the two-mode squeezed thermal state and has the form,
\begin{equation}
\label{eqn:Two_mode_state}
\sigma_{0}\,=\,
\begin{bmatrix}
D^{mn} & 0 & 0 & C^{mn} \\
0 & D^{nm} & C^{mn} & 0 \\
0 & C^{mn} & D^{mn} & 0 \\
C^{mn} & 0 & 0 & D^{nm}`
\end{bmatrix},\,\,\sigma_{E}\,=\,\bigoplus_{j\ne m,n}\lambda_{j}\,I,
\end{equation}
where we have introduced
\begin{subequations}
\begin{align}
D^{mn}&:=\lambda_{m}\cosh^{2}(r)+\lambda_{n}\sinh^{2}(r), \\
C^{mn}&:=(\lambda_{m}+\lambda_{n})\cosh(r)\sinh(r).
\end{align}
\end{subequations}

%final state
Exact elements of the the final state $\sigma(\epsilon)$ can be computed as
\begin{subequations}
\label{eqn:two_mode_XYsummed}
\begin{align}
\begin{split}X^{ij}&=D^{mn}(\ov{\A}^{im}\A^{jm}+\ov{\B}^{im}\B^{jm})-C^{mn}(\ov{\B}^{im}\A^{jn}+\ov{\A}^{im}\B^{jn})
+D^{nm}(\ov{\A}^{in}\A^{jn}+\ov{\B}^{in}\B^{jn})\\
&-C^{nm}(\ov{\B}^{in}\A^{jm}+\ov{\A}^{in}\B^{jm})
+\sum_{a\neq m,n}\lambda_{a}(\ov{\A}^{ia}\A^{ja}+\ov{\B}^{ia}\B^{ja}),
\end{split}\\
\begin{split}Y^{ij}&=-D^{mn}(\ov{\B}^{im}{\ov{\A}}^{jm}+\ov{\A}^{im}{\ov{\B}}^{jm})+C^{mn}(\ov{\A}^{im}{\ov{\A}}^{jn}+\ov{\B}^{im}{\ov{\B}}^{jn})
-D^{nm}(\ov{\B}^{in}{\ov{\A}}^{jn}+\ov{\A}^{in}{\ov{\B}}^{jn})\\
&+C^{nm}(\ov{\A}^{in}{\ov{\A}}^{jm}+\ov{\B}^{in}{\ov{\B}}^{jm})-\sum_{a\neq m,n}\lambda_{a}(\ov{\B}^{ia}{\ov{\A}}^{ja}+\ov{\A}^{ia}{\ov{\B}}^{ja}),
\end{split}
\end{align}
\end{subequations}
where $i,j=m,n$.

\subsection{Zero initial temperature}

In the two-mode case, the quantum Fisher information for zero temperature was computed using Eq.~\eqref{GeneralQFI} and is given as a series expansion in $\epsilon$. The resulting expressions are computable but considerably more involved. Here we present the results for the zero and first order contributions in $\epsilon$. In the linear contribution, we present only the case of zero squeezing. The formula for non-zero squeezing is too long and therefore, we have chosen to focus on the quantitative behaviour.
\begin{subequations}
\begin{align}
\label{eqn:two_mode_qfi_zero_temperature_zeroth_order}
H_{2}^{(0)}\,=&\,(X^{(2)mm}+X^{(2)nn})\cosh(2r)\nonumber\\
&\,-4|\beta^{(1)mn}|^{2}-2\,\mathrm{Re}[G^{m}G^{n}Y^{(2)mn}]\sinh(2r)\nonumber\\
&\,-4\,(|\alpha^{(1)mn}|^{2}+\mathrm{Im}[G^{m}\,\overline{\beta}^{(1)mn}]^{2})\sinh^{2}(2r), \\
\label{eqn:two_mode_qfi_zero_temperature_first_order}
H_{2}^{(1)}\big|_{r\,=\,0}\,=\,&\frac{2}{3}\,\Big(6\,\mathrm{Re}[G^{n}\beta^{(1)mn}Y^{(2)mn}]+X^{(3)mm}+X^{(3)nn}]\Big).
\end{align}
\end{subequations}
At zeroth order, the quantum Fisher information depends on the squeezing parameter in the same way as in the single mode channels studied in the previous section. However, at first order, particle creation terms  $\beta^{(1)mn}$ appear generating entanglement in the system. Therefore, we conclude that in this case entanglement does not provide an important improvement in precision. A highly squeezed single mode probe state could be enough to enable a good measurement strategy for the estimation of parameters uncoded in Bogoliubov transformations. Single mode states are usually more accessible in realistic experiments and this could provide an important advantage in quantum metrology for quantum fields.

%\subsection{Small initial temperature}

%We now compute the ``small" temperature contribution to the QFI for two mode probe states by expanding the thermal parameters as $\lambda_{m}\,=\,1+2\,q_{m}Z^{2}+\mathcal{O}(Z^{3})$ where $q_{m}$ is a fixed constant. We find that for zero squeezing, the first contribution to the QFI in the small temperature regimes has the following form,
%\begin{equation}
%\label{eqn:two_mode_small_temperature}
%H_2(\epsilon)\,\approx\, H_{2}^{(0)}-4\,(q_{m}+q_{n})\frac{Z^{2}}{\epsilon^{2}},
%\end{equation}
%where the zeroth order $H_{2}^{(0)}$ coincides with Eq.~\eqref{eqn:two_mode_qfi_zero_temperature_zeroth_order}. It should be noted that due to the complexity of the two-mode non-zero squeezing expression, the first correction term in Eq.~\eqref{eqn:two_mode_small_temperature} has only been verified for zero squeezing.

\subsection{Large initial temperature}

As in the single mode case, the thermal parameters take values strictly greater than unity, i.e. $\lambda_{j}>1$. The components of the state can be exactly computed and are given in Eq.~\eqref{eqn:two_mode_XYsummed}. We find that the quantum Fisher information, including linear contributions, is given by $H_2(\epsilon)\,=\,H^{(0)}_{2}+H^{(1)}_{2}\epsilon+\mathcal{O}(\epsilon^{2})$, with coefficients,
\begin{equation}
\label{eqn:two_mode_QFI_zeroth_order}
\begin{split}
H_{2}^{(0)}&\,=\,h_{00}+h_{02}\,\sinh^{2}(2r), \\
h_{00}&\,=\,\frac{2\,(\lambda_{m}-\lambda_{n})^{2}|\alpha^{(1)mn}|^{2}}{\lambda_{m}\lambda_{n}-1}+\frac{2\,(\lambda_{m}+\lambda_{n})^{2}|\beta^{(1)mn}|^{2}}{\lambda_{m}\lambda_{n}+1}, \\
h_{02}&\,=\,\frac{2\,(\lambda_{m}+\lambda_{n})^{2}((\lambda_{m}\lambda_{n}-1)^{2}+\lambda_{m}^{2}+\lambda_{n}^{2}-2)|\alpha^{(1)mn}|^{2}}{(\lambda_{m}^{2}+1)(\lambda_{n}^{2}+1)(\lambda_{m}\lambda_{n}-1)}
+\frac{2\,(\lambda_{m}+\lambda_{n})^{2}\mathrm{Im}[G^{m}\overline{\beta}^{(1)mn}]^{2}}{\lambda_{m}\lambda_{n}+1}.
\end{split}
\end{equation}
\begin{equation}
\begin{split}
H_{2}^{(1)}&\,=\,h_{10}+h_{11}\,\sinh(2r)+h_{12}\,\sinh^{2}(2r), \\
h_{10}&\,=\,4\,\Bigg(\frac{\lambda_{n}-\lambda_{m}}{\lambda_{m}\lambda_{n}-1}\,\mathrm{Re}[G_{n}\overline{\alpha}^{(1)mn} \overline{X}^{(2)mn}]-\frac{\lambda_{m}+\lambda_{n}}{\lambda_{m}\lambda_{n}+1}\,\mathrm{Re}[G_{n}\beta^{(1)mn}Y^{(2)mn}]\,\cosh(2r)\Bigg), \\
h_{11}&\,=\, \frac{2\,(\lambda_{m}+\lambda_{n})\mathrm{Re}[G^{m}\overline{\beta}^{(1)mn}]}{\lambda_{m}\lambda_{n}+1}(X^{(2)mm}+X^{(2)nn})\\
&-\frac{16\,\lambda_{m}\lambda_{n}(\lambda_{m}^{2}-\lambda_{n}^{2})^{2}|\alpha^{(1)mn}|^{2}\mathrm{Re}[G^{m}\overline{\beta}^{(1)mn}]}{(\lambda_{m}^{2}+1)(\lambda_{n}^{2}+1)(\lambda_{m}^{2}\lambda_{n}^{2}-1)}\cosh(2r) \\
&-\frac{2(\lambda_{m}+\lambda_{n})(\lambda_{m}\lambda_{n}+1)}{(\lambda_{m}^{2}+1)(\lambda_{m}^{2}+1)}\mathrm{Re}[G^{n}\overline{\alpha}^{(1)mn}(\overline{G}^{m}\overline{G}^{n} \overline{Y}^{(2)mm}-G^{m}G^{n}Y^{(2)nn})]\\
&+\frac{2(\lambda_{m}-\lambda_{n})(\lambda_{m}+\lambda_{n})^{2}}{(\lambda_{m}^{2}+1)(\lambda_{n}^{2}+1)(\lambda_{m}\lambda_{n}-1)}\mathrm{Re}[G^{n}\overline{\alpha}^{(1)mn}(\overline{G}^{m}\overline{G}^{n} \overline{Y}^{(2)mm}\!\!+\!G^{m}G^{n}Y^{(2)nn})]\cosh(2r),\\
h_{12}&\,=\,\frac{4\,(\lambda_{n}-\lambda_{m})(\lambda_{n}+\lambda_{m})^{2}\mathrm{Re}[G^{n}\overline{\alpha}^{(1)mn}\,\overline{X}^{(2)mn}]}{(1+\lambda_{m}^{2})(1+\lambda_{n}^{2})(\lambda_{m}\lambda_{n}-1)}.
\end{split}
\end{equation}
In general, the coefficients $\alpha^{(1)mn}$,  $\beta^{(1)mn}$ and the covariance matrix are time dependent. In the large temperature regime, to zeroth order, the quantum Fisher information for two-mode probe states is non-zero. This result is in contrast with the single-mode case. Even when the probe state  has zero squeezing, the quantum Fisher information is non-zero and it is proportional to the number of particles created after the Bogoliubov transformation~\cite{Bruschi2013c}. We can also analyse the effect of temperature on the quantum Fisher information by varying the parameters $\lambda_{m}$ and $\lambda_{n}$.
We note that at when estimating around the point $\epsilon\,=\,0$, the zero order expressions for the quantum Fisher information are exact, $H(0)=H^{(0)}$.

\section{Example: Estimation of the proper acceleration}

\begin{figure}[t!]
\centering
\includegraphics[width=0.8\linewidth]{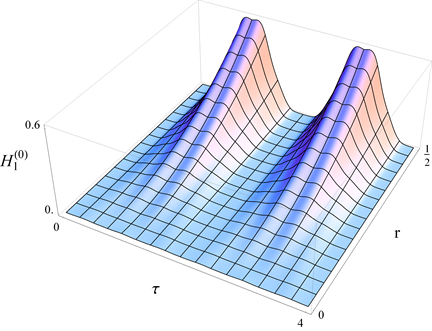}
\caption{The zeroth order of the quantum Fisher information for the estimation of the acceleration parameter $a$ using a one-mode squeezed state with initial zero temperature. Calculated using Eq.~\eqref{eqn:one_mode_qfi_zero_temp} and $m=1$ (using a Fock space corresponding to the first excited state within a cavity). The graph shows that to achieve the highest possible precision in estimation it is appropriate to measure at certain times ($\tau=1,3,5,\dots$). This periodic behavior is due to the fact that the information about the parameter moves into the modes we cannot access -- the environment -- and back. For times when the quantum Fisher information is the highest the estimation precision grows exponentially with the squeezing parameter $r$.}\label{fig:one-modeQFI}
\end{figure}

\begin{figure}[t!]
\centering
\includegraphics[width=1\linewidth]{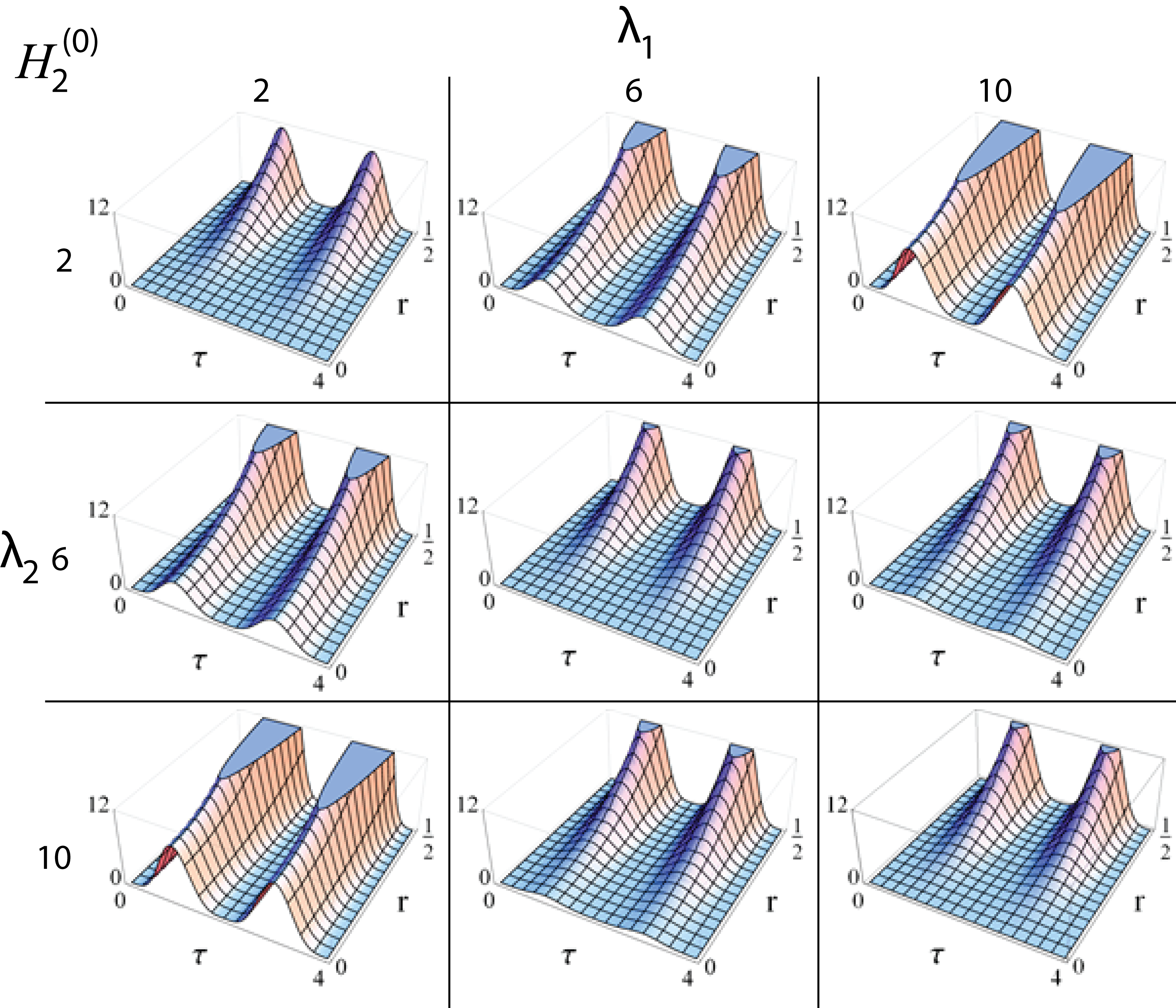}
\caption{The zeroth order of the quantum Fisher information for the estimation of the acceleration parameter $a$ using a two-mode squeezed thermal state with initial ``large'' temperature. Calculated using Eq.~\eqref{eqn:two_mode_QFI_zeroth_order} and $m=1$, $n=2$ (Fock spaces corresponding to the first and the second excited state within the cavity). Different combinations of initial temperatures are used, $\nu_{m,n}=2,6,10$. Similarly to the one-mode scenario, it is appropriate to measure at certain times ($\tau=1,3,5,\dots$) when the estimation precision grows exponentially with the squeezing parameter $r$. Moreover, the graph shows that the highest precision in estimation is achieved with large temperature difference between the modes, i.e., when $\nu_1=2$ and $\nu_2=10$, or $\nu_1=10$ and $\nu_2=2$. The diagonal $\nu_1=\nu_2\rightarrow\infty$ quickly converges to the double of the two-mode squeezed vacuum value given by $\nu_1=\nu_2=1$. An opportunity of using temperature difference between the modes is not the only advantage of using the two-mode states. In contrast to the one-mode states, two-mode states also achieve one order higher precision with the same amount of squeezing.}\label{fig:two-modeQFI}
\end{figure}

To illustrate the power of the derived formulae, we calculate the bound on the estimation of the proper acceleration using cavities. This section is a continuation of example~\ref{ex:accelerated_cavity}. Assume a quantum state inside of a non-moving cavity. Starting at proper time $\tau_0=0$, the cavity goes through a period $\tau$ of the proper acceleration $a$ (as measured in the centre of the cavity) and period $\tau$ of retardation $-a$, stopping again at time $2\tau$. We wish to estimate the proper acceleration $a$, we thus identify $a\equiv \epsilon$. The proper length of the cavity $L=1$ is considered constant during the whole procedure. Bogoliubov transformation of the state in this scenario has been calculated using a continuous perturbative expansion in the small parameter $a$ and is given by Eq.~\eqref{eq:bogos_acceleration}. Using these transformations, we calculate the zeroth order quantum Fisher information for a one-mode squeezed vacuum and a two-mode squeezed thermal state as shown on figures~\ref{fig:one-modeQFI} and~\ref{fig:two-modeQFI}.

\section{Conclusion and Discussion}\label{sec:QFT_conclusion}

In order to provide a general framework for estimating space-time parameters using quantum metrology, we have extended previous pure state analysis to the mixed case. The main motivation is that, for any practical and experimental purposes, quantum systems are always mixed. We have restricted our analysis to Gaussian probe states as the covariance matrix formalism provides a simple mathematical description. In particular, Gaussian states are also straightforward to prepare in quantum optical laboratories. We have computed general and exact expressions for the quantum Fisher information for one- and two- mode mixed Gaussian probe states undergoing arbitrary Bogoliubov transformations, and illustrated their use for the estimation of proper acceleration.

By expanding the Bogoliubov coefficients around the point $\epsilon_{0}=0$, we were able to evaluate the quantum Fisher information for the case of one- and two-mode Gaussian probe states. We obtained exact expressions for the quantum Fisher information at point $\epsilon\,=\,0$, and perturbative expressions for $\epsilon\neq 0$. In the single mode case, for a finite temperature, the quantum Fisher information is identically zero at $\epsilon=0$. This implies that for states which are at some temperature other than absolute zero, one cannot distinguish between two states in the neighbourhood of $\epsilon=0$. This can be explained in the following way: the assumption~\eqref{eqn:simplifying_condition} says that there is no change in the same mode up to the first order in $\epsilon$, i.e., the Bogoliubov transformation is a purely mode-entangling channel up to the first order in $\epsilon$. In general, the quantum Fisher information depends only on the first derivatives of the parameter. Since these derivatives are zero for a single mode state, the zeroth order of the quantum Fisher information is also zero. The only exception is for pure states because, as explained in section~\ref{sec:pure_states_QM}, there is a contribution from the change of purity given by the second derivative of the symplectic eigenvalues. This is why the zeroth order quantum Fisher information is non-zero in the zero and small temperature regime. For larger values of $\epsilon$, the quantum Fisher information is non-zero for all cases and the quantum Cram\'{e}r-Rao bound is finite. In the case of a thermal two-mode squeezed state there is, however, always the possibility of distinguishing between infinitesimally close states in the neighbourhood of $\epsilon=0$.

Higher squeezing and a high temperature difference of modes significantly improves the precision in estimation. We observed a similar behaviour in the estimation of unitary channels as studied in the previous chapter. Squeezed states are generally more sensitive to rotations and mode-mixing as well as particle creation when the squeezed state is appropriately oriented. The difference in temperature also helps because any mode-mixing channel given by a non-trivial passive coefficient $\alpha$ will, in general, mix temperatures of different modes. This effect vanishes when the modes have the same temperature. Ultimately, this is due to the fact that squeezed thermal states have high variance in energy which, according to the general equation for the quantum Fisher information~\eqref{pureK}, leads to a greater precision in estimation.

%This behaviour is due to the entanglement generated by the Bogoliubov transformation, which is related to particle creation. It is therefore clear that parameter estimation improves if particles are generated during a Bogoliubov transformation. The results can be interpreted in the following way: i) In the absence of initial excitations or thermal fluctuations, the dominating effect contributing to the quantum Fisher information is a result of the particle creation. This is because the only way to generate changes in the state is by generating excitations in different modes. ii) When there are either excitations from squeezing or thermal fluctuations, both pure particle creation \emph{and} passive transformation effects contribute to the quantum Fisher information. These passive transformations result from non-zero $\boldsymbol{\alpha}$ coefficients and allow for excitations to be shifted from one mode to another, therefore, allowing for a greater distinction between the initial state and the final state.

Our results will enable researchers to evaluate how well space-time parameters, such as the amplitude of gravitational waves, accelerations and local gravitational fields, can be estimated in the presence of background temperature~\cite{Sabin2014a,Sabin2015a}. We observe that strategies involving one- and two-mode probe states exhibit the same exponential gain for large squeezing. However, single mode thermal states do not perform well in the scenario when the channel is mostly mode-entangling, which is a common case in the literature~\cite{Bruschi2012a,Bruschi2013b,Friis2013a,Sabin2014a}.

Our results lead naturally to other important questions. The quantum Fisher information is the optimisation of the classical Fisher information over all possible measurements. One can therefore ask: ``What is the optimal measurement for our scheme?'' An analysis of the symmetric logarithmic derivative would certainly shed light on this and general knowledge in this direction has already been developed~\cite{Monras2013a,Gao2014a}, Eq.~\eqref{eq:SLD_Gaussian}. Furthermore, if the optimal measurement is found to be impractical then an analysis of more realistic measurements, such as homodyne and heterodyne measurements for Gaussian states, could prove fruitful. These questions are left for future work.

\begin{comment}
QUANTUM REFERENCE FRAMES
-quantum reference frames
    -the integral and review
    -reference frames in metrology
    -derived formulas in a simple language and examples, adding what hasn't been added
    -future guesses and possible directions (theorem about boob-shaped things, why everything goes to the same shape?), possible generalizations (remember what sam braunstein said??, look at the notes we did together)
\end{comment}
%% citations, conclusions,

%%%%%%%%%%%%%%%%%%%%%%%%%%%%%%
\chapter{Quantum metrology with imperfect reference frames}\label{chap:reference_frames}
%%%%%%%%%%%%%%%%%%%%%%%%%%%%%%

%introduction
The last chapter of this thesis will be slightly different from the previous in scope. This chapter is not primarily focused on Gaussian states and the operations studied here are not Gaussian operations. However, we will use many tools of quantum metrology as introduced in chapter~\ref{chap:QM}. In this chapter we show how not sharing a common reference frame between two parties affects the estimation precision of a parameter encoded in a quantum state which was sent from party A to party B. Moreover, we will show how this estimation precision can be improved using quantum reference frames. This chapter is mostly based on a paper we published in~\cite{safranek2015quantum} which also contains full proofs. %with space-based experiments

%scenario
Consider the following scenario: Alice encodes the parameter of interest into a known quantum state. She sends this state to Bob and it is Bob's task to decode the parameter. This known quantum state has been, however, defined with respect to Alice's reference frame, i.e., with respect to the measurement basis Alice chose to use. Bob's reference frame may be different but if he knows how to relate his reference frame to Alice's, he can simply rearrange his measurement basis to match hers and measure the parameter with the same precision. The problem begins when Bob does not know the relative orientation of his reference frame with respect to Alice's reference frame. Then his knowledge of the quantum state of interest is only partial and some or all information about the quantum state is lost. Therefore, Bob's precision in estimation of the parameter will be lower than Alice's. Nevertheless, there is a way to counter that. Because Bob does not share a reference frame with Alice, Alice can send her original state of interest along with another quantum state which represents a piece of her reference frame. This additional quantum state is called the quantum reference frame~(QRF). Bob will then perform the measurement on this composite system, in other words, he will perform a composite measurement which can be interpreted as the measurement on the original quantum state relative to the attached quantum reference frame. Because the parameter of interest is now encoded in the internal degrees of freedom of the composite system some precision in estimation of this parameter is retrieved.

%quantum reference frames
Quantum reference frames had been first introduced in~\cite{Peres2001Transmission,Peres2001entangled,bartlett2007reference,angelo2011physics}. A QRF is different from its classical counterpart in two ways: first, due to its quantum nature, it has an inherent uncertainty and the measurement results are only an approximation of what would be obtained using a classical reference frame.  For instance, if the reference frame describes a continuum of orientations in space, then states with different orientations are not perfectly distinguishable. Second, each time the QRF is used, it suffers a back-action, which causes future measurements to be less accurate. There has been extensive literature published on QRFs. We mention phase measurements using a QRF~\cite{bartlett2006degradation}, degradation of a directional QRF~\cite{bartlett2006degradation,poulin2007dynamics,ahmadi2010dynamics}, or a QRF considered as the resource state~\cite{gour2008the,gour2009measuring,skotiniotis2012alignment,narasimhachar2014phase}. QRFs enable us to achieve quantum information processing tasks without first establishing a shared reference frame. A QRF allows us to perform tasks in the absence of a common classical reference frame, in the same way that entangled states allow for the possibility of performing non-local quantum operations.

%content
In this chapter, we utilize the powerful machinery of quantum metrology to study the ultimate precision bounds in measurement of physical parameters with respect to QRFs. First we investigate how the ultimate precision in measurement of a parameter decreases due to inaccessibility of a perfect classical reference frame. We analyze the decrease in quantum Fisher information as a result. In particular, we provide necessary and sufficient conditions for two extreme cases that can occur in quantum parameter estimation. The first case is when the absence of a perfect reference frame does not affect the precision and the second case is when measurement of the parameter is no longer possible due to lack of access to a classical reference frame. We split the problem into two subproblems: the first case is when the encoding operator commutes with the operator representing the noise and the second is when it does not. Counter-intuitively, we show that the non-commuting case has some advantages over the commuting one. While the existence of ``decoherence-free subspaces'' is essential for encoding information in the commuting case~\cite{lidar1998decoherence,lidar2001decoherence,bacon2001coherence,bartlett2007reference}, for non-commuting operators the estimation is possible even in the absence of such subspaces. The trade-off is, however, that the precision will in general depend on the parameter to be estimated. %In addition, we explain the connection between noisy quantum metrology and alignment-free quantum communication.
Finally, we present three examples to further clarify different aspects of quantum metrology with imperfect frames of reference.

\section{Mathematical framework of quantum reference frames}
%Using quantum reference frames }

Consider $g\in G$ to be the group element that describes the passive transformation from Alice's to Bob's reference frame. Alice prepares a state ${\R}_{A}$ relative to her local reference frame. In the Bob's reference frame this state looks as $\hat{U}(g){\R}_A \hat{U}(g)^\dag$, where $\hat{U}$ is a unitary operator parametrized by a group element $g$. However, if Bob is completely unaware of the relation between his local reference and Alice's local reference frame, we can assume that the group element $g$ is completely unknown. It follows that relative to Bob's reference frame this state is seen as\footnote{We will restrict our attention to Lie-groups that are compact, so that they possess a group-invariant (Haar) measure ${\mathrm{d}} g$. We refer the readers for more details to \cite{bartlett2007reference}.} \cite{bartlett2007reference}
\[
\label{gtwirl}
{\R}_{B}={\cal{G}}[{\R}_{A}]=\int {\mathrm{d}} g \hat{U}(g) {\R} _{A}\hat{U}(g)^{\dag}.
\]
Therefore, lacking such a shared reference frame is equivalent to having a noisy completely positive trace-preserving map which is known as the ``\textsl{g-twirling map}'', i.e. $\G({\R}_{A})$. The resulting density matrix $\rho_B$ then represents the state of knowledge Bob has about the physical system. It follows from the construction that such state must be less or equally pure to the state of Alice, and no information is lost only if the $\R_A$ is invariant under the g-twirling map. The integral above is over the Haar measure ${\mathrm{d}} g$ which in case of compact Lie groups is group-invariant.

Now we formalize the metrological scenario discussed in the introduction: We assume Alice sends a state ${\R}_\epsilon$ which depends on the parameter of interest $\epsilon$. This state can represent either solely the original state with the encoded parameter ${\R}_\epsilon={\R}_{A\epsilon}$ or the original state plus the attached quantum reference frame, ${\R}_\epsilon={\R}_{A\epsilon}\otimes{\R}_{QRF}$. We moreover assume that the unitary operator which connects states from the Alice's reference frame and the Bob's reference frame is a one-parameter unitary group $\hat{U}(t)=e^{-i\hat{G}t}$, where $\hat{G}$ is a Hermitian operator. Eq.~\eqref{gtwirl} becomes
\begin{equation}\label{Htwirl}
{\R}_{B}=\G[{\R}_{\epsilon}]=\lim_{T\rightarrow\infty} \frac{1}{T}\int_{0}^{T} dt\ \hat{U}(t){\R}_{\epsilon}\hat{U}(t)^{\dag}
\end{equation}
Parameter $t$ is the element of the group and $\hat{G}$ is the generator of translations in this parameter. For example, if two parties do not have synchronized clocks, i.e. they do not share a time reference frame, the parameter $t$ represents time and generator of translation in time $\hat{G}$ is the Hamiltonian of the system. $\hat{G}$ and $t$ are thus determined by the type of reference frame that is lacking.

Assuming the spectral decomposition $\hat{G}=\sum_{i}G_{i}\hat{P}_{i}$, where the $\hat{P}_{i}$s are the projectors into subspaces with eigenvalues $G_{i}$ and $\sum_{i}\hat{P}_{i}=I$, one can easily check that the state $\G[{\R}_{\epsilon}]$ in~\eqref{Htwirl} can be written as
\[\label{rhoB}
\G[{\R}_{\epsilon}]=\sum_{i}\hat{P}_i{\R}_{\epsilon}\hat{P}_i.
\]
The full aim of this chapter motivated by quantum reference frames is quite simple: It is to understand how the quantum Fisher information behaves under the transformation~\eqref{rhoB}. Because of the computational complexity we consider only states which are pure in Alice's reference frames, ${\R}=\ket{\psi_\epsilon}\bra{\psi_\epsilon}$. From now on we also drop the lower index $\epsilon$.

\section{Loss of the estimation precision}

Assuming ${\R}=\ket{\psi}\bra{\psi}$ is a pure state the quantum Fisher information is given by Eq.~\eqref{eq:QFI_pure},
\[\label{eq:QFI_pure2}
H({\R})=4(\braket{\partial_{\epsilon}\psi}{\partial_{\epsilon}\psi}-\norm{\braket{\psi}{\partial_{\epsilon}\psi}}^2).
\]
Using Eq.~\eqref{QFI} for the quantum Fisher information, the properties of the quantum channel $\G$~\eqref{rhoB} and the Parseval identity (see appendix~\ref{app:FirstFormula}) we derive Bob's quantum Fisher information as
\[\label{HrhoB}
H(\G[{\R}])=4\braket{\partial_\epsilon\psi}{\partial_\epsilon\psi}-
4\sum_i\frac{(\Im\bra{\psi}\hat{P}_i\ket{\partial_\epsilon\psi})^2}{\bra{\psi}\hat{P}_i\ket{\psi}},
\]
where the summation is over the indices $i$ for which $p_i=\bra{\psi_{\epsilon}}\hat{P}_i\ket{\psi_{\epsilon}}\neq0$. Note that we will use this convention throughout the rest of this chapter.

As mentioned before, because Bob does not share a reference frame with Alice, his precision in estimation in estimation should be lower than Alice's. We formalise this statement in the theorem below together with the necessary and sufficient conditions for two extreme cases. The first case is where the precision in measurement of $\epsilon$ remains the same both in the absence or the presence of a perfect reference frame and the second case is where the measurement of $\epsilon$ is not possible anymore due to inaccessibility of such reference frames. The full proof can be found in appendix~\ref{app:nonnegQFIl} or~\cite{safranek2015quantum}.
\begin{theorem}\label{theorem}
Let ${\R}=\pro{\psi}{\psi}$ be a pure initial state. Then the quantum Fisher information ${\R}_B=\G[{\R}]$ is bounded,
$0\leq H(\G[{\R}])\leq H({\R})$.

No precision in the estimation is lost (no loss), i.e. $H(\G[{\R}])= H({\R})$, if and only if
\[\label{cnoloss}
\forall i,\ \Im\bra{\psi}\hat{P}_i\ket{\partial_\epsilon\psi}=-i\PI\braket{\psi}{\partial_\epsilon\psi}.
\]

$\epsilon$ cannot be estimated anymore (maximum loss), i.e. $H(\G[{\R}])=0$, if and only if
\[\label{maxlosstheorem}
\forall i,\ \Re\bra{\psi}\hat{P}_i\ket{\partial_\epsilon\psi}=0\ \wedge\ \forall \ket{\phi_j},\ \braket{\phi_j}{\partial_\epsilon\psi}=0,
\]
where $\{\frac{\hat{P}_i\ket{\psi}}{\sqrt{p_i}},\ket{\phi_j}\}_{i,j}$ forms an orthonormal basis of the Hilbert space.
\end{theorem}
Without loss of generality in Eq.~\eqref{cnoloss}, we can restrict our analysis to the terms for which  $p_i=\PI\neq0$, since using the Cauchy-Schwarz inequality it can be checked that the condition~\eqref{cnoloss} holds trivially if $p_i=0$. Also, the set of states $\{\ket{\phi_j}\}_j$ are orthonormal states which together with the set of normalised states $\{\frac{\hat{P}_i\ket{\psi}}{\sqrt{p_i}}\}_i$ form a complete basis. We can always find the set of states $\{\ket{\phi_j}\}_j$ via the Gram-Schmidt process for orthonormalisation of a set of vectors. Assuming the eigenvectors of $\hat{G}$ span the whole Hilbert space, $\{\ket{\phi_j}\}_j$ is exactly the set of the eigenvectors of $\G({\R})$ with the respective eigenvalue $0$.\\

%From Eq.~\eqref{cnoloss} we can also obtain necessary condition for the scenario when no precision in estimation is lost. Summing over all projectors $\hat{P}_i$ gives

Using similar analysis we can find the symmetric logarithmic derivative as
\[\label{NSLD}
L(\G({\R}))=\sum_i\pro{\varphi_i}{\psi_i}+\pro{\psi_i}{\varphi_i},
\]
where $|\psi_{i}\rangle$ and $|\varphi_{i}\rangle$ are defined as
\begin{equation}
\ket{\psi_i}=\frac{\hat{P}_i\ket{\psi}}{\sqrt{p_i}},\quad
\ket{\varphi_i}=\frac{1}{\sqrt{p_i}}\left(2\hat{P}_i\ket{\partial_\epsilon\psi}-
\langle{\psi_i}\ket{\partial_\epsilon\psi}\ket{\psi_i}\right).
\end{equation}
As discussed in the first chapter, eigenvectors of the symmetric logarithmic derivative give the optimal POVM for the estimation of the parameter.

\section{Unitary encoding operations}

In this section we will assume the parameter of interest has been encoded via one-parameter unitary group, $\ket{\psi}=e^{-i\hat{K}\epsilon}\ket{\psi_0}$.
We will analyse the quantum Fisher information, the no-loss and the and maximum-loss conditions in terms of the hermitian operator $\hat{K}$. We also remind that the Hermitian operator $\hat{G}$ is the generator of the the noisy channel $\G$. This way we split the problem into two different cases. The first case is where the encoding process in general does not commute with the noisy channel, i.e. $[\hat{K},\hat{G}]\neq0$. We call such noise non-commutative. The second is when when the noise is commutative. If the noise is commutative, it simply means that the noisy channel \eqref{Htwirl} commutes with the encoding process, i.e. $[\hat{K},\hat{G}]=0$. In that case our results can be also applied on systems where the noise \eqref{rhoB} precedes the encoding operation.

Suprisingly, for the parameter encoded via one-parameter unitary group the quantum Fisher information takes a very elegant form. This form follows directly from Eq.~\eqref{HrhoB} and is formalized in the following theorem.
\begin{theorem}\label{nicetheorem}
For an initial pure state $\ket{\psi_0}$, a generator $\hat{K}$ of a unitary encoding operator $\hat{U}(\epsilon)=\text{exp}(-i\hat{K}\epsilon)$ and projectors $\hat{P}_i$ of the g-twirling map in \eqref{rhoB}, the quantum Fisher information of the state $\G[{\R}_{\epsilon}]$ is
\begin{equation}\label{BobQFI}
\begin{split}
 H(\G[{\R}_{\epsilon}])
=4\mathrm{Var}_{{\R}_{\epsilon}}(\hat{K})-4\sum_i
p_i\left[\mathrm{Cov}_{{\R}_{\epsilon}}\left(\frac{\hat{P}_i}{p_i},\hat{K}\right)\right]^2,
\end{split}
\end{equation}
where ${\R}_{\epsilon}=|\psi_{\epsilon}\rangle\langle\psi_{\epsilon}|$, $\ket{\psi_{\epsilon}}=\hat{U}(\epsilon)\ket{\psi_0}$ and $p_i=\mean{\hat{P_i}}_{{\R}_\epsilon}$. The covariance between two operators $\hat{A}$ and $\hat{B}$ is defined as $\mathrm{Cov}_{\R}(\hat{A},\hat{B})=\frac{1}{2}\mean{\{\hat{A},\hat{B}\}}_{\R}-\mean{\hat{A}}_{\R}\mean{\hat{B}}_{\R}$ and the variance can be written as $\mathrm{Var}_{\R}(\hat{A})=\mathrm{Cov}_{\R}(\hat{A},\hat{A})$. Note that the first part of the expression is equal to the quantum Fisher information in Alice's reference frame, $H({\R})=4\mathrm{Var}_{{\R}_{\epsilon}}(\hat{K})$.

No precision in the estimation is lost (no loss), i.e. $H(\G[{\R}])= H({\R})$, if and only if
\[\label{cnolossKG}
\forall i,\ \mathrm{Cov}_{\R}(\hat{P}_i,\hat{K})=0.
\]

$\epsilon$ cannot be estimated anymore (maximum loss), i.e. $H(\G[{\R}])=0$, if and only if
\[\label{maxlosstheoremKG}
\forall i,\ \bra{\psi}[\hat{P}_i,\hat{K}]\ket{\psi}=0\ \wedge\ \forall \ket{\phi_j},\ \bra{\phi_j}\hat{K}\ket{\psi}=0,
\]
where $\{\frac{\hat{P}_i\ket{\psi}}{\sqrt{p_i}},\ket{\phi_j}\}_{i,j}$ forms an orthonormal basis of the Hilbert space.
\end{theorem}

From Eq.~\eqref{BobQFI} we deduce that the decrease in the quantum Fisher information is proportional to the mean of squared covariances between the normalized\footnote{$\mean{\hat{P}_i/{p_i}}_{{\R}}=1$} projectors ${\hat{P}_i}/{p_i}$ and the encoding operator $\hat{K}$. This means that  the more projectors ${\hat{P}_i}/{p_i}$ are correlated with the encoding operator $\hat{K}$, the more precision is lost. Roughly speaking, in order to lose the minimum amount of precision one should choose an encoding operator $\hat{K}$ which is less correlated with the decoherence caused by the noisy channel $\G$. Explicit examples will be presented in the next section.

Another useful theorem will show that in certain common scenarios the estimation of the parameter is impossible.
\begin{theorem}\label{thirdtheorem}
Let $\hat{G}$ have a non-degenerate spectrum, i.e., each projector $\hat{P}_i$ has rank $1$. If $[\hat{K},\hat{G}]=0$, then $H(\G[{\R}])=0$, i.e., parameter $\epsilon$ cannot be estimated anymore.
\end{theorem}
Because $\hat{K}$ and $\hat{G}$ commute if and only if $\hat{K}$ commutes with every projector $\hat{P}_i$ in the spectral decomposition of $\hat{G}$, the proof immediately follows from Eq.~\eqref{maxlosstheoremKG}. Physically, this theorem shows that non-degenerate simply wipe out all information about the parameter in the system. The above theorem can be also explained as follows. To be able to estimate the parameter in the presence of a commutative noise it is necessary to have decoherence-free subspaces in which the information about the parameter is stored. Decoherence-free subspaces are given by projectors associated with degenerate eigenvalues of $\hat{G}$. These subspaces are spanned by the eigenvectors of these projectors associated with the eigenvalue 1, i.e., by the states which are invariant under the projection. However, decoherence-free subspaces are not needed for the parameter estimation in the presence of the non-commutative noise.

From Eq.~\eqref{cnolossKG} we can also derive a necessary condition for the scenario when no precision in estimation is lost. Summing over all projectors $\hat{P}_i$ yields
\[\label{noloss}
H(\G[{\R}])=H({\R})\ \Rightarrow\ \mathrm{Cov}_{\R}(\hat{G},\hat{K})=0.
\]
This means that if operators $\hat{K}$ and $\hat{G}$ are correlated with respect to the pure initial state ${\R}$, i.e. $\mathrm{Cov}_{\R}(\hat{G},\hat{K})\neq0$, then some precision is lost due to the misalignment of reference frames. It is worth emphasising that this condition is not sufficient. As an example consider the operators $\hat{K}=\pro{2}{2}$, $\hat{G}=6\pro{0}{0}+3\pro{1}{1}+4\pro{2}{2}$, and the initial state $\ket{\psi_0}=\frac{1}{\sqrt{6}}\ket{0}+\frac{1}{\sqrt{3}}\ket{1}+\frac{1}{\sqrt{2}}\ket{2}$. In this example the covariance between $\hat{G}$ and $\hat{K}$  is zero, nevertheless, since $\hat{K}$ and $\hat{G}$ commute and the fact that no non-degenerate subspace exists, according to theorem~\ref{thirdtheorem} we will not be able to extract any information about $\epsilon$.

\section{Examples}~\label{sec:Ex}

In the previous sections we analysed how the quantum Fisher information changes when of the party which decodes the parameter of interest does not share a local reference frame with the party which encoded the parameter. Now we will present some explicit examples. In the first example we show the scenario where the encoding operator $\hat{K}$ commutes with the generator connecting the two misaligned reference frames $\hat{G}$. In such scenario the decoherence-free subspaces are necessary. In the second example, however, due to the interaction between the system and the QRF $\hat{K}$ and $\hat{G}$ no longer commute. The third example is somewhat different. There we present the case where Alice and Bob share only one axis of their reference frames but not the others. We also find the optimal encoding operator $\hat{K}$ which maximizes the amount of information extracted by Bob.

\begin{figure}[t]
\centering
\includegraphics[width=\linewidth]{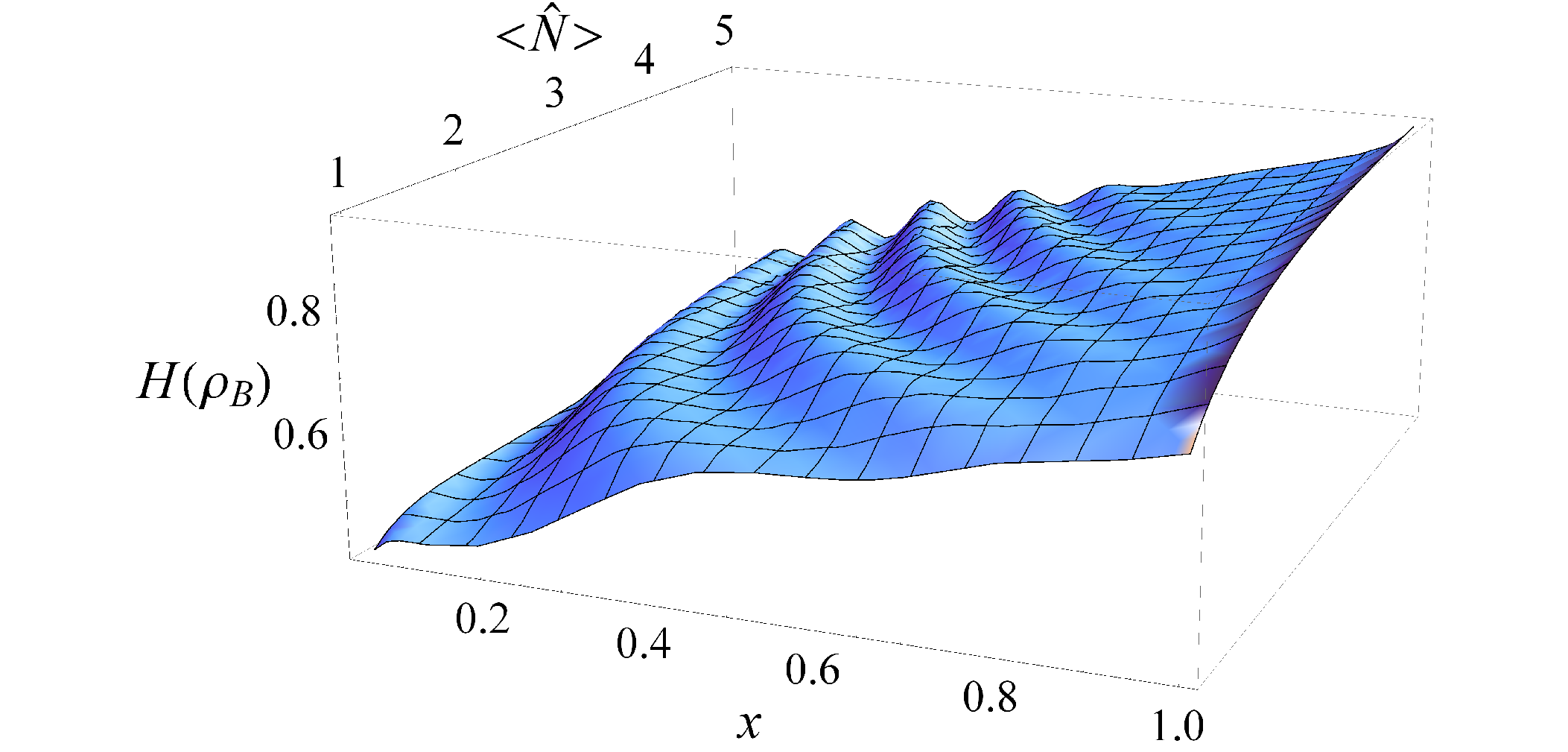}
\caption{Bob's quantum Fisher information in terms of mean photon number ${\mean{\hat{N}}}$ and $x$ for a squeezed, displaced vacuum state, i.e. $|\psi_{QRF}\rangle=|\alpha,r\rangle$, as the initial state of the QRF. Parameter $x$
denotes the fraction of mean energy due to displacing the vacuum, i.e. $x=\frac{\alpha^2}{\mean{\hat{N}}}$.}\label{SQD}
\end{figure}

\subsection{Two non-interacting quantum harmonic oscillators}\label{sec:two_nonint_osc}

The scenario that we consider in this example is as follows. Alice and Bob do not have access to synchronised clocks, i.e. they do not share a time reference frame. Alice prepares a state $|\psi_{\epsilon}\rangle= \hat{U}_{\epsilon}|\psi_{0}\rangle$, where $\hat{U}_{\epsilon}=e^{-i \hat{K}\epsilon}$. Since the local clocks of the parties are not synchronised, in Bob's frame the state of the system is given by Eq.~\eqref{gtwirl}, where $\hat{U}(t)=e^{-i\hat{H}t}$ and $\hat{G}\equiv\hat{H}$ is the Hamiltonian of the qubit and the $QRF$. The operators $\hat{P}_{i}$ are the projectors into subspaces with total energy $E_{i}$. We analyse the quantum Fisher information of the state ${\R}_{B}=\G[{\R}]$ which shows how precise Bob will be able to measure $\epsilon$.\\

Let us consider the example of two non-interacting quantum harmonic oscillators with the Hamiltonian $H=\hbar \omega (\hat{a}^{\dag}\hat{a}+\hat{b}^{\dag}\hat{b})$,  where $\hat{a}$ and $\hat{a}^{\dag}$ are the creation and annihilation operators corresponding to the first quantum harmonic oscillator and $\hat{b}$ and $\hat{b}^{\dag}$ to the second respectively. One harmonic oscillator will serve as a carrier of the information about the parameter while the other will represent a quantum reference frame. The initial state is therefore of the product form $|\psi_0\rangle=|\psi_q\rangle \otimes |\psi_{QRF}\rangle$, where $|\psi_q\rangle=\frac{1}{\sqrt{2}}(|0\rangle+|1\rangle)$ and $|0\rangle$ and $|1\rangle$ are the eigenstates of number operator $\hat{N_q}=\hat{a}^{\dag}\hat{a}$ with eigenvalues $0$ and $1$ respectively. We choose the generator of the unitary channel $\hat{U}_{\epsilon}$ to be number operator associated with the first harmonic oscillator, $\hat{K}=\hat{a}^{\dag}\hat{a}$. It is worth emphasising at this point that in this example $[\hat{K},\hat{H}]=0$. Note that this example is similar to the quantum communication scheme between two parties when they do not have a common phase reference frame as was considered in~\cite{bartlett2009quantum_communication}.

Using Eq.~\eqref{eq:QFI_pure2}, it is straightforward to find the quantum Fisher information in Alice's frame as $H({\R})=1$. Note that Alice's quantum Fisher information is independent of the state of the QRF. On the other hand, if we consider the state $|\psi_{QRF}\rangle=\sum_{n=0}^{N-1}c_{n}|n\rangle$, then using either Eq.~\eqref{BobQFI} or Eq.~\eqref{HrhoB}, we find the quantum Fisher information in Bob's frame as
\[\label{foranystate}
H({\R}_B)=2\sum_{n=0}^{N-2}\frac{\abs{c_n}^2\abs{c_{n+1}}^2}{\abs{c_n}^2+\abs{c_{n+1}}^2}.
\]
If Alice chooses a uniform superposition of Fock states, i.e. the state $|\psi_{US}\rangle=\frac{1}{\sqrt{N}}\sum_{n=0}^{N-1}|n\rangle$, then using~\eqref{foranystate} we can easily compute Bob's quantum Fisher information as $H({\R}_B)=1-\frac{1}{N}$. Using Eq.~\eqref{SLD}, we find the elusive symmetric logarithmic derivative for this case as
\[
L({\R}_{B,US})=\sum_{n=1}^{N-1}ie^{i\epsilon}\ket{0}\pro{n}{n-1}\bra{1}-ie^{-i\epsilon}\ket{1}\pro{n-1}{n}\bra{0}.
\]

\begin{figure}[t]
\centering
\includegraphics[width=\linewidth]{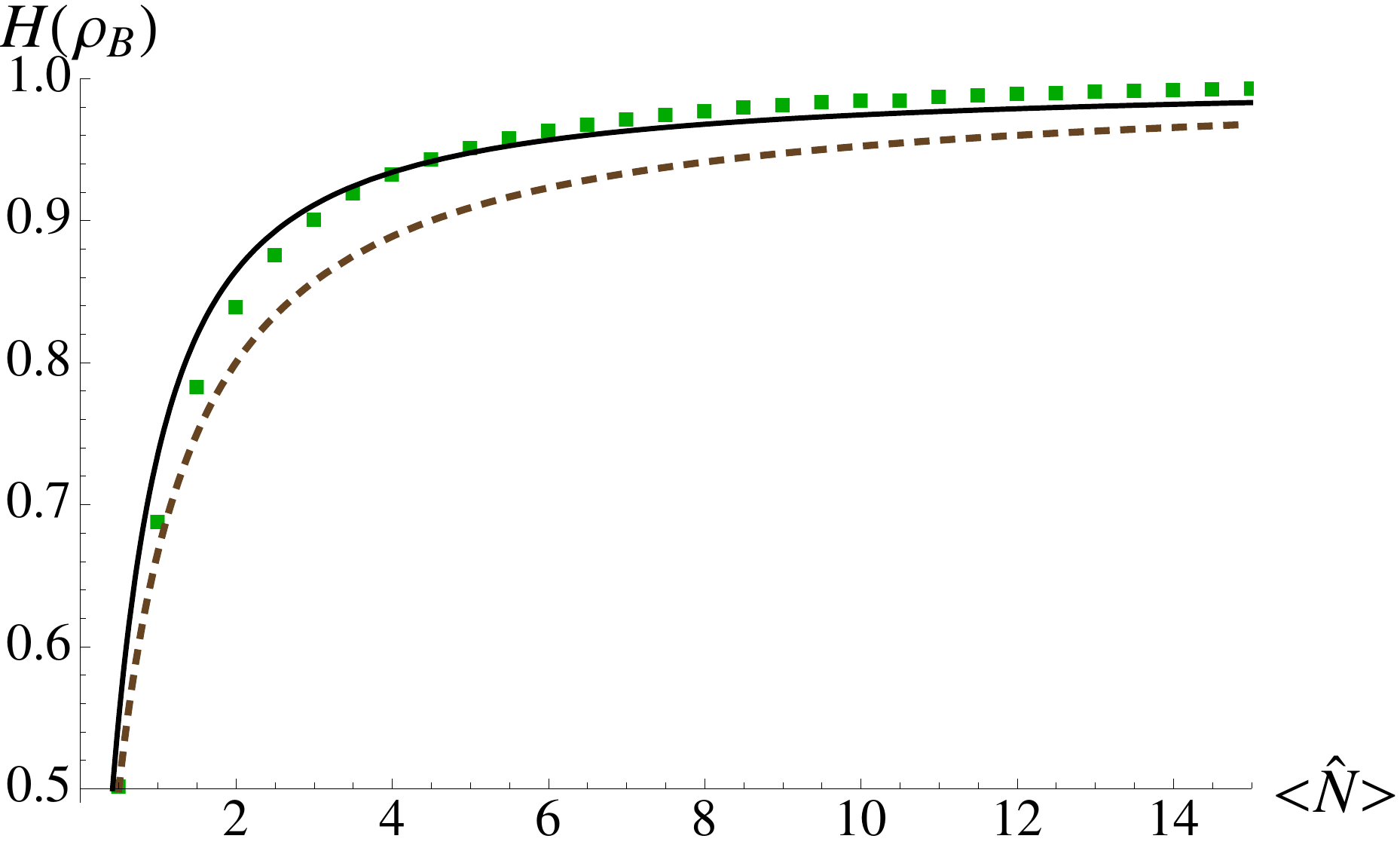}
\caption{Bob's quantum Fisher information in terms of mean photon number in the initial state of the QRF for three different states. The solid-black, dashed-brown and dotted-green curves correspond to coherent state, uniform superposition state $|\psi_{\tiny{US}}\rangle$, and the optimal state with finite cut-off $N=2\mean{\hat{N}}+1$.}\label{comp}
\end{figure}

Let us next consider a squeezed, displaced vacuum state~\cite{gerry2005introductory} $|\alpha,r\rangle=D(\alpha)S(r)\ket{0}$ as the state of the QRF.
The mean energy of this state is equal to $\mean{\hat{N}}=\alpha^2+\sinh^{2}r$. We define parameter $x$ as the fraction of initial mean energy due to displacing the vacuum, i.e. $x=\frac{\alpha^2}{\mean{\hat{N}}}$. Note that with this definition, $x=0$ and $x=1$ represent a squeezed state and a coherent state respectively. In particular, noticing that in the Fock basis a squeezed state is of the form $|r\rangle=\sum_{n}c_{n}|2n\rangle$ together with Eq.~\eqref{gtwirl} we find that $H(\G[\ket{r}\bra{r}])=0$, i.e. Bob will not be able to decode $\epsilon$ if Alice prepares the QRF in a squeezed state.\\

In figure \ref{SQD} we have plotted Bob's quantum Fisher information for the state $|\alpha,\xi\rangle$ in terms of $x$ and the mean energy $\mean{\hat{N}}$. As can be seen in this figure, if we fix the mean energy of the QRF, then it is optimal to have zero squeezing in the initial state of the QRF, i.e. $x=1$. This corresponds to preparing the QRF in a coherent state. Using Eq.~\eqref{foranystate} we find Bob's quantum Fisher information for a coherent state as
\begin{equation}\label{CohQFI}
H({\R}_B)=2\frac{\abs{\alpha}^2}{1+\abs{\alpha}^2}M\left(1,2+\abs{\alpha}^2,-\abs{\alpha}^2\right),
\end{equation}
where $M(a,b,z)$ is a confluent hypergeometric function. We derive the asymptotic expression for the limit of large
mean energy, i.e. $\abs{\alpha}^2\rightarrow\infty$, as
\[\label{CoherentH}
H({\R}_B) \approx 1-\frac{1}{4(\abs{\alpha}^2+1)}.
\]

In figure \ref{comp}, we compare Bob's quantum Fisher information for different QRFs. This figure shows that a coherent state outperforms the uniform superposition of Fock states. This is in complete agreement with the results of~\cite{ahmadi2013the_Wigner} where it is shown that if Bob chooses the Maximum-likelihood estimation process to decode $\epsilon$, then choosing a coherent state as the initial state of the QRF instead of the state $|\psi_{US}\rangle$ improves the efficiency of the communication protocol. We also optimize over the quantum reference frames numerically by maximizing Eq.~\eqref{foranystate} over amplitudes $c_1,\dots,c_{N-1}$.\footnote{We found that the resulting optimized states, $\ket{\psi_{OPT}}=\sum_{n=0}^{N-1}c_{n}|n\rangle$, representing the quantum reference frames have always symmetric amplitudes around the value $\mean{\hat{N}}=\frac{N-1}{2}$.} The green square-shaped dots in figure \ref{comp} represent the quantum Fisher information for this optimal state. As can be seen from the figure the coherent state is nearly optimal in this case.

\subsection{Two interacting quantum harmonic oscillators}\label{sec:two_int_osc}

In this example we again consider Alice and Bob not having an access to synchronised clocks. However, in this example we assume that the QRF interacts with the original state.

Let us consider the example of two interacting quantum harmonic oscillators with the total Hamiltonian
\[
\hat{H}=\hbar\omega(\hat{a}^{\dagger}\hat{a}+\hat{b}^{\dagger}\hat{b})+\hbar \kappa(\hat{a}^{\dagger}\hat{b}+\hat{b}^{\dagger}\hat{a}),
\]
where $\kappa$ is the interaction strength. Similar to the example of two non-interacting quantum harmonic oscillators, we consider the generator of the unitary channel to be the number operator, i.e. $\hat{K}=\hat{a}^{\dag}\hat{a}$. Note that the two operators $\hat{K}$ and $\hat{H}$ do not commute in this case, $[\hat{K},\hat{H}]=\kappa (\hat{a}^{\dag}\hat{b}-\hat{a}\hat{b}^{\dag})$. As mentioned earlier whenever these two operators do not commute, even in the absence of degenerate subspaces of total energy, we may still be able to estimate the parameter. For simplicity we also assume that frequency $\omega$ is not a fraction of the interaction strength $\kappa$, i.e.
\[
\forall P,R\in \mathbb{Z},\ P\omega\neq R\kappa.
\]
This assumption ensures that the hamiltonian $\hat{H}$ does not possess any degenerate eigenvalues.  In order to make the computations easier, we change the basis by defining a new set of annihilation operators as \cite{estes1968quantum}
\[
\hat{A}=\frac{1}{\sqrt{2}}(\hat{a}+\hat{b}),\ \ \hat{B}=\frac{1}{\sqrt{2}}(\hat{a}-\hat{b}).
\]
This change of basis allows us to write the Hamiltonian as
$\hat{H}=\hbar(\omega+\kappa)\hat{A}^{\dagger}\hat{A}+\hbar(\omega-\kappa)\hat{B}^{\dagger}\hat{B}$
with the eigenvectors
\[
\ket{\widetilde{m,n}}\!=\!\frac{(\hat{A}^{\dagger})^m}{\sqrt{m!}}\frac{(\hat{B}^{\dagger})^n}{\sqrt{n!}}\ket{\widetilde{0,0}}\!,
\]
which using $\ket{\widetilde{0,0}}=\ket{0,0}$ and applying the creation operators can be written in terms of the Fock basis as
\[
\ket{\widetilde{m,n}}=\sum_{k=0}^m\sum_{l=0}^n{{m}\choose{k}}{{n}\choose{l}}\sqrt{\frac{(k+l)!(m+n-k-l)!}{2^{m+n}m!n!}}\ket{k+l,m+n-k-l}.
\]

\begin{figure}[t]
\centering
\includegraphics[width=\linewidth]{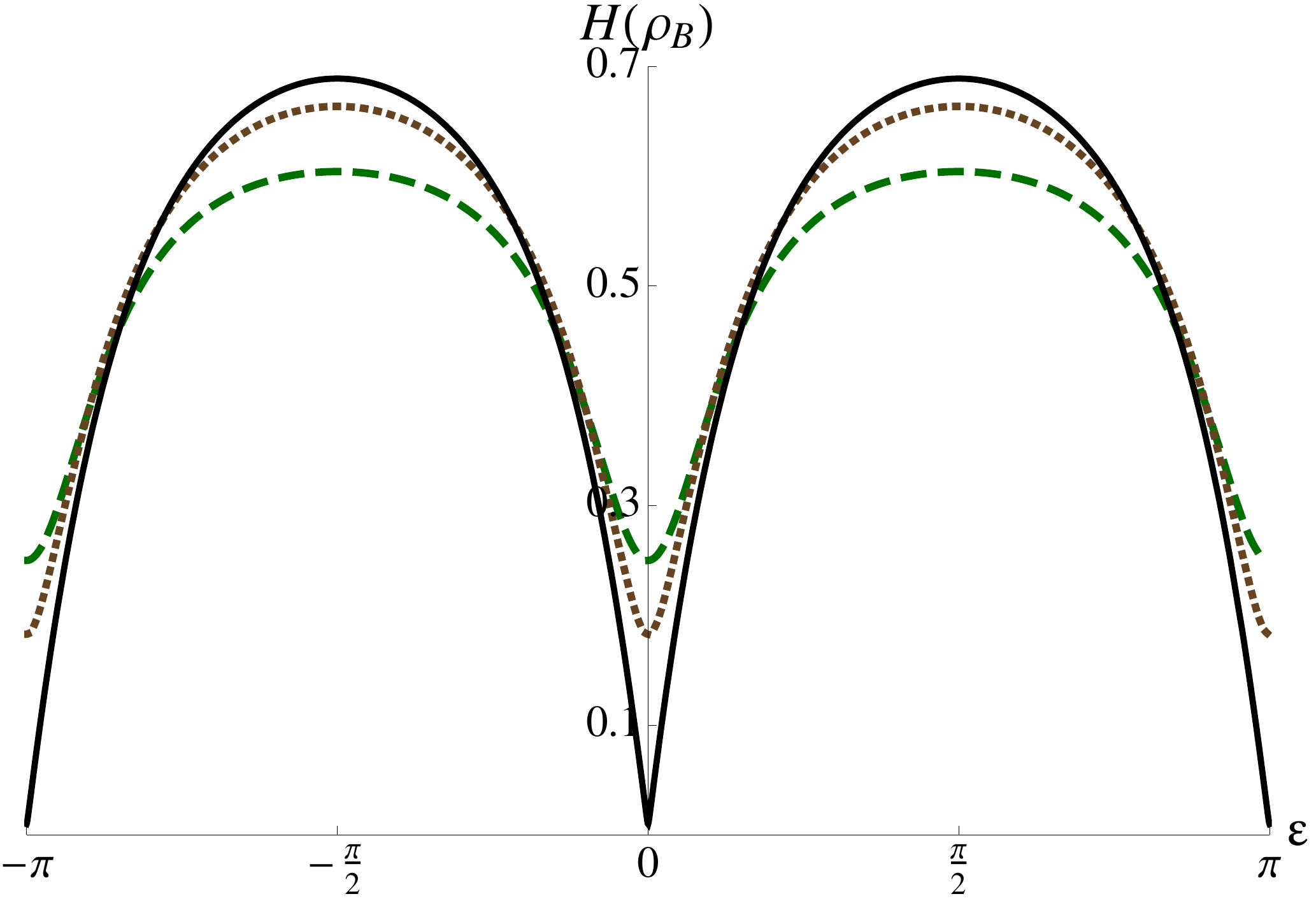}
\caption{Bob's quantum Fisher information vs. $\epsilon$ for two interacting quantum harmonic oscillators. The initial state is considered as $|\psi_{0}\rangle= \frac{1}{\sqrt{2}}(|0\rangle+|1\rangle)\otimes|\psi_{\tiny{US}}\rangle$. The dashed(green), dotted(brown) and solid(black) curves correspond to $N=4,10$ and $300$ respectively.}\label{nonvsint}
\end{figure}

Now let us consider that the QRF is initially prepared in the uniform superposition of Fock states. Using Eq.~\eqref{HrhoB}, we derive the quantum Fisher information of the averaged state as
\[
%\begin{split}
H({\R}_B)=2-\frac{8}{N}\Bigg(\sum_{m=1}^{\lfloor\frac{N}{2}\rfloor}\sum_{n=0}^{m-1}c_{m,n}(1-d_{m,n}(\epsilon))
-\!\!\!\!\sum_{m=\lfloor\frac{N}{2}\rfloor+1}^{N}\!\sum_{n=0}^{N-m}c_{m,n}-\!\!\!\!\sum_{m=\lfloor\frac{N}{2}\rfloor+1}^{N-1}\!\!\!\!\sum_{n=0}^{N-m-1}c_{m,n}d_{m,n}(\epsilon)\Bigg),
%\end{split}
\]
where $d_{m,n}(\epsilon)=\frac{(m+n)((m-n)^2+m+n)\sin^2\epsilon}{((m-n)^2-m-n)^2+4(m+n)(m-n)^2\sin^2\epsilon}$, $c_{m,n}=\frac{(m+n-1)!(m-n)^2}{2^{m+n+1}m!n!}$, and $\lfloor\cdot\rfloor$ is the floor function.

Because $\hat{K}$ and $\hat{G}$ do not commute in this example, $H(\G[{\R}])$ is $\epsilon$-dependent as opposed to the first example where Bob's quantum Fisher information was independent of the encoded parameter $\epsilon$. In figure \ref{nonvsint} we have plotted the quantum Fisher information $H(\G[{\R}])$ in terms of $\epsilon$ for increasing values of the mean energy in the state of the QRF. The maximum and minimum of the quantum Fisher information occurs at $\epsilon=\pm \frac{\pi}{2}$ and $\epsilon=0,\pm\pi$ respectively. Note that, even for very large $N$, the quantum Fisher information converges but it does not approach the ideal case. In other words, even in the limit of very large mean energy in the initial state of the quantum clock, we can not estimate the phase parameter $\epsilon$ as precise as we could if we had access to a classical clock. This can be proved using the necessary conditions~\eqref{noloss}. One can easily check that $\mathrm{Cov}_{\R}(\hat{G},\hat{K})=\frac{\hbar\omega}{4}$, which means that independent of $N$ and $\epsilon$, the quantum Fisher information is always smaller than one, i.e. $H(\G[{\R}])<1$. %Similarly, since $\mean{[\hat{K},\hat{G}]}_{\R}\approx\frac{2i\hbar\kappa}{3}\sqrt{N}\sin\epsilon$, we can deduce that independent of $N$ for $\epsilon\neq-\pi,0,\pi$, the quantum Fisher information is always positive.

\begin{figure}[t]
\centering
\includegraphics[width=0.30\textwidth]{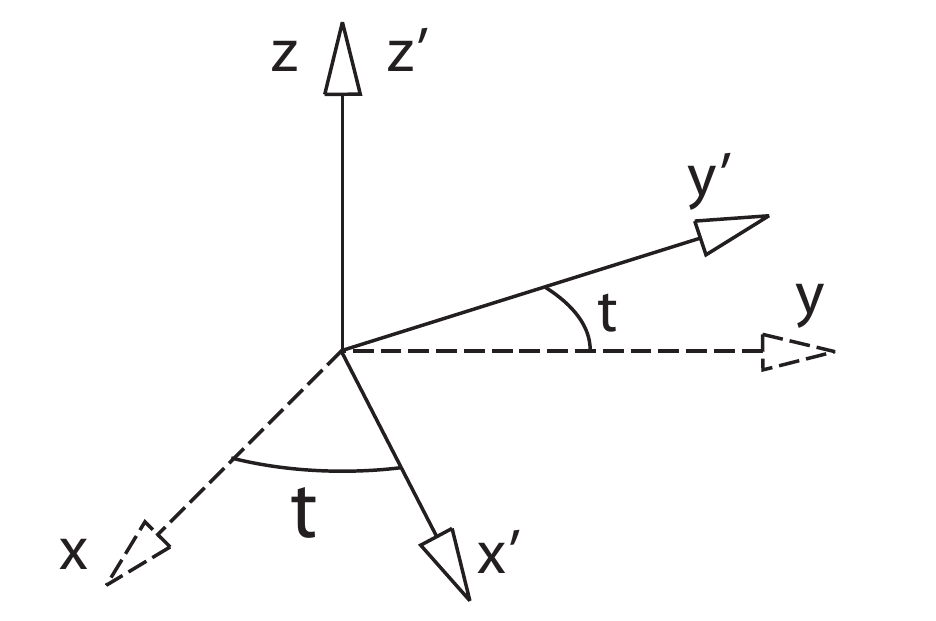}
\caption{Bob only shares his z-axis with Alice, i.e., he lacks the knowledge about the angle $t$ that relates his other two axes to Alice's axes.}\label{BobRotated}
\end{figure}

\subsection{Direction indicator}\label{RotatingObserver}

In the first two examples we observed how a QRF can help the estimation in the scenario when Alice and Bob do not have synchronized clocks. Here, we investigate the precision of estimation of an angle encoded in a qubit when Bob's measurement apparatus is rotated by an unknown angle with respect to Alice's measurement apparatus.

Let us start with the case where Alice wishes to both encode and decode a parameter herself. She chooses a spin-$\frac{1}{2}$ particle as the physical system to encode a parameter $\epsilon$ and then she encodes this parameter using a unitary channel with the generator
\[
\hat{K}=\frac{1}{2}\vec{n}\cdot\vec{\sigma}=\frac{1}{2}(x\sigma_x+y\sigma_y+z\sigma_z).
\]
This is the generator of a general rotation in the Bloch sphere around the axis $\vec{n}=(x,y,z)$, where $x^2+y^2+z^2=1$ and $x,y,z$ are real parameters.
For simplicity we choose the initial state to be the eigenstate of $\sigma_z$ with eigenvalue $1$, i.e. $\ket{\psi_0}=\ket{0}$.

\begin{figure}[t]
\includegraphics[width=\linewidth]{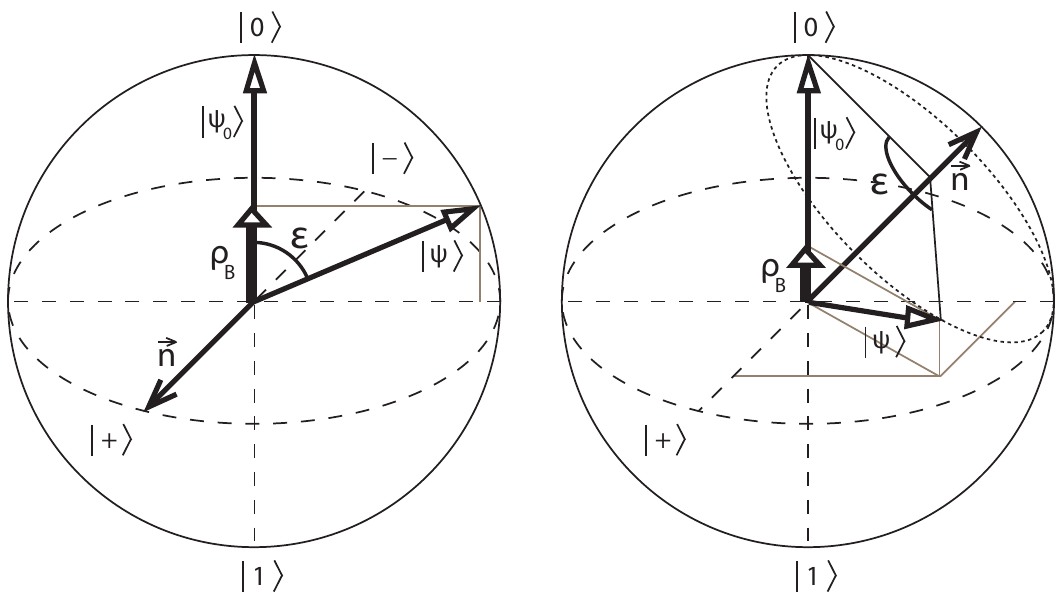}
\caption{Encoding $\epsilon$ via rotating the fiducial state $\ket{\psi_0}=\ket{0}$ around the unit vector $\vec{n}$. For $\vec{n}=(1,0,0)$ the state of the qubit in Bob's frame is ${\R}_B=\cos^2(\frac{\epsilon}{2})\pro{0}{0}+\sin^2(\frac{\epsilon}{2})\pro{1}{1}$ and Bob's quantum Fisher information is the same as Alice's. For $\vec{n}=(0,\frac{1}{\sqrt{2}},\frac{1}{\sqrt{2}})$, ${\R}_B$ is ${\R}_B=(1-\frac{1}{2}\sin^2(\frac{\epsilon}{2}))\pro{0}{0}+\frac{1}{2}\sin^2(\frac{\epsilon}{2})\pro{1}{1}$. Note that ${\R}_{B}$ is the projection of $|\psi\rangle$ onto the z-axis. Also note that in the latter case Bob's quantum Fisher information is $\epsilon$-dependent(See figure \ref{BobsQFIexample3}).}\label{BlochSpheres}
\end{figure}

Using Euler's formula for Pauli matrices $e^{-i\hat{K}\epsilon}=\cos(\frac{\epsilon}{2})\mathbb{I}-i\sin(\frac{\epsilon}{2})(\vec{n}\cdot\vec{\sigma})$ we can write the state Alice prepared as
\begin{equation}\label{EX3state}
\ket{\psi_{\epsilon}}=\left(\cos\left(\tfrac{\epsilon}{2}\right)-iz\sin\left(\tfrac{\epsilon}{2}\right)\right)\ket{0}+
(y-ix)\sin\left(\tfrac{\epsilon}{2}\right)\ket{1}.
\end{equation}

Then using Eq.~\eqref{eq:QFI_pure2}, the quantum Fisher information in Alice's frame reads as
\[\label{example3AlicesH}
H({\R})=1-z^2 .
\]
Note that for $z=1$, the corresponding generator is $\hat{K}=\frac{1}{2}\sigma_z$ which leaves the initial state invariant, i.e. $\text{exp}(-i\frac{\sigma_{z}}{2})|0\rangle=|0\rangle$. Since the encoding process is not successful, the quantum Fisher information $H({\R})$ vanishes which simply means that a different generator needs to be used at the preparation stage. The quantum Fisher information achieves its maximum when when the parameter $\epsilon$ is encoded via a rotation around any vector in the $xy$-plane, i.e. when $z=0$.

\begin{figure}[t]
\includegraphics[width=\linewidth]{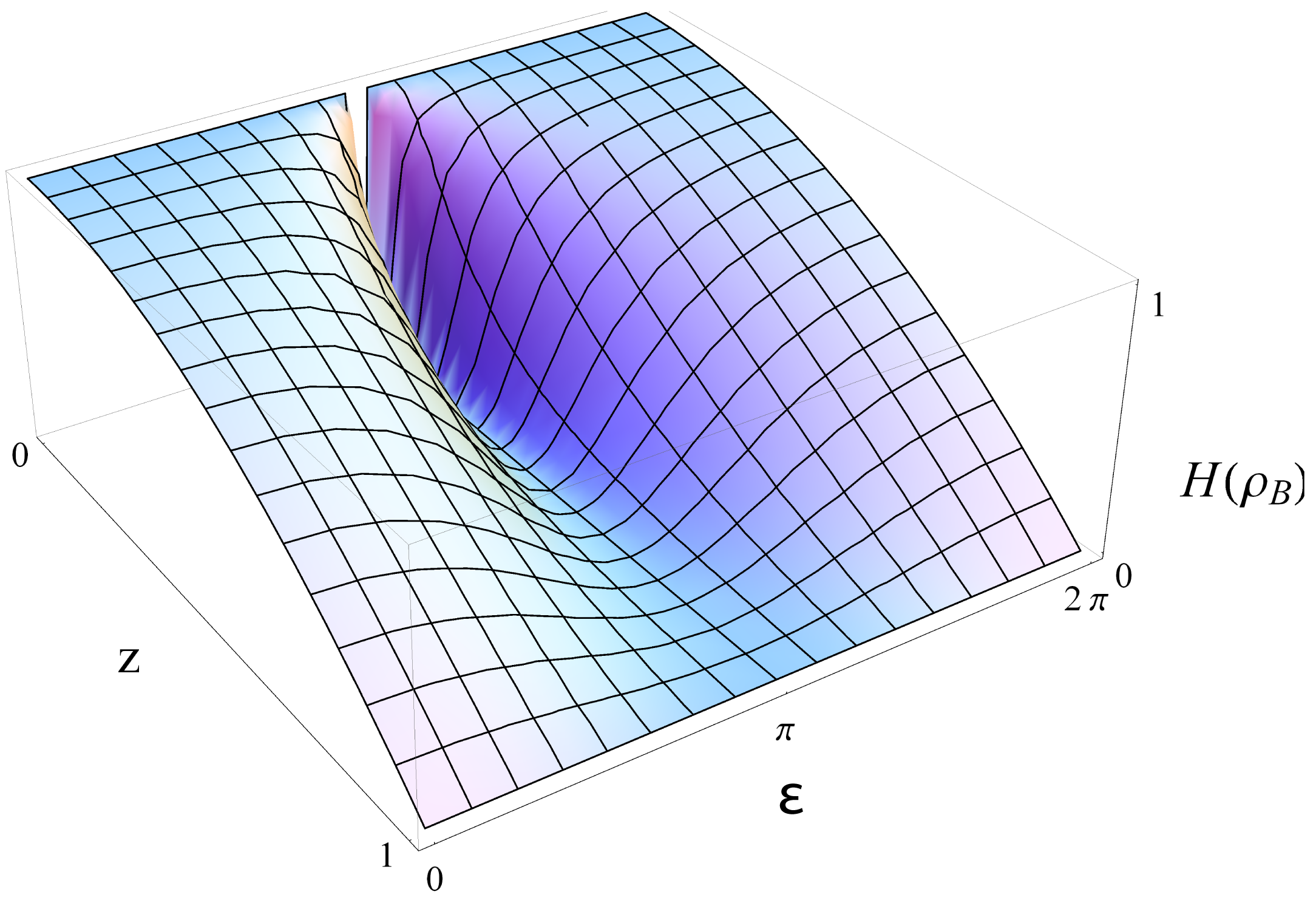}
\caption{Bob's quantum Fisher information in terms of $\epsilon$ and $z$ for general $\vec{n}=(x,y,z)$.}\label{BobsQFIexample3}
\end{figure}

Now suppose that Alice and Bob only share their $z$-axis, i.e. Bob is completely unaware of the relative angle $t$ between his other two axes and Alice's, as depicted in figure \ref{BobRotated}. In this case, $\hat{G}$ is the generator of rotations around $z$-axis, i.e. $\hat{G}=\frac{1}{2}\sigma_z$. Using Eq.~\eqref{BobQFI}, the quantum Fisher information in Bob's frame can be written as
\[\label{example3BobsH}
H({\R}_B)=\frac{1-z^2}{1+z^2\tan^2\left(\tfrac{\epsilon}{2}\right)}=\frac{H({\R})}{1+z^2\tan^2\left(\tfrac{\epsilon}{2}\right)}.
\]

Again note that for $z=1$, the quantum Fisher information is zero in Bob's frame. This is expected, since Bob lacks some information with respect to Alice, therefore Alice's inability in extracting information about $\epsilon$ means that Bob will not be able to decode the message either, i.e. $H({\R}_{B})=0$. On the other hand, as can be seen from \eqref{example3BobsH}, when $z=0$ the quantum Fisher information is the same in Alice's frame and Bob's frame. Figure \ref{BlochSpheres} depicts the two cases of $\vec{n}=(1,0,0)$ and $\vec{n}=(0,\frac{1}{\sqrt{2}},\frac{1}{\sqrt{2}})$. For the former case, the efficiency of communication is $\epsilon$-independent, whereas for the latter case it is $\epsilon$-dependent, as can be seen in figure \ref{BobsQFIexample3}. In this figure, we have plotted Bob's quantum Fisher information in terms of $\epsilon$ and $z$ for general direction $\vec{n}=(x,y,z)$. We observe that as $ \epsilon$ approaches the value $\pi$, the quantum Fisher information approaches its minimum value, i.e.
$H({\R}_B)\rightarrow 0$. In other words, for the chosen encoding operator $\hat{K}$ and the fiducial state $|0\rangle$, Bob will not be able to distinguish ${\R}_{\pi}$ form its neighbouring states ${\R}_{\pi\pm\de}$, where $\de$ is a very small change in $\epsilon$.

Also after some algebra and with the aid of Eq.~\eqref{NSLD}, we find the symmetric logarithmic derivative that achieves the quantum Fisher information in~\eqref{example3BobsH} as
\[
L({\R}_B)=\frac{(z^2-1)\tan\left(\tfrac{\epsilon}{2}\right)}{1+z^2\tan^2\left(\tfrac{\epsilon}{2}\right)}\pro{0}{0}+\cot\left(\tfrac{\epsilon}{2}\right)\pro{1}{1}.
\]
Again the optimal POVM can be constructed from the eigenvectors of this operator, i.e. $\{\ket{0}\langle 0|,\ket{1}\langle 1|\}$. This simply means that the most informative measurement for Bob is the number-counting measurement.

\section{Conclusion}

In quantum metrological schemes the existence of a perfect classical reference frame is often assumed. In this chapter we analyzed how the ultimate limits of precision change due to the absence of such frames of reference, and how attaching a quantum reference frame can improve the estimation precision.

We considered effects of noise due to lack of a certain reference frame. %In our examples, commutative noise corresponds to a QRF which does not interact with the original state, while non-commutative noise is caused by interacting QRF.
We have shown that more precision is lost when the encoding process resembles the nature of the noise. We observed two qualitatively different scenarios. The first scenario is when the the encoding operator commutes with the noise, and the second is when it does not. Interestingly, we demonstrated that choosing an encoding operator which does not commute with the generator of the noise may be advantageous in certain situations. For example when the noise is non-degenerate using a commutative encoding operation would lead to the complete inability of extracting the parameter. Moreover, we derived necessary and sufficient conditions for two extreme cases. One in which the parameter can no longer be estimated due to the lack of a reference frame, and the second in which the parameter can be extracted with the maximal possible precision.

We proved that the use of quantum reference frames is a feasible strategy for the estimation of quantum parameters in the scenario when a common reference frame is lacking. In certain cases, not using quantum reference frames would lead to the complete loss of precision in estimation. However, when using a QRF of large energy, it is often possible to extract the parameter with the highest possible precision.

%%%%%%%%%%%%%%%%%%%%%%%%%%%%%%%%%%%%%%%%%%%%%%%%%%%%%%%%%
%%%%%%%%%%%%%%%%%%%%%%%%%%%%%%%%%%%%%%%%%%%%%%%%%%%%%%%%%
\chapter{Final remarks}\label{chap:final_remarks}
%%%%%%%%%%%%%%%%%%%%%%%%%%%%%%%%%%%%%%%%%%%%%%%%%%%%%%%%%
%%%%%%%%%%%%%%%%%%%%%%%%%%%%%%%%%%%%%%%%%%%%%%%%%%%%%%%%%

In this last chapter we discuss possible future directions and open questions. Then we conclude this thesis.

\section{Future research directions and open questions}\label{sec:open_questions}

%In this last section we give a list of possible future research directions and open problems with small explanations and ideas of how to solve them.

%During work on the quantum metrology of Gaussian states, and in the quantum metrology in general, several questions naturally appear. The following list is definitely not the entirely list of possible

The following is a list of questions which arose during our work in quantum metrology. It is by no means a complete list but rather, in our view, a collection of the most interesting or most important questions to be answered.

\begin{itemize}[leftmargin=*]
\item \emph{Full parametrization of Gaussian states.} In chapter~\ref{chap:GS} we managed to fully parameterized one-, two-, and three-mode Gaussian states. To do this we have used the parametrization of unitary matrices up to size $3\times 3$. However, a general parametrization of unitary matrices is not known. The natural question arises: how to parametrize Gaussian states consisting of $N\geq4$ number of modes? Of course, it is possible to parametrize the covariance matrix directly~\eqref{def:covariance_matrix} as its submatrices are Hermitian and symmetric matrices. However, there would be many conditions on these parameters~\eqref{eq:sigma_K_positivity} in order for the matrix $\sigma$ to be a valid covariance matrix of a Gaussian state. Answering this question would be especially useful in finding optimal probe states for the multi-mode channels, or for the multi-parameter estimation.
\item \emph{Are Gaussian states extremal on the set of all quantum probe states?} Let us consider an arbitrary encoding channel $C_\epsilon$ on a bosonic Fock space $\HS$. For any quantum state in the Fock space it is possible to calculate the first and second moments of the field operators~\eqref{def:covariance_matrix}. The proposed statement is the following: from all quantum states $\R$ with a given displacement vector $\bd$ and covariance matrix $\sigma$, it is the Gaussian state $\R_G$ with those moments which performs the worst as a probe, i.e., the inequality for the quantum Fisher information holds,
    \[
    H(C_\epsilon(\R))\geq H(C_\epsilon(\R_G)).
    \]
    We strongly believe that such a statement holds for Gaussian encoding operations $C_\epsilon$. This is because a Gaussian probe state is fully described by the first and the second moments even after the parameter $\epsilon$ is encoded, therefore the higher moments of this state cannot provide any additional information about the parameter. In contrast, the first and the second moments of a non-Gaussian states change the same way as the moments of the Gaussian state but the higher moments could provide some additional information about the parameter. A similar statement about extremality of Gaussian states has been proven for other quantum information measures such as entropy~\cite{wolf2006extremality}. The proof there is based on the super-additivity of such measures, however, the quantum Fisher information is not super-additive~\cite{hansen2007wigner}.
\item \emph{Are squeezed states the optimal probe states?} In all examples of probing Gaussian unitary channels we found that the optimal scaling with the number of particles is always achieved by some multi-mode squeezed state when restricting ourselves to Gaussian probe states. Is this always the case for any Gaussian channel? This topic was partially discussed in section~\ref{sec:are_pure_optimal}.
%\item \emph{Finding optimal probe states for non-unitary channels}
\item \emph{The quantum Fisher information of non-Gaussian states in terms of their moments.} We have derived various formulae for the quantum Fisher information of Gaussian states in terms of the first and second moments in the field operators. Is it possible to write the quantum Fisher information of non-Gaussian states in terms of their (possibly higher) moments? If possible, would the contribution to the quantum Fisher information from the first and second moments correspond to the terms in the known formulae for Gaussian states? Answering this question could also resolve the open problem introduced earlier -- the extremality of Gaussian probe states.
\item \emph{Optimal measurements.} In this thesis we focused on finding optimal Gaussian probe states. It is known that the optimal measurement is given by the symmetric logarithmic derivative~\eqref{eq:SLD_Gaussian}. Diagonalizing this quantity usually shows that the optimal measurement is a projective measurement on infinitely squeezed states. However, such states contain an infinite amount of energy and are thus unattainable in a lab. Is it possible to find some realistic measurements which achieve the ultimate limit of precision given by the quantum Fisher information? The first step towards this direction has been taken in~\cite{Monras2013a} where it was proven that homodyne detection is the optimal measurement in case of measuring isothermal Gaussian states. However, the general result is still missing.
\item \emph{Universality of quantum reference frames.} In section~\ref{sec:Ex}, we showed examples in which a quantum reference frame that does not interact with the original state~\ref{sec:two_nonint_osc}, and a quantum reference frame interacting with the original state~\ref{sec:two_int_osc}. We observed a certain universality of QRFs. In both examples, the uniform superposition of Fock states and the coherent state used as a QRF ultimately led to the same quantum Fisher information as energy grows to infinity. In the non-interacting scenario the quantum Fisher information converges to a constant function, while in the interacting scenario the limit is a function depicted on Fig.~\ref{nonvsint}. We also observed that not every quantum reference frame is useful. For example, the squeezed state did not lead to any improvement in precision in the non-interacting scenario. We therefore pose the following questions: is it possible to classify quantum reference frames in some practical way? Are there types of quantum reference frames that are in some sense universal? In other words, is it possible to always use only one type of a QRF and achieve the same precision as with any other ``equally good'' QRF? We formalize these questions in appendix~\ref{app:QRF_questions}.
\item \emph{Generalizations of not sharing a reference frame.} There are several possible way of how to generalize what it means not to share a reference frame. The first generalization is quite straigtforward: in Eq.~\eqref{Htwirl} we considered only the complete lack of knowledge about the parameter of the reference frame. However, we could relatively easily incorporate some weight $p(t)$ representing the knowledge about this parameter,
    \[
    \G[{\R}_{\epsilon}]=\int_{-\infty}^{+\infty}dt\ p(t)\hat{U}(t){\R}_{\epsilon}\hat{U}(t)^{\dag}=\sum_{i,j}\hat{p}\Big(\tfrac{G_i-G_j}{2\pi}\Big)\hat{P}_i{\R}_{\epsilon}\hat{P}_j.
    \]
    $\hat{p}$ denotes the Fourier transform of the function $p$. When $p$ is a Gaussian function, transformation $\G$ is a Gaussian transformation, and it should be possible to derive how such a $\G$-twirling map translates into the phase-space formalism of Gaussian states, which gives the second possible generalization. The third generalization could be possible in the direction of quantum field theory in curved space-time. However, it is not entirely clear what this generalization should look like. This is because there is no general consensus of how to define a reference frame in general relativity, especially when combined with quantum physics. The simplest scenario involves constructing the $\G$-twirling map by averaging over the group elements of the Lorenz or the Poincar\'e group as suggested in~\cite{bartlett2007reference}. The first steps towards this direction has been taken in~\cite{ahmadi2015communication}. Other approaches could involve averaging over histories in the path-integral formulation of quantum field theory, averaging over certain local coordinates associated with an observer in general relativity, or averaging over tetrads in the tetrad formalism of general relativity.
\end{itemize}

\section{Conclusion}
%unique: complex form of the covariance matrix
%study of temperature:
%practical formalism which has been immensely difficult before
%many new formulae
%Finally new
%realistic gravitational detection
%bound states

In this thesis we combined tools of many different fields: quantum metrology, Gaussian states, quantum field theory, and quantum reference frames. We followed on from the current state-of-the-art quantum metrology and derived new formulae for the optimal estimation of parameters encoded in Gaussian states. These new formulae are such powerful tools for the treatment of Gaussian states that by using them we have managed to single-handedly generalize all previous bounds on the precision with which Gaussian unitary channels can be estimated. Moreover, these formulae provide a deep insight into the structure of estimation: they show which combinations of initial squeezing, displacement, and temperature of the probe state leads to the highest possible sensitivity. In other words, these formulae can be used to design the core of future quantum detectors.

The application of such mathematical framework can also provide strategies for the estimation of space-time parameters. Previous studies in this direction considered almost exclusively pure initial probe states. However, in real scenarios probe states are mixed. We derived the limits of precision in the estimation of a general space-time parameter using these realistic probes. With such expressions it is possible to determine which space-time parameters are within experimental reach of current and future technology, and which experimental paths are worth pursuing. The ultimate aim is to measure the elusive predictions of quantum field theory in curved space-time such as the Unruh effect, Hawking radiation, or the dynamical Casimir effect. These phenomena have been confirmed so far only in the analogue systems~\cite{Wilson2011a,weinfurtner2011measurement}. Measuring these effects could either validate the theory or possibly lead to a new theory of quantum gravity.

%To construct these detectors it might not be possible to do it on earth because of noise or whatever but we have to put it into space

To achieve the precision required to measure these space-time effects it may be necessary to construct such detectors in space. By doing so we can greatly increase the scale of such experiments, as well as avoiding various sources of noise. Such design could involve transmission of quantum parameters between the orbiting detector and the control centre on Earth. However, due to the motion of such space-based detectors, the reference frames of the detector and the operator on Earth can become easily misaligned. Such misalignment leads to the loss of precision in the estimation of the parameters. We discovered that attaching a quantum reference frame provides a feasible strategy for preventing this loss. Moreover, these results could help in designing novel ways of storage of quantum parameters for quantum information protocols, or develop satellite-based quantum key distribution.

%Generally what we did
%In this thesis we derived new formulae for the optimal estimation of parameters encoded in Gaussian states. We developed a practical formalism for finding optimal probe states for the estimation of Gaussian unitary channels, and discussed how different characteristics of the probe state affect the estimation precision. We demonstrated this method on several examples. We also used the derived formulae to study how Gaussian states perform as probes for the estimation of space-time parameters. Finally, we derived how the precision in estimation changes when two parties exchanging a quantum state with an encoded parameter do not share a reference frame. Moreover, we have shown that using quantum reference frames in that scenario can effectively counter this loss of precision.

%How what we did translates to the previous knowledge: putting it into perspective.

%What we did contretely
We consider the following results as the main results of this thesis: connection between the quantum Fisher information matrix and the Bures metric at the boundary of the space of density matrices~\eqref{eq:connection_between_Hc_and_H}; the quantum Fisher information matrix in the scenario when the Williamson's decomposition of the covariance matrix is known~\eqref{eq:multimode_QFI}; the limit formula for the quantum Fisher information matrix of Gaussian states~\eqref{eq:limit_formula}; a general method for finding optimal Gaussian probe states, section~\ref{sec:general_method}; finding that effects of the temperature of the probe state on the estimation are generic, section~\ref{sec:effects_of_temperature}; % i.e., for any Gaussian channel the temperature dependence always come in multiplicative factors of four types,
expression for the quantum Fisher information for the estimation of space-time parameters using two-mode squeezed thermal states~\eqref{eqn:two_mode_QFI_zeroth_order};
and finally Eq.~\eqref{BobQFI} which shows how to encode the parameter in a quantum state such that the loss of precision in the estimation due to the lack of a shared reference frame is minimized.

%what can it be useful for
Our results can lead to applications in existing gravitational wave detectors~\cite{Abbott2004a,Caron1995a}, they may be useful for designing new gravimeters~\cite{Snadden1998a,Altin2013a,Sabin2014a}, climate probes~\cite{Tapley2004a}, or sensors for the estimation of space-time parameters~\cite{Danzmann1996a,Everitt2011a,Bruschi2014a}. Our results will enable researchers to evaluate how well space-time parameters, such as the amplitude of a gravitational wave, accelerations, and local gravitational fields, can be estimated in the presence of background temperature~\cite{Sabin2014a,Sabin2015a}. Our studies on metrology with misaligned reference frames could prove useful for future space-based experiments such as the gravitational wave detector eLISA~\cite{amaro2012low} or miniaturized satellites~\cite{bedington2015deploying}. Finally, our results could help the efforts in bringing the new era of quantum technologies to the general public.

\appendix

\chapter{Appendix}
%\addcontentsline{toc}{chapter}{Appendix}

\section{The full statement of the Cram\'er-Rao bound}\label{app:CR}

In this section we state the Cram\'er-Rao bound in full with all assumptions which are not usually mentioned in the quantum metrology literature. Moreover, we show the proof for this theorem for a finite sample space while taking particular care about points where $p(x|\epsilon)=0$. Then we show that by plugging the measurements given by the eigenvectors of the symmetric logarithmic derivative into the obtained expression will lead to the definition of the quantum Fisher information~\eqref{def:H_using_L}. This is useful from the theoretical perspective because by doing that we show that the (possibly discontinuous) quantum Fisher information~\eqref{def:H_using_L} is the correct figure of merit in the quantum Cram\'er-Rao bound, while similar theorem might not hold for the continuous quantum Fisher information~\eqref{eq:connection_between_Hc_and_H}. For more details on the (non-quantum) estimation theory see for example~\cite{lehmann2006theory,amari2007methods}.

\begin{theorem} (Cram\'er-Rao)
Let $p(x|\epsilon)$ be a probability distribution, $\Omega$ a sample space, and $\hat{\epsilon}$ a locally unbiased estimator. Let the following conditions hold:
\begin{enumerate}
  \item The Fisher information~\eqref{eq:the_Fisher_information} is always defined; equivalently, for all $x\in\Omega$ such that $p(x|\epsilon)>0$,
      \[
      \frac{\partial_\epsilon p(x|\epsilon)}{p(x|\epsilon)}
      \]
      exists and is finite.
  \item The operations of integration with respect to $x$ with respect to $\epsilon$ can be interchanged in the expectation value of $\hat{\epsilon}$, i.e.,
      \[
      \partial_\epsilon\left(\int_\Omega\!\!\dif{x}\ \hat{\epsilon}(x)p(x|\epsilon)\right)=\int_\Omega\!\!\dif{x}\ \hat{\epsilon}(x)\partial_\epsilon p(x|\epsilon),
      \]
      whenever the right-hand side is finite.
\end{enumerate}
Then
\[
\mean{\Delta\hat{\epsilon}^2}\geq\frac{1}{F(\epsilon)}.
\]
\end{theorem}
Note that $N$ in the equation~\eqref{eq:Cramer_Rao} comes from the fact that the Fisher information is additive under independent random variables.

\begin{proof} We will prove the theorem for the finite sample space $\Omega$. We stick to the following notation: the subscript $x$ under the sum means that the sum goes over all $x\in\Omega$, while subscript $p(x|\epsilon)>0$ means that the sum goes over all elements $x$ for which $p(x|\epsilon)>0$.

The estimator is locally unbiased which by definition means
\[
\sum_x(\hat{\epsilon}(x)-\epsilon)p(x|\epsilon)=0.
\]
We differentiate this expression with respect to $\epsilon$. We can swap the derivative $\partial_\epsilon$ and the sum, because the summation goes only over a finite amount of elements. Therefore $1=\sum_x\big(\hat{\epsilon}(x)-\epsilon\big)\partial_\epsilon p(x|\epsilon)$. The following identities and inequalities hold:
\begin{equation}
\begin{split}
    1&=\sum_x\big(\hat{\epsilon}(x)-\epsilon\big)\partial_\epsilon p(x|\epsilon)
    =\sum_{p(x|\epsilon)>0}\big(\hat{\epsilon}(x)-\epsilon\big)\partial_\epsilon p(x|\epsilon) \\
     &=\sum_{p(x|\epsilon)>0}\big(\hat{\epsilon}(x)-\epsilon\big)\sqrt{p(x|\epsilon)}\frac{\partial_\epsilon p(x|\epsilon)}{\sqrt{p(x|\epsilon)}} \\
     &\leq\sqrt{\sum_{p(x|\epsilon)>0}\big(\hat{\epsilon}(x)-\epsilon\big)^2p(x|\epsilon)}\sqrt{\sum_{p(x|\epsilon)>0}\frac{(\partial_\epsilon p(x|\epsilon))^2}{p(x|\epsilon)}} \\
     &=\sqrt{\sum_{x}\big(\hat{\epsilon}(x)-\epsilon\big)^2p(x|\epsilon)}\sqrt{F(\epsilon)}=\sqrt{\mean{\Delta\hat{\epsilon}^2}}\sqrt{F(\epsilon)}.
\end{split}
\end{equation}
The second identity follows from the fact that because $p(x|\epsilon)\geq0$, the function $p(x|\epsilon)$ achieves the local minimum at point $\epsilon$ defined by $p(x|\epsilon)=0$. Therefore $\partial_\epsilon p(x|\epsilon)=0$ at such point $\epsilon$. The inequality is the usual Cauchy-Schwartz inequality for vectors.
\end{proof}

Notice that in the definition of the Fisher information derived in the proof we sum only over such values of $x$ such that $p(x|\epsilon)>0$. Now, accordingly to Eq.~\eqref{def:prob_distribution}, we set
\begin{subequations}
\begin{align}
p(x|\epsilon)&=\tr[P_x\R_\epsilon],\\
\partial_\epsilon p(x|\epsilon)&=\tr[P_x\partial_\epsilon\R_\epsilon],
\end{align}
\end{subequations}
where $P_x$ are the projectors from the spectral decomposition of the symmetric logarithmic derivative~\eqref{def:SLDsolution}, $L=\sum_xl_xP_x$. Then
\begin{equation}
\begin{split}
    F(\epsilon)&=\sum_{p(x|\epsilon)>0}\frac{\tr[P_x(L\R_\epsilon+\R_\epsilon L)]^2}{4\tr[P_x\R_\epsilon]}
    =\sum_{p(x|\epsilon)>0}\frac{\tr[2l_xP_x\R_\epsilon]^2}{4\tr[P_x\R_\epsilon]}\\
    &=\sum_{p(x|\epsilon)>0}\frac{l_x^2\tr[P_x\R_\epsilon]^2}{\tr[P_x\R_\epsilon]}=\sum_{p(x|\epsilon)>0}l_x^2\tr[P_x\R_\epsilon]=\sum_{x}l_x^2\tr[P_x\R_\epsilon]\\
    &=\tr[\sum_{x}l_x^2P_x\R_\epsilon]=\tr[L^2\R_\epsilon]=H(\epsilon).
\end{split}
\end{equation}
This shows that the measurement given by the eigenvectors of the symmetric logarithmic derivative truly results in the Fisher information to reach the upper limit given by the quantum Fisher information. Moreover, this also shows that the quantum Fisher information given by the symmetric logarithmic derivative is the correct figure of merit for which the quantum Cram\'er-Rao bound holds, in contrast to the continuous quantum Fisher information~\eqref{eq:connection_between_Hc_and_H} for which such a theorem is not known.

\section{Discontinuity of the quantum Fisher information matrix and the Bures metric}\label{app:discontinuity_of_QFI}

In this section we derive the general expression for the Bures metric defined via Eq.~\eqref{eqn:bures}. To do that, we modify the proof from~\cite{hubner1992explicit} to include the case where the density matrix is not invertible and generalize that result. Then we show that we could get the same expression (up to a multiplicative constant) by redefining the quantum Fisher information matrix in a way which makes it a continuous function in the sense explained in section~\ref{QMsec:multi}. By that we prove not only that the Bures metric coincides with this continuous quantum Fisher information, but also that it shares the same properties, i.e., it is also continuous in the same sense. Finally, we show a simple example to illustrate the difference between the quantum Fisher information and its continuous version.

%notation
We introduce the following notation: if a symbol of an index appear under the sum, the sum goes over all values of the index such that the property is satisfied. For example, $\sum_{p_i>0}$ means that the sum goes over all $i$ such that $p_i>0$. If there is no condition present, the sum goes over all parameters. We will also drop noting the explicit dependence on the vector of parameters $\be$ unless the argument is not only $\be$. For example, instead of $p_i(\be)$ we write only $p_i$, but for $p_i(\be+\bdeps)$ we write the full form. In rare cases we also write the full form when we want to stress out the dependence on $\be$. We denote the partial derivatives as $\partial_i=\partial_{\epsilon_i}$ and $\partial_{ij}=\partial_{\epsilon_i}\partial_{\epsilon_j}$. Derivatives with respect to elements of $\bdeps$ will be denoted as $\partial_{\de_i}$ for the first derivatives, and $\partial_{\de_i\de_j}$ for the second derivatives. We also write the spectral decomposition of the density matrix as $\R=\sum_kp_k\pro{k}{k}$.

%expression for the Bures metric
We combine the defining relation for the Bures metric~\eqref{eqn:bures}, the definition of the Bures distance~\eqref{def:bures_distance}, and the definition of the Uhlmann fidelity~\eqref{def:Uhlmann_Fidelity}, by which we obtain the expression for the Bures metric,
\[\label{eq:bures_metric_fid}
\sum_{i,j}g^{ij}(\be)\mathrm{d}\epsilon_i\mathrm{d}\epsilon_j=2\Big(1-\tr\Big[\sqrt{\sqrt{\R_{\be}}\R_{\be+\bdeps}\sqrt{\R_{\be}}}\Big]\Big).
\]
We define the operator $\hat{O}(\bdeps):=\sqrt{\R_{\be}}\R_{\be+\bdeps}\sqrt{\R_{\be}}$. We omit writing the explicit dependence on $\be$. Because this operator is in between two matrices $\sqrt{\R}=\sum_k\sqrt{p_k}\pro{k}{k}$, it can be written as
\[
\hat{O}(\bdeps)=\sum_{p_k>0,\,p_l>0}o_{kl}(\bdeps)\pro{k}{l}
\]
As a result, this operator clearly belongs to the subspace of linear operators acting on the Hilbert space spanned by the eigenvectors associated with non-zero eigenvalues $p_k$, i.e., $\hat{O}\in\mathcal{L}(\HS_{>0})$, where $\HS_{>0}:=\mathrm{span}\{\ket{k}\}_{p_k>0}$. Now let us define its square root as $\hat{A}(\bdeps)$,
\[\label{eq:A_definition}
\hat{A}(\bdeps)\hat{A}(\bdeps)=\hat{O}(\bdeps).
\]
Because $\hat{O}\in\mathcal{L}(\HS_{>0})$, also $\hat{A}\in\mathcal{L}(\HS_{>0})$ together with its all derivatives. To show that, we assume that $\hat{O}$ has a spectral decomposition\footnote{Spectral decomposition exists, because $\hat{O}$ is a Hermitian operator.} $\hat{O}(\bdeps)=\sum_mo_m^{\mathrm{diag}}(\bdeps) P_m(\bdeps)$, where $P_m(\bdeps)=\sum_{p_k>0,\,p_l>0}c^{(m)}_{kl}(\bdeps)\pro{k}{l}$ (this is possible because the operator $\hat{O}$ lies in the previously mentioned subspace). The square root is then given by
\[
\hat{A}(\bdeps)=\sum_m\sqrt{o_m^{\mathrm{diag}}(\bdeps)} P_m(\bdeps)=\sum_{p_k>0,\,p_l>0}\left(\sum_{m}\sqrt{o_m^{\mathrm{diag}}(\bdeps)} c^{(m)}_{kl}(\bdeps)\right)\pro{k}{l}.
\]
Clearly, any derivatives of $\hat{A}(\bdeps)$ with respect to $\de_i$ will change only the factors, so the resulting operator will still remain in the same subspace $\mathcal{L}(\HS_{>0})$.
From Eq.~\eqref{eq:bures_metric_fid} we have
\[
\sum_{i,j}g^{ij}(\be)\mathrm{d}\epsilon_i\mathrm{d}\epsilon_j=2(1-\tr[A(\bdeps)]).
\]
which gives the expression for the elements of the Bures metric,
\[
g^{ij}(\be)=-\tr[\partial_{\de_i\de_j}\!A(0)],
\]
if the second derivative exists. For that reason we assume that $\R\in C^{(2)}$, i.e., the second derivatives of $\R$ exist and are continuous.\footnote{Actually, the assumption can be slightly weakened. We can assume that the second derivatives exists, but may not be necessarily continuous. But the continuity of the second derivatives implies $\partial_{ij}\R=\partial_{ji}\R$ which will be useful later when discussing the continuity of the Bures metric.} To obtain these second partial derivatives we rewrite Eq.~\eqref{eq:A_definition} while expanding $\R_{\be+\bdeps}$ around point $\be$,
\[\label{eq:A_definition2}
\hat{A}(\bdeps)\hat{A}(\bdeps)=\sqrt{\R}\bigg(\R+\sum_k\partial_k\R\ \de_k + \frac{1}{2} \sum_{k,l}\partial_{kl}\R\ \de_k \de_l\bigg)\sqrt{\R}.
\]
By differentiating with respect to $\de_i$  and setting $\bdeps=0$ we obtain
\[
\partial_{\de_i}\!\hat{A}(0)\,\R+\R\,\partial_{\de_i}\!\hat{A}(0)=\sqrt{\R}\partial_i\R\sqrt{\R},
\]
because $\hat{A}(0)=\R$. By applying $\bra{k}\ \ket{l}$ for $p_k>0$, $p_l>0$ we obtain the matrix elements of $\partial_{\de_i}\!\hat{A}(0)$,
\[\label{eq:elem_of_Adot}
\bra{k}\partial_{\de_i}\!\hat{A}(0)\ket{l}=\frac{\sqrt{p_kp_l}\bra{k}\partial_i\R\ket{l}}{p_k+p_l}.
\]
Elements $\bra{k}\partial_{\de_i}\!\hat{A}(0)\ket{l}$ for which $p_k=0$ or $p_l=0$ are identically zero, because as we proved earlier all derivatives of $\hat{A}$ lie in the subspace $\mathcal{L}(\HS_{>0})$. Differentiating Eq.~\eqref{eq:A_definition2} for the second time and setting $\bdeps=0$ yields
\[
\partial_{\de_i\de_j}\!\hat{A}(0)\, \R+\{\partial_{\de_i}\!\hat{A}(0),\partial_{\de_j}\!\hat{A}(0)\}+\R\, \partial_{\de_i\de_j}\!\hat{A}(0)=\sqrt{\R}\partial_{ij}\R\sqrt{\R}.
\]
Now, restricting ourselves to the subspace $\mathcal{L}(\HS_{>0})$ the density matrix has the inverse matrix $\R^{-1}$. We multiply the above equation by this matrix and perform the trace on this subspace,
\[
2\tr_{\mathcal{L}(\HS_{>0})}[\partial_{\de_i\de_j}\hat{A}(0)]+\tr_{\mathcal{L}(\HS_{>0})}[\rho^{-1}\{\partial_{\de_i}\hat{A}(0),\partial_{\de_j}\hat{A}(0)\}]
=\tr_{\mathcal{L}(\HS_{>0})}[\partial_{ij}\R].
\]
Because all derivatives of $\hat{A}$ lie in the subspace $\mathcal{L}(\HS_{>0})$, the traces of such operators are identical on both the subspace and the full space, $\tr_{\mathcal{L}(\HS_{>0})}[\partial_{\de_i\de_j}\hat{A}(0)]=\tr[\partial_{\de_i\de_j}\hat{A}(0)]$. However that is not necessarily true for the last element for which $\tr_{\mathcal{L}(\HS_{>0})}[\partial_{ij}\R]=\tr[\hat{P}_{\HS_{>0}}\partial_{ij}\R]$, where $\hat{P}_{\HS_{>0}}$ denotes the projector on the Hilbert space $\HS_{>0}$. Because $\tr[\partial_{ij}\R]=0$, this term can be equivalently written as $\tr[\hat{P}_{\HS_{>0}}\partial_{ij}\R]=-\tr[\hat{P}_0\partial_{ij}\R]$, where projector $\hat{P}_0:=\hat{I}-\hat{P}_{\HS_{>0}}$ projects onto the subspace spanned by the eigenvectors of the density matrix $\R$ associated with the zero eigenvalue.
Therefore we have,
\[\label{eq:g_in_terms_of_A}
g^{ij}=-\tr[\partial_{\de_i\de_j}\!\hat{A}(0)]=\frac{1}{2}\big(\tr_{\mathcal{L}(\HS_{>0})}[\rho^{-1}\{\partial_{\de_i}\!\hat{A}(0),\partial_{\de_j}\!\hat{A}(0)\}]
+\tr[\hat{P}_0\partial_{ij}\R]\big).
\]
The first term of the right hand side can be readily computed from Eq.~\eqref{eq:elem_of_Adot} while the antisymmetric part vanishes under the sum,
\[\label{eq:first_term}
\tr_{\mathcal{L}(\HS_{>0})}[\rho^{-1}\{\partial_{\de_i}\!\hat{A}(0),\partial_{\de_j}\!\hat{A}(0)\}]=
\!\!\!\!\sum_{p_k>0,\,p_l>0}\!\!\!\!\frac{\Re(\bra{k}\partial_i\R\ket{l}\bra{l}\partial_j\R\ket{k})}{p_k+p_l}.
\]
Now we compute the second term. The second derivative of $\R$ is given by
\[\label{eq:second_derivative_rho}
\begin{split}
   \partial_{ij}\R =\sum_{k} &\partial_{ij}p_k\pro{k}{k} + p_k\big(\pro{\partial_i k}{\partial_j k}+\pro{\partial_j k}{\partial_i k}\big)+ p_k\big(\pro{\partial_{ij} k}{k} + \pro{k}{\partial_{ij} k}\big)\\
     + &\partial_jp_k\big(\pro{\partial_i k}{k}+\pro{k}{\partial_i k}\big)+\partial_ip_k\big(\pro{\partial_j k}{k}+\pro{k}{\partial_j k}\big).
\end{split}
\]
We stress out that the summation goes over all values $k$, even over those for which $p_k=0$. When using the above equation to calculate $\tr[\hat{P}_0\partial_{ij}\R]$ we find that many terms vanish:
\begin{itemize}
  \item for $k$ such that $p_k>0$, $P_0\ket{k}=0$,
  \item for $k$ such that $p_k=0$, also $\partial_ip_k=\partial_jp_k=0$, because $p_k$ reaches the local minimum at point $\epsilon$ such that $p_k(\epsilon)=0$.
\end{itemize}
Only parts of the first two terms of Eq.~\eqref{eq:second_derivative_rho} remain,
\[\label{eq:second_term}
\begin{split}
\tr[\hat{P}_0\partial_{ij}\R]&=\sum_{p_k=0}\partial_{ij}p_k+2\!\!\!\!\!\!\sum_{p_k>0,p_l=0}\!\!\!\!\!p_k\ \Re(\braket{l}{\partial_i k}\braket{\partial_j k}{l})\\
&=\sum_{p_k=0}\partial_{ij}p_k+2\!\!\!\!\!\!\sum_{p_k>0,p_l=0}\!\!\!\!\!\!\frac{\Re(\bra{k}\partial_i\R\ket{l}\bra{l}\partial_j\R\ket{k})}{p_k+p_l}.
\end{split}
\]
Combining Eqs.~\eqref{eq:g_in_terms_of_A}, \eqref{eq:first_term}, \eqref{eq:second_term}, and the expression for the quantum Fisher information~\eqref{QFI_multi} we derive
\[\label{eq:Bures_metric_app}
g^{ij}=\frac{1}{2}\sum_{p_k+p_l> 0}\!\!\!\!\!\!\frac{\Re(\bra{k}\partial_i\R\ket{l}\bra{l}\partial_j\R\ket{k})}{p_k+p_l}
+\frac{1}{2}\sum_{p_k=0}\!\!\partial_{ij}p_k
=\frac{1}{4}\Big(H^{ij}+2\!\!\!\sum_{\,p_k=0}\!\!\!\partial_{ij}p_k\Big),
\]
which proves Eq.~\eqref{eq:bures_metric}.

%The continuous quantum Fisher information
Now we show that by redefining possible problematic points of the quantum Fisher information to make the quantum Fisher information a continuous function (in certain sense which will become clear later), we obtain the same expression as four-times the Bures metric. In other words, the Bures metric gives the continuous version of the quantum Fisher information. To show that, we will study the neighborhood of the quantum Fisher information matrix, i.e., we will study function $H^{ij}(\be+\bdeps)$.  We are going to expand this function around point $\be$ and define any problematic value of $H^{ij}(\be)$ as a limit of the values given by the neighborhood of $\be$. Eq.~\eqref{QFI_multi} yields
\begin{equation}\label{eq:QFI_in_the_neighborhood}
\begin{split}
H^{ij}(\be+\bdeps) &=2\!\!\!\!\!\!\sum_{p_{k\,\be+\bdeps}+p_{l\,\be+\bdeps}> 0}\!\!\!\!\!\!\frac{\Re(\bra{k_{\be+\bdeps}}\partial_i\R_{\be+\bdeps}\ket{l_{\be+\bdeps}}\bra{l_{\be+\bdeps}}\partial_j\R_{\be+\bdeps}\ket{k_{\be+\bdeps}})}{p_{k\,\be+\bdeps}+p_{l\,\be+\bdeps}} \\
     &=2\!\!\!\!\!\!\sum_{p_{k\,\be}+p_{l\,\be}> 0}\!\!\!\!\!\!
     \frac{\Re(\bra{k_{\be}}\partial_i\R_{\be}\ket{l_{\be}}\bra{l_{\be}}\partial_j\R_{\be}\ket{k_{\be}}+\mathcal{O}(\bdeps))}{p_{k\,\be}+p_{l\,\be}+\mathcal{O}(\bdeps)} \\
     &+2\!\!\!\!\!\!\sum_{p_{k\,\be}+p_{l\,\be}= 0}\!\!\!\!\!\!\frac{\Re(\bra{k_{\be+\bdeps}}\partial_i\R_{\be+\bdeps}\ket{l_{\be+\bdeps}}\bra{l_{\be+\bdeps}}\partial_j\R_{\be+\bdeps}\ket{k_{\be+\bdeps}})}{p_{k\,\be+\bdeps}+p_{l\,\be+\bdeps}}
\end{split}
\end{equation}
Assuming $\R_{\be}\in C^{(2)}$ we can write expansions,
\begin{subequations}
\begin{align}
p_{k\,\be+\bdeps}&=p_{k}+\sum_m\partial_mp_{k}\de_m+\frac{1}{2}\sum_{m,n}\partial_{mn}p_{k}\de_m\de_n+\mathcal{O}(\bdeps^3),\\
\R_{\be+\bdeps}&=\R+\sum_m\partial_m\R\de_m+\frac{1}{2}\sum_{m,n}\partial_{mn}\R\de_m\de_n+\mathcal{O}(\bdeps^3),\\
\ket{k_{\be+\bdeps}}&=\ket{k}+\sum_m\ket{\partial_mk}\de_m+\frac{1}{2}\sum_{m,n}\ket{\partial_{mn}k}\de_m\de_n+\mathcal{O}(\bdeps^3).
\end{align}
\end{subequations}
Using these expansions, Eq.~\eqref{eq:second_derivative_rho}, and assuming $p_{k\,\be}+p_{l\,\be}= 0$ we have $\partial_ip_{k\,\be}+\partial_ip_{l\,\be}=0$ and
\begin{subequations}\label{eq:elements_p_and_kl}
\begin{align}
p_{k\,\be+\bdeps}+p_{l\,\be+\bdeps}&=\frac{1}{2}\sum_{m,n}(\partial_{mn}p_{k}+\partial_{mn}p_{l})\de_m\de_n+\mathcal{O}(\bdeps^3),\\
\bra{k_{\be+\bdeps}}\partial_i\R_{\be+\bdeps}\ket{l_{\be+\bdeps}}&=
\sum_m\big(\bra{\partial_mk}\partial_i\R\ket{l}+\bra{k}\partial_i\R\ket{\partial_ml}+\bra{k}\partial_{im}\R\ket{l}\big)\de_m+\mathcal{O}(\bdeps^2)\nonumber\\
&=\delta_{kl}\sum_m\partial_{im}p_k\de_m+\mathcal{O}(\bdeps^2),
\end{align}
\end{subequations}
where we used $\braket{k}{\partial_i j}=-\braket{\partial_i k}{j}$ which comes from the orthonormality condition. Inserting expressions~\eqref{eq:elements_p_and_kl} to Eq.~\eqref{eq:QFI_in_the_neighborhood} reads
\[\label{eq:H_expanded}
\begin{split}
H^{ij}(\be+\bdeps)&=2\!\!\!\!\!\!\sum_{p_{k}+p_{l}> 0}\!\!\!\!\!\!
     \frac{\Re(\bra{k}\partial_i\R\ket{l}\bra{l}\partial_j\R\ket{k})+\mathcal{O}(\bdeps)}{p_{k}+p_{l}+\mathcal{O}(\bdeps)}\\ &+2\!\sum_{p_{k}= 0}\!\frac{\big(\sum_{m}\partial_{im}p_k \de_m\big) \big(\sum_{n}\partial_{jn}p_k\de_n\big)+\mathcal{O}(\bdeps^3)}{\sum_{s,t}\partial_{st}p_{k}\de_s\de_t+\mathcal{O}(\bdeps^3)}.
\end{split}
\]
%Redefinition of the points.
The first part of this expression goes to the limit of the quantum Fisher information as $\bdeps$ goes to zero. However, the second part also contributes. In contrast to the first part, the second part in general does not have a well-defined limit at point $\be$. This is because for two differently chosen vectors $\bdeps:=\bdeps_1,\bdeps_2$ from the neighborhood of $0$, as both $\bdeps_1$, $\bdeps_2$ goes to zero, the second part goes to different values depending on these vectors particular structure. For example, choosing $\bdeps_1=(1,0,\dots)\de$ leads to $2\sum_{p_{k}= 0}\frac{\partial_{i1}p_k\partial_{j1}p_k}{\partial_{11}p_k}$ in the second part, while choosing $\bdeps_2=(1,1,0,\dots)\de$ leads to $2\sum_{p_{k}= 0}\frac{(\partial_{i1}p_k+\partial_{i2}p_k)(\partial_{j1}p_k+\partial_{j2}p_k)}{\partial_{11}p_k+2\partial_{12}p_k+\partial_{22}p_k}$. Therefore, the quantum Fisher information is not in general a continuous function at points $\be$ such that $p(\be)=0$. Moreover, it cannot be made continuous at point $\be$ even after certain redefinition of those points, because the limit does not exist. On the other hand, despite the fact that this function cannot be continuous in the topology of multiple parameters $\be$, it can be continuous in the topology of a single parameter. In other words, it can be defined in a way that for each element of the quantum Fisher information matrix, it is continuous for some parameters while not in others. Namely, choosing $\bdeps=(0,\dots,0,\de,0,\dots)$ where $\de$ is in the $i$-th place gives a limit $2\sum_{p_{k}= 0}\partial_{ij}p_k$ of the second part. Similarly, choosing $\bdeps=(0,\dots,0,\de,0,\dots)$ where $\de$ is in the $j$-th place gives the same limit. Together, we define the continuous quantum Fisher information as
\[\label{eq:construction_of_continuous_QFI}
H_c^{ij}(\be):=\!\lim_{\de_i\rightarrow 0}\lim_{\de_j\rightarrow 0}\lim_{\overset{\de_r\rightarrow 0}{r\neq i,j}}H^{ij}(\be+\bdeps),
\]
%By redefining the quantum Fisher information~\eqref{eq:H_expanded} by this limit at the problematic points we ensure that this new function $H_c^{ij}$ will be continuous in $\epsilon_i$ and $\epsilon_j$.
and by performing the above limit with the use of Eq.~\eqref{eq:H_expanded} we show
\[
H_c^{ij}(\be)=H^{ij}(\be)+2\!\!\!\!\sum_{p_{k}(\be)= 0}\!\!\!\!\partial_{ij}p_k(\be),
\]
which is a continuous function in both parameter $\epsilon_i$ when keeping all other parameters fixed, and a continuous function in parameter $\epsilon_j$ when keeping all other parameters fixed respectively. The above expression is identical to the four times the Bures metric given by Eq.~\eqref{eq:Bures_metric_app}, which is what we wanted to prove.

\begin{figure}[t!]
  \centering
\includegraphics[width=0.49\linewidth]{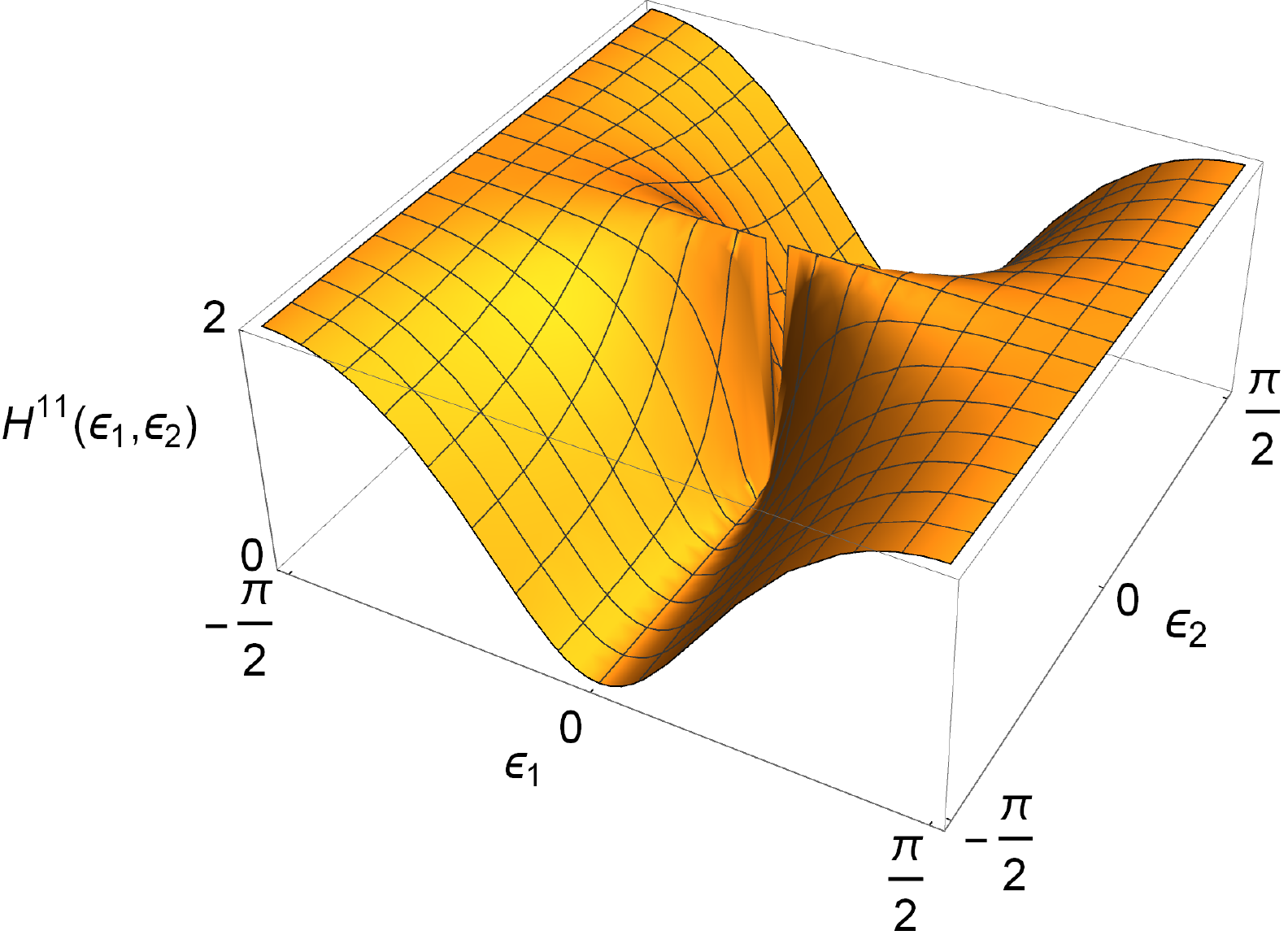}
  \includegraphics[width=0.49\linewidth]{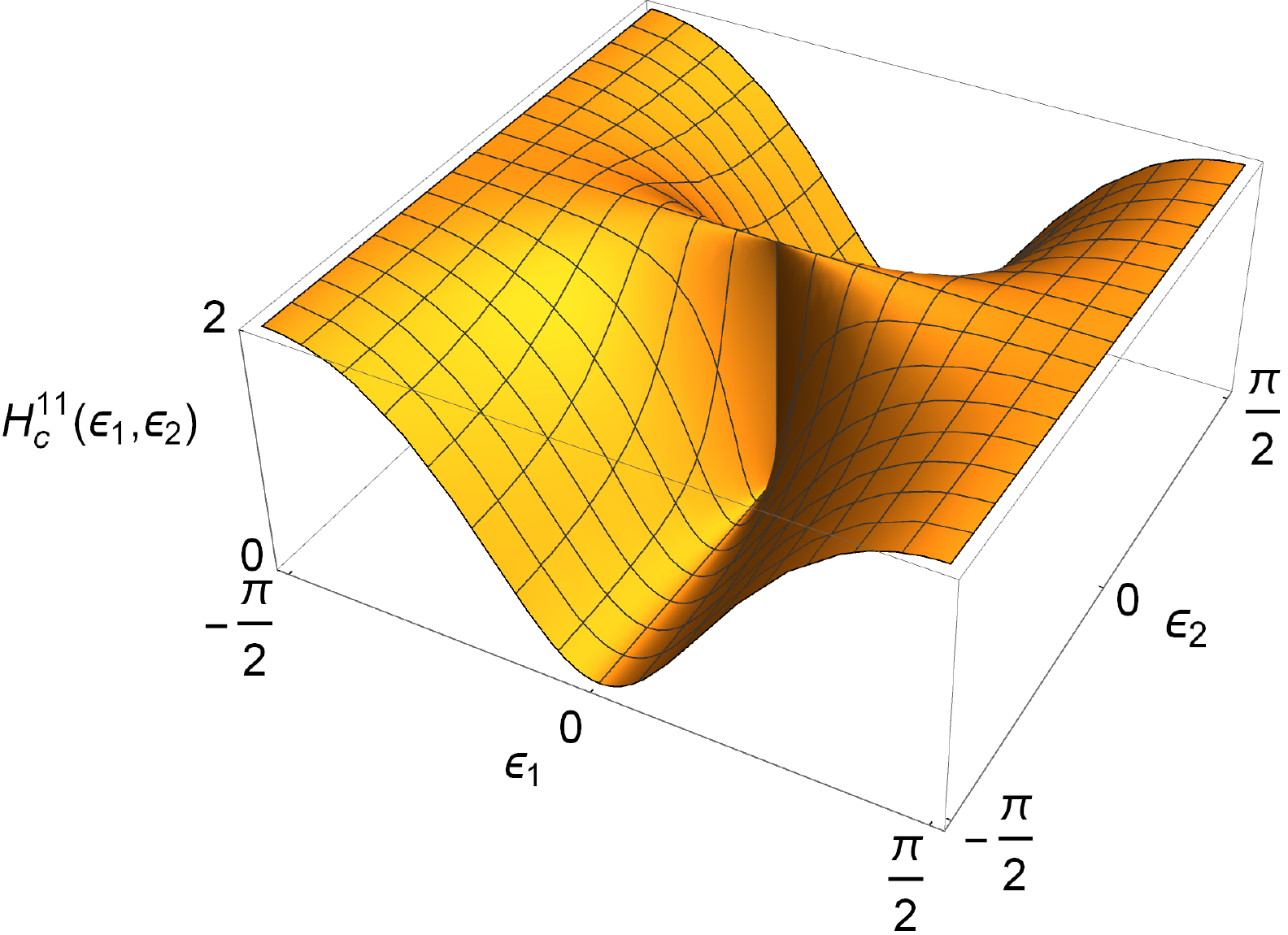}
  \caption{Graphs of the first element of the quantum Fisher information matrix, and the continuous quantum Fisher information matrix respectively, for the estimation of parameters of the density matrix~\eqref{eq:discontinuous_example}. These graphs are identical everywhere apart from the point $(\epsilon_1,\epsilon_2)=(0,0)$. Clearly, neither function is a continuous function in both parameters at the same time, however, $H_c^{11}$ is guaranteed to be a continuous function in $\epsilon_1$ for any $\epsilon_2$.}\label{fig:Continuous_QFI}
\end{figure}

%An example
Let us illustrate the discontinuity of the quantum Fisher information matrix on the following example. We consider a quantum state depending on two parameters,
\[\label{eq:discontinuous_example}
\R_{\be}=\frac{1}{2}(\sin^2 \epsilon_1+\sin^2 \epsilon_2)\pro{0}{0}+\frac{1}{2}\cos^2 \epsilon_1\pro{1}{1}+\frac{1}{2}\cos^2 \epsilon_2\pro{2}{2}.
\]
We are going to study the first element of the quantum Fisher information matrix $H^{11}$ which measures the mean squared error in estimating parameter $\epsilon_1$. While the expression for the quantum Fisher information matrix~\eqref{QFI_multi} assigns value $H^{11}(0,0)=0$ to the problematic point $\be=(0,0)$, the continuous quantum Fisher information assigns value $H^{11}(0,0)=2$, which makes this function a continuous function in $\epsilon_1$, but not necessarily in $\epsilon_2$. Graphs of these functions are shown on Fig.~\ref{fig:Continuous_QFI}.

%Regularization procedure for any state.
In the last paragraph of this section we show that the quantum Fisher information of any state can be calculate as a limit of the quantum Fisher information of a completely mixed state. We define a density matrix
\[
\R_{\be,\nu}:=(1-\nu)\R_{\be}+\nu\R_{0},
\]
where $0<\nu<1$ is a real parameter and $\R_{0}$ is any $\be$-independent full-rank density matrix that is diagonal in the eigenbasis of the density matrix $R_\epsilon$. This can be for example a multiple of identity, $\R_0=\frac{1}{\mathrm{dim}{\HS}}\hat{I}$, or a thermal state $\R_0=\R_{\mathrm{th}}$. The resulting matrix $\R_{\be,\nu}$ is also a full-rank matrix, and by inserting this full-rank matrix into Eq.~\eqref{QFI_multi} it is easy to show that
\[\label{eq:regularization_procedure_density_matrix}
H, H_c(\be)=\lim_{\nu\rightarrow 0}H(\R_{\be,\nu})+c\!\!\!\!\sum_{p_{k}(\be)= 0}\!\!\!\!\mathcal{H}_k(\be).
\]
$c$ is the real parameter defined as follows: to obtain the expression for $H$ we set $c=0$, and to obtain the expression for $H_c$ we set $c=1$. $\mathcal{H}_k^{ij}(\be):=\partial_{ij}p_k(\be)$ is the Hessian matrix of the eigenvalue $p_k$. This process that we call the regularization procedure will be introduced again for Gaussian states in Eq.~\eqref{eq:regularization_procedure}.

\section{The real form of the covariance matrix and its relation to the complex form}\label{app:real_covariance}

In this section we describe the structure of covariance matrices and displacement. We introduce the real form covariance matrix formalism and find how covariance matrices in the real form transform into its complex form.

The complex form of the covariance matrix and displacement was defined by equations \eqref{def:commutation_relation} and \eqref{def:covariance_matrix}. From the definition we can observe the following structure of the first and second moments:
\[\label{def:first_and_second_moments2}
\bd=
\begin{bmatrix}
\boldsymbol{\tilde{d}} \\ \overline{\boldsymbol{\tilde{d}}}
\end{bmatrix},\quad
\sigma\,=\,\begin{bmatrix}
X & Y \\
\overline{Y} & \overline{X}
\end{bmatrix}
\]
and $\sigma^\dag=\sigma$, i.e., $X^\dag=X$ and $Y^T=Y$. Using the Williamson's theorem\cite{Williamson1936a,Arvind1995a,Simon1998a} we can write $\sigma=SDS^\dag$, where $SKS^\dag=K$. Although the symplectic matrices $S$ are not necessarily Hermitian, they follow the same structure as $\sigma$ which is expressed by Eq.~\eqref{def:structure_of_S}.

Construction of the real form of the covariance matrix is analogous to the complex form described in the introduction. It is usually defined with respect to the collection of position and momenta operators $\boldsymbol{\hat{Q}}:=(\hat{x}_1,\dots,\hat{x}_N,\hat{p}_1,\dots,\hat{p}_N)^T$, where $\hat{x}_i:=\frac{1}{\sqrt{2}}(\hat{a}^\dag_i+\hat{a}_i)$, $\hat{p}_i:=\frac{i}{\sqrt{2}}(\hat{a}^\dag_i-\hat{a}_i)$. The canonical commutation relations of these operators can be conveniently expressed as
\begin{equation}
[\hat{{Q}}^{m},\hat{{Q}}^{n}]\,=+i\,\Omega_R^{mn}\,\mathrm{id}\quad\Rightarrow\quad \Omega_R=
\begin{bmatrix}
0 & I \\
-I & 0
\end{bmatrix}.
\end{equation}
Properties of $\Omega_R$ are $-\Omega_R^2=I$ and $\Omega_R^T=-\Omega_R$, in contrast to the complex form version $K$. In the real form, the definitions of the first and second moments are
\begin{subequations}\label{def:covariance_matrix_real}
\begin{align}
\bd_R^i&=\mathrm{tr}\big[\hat{\rho}\boldsymbol{\hat{Q}}^i\big],\\
\sigma_R^{ij}&=\mathrm{tr}\big[\hat{\rho}\,\{\Delta\boldsymbol{\hat{Q}}^i,\Delta\boldsymbol{\hat{Q}}^j\}\big],
\end{align}
\end{subequations}
where $\Delta\boldsymbol{\hat{Q}}:=\boldsymbol{\hat{Q}}-\bd_{R}$, with internal structure
\[
\bd_{R}=\begin{bmatrix}\boldsymbol{x} \\ \boldsymbol{p}\end{bmatrix},\quad
\sigma_{R}=\begin{bmatrix}X_{R} & Y_{R} \\ Y^{T}_{R} & Z_{R} \end{bmatrix}.
\]
The real covariance matrix is symmetric, i.e. $X_R=X_R^T$ and $Z_R=Z_R^T$. The corresponding real symplectic matrices are given by $\sigma_R=S_RD_RS_R^T$, where $S\Omega_R S^T=\Omega_R$, which is a defining relation of the real symplectic group $Sp(2N,\mathbb{R})$.

Since the change between real and complex form of the covariance matrix is a simple basis transformation, $\boldsymbol{\hat{Q}}\rightarrow\boldsymbol{\hat{A}}$, we can relate these two using the unitary matrix $U$,
\[\label{def:U_transform_real_cov}
\boldsymbol{\hat{A}}=U\boldsymbol{\hat{Q}},\ \ U=\frac{1}{\sqrt{2}}\,\begin{bmatrix}I & +iI \\ I & -iI\end{bmatrix}.
\]
The resulting transformation between real and complex covariance matrices and displacement are
\[
\bd=U\bd_{R},\ \ \sigma=U{\sigma}_{R}U^\dag,
\]
and the transformations related to the Williamson's decomposition are
\[
S=US_RU^\dag,\ \ D=UD_RU^\dag=D_R,\ \ K=Ui\Omega_R U^\dag.
\]
We explicitly write the connection between real and complex form of symplectic matrix,
\[
S_{R}=
\begin{bmatrix}
\alpha_R & \beta_R \\
\gamma_R & \delta_R
\end{bmatrix}=
\begin{bmatrix}
\operatorname{Re}\,[\A+\B] & -\operatorname{Im}\,[\A-\B] \\
\operatorname{Im}\,[\A+\B] & \operatorname{Re}\,[\A-\B]
\end{bmatrix}.
\]
Consequently, $\A$ and $\B$ can be expressed in the real form symplectic matrix elements as
\begin{subequations}
\begin{align}
\A&=\frac{1}{2}(\alpha_R+\delta_R+i\gamma_R-i\beta_R),\\
\B&=\frac{1}{2}(\alpha_R-\delta_R+i\gamma_R+i\beta_R).
\end{align}
\end{subequations}

Since all important matrices are related via this unitary transformation and traces and determinants are invariant under such transformations, it is clear that every formula we derived can be easily rewritten in the real form formalism by formal substitution $\sigma\rightarrow\sigma_R$ and $K\rightarrow i\Omega_R$. On the other hand, the complex form provides much more elegant structure and exposes the inner symmetries of symplectic and covariance matrices in more detail. Also, it is much easier to work with $K$ since it is diagonal, unitary, and Hermitian in contrast to non-diagonal skew-Hermitian matrix $\Omega_R$, providing much more convenient language.

\section{Derivation of Gaussian unitary transformations in the phase-space formalism}\label{app:derivation_of_S_and_b}

Let us consider the most general Gaussian unitary from Eq.~\eqref{def:Gaussian_unitary}. A Gaussian state $\R$ transforms under such unitary as
\[
\R'=\hat{U} \R \hat{U}^\dag.
\]
The first and the second moments transform according to the rule,
\begin{subequations}\label{def:covariance_matrix_app}
\begin{align}
{\bd'}^{i}&=\mathrm{tr}\big[\R'\boldsymbol{\hat{A}}^{i}\big]=\mathrm{tr}\big[{\rho}\boldsymbol{\hat{A}'}^{i}\big],\\
{\sigma'}^{ij}&=\mathrm{tr}\big[\R'\,\{\Delta\boldsymbol{\hat{A}}^{i},\Delta\boldsymbol{\hat{A}}^{\dag j}\}\big]=\mathrm{tr}\big[\R\,\{\Delta\boldsymbol{\hat{A}'}^{i},\Delta\boldsymbol{\hat{A}'}^{\dag j}\}\big].
\end{align}
\end{subequations}
We switched to the Heisenberg picture by defining the transformed vector of creation and annihilation operators,
\[\label{def:Heisenberg_picture}
\boldsymbol{\hat{A}'}^{i}=\hat{U}^\dag \bA^i \hat{U}.
\]

Now we derive the transformation of the first and the second moments in the phase-space formalism, i.e., we derive how we can write the transformed covariance matrix $\sigma'$ in terms of $\sigma$. The unitary transformation given by Eq.~\eqref{def:Gaussian_unitary} depends on the matrix $W$ and the vector $\bg$. In the following, we generalize the proof from~\cite{Luis1995a} which has been done so far only for $\bg=0$. We are going to use the identity
\[\label{eq:BHC_formula_corollary}
e^{\hat{X}}\bA^ie^{-\hat{X}}=\sum_{n=0}^\infty\frac{1}{n!}[\hat{X},\bA^i]_n,
\]
where $[\hat{X},\bA^i]_n=[\hat{X},[\hat{X},\bA^i]_{n-1}]$, $[\hat{X},\bA^i]_0=\bA^i$. Denoting $\hat{X}=-\tfrac{i}{2}\bA^\dag W \bA-\bA^\dag K \bg$, and using commutation relations
\[
[\hat{X},\bA^i]=(KW\bA)^i+\bg^i,
\]
we derive by induction
\[\label{eq:nth_element_of_BHC}
[\hat{X},\bA^i]_n=\big((iKW)^n\bA+(iKW)^{n-1}\bg\big)^i.
\]
Combining Eqs.~\eqref{def:Heisenberg_picture}, \eqref{eq:BHC_formula_corollary}, and~\eqref{eq:nth_element_of_BHC} we derive
\[\label{eq:vector_of_ann_op_transformation}
\bA'=S\bA+\bb,
\]
where
\begin{subequations}
\begin{align}
S&=e^{iKW},\\
\bb&=\sum_{n=0}^\infty\frac{(iKW)^n}{(n+1)!}\bg=\Big(\!\int_0^1e^{iKWt}\mathrm{d}t\!\Big)\ \!\bg.
\end{align}
\end{subequations}
For invertible $W$ we can also write
\[
\bb=(iKW)^{-1}\big(e^{iKW}-I\big)\bg.
\]
It is easy to prove that matrix $S$ is symplectic, i.e., it satisfies conditions~\eqref{def:structure_of_S}. Inserting Eq.~\eqref{eq:vector_of_ann_op_transformation} into Eq.~\eqref{def:covariance_matrix_app} we derive the transformation of the first and the second moments,
\begin{subequations}\label{def:covariance_matrix_app}
\begin{align}
{\bd'}&=S\bd+\bb,\\
{\sigma'}&=S\sigma S^\dag.
\end{align}
\end{subequations}

\section{Derivation of the quantum Fisher information for mixed Gaussian states, the real form expression, and the symmetric logarithmic derivative}\label{app:mixed_state}
Here we use the general result of~\cite{Gao2014a} to derive Eq.~\eqref{eq:mixed_QFI}. According to~\cite{Gao2014a} while using the Einstein's summation convention, the quantum Fisher information for $N$-mode Gaussian state can be calculated as
\[\label{eq:Gao_unpolished}
H_{i,j}(\be)=\frac{1}{2}\mathfrak{M}_{\alpha\beta,\mu\nu}^{-1}\partial_j\Sigma^{\alpha\beta}\partial_i\Sigma^{\mu\nu}+\Sigma_{\mu\nu}^{-1}\partial_j\boldsymbol{\lambda}^\mu\partial_i\boldsymbol{\lambda}^\nu.
\]
The displacement vector and the covariance matrix are defined as
$\boldsymbol{\lambda}^m=\mathrm{tr}\big[\hat{\rho}\boldsymbol{\hat{A}}_{G}^m\big]$ and $\Sigma^{mn}=\mathrm{tr}\big[\hat{\rho}\,\{(\boldsymbol{\hat{A}}_G-\boldsymbol{\lambda})^m,(\boldsymbol{\hat{A}}_G-\boldsymbol{\lambda})^n\}\big]$, $\bA_G=(\hat{a}_1,\hat{a}_1^\dag,\dots,\hat{a}_N,\hat{a}_N^\dag)^T$, and the symplectic form is given by $[\boldsymbol{\hat{A}}_{G}^m,\boldsymbol{\hat{A}}_{G}^{n}]=:\Omega^{mn}\mathrm{id}$. The inverse of the tensor $
\mathfrak{M}_G^{\alpha\beta,\mu\nu}=\Sigma^{\alpha\mu}\Sigma^{\beta\nu}+\frac{1}{4}\Omega^{\alpha\mu}\Omega^{\beta\nu}$ is defined via
\[\label{eq:inverse_matrix}
(\mathfrak{M}_G^{-1})_{\tilde{\mu}\tilde{\nu},\alpha\beta}\mathfrak{M}_G^{\alpha\beta,\mu\nu}=\delta_{\tilde{\mu}}^\mu\delta_{\tilde{\nu}}^\nu,
\]
where $\delta_{\tilde{\mu}}^\mu$ denotes the Kronecker delta. Considering the above definition, we can find a matrix form to Eq.~\eqref{eq:Gao_unpolished},
\[\label{eq:Gao_polished}
H^{ij}(\be)=\frac{1}{2}\vectorization{\partial_i\Sigma}^T\mathfrak{M}_G^{-1}\vectorization{\partial_j\Sigma}+\partial_i\boldsymbol{\lambda}^T\Sigma^{-1}\partial_j\boldsymbol{\lambda},
\]
where $\mathfrak{M}_G=\Sigma\otimes\Sigma+\frac{1}{4}\Omega\otimes\Omega$, $\otimes$ denotes the Kronecker product, and $\vectorization{\cdot}$ is a vectorization of a matrix.

To obtain the result in our notation we need to consider transformation relations
\[
\begin{split}
\sigma&=2P\Sigma XP^T,\\
K&=P\Omega XP^T,\\
\boldsymbol{d}&=P\boldsymbol{\lambda},
\end{split}
\]
where $X=\bigoplus_{i=1}^N\sigma_x$ ($X$ is real and $X^2=I$) and $P$ is a certain permutation matrix ($P$ is real and $PP^T=I$). Using properties
\[
\begin{split}
X\Sigma X&=\ov{\Sigma},\\
X\Omega X&=-\Omega,
\end{split}
\]
the fact that $\Omega$ is real, and identities
\[\label{id:Kronecker_product_ids}
\begin{split}
(ABC)\otimes(A'B'C')&=(A\otimes A')(B\otimes B')(C\otimes C'),\\
(C^T\otimes A)\vectorization{B}&=\vectorization{ABC},%\\
%\vectorization{B}^\dag(\ov{C}\otimes A^\dag)&=\vectorization{ABC}^\dag,
\end{split}
\]
we derive
\[\label{eq:app_mixed_states}
H^{ij}(\be)=\frac{1}{2}\vectorization{\partial_i\sigma}^\dag\mathfrak{M}^{-1}\vectorization{\partial_j\sigma}+2\partial_i\boldsymbol{d}^\dag\sigma^{-1}\partial_j\boldsymbol{d},
\]
where $\mathfrak{M}=\ov{\sigma}\otimes\sigma-K\otimes K$.

Using definition of the real form covariance matrix~\eqref{def:covariance_matrix_real}, transformation relations~\eqref{def:U_transform_real_cov}, Eqs.~\eqref{id:Kronecker_product_ids} and~\eqref{eq:app_mixed_states} we derive the quantum Fisher information in the real form,
\[
H^{ij}(\be)=\frac{1}{2}\vectorization{\partial_i\sigma_R}^T\mathfrak{M}_R^{-1}\vectorization{\partial_j\sigma_R}+2\partial_i\boldsymbol{d}_R^T\sigma_R^{-1}\partial_j\boldsymbol{d}_R,
\]
where $\mathfrak{M}_R=\sigma_R\otimes\sigma_R-\Omega_R\otimes \Omega_R$.

With a similar approach we can rewrite expressions for the symmetric logarithmic derivatives~\cite{Gao2014a} in an elegant matrix form,
\[\label{eq:SLD_Gaussian}
\mathcal{L}_i=\Delta \bA^\dag\mathcal{A}_i\Delta \bA-\frac{1}{2}\tr[\sigma\mathcal{A}_i]+2\Delta\boldsymbol{A}^\dag\sigma^{-1}\partial_i{\boldsymbol{d}},
\]
where $\Delta \bA:=\bA-\bd$, $\vectorization{\mathcal{A}_i}:=\mathfrak{M}^{-1}\vectorization{\partial_i\sigma}$. The quantum Fisher information matrix is then defined as $H^{ij}(\boldsymbol{\epsilon})=\frac{1}{2}\tr[\hat{\rho}\{\mathcal{L}_i,\mathcal{L}_j\}]$.

\section{Derivation of the quantum Fisher information in the case when the Williamson's decomposition is known}\label{app:Williamson's_QFI}
Here we use Eq.~\eqref{eq:mixed_QFI} to derive Eq.~\eqref{eq:multimode_QFI}. Using the Williamson's decomposition $\sigma=SDS^\dag$, identities~\eqref{def:structure_of_S} and~\eqref{id:Kronecker_product_ids} we derive
\[
\begin{split}
\mathfrak{M}^{-1}&=(\!(S^{-1}\!)^T\!\otimes\! (\!KSK\!)\!)(\!D\!\otimes\! D\!-\!K\!\otimes\! K)^{-1}(\!(\!KSK\!)^T\!\otimes\! S^{-1}\!),\\
\partial_i\sigma&=\partial_iS DKS^{-1}K+S\partial_iD KS^{-1}K-SDKS^{-1}\partial_iSS^{-1}K.
\end{split}
\]
The first part of Eq.~\eqref{eq:mixed_QFI} then reads
\begin{multline}\label{eq:P1_first_part}
\frac{1}{2}\vectorization{\partial_i\sigma}^\dag\mathfrak{M}^{-1}\vectorization{\partial_j\sigma}=\\
\vectorization{\!P_{i}D\!-\!DKP_{i}K\!+\!\partial_iD\!}^\dag\mathfrak{M}_{diag}^{-1}\vectorization{\!P_{j}D\!-\!DKP_{j}K\!+\!\partial_jD\!},
\end{multline}
where $\mathfrak{M}_{diag}=D\!\otimes\! D\!-\!K\!\otimes \!K$. Solving Eq.~\eqref{eq:inverse_matrix} for the diagonal matrix $\mathfrak{M}_{diag}$ we find
\[
(\mathfrak{M}_{diag}^{-1})_{\tilde{\mu}\tilde{\nu},\mu\nu}=\frac{\delta_{\tilde{\mu}}^\mu\delta_{\tilde{\nu}}^\nu}{D^{\mu\mu}D^{\nu\nu}-K^{\mu\mu}K^{\nu\nu}}.
\]
Changing to element-wise notation and using Einstein's summation convention ($\mu,\nu\in\{1,2N\}$, $k,l\in\{1,N\}$) we expand Eq.~\eqref{eq:P1_first_part},
\[
\begin{split}
&\frac{1}{2}\vectorization{\partial_i\sigma}^\dag\mathfrak{M}^{-1}\vectorization{\partial_j\sigma}\\
&=\frac{1}{2}(\ov{P_{i}}D-DK\ov{P_{i}}K+\partial_jD)^{\tilde{\mu}\tilde{\nu}}
\frac{\delta_{\tilde{\mu}}^\mu\delta_{\tilde{\nu}}^\nu}{D^{\mu\mu}D^{\nu\nu}-K^{\mu\mu}K^{\nu\nu}}
(P_{j}D-DKP_{j}K+\partial_jD)^{\mu\nu}\\
&=\frac{1}{2}
\frac{(\ov{P_{i}}^{\mu\nu}D^{\nu\nu}\!-\!D^{\mu\mu} K^{\mu\mu}\ov{P_{i}}^{\mu\nu}K^{\nu\nu}\!+\!\partial_iD^{\mu\mu}\delta^{\mu\nu})(P_{j}^{\mu\nu}D^{\nu\nu}\!-\!D^{\mu\mu} K^{\mu\mu} P_{j}^{\mu\nu}K^{\nu\nu}\!+\!\partial_jD^{\mu\mu}\delta^{\mu\nu})}{D^{\mu\mu}D^{\nu\nu}-K^{\mu\mu}K^{\nu\nu}}\\
&=\frac{1}{2}\bigg(
\frac{1}{\lambda_k\lambda_l-1}
(\ov{R_{i}}^{kl}\lambda_l-\lambda_k\ov{R_{i}}^{kl}+\partial_i\lambda_k\delta^{kl})(R_{j}^{kl}\lambda_l-\lambda_k R_{j}^{kl}+\partial_j\lambda_k\delta^{kl})\\
&+\frac{1}{\lambda_k\lambda_l+1}(\ov{Q_{i}}^{kl}\lambda_l+\lambda_k \ov{Q_{i}}^{kl})(Q_{j}^{kl}\lambda_l+\lambda_k Q_{j}^{kl})\\
&+\frac{1}{\lambda_k\lambda_l-1}
({R_{i}}^{kl}\lambda_l-\lambda_k{R_{i}}^{kl}+\partial_i\lambda_k\delta^{kl})(\ov{R_{j}}^{kl}\lambda_l-\lambda_k \ov{R_{j}}^{kl}+\partial_j\lambda_k\delta^{kl})\\
&+\frac{1}{\lambda_k\lambda_l+1}(Q_{i}^{kl}\lambda_l+\lambda_k Q_{i}^{kl})(\ov{Q_{j}}^{kl}\lambda_l+\lambda_k \ov{Q_{j}}^{kl})\bigg)\\
&=\frac{1}{2}\bigg(
\frac{1}{\lambda_k\lambda_l\!-\!1}
((\lambda_l\!-\!\lambda_k)\ov{R_{i}}^{kl}+\partial_i\lambda_k\delta^{kl})((\lambda_l\!-\!\lambda_k)R_{j}^{kl}+\partial_j\lambda_k\delta^{kl})+\frac{1}{\lambda_k\lambda_l\!+\!1}
(\lambda_l\!+\!\lambda_k)^2\ov{Q_{i}}^{kl}Q_{j}^{kl}\\
&\ \ \!\ +\frac{1}{\lambda_k\lambda_l\!-\!1}
((\lambda_l\!-\!\lambda_k){R_{i}}^{kl}+\partial_i\lambda_k\delta^{kl})((\lambda_l\!-\!\lambda_k)\ov{R_{j}}^{kl}+\partial_j\lambda_k\delta^{kl})+\frac{1}{\lambda_k\lambda_l\!+\!1}
(\lambda_l\!+\!\lambda_k)^2Q_{i}^{kl}\ov{Q_{j}}^{kl}\bigg)\\
&=\frac{(\lambda_l-\lambda_k)^2}{\lambda_k\lambda_l-1}\Re[\ov{R_{i}}^{kl}R_{j}^{kl}]+\frac{(\lambda_l+\lambda_k)^2}{\lambda_k\lambda_l+1}\Re[\ov{Q_{i}}^{kl}Q_{j}^{kl}]
+\frac{\partial_i\lambda_k\partial_j\lambda_k}{\lambda_k^2-1},\\
\end{split}
\]
which in combination with term $2\partial_i\boldsymbol{d}^\dag\sigma^{-1}\partial_j\boldsymbol{d}$ gives Eq.~\eqref{eq:multimode_QFI}.

\section{Derivation of the bound on the remainder in the limit formula for the quantum Fisher information}\label{app:Remainder}
Here we prove the bound on the remainder of the general multi-mode formula. We consider the Williamson's decomposition $\sigma=SDS^\dag$. An element of the sum Eq.~\eqref{eq:limit_formula} can be written as
\[
a_n^{ij}=\tr\big[A^{-n}\partial_i{A}A^{-n}\partial_j{A}]=\tr\big[{D^{-n}B_iD^{-n}B_j}\big],
\]
where $B_i=S^\dag\partial_i{A}(S^\dag)^{-1}K^{-n-1}$. We can derive the inequalities
\[\label{eq:Remainder1}
\begin{split}
\norm{a_n^{ij}}&=\norm{\sum_{k,l}\frac{1}{\lambda_k^n\lambda_l^n}B_i^{kl}B_j^{lk}}
\leq\norm{\sum_{k,l}\frac{1}{\lambda_k^{n}\lambda_l^{n}}\norm{B_i^{kl}}\norm{B_j^{lk}}}
\leq\frac{1}{\lambda_{\mathrm{min}}^{2n}}\norm{\sum_{k,l}\norm{B_i^{kl}}\norm{B_j^{lk}}}\\
&\leq\frac{1}{\lambda_{\mathrm{min}}^{2n}}\sqrt{\sum_{k,l}\norm{B_i^{kl}}^2}\sqrt{\sum_{k,l}\norm{B_j^{lk}}^2}
=\frac{1}{\lambda_{\mathrm{min}}^{2n}}\sqrt{\tr[B_i^\dag B_i]}\sqrt{\tr[B_j^\dag B_j]},
\end{split}
\]
where the last inequality is the Cauchy-Schwarz inequality between $B_i^{kl}$ and $B_j^{lk}$ considered as vectors with $2N\times2N$ entries where $N$ is number of modes, $\lambda_{\mathrm{min}}:=\min_{k}\{\lambda_k\}$ is the smallest symplectic eigenvalue. Defining the Hermitian matrix $C_i:=S^\dag \partial_i{A}KS$ we have
\[\label{eq:Remainder2}
\tr[(A\partial_i{A})^2]=\tr[C_i^\dag D C_i D]=\sum_{k,l}\norm{C_i^{kl}}^2\lambda_k\lambda_l\geq\lambda_{\mathrm{min}}^2\tr[C_i^\dag C_i]=\lambda_{\mathrm{min}}^2\tr[B_i^\dag B_i].
\]
Combining \eqref{eq:Remainder1} and \eqref{eq:Remainder2} gives
\[
\norm{a_n^{ij}}\leq\sqrt{\tr\big[(A\partial_i{A})^2\big]}\sqrt{\tr\big[(A\partial_j{A})^2\big]}\lambda_{\mathrm{min}}^{-2n-2}.
\]
For $\lambda_{\mathrm{min}}>1$ we can estimate the remainder,
\[
\norm{R_M^{ij}}\leq \frac{\sqrt{\tr\big[(A\partial_i{A})^2\big]}\sqrt{\tr\big[(A\partial_j{A})^2\big]}}{2}\!\!\sum_{n=M+1}^\infty\!\!\!\!{\lambda_{\mathrm{min}}^{-2n-2}}
=\frac{\sqrt{\tr\big[(A\partial_i{A})^2\big]}\sqrt{\tr\big[(A\partial_j{A})^2\big]}}{2\lambda_{\mathrm{min}}^{2M+2}(\lambda_{\mathrm{min}}^2-1)}.
\]

\section{Proof of the theorem about Taylor expansion of the quantum Fisher information}\label{app:not_existing_form}
In this section we are going to prove theorem~\ref{not_existing_form}. We are going to prove that if the function $\sigma(\lambda,\epsilon)$ is at least $C^{(3)}$ at the point $(p,\epsilon)$, then the Taylor expansion does not exist. To do that, we prove $\partial_p H(\sigma(p,\epsilon))=-\infty$ from which the statement immediately follows. We also use the first definition of the problematic points given by Eq.~\eqref{def:problematic_points}. The proof for the second definition is analogous. We are going to use the general formula
\[
H(p,\epsilon)=H_0(p,\epsilon)+\sum_{i=1}^N\frac{(\partial_\epsilon{\lambda_i})^2}{\lambda_i^2-1},
\]
where
\[
H_0(p,\epsilon)=\sum_{k,l=1}^N\frac{(\lambda_k-\lambda_l)^2}{\lambda_k\lambda_l-1}\norm{R^{kl}}^2
+\frac{(\lambda_k+\lambda_l)^2}{\lambda_k\lambda_l+1}\norm{Q^{kl}}^2+2\dot{d}^\dag\sigma^{-1}\dot{d}
\]
First we differentiate $H_0$ and show that the derivative exists. Since $\sigma\in C^{(3)}$, the derivatives of $R$, $Q$, and $\frac{(\lambda_k+\lambda_l)^2}{\lambda_k\lambda_l+1}$ the derivatives at point $(p,\epsilon)$ exist and are continuous. The only terms which may be problematic in $H_0$ is $\frac{(\lambda_k-\lambda_l)^2}{\lambda_k\lambda_l-1}$ when both $\lambda_k(p,\epsilon)=\lambda_l(p,\epsilon)=1$. For such cases we can write the Taylor expansions
\begin{align}
%\begin{subequations}
\lambda_k(p+\mathrm{d} p,\epsilon)&=\lambda_k(p,\epsilon)+\partial_p\lambda_k(p,\epsilon)\mathrm{d} p+\cdots=1+\partial_p\lambda_i(p,\epsilon)\mathrm{d} p+\cdots\\
\partial_\epsilon\lambda_k(p+\mathrm{d} p,\epsilon)&=\partial_\epsilon\lambda_k (p,\epsilon)+\partial_{p\epsilon}\lambda_k(p,\epsilon)\mathrm{d} p+\dots=0+\partial_{p\epsilon}\lambda_i(p,\epsilon)\mathrm{d} p+\cdots.
%\end{subequations}
\end{align}
Then we have
\[
\begin{split}
\partial_p\frac{(\lambda_k-\lambda_l)^2}{\lambda_k\lambda_l-1}(p,\epsilon)
&=\lim_{\mathrm{d} p\rightarrow 0}\frac{\frac{(\lambda_k-\lambda_l)^2}{\lambda_k\lambda_l-1}(p+\mathrm{d} p,\epsilon)-0}{\mathrm{d} p}
=\lim_{\mathrm{d} p\rightarrow 0}\frac{\frac{(\partial_p\lambda_k(p,\epsilon)-\partial_p\lambda_l(p,\epsilon))^2\mathrm{d} p^2}{(\partial_p\lambda_k(p,\epsilon)+\partial_p\lambda_l(p,\epsilon))\mathrm{d} p}}{\mathrm{d} p}\\
&=\frac{(\partial_p\lambda_k(p,\epsilon)-\partial_p\lambda_l(p,\epsilon))^2}{\partial_p\lambda_k(p,\epsilon)+\partial_p\lambda_l(p,\epsilon)},
\end{split}
\]
which is a finite number. If both $\partial_p\lambda_k(p,\epsilon)=\partial_p\lambda_l(p,\epsilon)=0$, we can use the Taylor expansion to the higher order.
We thus conclude that $\partial_p H_0$ is a finite number.

Now, let us take a look at the problematic term $\sum_{k=1}^N\frac{(\partial_\epsilon{\lambda_k})^2}{\lambda_k^2-1}$. For $\lambda_k(p,\epsilon)=1$ we have
\[
\begin{split}
\partial_p\frac{(\partial_\epsilon{\lambda_k})^2}{\lambda_k^2-1}&=\lim_{\mathrm{d} p\rightarrow 0}\frac{\frac{(\partial_\epsilon{\lambda_k})^2}{\lambda_k^2-1}(p+\mathrm{d} p,\epsilon)-\partial_{\epsilon\epsilon}\lambda_k(p,\epsilon)}{\mathrm{d} p}\\
&=\frac{(\partial_{p\epsilon}\lambda_k(p,\epsilon))^2}{2\partial_p\lambda_k(p,\epsilon)}-\lim_{\mathrm{d} p\rightarrow 0}\frac{\partial_{\epsilon\epsilon}\lambda_k(p,\epsilon)}{\mathrm{d} p}=-\infty,
\end{split}
\]
because $\partial_{\epsilon\epsilon}\lambda_k(p,\epsilon)\neq 0$ and because symplectic eigenvalues must be larger than one, it must be positive, $\partial_{\epsilon\epsilon}\lambda_k(p,\epsilon)>0$.
We proved that while part $H_0$ has a finite derivative while the other part $\sum_{k=1}^N\frac{(\partial_\epsilon{\lambda_k})^2}{\lambda_k^2-1}$ does not, we conclude that $H$ does not have the finite derivative which proves the theorem.\qed

\section{Derivation of quantum Fisher information in the absence of perfect reference frames}\label{app:FirstFormula}
From Eq.~\eqref{rhoB}, we immediately observe that eigenvalues of transformed density matrix $\rho_B$ are $p_i=\langle\psi_{\lambda}|\hat{P}_i|\psi_{\lambda}\rangle$ with respective normalised eigenvectors $\frac{\hat{P_i}\ket{\psi_{\lambda}}}{\sqrt{p_i}}$. Let $\{\ket{\phi_j}\}_j$ be a set orthonormal eigenvectors of $\rho_B$ with respective eigenvalue $0$. %Eigenvectors of $\rho_B$ form an orthonormal basis of the Hilbert space, $\mathcal{H}=\mathrm{span}\{\frac{\hat{P_i}\ket{\psi}}{\sqrt{p_i}},\ket{\phi_j}\}_{i,j}$. %$span\{\frac{\hat{P_i}\ket{\psi}}{\sqrt{p_i}},\ket{\phi_j}\}$.
Using Eq.~(\ref{QFI}) we have
\begin{equation}
\begin{split}
H(\rho_B)&=\ \ \ 2\!\!\!\!\!\!\!\!\!\!\sum_{i,j,p_i\neq0,p_j\neq0}\!\!\!\!\frac{\big|\frac{\bra{\psi}\hat{P_i}}{\sqrt{p_i}}\sum_k\hat{P}_k\partial_\lambda\rho \hat{P}_k\frac{\hat{P}_j\ket{\psi}}{\sqrt{p_j}}\big|^2}{p_i+p_j}+4\sum_{i,j}\frac{\big|\frac{\bra{\psi}\hat{P_i}}{\sqrt{p_i}}\sum_k \hat{P}_k\partial_\lambda\rho \hat{P}_k\ket{\phi_j}\big|^2}{p_i}\\
&=\sum_i\frac{\big|\frac{\bra{\psi}\hat{P_i}}{\sqrt{p_i}}\partial_\lambda\rho \frac{\hat{P_i}\ket{\psi}}{\sqrt{p_j}}\big|^2+4\sum_j\big|\frac{\bra{\psi}\hat{P_i}}{\sqrt{p_i}}\partial_\lambda\rho \hat{P_i}\ket{\phi_j}\big|^2}{p_i}
\end{split}
\end{equation}
and together with the Parseval identity, i.e.
\begin{equation}
\sum_j\Big|\frac{\bra{\psi}\hat{P}_i}{\sqrt{p_i}}\partial_\lambda\rho \hat{P}_i\ket{\phi_j}\Big|^2=
\Big|\!\Big|\hat{P}_i\partial_\lambda\rho\frac{\hat{P}_i\ket{\psi}}{\sqrt{p_i}}\Big|\!\Big|^2-\sum_j\Big|\frac{\bra{\psi}\hat{P}_i}{\sqrt{p_i}}\partial_\lambda\rho \hat{P}_i\frac{\hat{P}_j\ket{\psi}}{\sqrt{p_j}}\Big|^2
\end{equation}
we can remove the dependence on states $\ket{\phi_j}$. Then $H(\rho_{B})$ is
\[
H(\rho_B)\!=\!\!\sum_i\!\!\frac{4p_i\bra{\psi}\hat{P}_i\partial_\lambda\rho \hat{P}_{i}\partial_\lambda\rho \hat{P}_{i}\ket{\psi}\!-\!3\abs{\bra{\psi}\hat{P}_{i}\partial_\lambda\rho \hat{P}_{i}\ket{\psi}}^2}{p_i^3}.
\]
After substituting $\rho_{\lambda}=\pro{\psi_{\lambda}}{\psi_{\lambda}}$
\[\label{prefinal}
H(\rho_B)\!=\!\!\!\!\sum_{i,p_i\neq0}\!\!\!4\bra{\partial_\lambda\psi}\hat{P}_{i}\ket{\partial_\lambda\psi}+
\frac{(\bra{\psi}\hat{P}_{i}\ket{\partial_\lambda\psi}\!-\!\bra{\partial_\lambda\psi}\hat{P}_{i}\ket{\psi})^2}{p_i},
\]
where $\ket{\partial_\lambda\psi}=\sum_k(\partial_\lambda\alpha_k)\ket{k}$ for $\lambda$-independent basis $\{\ket{k}\}$. The sum in (\ref{prefinal}) consists only of elements where $p_i\neq0$, however, by differentiating $p_i=\bra{\psi}\hat{P}_{i}\ket{\psi}=0$ and using Cauchy-Schwarz inequality on $\bra{\partial_{\lambda\lambda}\psi}\hat{P}_{i}\ket{\psi}$ we get $\bra{\partial_\lambda\psi}\hat{P}_{i}\ket{\partial_\lambda\psi}=0$. Now summing over all $i$ and using the completeness relation $\sum_i\hat{P}_{i}=\mathbf{1}$ we get
\[\label{almostdone}
H(\rho_B)=4\braket{\partial_\lambda\psi}{\partial_\lambda\psi}-
4\sum_i\frac{(\mathfrak{Im}\bra{\psi}\hat{P}_{i}\ket{\partial_\lambda\psi})^2}{\bra{\psi}\hat{P}_{i}\ket{\psi}}.
\]
\emph{Symmetric logarithmic derivative} \eqref{NSLD} can be derived analogously, where instead of Parseval identity we use completeness relation
$\sum_j\pro{\phi_j}{\phi_j}=\mathbb{I}-\sum_i\frac{P_i\pro{\psi}{\psi}P_i}{p_i}$.

\section{Proof of the theorem about the loss of estimation precision due to a lack of reference frames}\label{app:nonnegQFIl}
Here we prove that $0\leq H(\G[\rho])\leq H(\rho)$ and the equality conditions. First, we define the \emph{quantum Fisher information loss} as
\begin{equation}\label{QFIlossdef}
l(\rho,\hat{G})=H(\rho)-H(\G[\rho]).
\end{equation}
This operational measure enables us to analyze how much information is lost due to a lack of a reference frame. Instead of the original statement, we prove the equivalent $0\leq l(\rho,G)\leq H(\rho)$ together with the equality conditions. $l(\rho,G)\leq H(\rho)$ follows immediately from definition \eqref{QFIlossdef}. Let us prove $l(\rho,G)\geq 0$. Looking at the expression for the quantum Fisher information loss, i.e. Eq.~\eqref{QFIlossdef}, we need to prove that
\[\label{lnonnegativestart}
\sum_i\frac{(\mathfrak{Im}\bra{\psi}\hat{P}_{i}\ket{\partial_\lambda\psi})^2}{\bra{\psi}\hat{P}_{i}\ket{\psi}}\geq|\braket{\psi}{\partial_\lambda\psi}|^2.
\]
First, let us define $\ket{\widetilde{\partial_\lambda\psi}}:=\sum_i\frac{\mathfrak{Im}\bra{\psi}\hat{P}_{i}\ket{\partial_\lambda\psi}}{\PI}\hat{P}_{i}\ket{\psi}$.
Then using the fact that the state $|\psi\rangle$ is normalized, i.e. $\norm{\ket{\psi}}=1$, the Cauchy-Schwarz inequality and that for any state $|\psi\rangle$, $p_i\equiv\bra{\psi}\hat{P}_{i}\ket{\psi}=0$ if and only if  $\hat{P}_{i}\ket{\psi}=0$ and therefore $\bra{\psi}\hat{P}_{i}\ket{\partial_\lambda\psi}=0$, together with the completeness relation $\sum_i\hat{P}_i=\mathbf{1}$, we have
\begin{equation}
\begin{split}
LHS&=\norm{\ket{\widetilde{\partial_\lambda\psi}}}^2=\norm{\ket{\widetilde{\partial_\lambda\psi}}}^2\norm{\ket{\psi}}^2\geq\abs{\braket{\psi}{\widetilde{\partial_\lambda\psi}}}^2\\
&=\bigg|\bra{\psi}\sum_{i,p_i\neq0}\frac{\mathfrak{Im}\bra{\psi}\hat{P}_{i}\ket{\partial_\lambda\psi}}{\PI}\hat{P}_{i}\ket{\psi}\bigg|^2\\
&=\Big|\mathfrak{Im}\bra{\psi}\sum_i\hat{P}_{i}\ket{\partial_\lambda\psi}\Big|^2=\abs{\mathfrak{Im}\braket{\psi}{\partial_\lambda\psi}}^2\\
&=\left|\braket{\psi}{\partial_\lambda\psi}\right|^2
\end{split}
\end{equation}
where for the last step we have used the fact that $\braket{\psi_{\lambda}}{\partial_\lambda\psi_{\lambda}}$ is purely imaginary (which is a consequence of the normalization condition).

Now, because Cauchy-Schwarz inequality is saturated if and only if there exists a complex number $c$ such that $\ket{\widetilde{\partial_\lambda\psi}}=c\ket{\psi}$, by re-writing the state $\ket{\psi}$ as $\sum_i\frac{p_i}{p_i}\hat{P}_{i}\ket{\psi}$ we find that Eq.~\eqref{lnonnegativestart} is saturated if and only if
\[
\sum_i\frac{\mathfrak{Im}\bra{\psi}\hat{P}_{i}\ket{\partial_\lambda\psi}-cp_i}{p_i}\hat{P}_{i}\ket{\psi}=0,
\]
which together with orthogonality condition for the projectors $\hat{P}_{i}$ leads to the no-loss condition \eqref{cnoloss}, i.e.
\[
l(\rho,G)=0\ \Leftrightarrow\ \exists c\in\mathbb{C},\ \forall i,\ \mathfrak{Im}\bra{\psi}\hat{P}_{i}\ket{\partial_\lambda\psi}=c\PI.
\]
Summing over all $i$'s we find that the only value $c$ can take is $c=\mathfrak{Im}\braket{\psi}{\partial_\lambda\psi}=-i\braket{\psi}{\partial_\lambda\psi}$.

Now let us derive the max-loss condition~\eqref{maxlosstheorem}. From the definition of quantum Fisher information loss and that $\abs{\braket{\psi}{\widetilde{\partial_\lambda\psi}}}=\abs{\braket{\psi}{\partial_\lambda\psi}}$, we can write
\begin{equation}\label{lossformula}
\begin{split}
l(\rho,G)&=4\braket{\widetilde{\partial_\lambda\psi}}{\widetilde{\partial_\lambda\psi}}-4\abs{\braket{\psi}{\widetilde{\partial_\lambda\psi}}}^2\\
&=4\braket{\widetilde{\partial_\lambda\psi}}{\widetilde{\partial_\lambda\psi}}-4\abs{\braket{\psi}{{\partial_\lambda\psi}}}^2.
\end{split}
\end{equation}
Therefore by comparing (\ref{lossformula}) and (\ref{eq:QFI_pure2}) we have
\[\label{nolossformulaappendix}
l(\rho,\hat{G})=H(\rho)\ \Leftrightarrow
\braket{\widetilde{\partial_\lambda\psi}}{\widetilde{\partial_\lambda\psi}}=\braket{{\partial_\lambda\psi}}{{\partial_\lambda\psi}}.
\]
Similar to the previous case we can write $\ket{\partial_\lambda\psi}$ in the complete orthonormal basis $\{\frac{\hat{P}_{i}\ket{\psi}}{\sqrt{p_i}},\ket{\phi_j}\}_{i,j}$ as
\[
\ket{\partial_\lambda\psi}=\sum_i\frac{\bra{\psi}\hat{P}_{i}\ket{\partial_\lambda\psi}}{\sqrt{p_i}}\frac{\hat{P}_{i}\ket{\psi}}{\sqrt{p_i}}
+\sum_j\braket{\phi_j}{\partial_\lambda\psi}\ket{\phi_j}.
\]
where $\ket{\phi_j}$ span the rest of the Hilbert space which is not spanned by vectors $\frac{\hat{P}_{i}\ket{\psi}}{\sqrt{p_i}}$.
After multiplying by $\bra{\partial_\lambda\psi}$ we get the Parseval identity, i.e.
\[
\braket{\partial_\lambda\psi}{\partial_\lambda\psi}=\sum_i\frac{\abs{\bra{\psi}\hat{P}_{i}\ket{\partial_\lambda\psi}}^2}{p_i}
+\sum_j\abs{\braket{\phi_j}{\partial_\lambda\psi}}^2.
\]
Comparing this with (\ref{nolossformulaappendix}) we get condition for max-loss as $l(\rho,\hat{G})=H(\rho)\ \Leftrightarrow$
\[
\forall i,\ \mathfrak{Re}\bra{\psi}\hat{P}_{i}\ket{\partial_\lambda\psi}=0\ \wedge\ \forall \ket{\phi_j},\ \braket{\phi_j}{\partial_\lambda\psi}=0.
\]

\section{Questions about the universality of quantum reference frames}\label{app:QRF_questions}

In this appendix we construct a way how to ask questions posed in section~\ref{sec:open_questions} about ``universality of quantum reference frames'' in the precise language of mathematics.

\begin{definition}
Let $\rho_\epsilon\in \mathcal{L}(\HS)$ be a quantum state with the encoded parameter $\epsilon$, $s=(\R_n)_{n=1}^\infty,\ \R_n\in\mathcal{L}(\HS_{QRF}),$ a sequence of quantum reference frames, and $\G$ a $\G$-twirling map on $\HS\otimes\HS_{QRF}$. We call two sequences of QRFs, $s_1=\big(\R_n^{(1)}\big)_{n=1}^\infty$ and $s_2=\big(\R_n^{(2)}\big)_{n=1}^\infty$, equivalent if and only if
\[
\forall \epsilon,\quad \lim_{n\rightarrow\infty}H\big(\G\big(\rho_\epsilon\otimes\R_n^{(1)}\big)\big)=\lim_{n\rightarrow\infty}H\big(\G\big(\rho_\epsilon\otimes\R_n^{(2)}\big)\big).
\]
\end{definition}
Using this equivalence we can define equivalence classes,
\[\label{eq:equivalence_classes}
[f]:=\{\ \!s\ |\ \forall \epsilon,\ \lim_{n\rightarrow\infty}H(\G(\rho_\epsilon\otimes\R_n))=f(\epsilon)\}
\]
We also define a partial ordering on these equivalence classes: a class of quantum reference frames $[s_2]$ is better than a class $[s_1]$, $[s_1]\leq [s_2]$, if and only if
\[
\forall \epsilon,\quad s_1(\epsilon)\leq s_2(\epsilon).
\]
We call $[H(\rho_\epsilon)]$ the class of perfect reference frames. This class contains all QRFs which in the limit of large $n$ retrieve the same precision as when two parties share a reference frame. Note that for all classes $[s]$, $[s]\leq[H(\rho_\epsilon)]$. We call class $[s]$ \emph{a good class for estimating $\epsilon$} if and only if
\[
\forall s_1,\quad s_1(\epsilon)\leq s(\epsilon).
\]
We call class $[s]$ \emph{the good class} when it is good for estimating all parameters $\epsilon$, i.e., when \[\forall [s_1],\quad[s_1]\leq [s].\]

It follows from the definition that the class of perfect reference frames is also the good class. Note however, that both the class of perfect reference frames and the good class might be empty. In other words, there might not be a quantum reference frame which serves as a classical reference frame in the limit of large $n$, and there might not exist any class which is better than all the other classes in estimating any parameter $\epsilon$.

In the first example~\ref{sec:two_nonint_osc} the perfect class was non-empty. Both superposition of uniform states and coherent states belonged to the perfect class. The second example has shown, as depicted on Fig.~\ref{nonvsint}, that there might be a quantum reference frames which perform better at the estimation of certain parameters $\epsilon$ while sacrificing the precision in the estimation of others.

Now we can phrase the questions posed in the beginning in a different manner: Does there always exist a perfect or a good class of quantum reference frames? Under which conditions do they exist? How large are these classes, i.e., how many types of different QRFs they contain? How many different classes are there? Is there some equivalent and more physically intuitive way of defining the equivalence classes? We leave answering these questions for future work to anyone who wants to tackle them.

\section{Notation}\label{app:notations}
Throughout the thesis the lower indices label different matrices, while upper indices represent the components of a matrix or a vector. We also systematically use the Planck units, $\hbar=c=k_B=1$. In this thesis we use the following notation.
\vspace{0.5cm}

\noindent\begin{tabular}{ r | l }
$\overline{\phantom{\alpha}}$, $^T$, $^\dag$  & Complex conjugate, transpose, conjugate transpose respectively.\\
${\Re}$ & the real value of a complex number\\
${\Im}$ & the imaginary value of a complex number\\
$\norm{\ \cdot\ }$ &  absolute value of a complex number, determinant of a matrix \\
$\bigotimes, \otimes$  & tensor product, Kronecker product of matrices \\
$\bigoplus,\oplus$ &  direct product \\
$\Sigma_{i}$ &  summation over $i$ \\
$[\hat{A},\hat{B}]$ &  commutator, $[\hat{A},\hat{B}]:=\hat{A}\hat{B}-\hat{B}\hat{A}$ \\
$\{\hat{A},\hat{B}\}$ & anti-commutator, $\{\hat{A},\hat{B}\}:=\hat{A}\hat{B}+\hat{B}\hat{A}$ \\
$M$ &  matrix \\
$M^{kl}$ & components of a matrix \\
$M_i$ & matrix labeled $i$ \\
$\vectorization{M}$ & vectorization of a matrix\\
$\boldsymbol{v}$ &  vector \\
$\boldsymbol{v}^{k}$ & component of a vector \\
$\HS$ & Hilbert space\\
$\hat{O}$ & Hilbert space operator \\
\end{tabular}

\noindent\begin{tabular}{ r | l }
$\epsilon,\be$ & parameter, vector of parameters we estimate\\
$\dot{~}, \partial_i$ & derivatives with respect to the parameter we want to estimate, $\dot{~}:=\frac{\partial}{\partial \epsilon}$,
$\partial_i:=\frac{\partial}{\partial \epsilon_i}$\\
$H(\epsilon)$ & the quantum Fisher information, Eq.~\eqref{def:H_using_L}\\
$H^{ij}(\be)$ & elements of the quantum Fisher information matrix, Eq.~\eqref{def:Information_matrix}\\
$H_c^{ij}(\be)$ & elements of the continuous quantum Fisher information matrix, Eq.~\eqref{eq:connection_between_Hc_and_H}\\
$K$ & the constant matrix defining a complex represenation of the real symplectic\\
& group defined called the symplectic form, Eq.~\eqref{def:commutation_relation} \\
$\bA$ & vector of annihilation and creation operators, Eq.~\eqref{def:commutation_relation}\\
$\bd$ & the displacement vector, Eq.~\eqref{def:covariance_matrix}\\
$\sigma$ & the covariance matrix, Eq.~\eqref{def:covariance_matrix}\\
$X, Y$ & submatrices of the covariance matrix $\sigma$, Eq.~\eqref{def:first_and_second_moments}\\
$A$ & $A:=K\sigma$, a multiple of the two previously mentioned matrices, Eq.~\eqref{def:A}\\
$S$ & The symplectic matrix, Eq.~\eqref{def:structure_of_S}\\
$\A, \B$ & Submatrices of the matrix $S$, Eq.~\eqref{def:structure_of_S}, the Bogoliubov coefficients, Eq.~\eqref{def:Bogoliubov_S}\\
$W,\bg$ & generators of a Gaussian unitary operator, Eq.~\eqref{def:Gaussian_unitary}\\
$D,L$ & diagonal matrices consisting of the symplectic eigenvalues, Eqs.~\eqref{def:Williamson's_decomposition}, \eqref{eq:exact_multimode_compact}\\
$\lambda_i$ & symplectic eigenvalue\\
$r_i$ & squeezing parameter\\
$N$ & number of modes of a Gaussian state, number of identical states, Ex.~\eqref{ex:Heisenberg_scaling},\\
    & number of performed mesurements, Eq.~\eqref{eq:Cramer_Rao}\\
$n$ & the mean total number of particles in a Gaussian state, Eq.~\eqref{eq:mean_number_of_particles}\\
$P,P_i$ & $P_i:=S^{-1}\partial_i{S}$, elements of the Lie algebra associated with the symplectic\\
& group, Eqs.~\eqref{def:P_1}, \eqref{def:P_i}\\
$R_i, Q_i$ & Submatrices of the matrix $P_i$, Eq.~\eqref{def:P_i}\\
\end{tabular}

%\printglossary[title=List of Symbols,style=mcolindex]

\renewcommand{\bibname}{References} % changes the header; default: Bibliography
\bibliographystyle{unsrt}
\bibliography{Optimal_probe_states_BiBTeX} % adjust this to fit your BibTex file

\end{document}